%% file: Fifty_Years_MIMO_Detection.tex
\documentclass[final, journal, twocolumn]{IEEEtran}

\usepackage{threeparttable} % to put notes at the end of a table
\usepackage{graphicx}
\usepackage{picinpar}
\usepackage{rotating}

\usepackage{psfrag}
\usepackage{placeins}
\usepackage{times}
\usepackage{amsmath}
\usepackage{amsfonts}
\usepackage{subfigure}

\usepackage{pifont}
\usepackage{amssymb}
\usepackage{setspace}
\usepackage[color]{changebar}
\cbcolor{red}
\usepackage{flafter}
\DeclareGraphicsExtensions{.pdf}
\usepackage{booktabs}
\usepackage{tabularx}
\usepackage{array}
\usepackage{color}

\newcommand{\PreserveBackslash}[1]{\let\temp=\\#1\let\\=\temp}
\newcolumntype{C}[1]{>{\PreserveBackslash\centering}p{#1}}
\newcolumntype{R}[1]{>{\PreserveBackslash\raggedleft}p{#1}}
\newcolumntype{L}[1]{>{\PreserveBackslash\raggedright}p{#1}}

\usepackage{mdwmath}
\usepackage{mdwtab}
\usepackage{dblfloatfix}
\usepackage{morefloats}
\usepackage{url}
\usepackage{cite}

\usepackage[style=list,nonumberlist]{glossaries}
\renewcommand{\glsgroupskip}{}
\input{glossaries.tex}
\makeglossaries

\begin{document}
\bibliographystyle{../../../shaoshi_bib/IEEEtran}
\title{Fifty Years of MIMO Detection: The Road to Large-Scale MIMOs}
\author{Shaoshi~Yang,~\IEEEmembership{Member,~IEEE} and~Lajos~Hanzo,~\IEEEmembership{Fellow,~IEEE}
\thanks{
% Copyright (c) 2011 IEEE. Personal use of this material is permitted. However, permission to use this material for any other purposes must
% be obtained from the IEEE by sending a request to pubs-permissions@ieee.org.

The financial support of the Research Councils UK (RCUK) under the India-UK Advanced Technology Center (IU-ATC), of the EU under the auspices of the Concerto project, and of the European Research Councils Advanced Fellow Grant is gratefully acknowledged.}
\thanks{The authors are with the School of Electronics and Computer Science, University of Southampton,
Southampton, SO17 1BJ, UK (e-mail: \{lh, shaoshi.yang\}@ecs.soton.ac.uk).

% T. Lv is with the School of Information and Communication
% Engineering, Beijing University of Posts and Telecommunications
% (BUPT), Beijing, 100876, China (e-mail: lvtiejun@bupt.edu.cn).
}}

\markboth{Accepted to appear on IEEE Communications Surveys \& Tutorials, 2015}%
{Shell \MakeLowercase{\textit{et al.}}: Bare Demo of IEEEtran.cls
for Journals}

\maketitle
\begin{abstract}
The emerging massive/large-scale multiple-input multiple-output (\gls{LS-MIMO}) systems that rely on very large antenna arrays have become a hot topic of  wireless communications. Compared to multi-antenna aided systems being built at the time of writing, such as the long-term evolution (\gls{LTE}) based fourth generation (\gls{4G}) mobile communication system which allows for up to eight antenna elements at the base station (BS), the LS-MIMO system entails an unprecedented number of antennas, say 100 or more, at the BS. The huge leap in the number of BS antennas opens the door to a new research field in communication theory, propagation and electronics, where random matrix theory begins to play a dominant role. Interestingly, LS-MIMOs also constitute a perfect example of one of the key philosophical principles of the Hegelian Dialectics, namely that ``quantitative change leads to qualitative change''. 

In this treatise, we provide a recital on the historic heritages and novel challenges facing LS-MIMOs from a detection perspective. Firstly, we highlight the fundamentals of MIMO detection, including the nature of co-channel interference (CCI), the generality of the MIMO detection problem, the received signal models of both linear memoryless MIMO channels and dispersive MIMO channels exhibiting memory, as well as the complex-valued versus real-valued MIMO system models. Then, an extensive review of the representative MIMO detection methods conceived during the past fifty years (\textit{1965-2015}) is presented, and relevant insights as well as lessons are inferred for the sake of designing complexity-scalable MIMO detection algorithms that are potentially applicable to LS-MIMO systems. Furthermore, we divide the LS-MIMO systems into two types, and elaborate on the distinct detection strategies suitable for each of them. The type-I LS-MIMO corresponds to the case where the number of active users is much smaller than the number of BS antennas, which is currently the mainstream definition of LS-MIMO. The type-II LS-MIMO corresponds to the case where the number of active users is comparable to the number of BS antennas. Finally, we discuss the applicability of existing MIMO detection algorithms in LS-MIMO systems, and review some of the recent advances in LS-MIMO detection. 

\end{abstract}


\makeatletter
\def\hlinewd#1{%
  \noalign{\ifnum0=`}\fi\hrule \@height #1 \futurelet
   \reserved@a\@xhline}
\makeatother

\IEEEpeerreviewmaketitle

{\footnotesize{\printglossaries}}

%%%%%%%%%%%%%%%%%%%%%%%%%%%%%%%%%%%%%%%%%%%%%%%%%%%%%%%%%%%%%%%%
\section{Introduction}\label{sec:intro}
%%%%%%%%%%%%%%%%%%%%%%%%%%%%%%%%%%%%%%%%%%%%%%%%%%%%%%%%%%%%%%%%
\subsection{Why are Large-Scale MIMOs Important?}
\begin{figure}[t]
\centering
\includegraphics[width=0.85\linewidth]{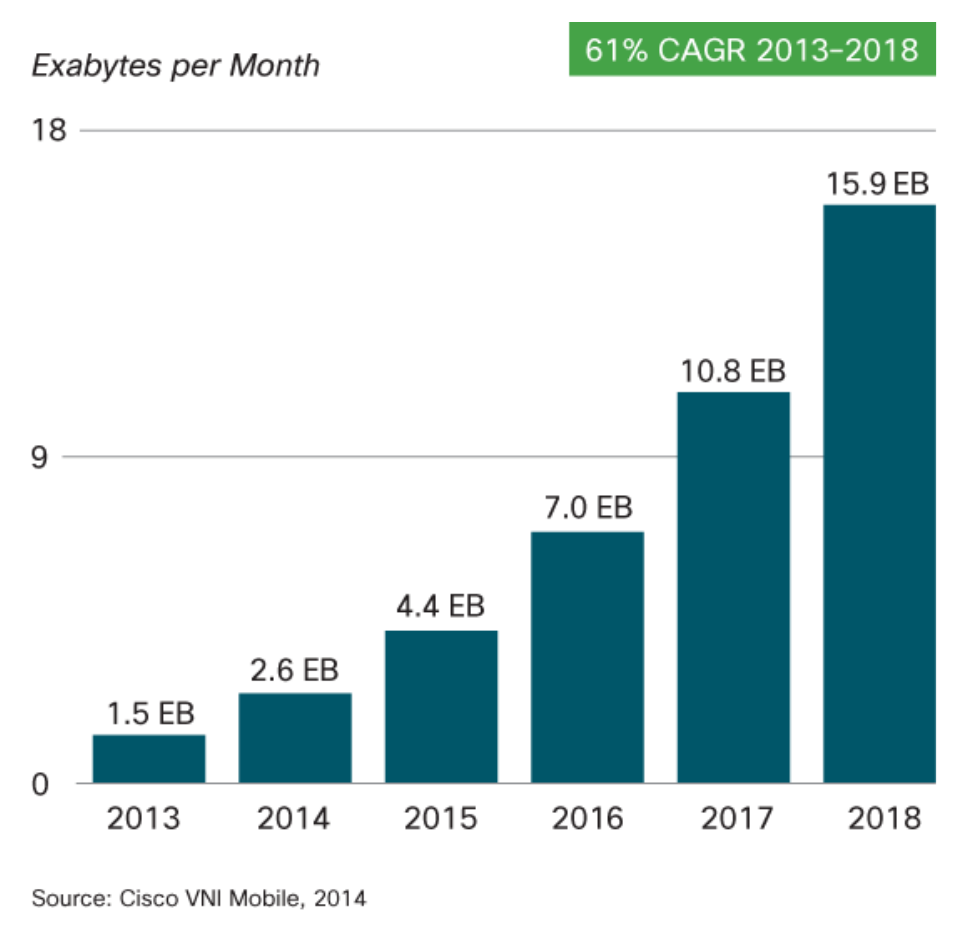}
\caption{Cisco VNI: global mobile data traffic forecast, 2013-2018.} \label{fig:mobile_data_growth_overall}
\end{figure}

\begin{figure*}[t]
\centering
\includegraphics[width=0.7\linewidth]{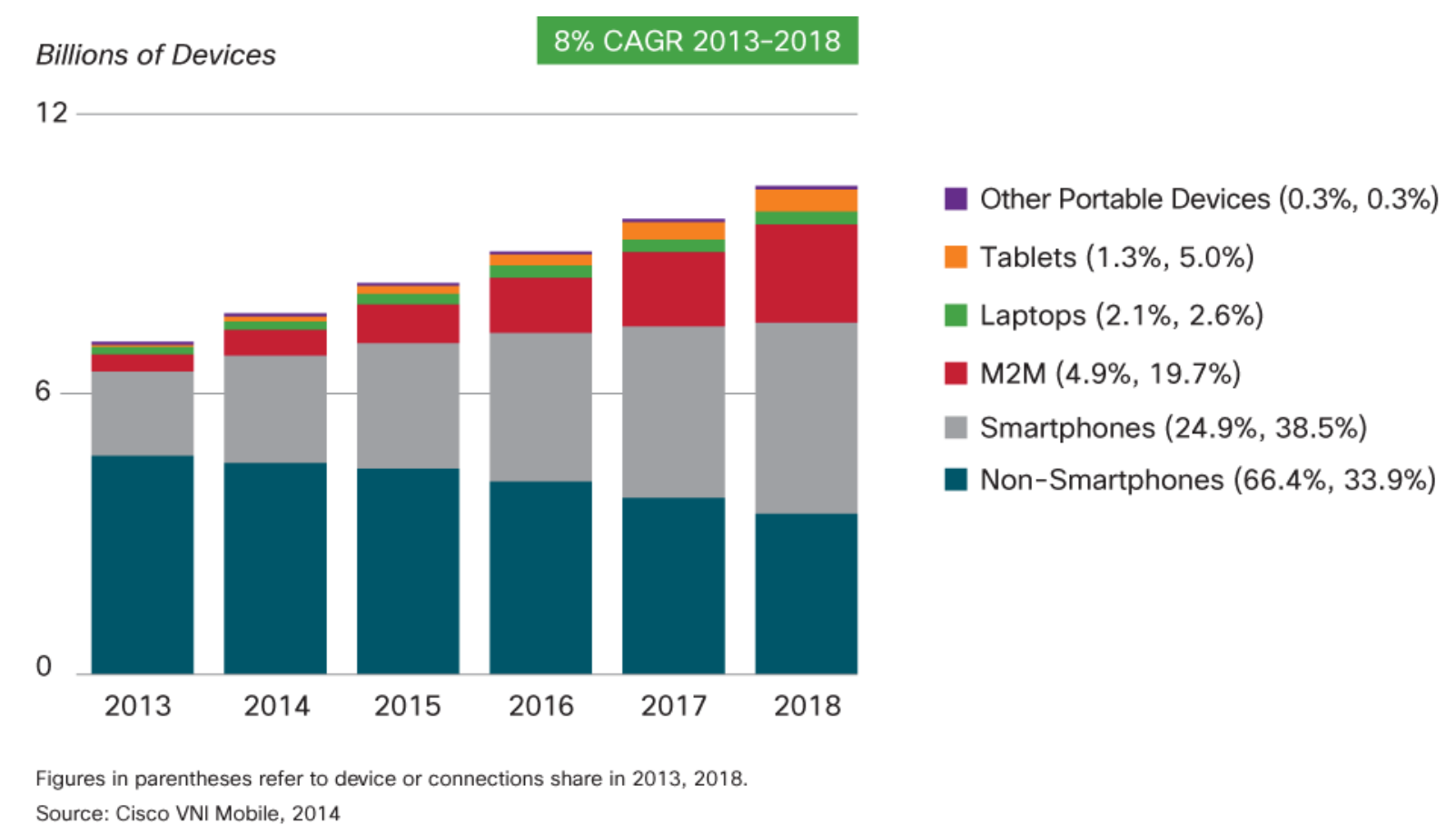}
\caption{Cisco VNI: global mobile devices and connections growth forecast, 2013-2018.} \label{fig:mobile_data_growth_device}
\end{figure*}

The multimedia data traffic conveyed by the global mobile networks has been soaring\cite{Mcqueen_2009:momentum_LTE, Lee_2010:Mobile_data_offloading_wifi, Sayed_2011:mobile_data_explosion_monetizing, Ranganathan_2011:data_explosion, Han_2012:mobile_data_offloading_opportunistic_social}, and this trend is set to continue, as indicated by Cisco's visual networking index (\gls{VNI}) forecast\cite{Cisco_2014:whitepaper_mobile_data_growth_forcast,Cisco_2014:VNI_project}. More specifically,\footnote{Fig. \ref{fig:mobile_data_growth_overall} and Fig. \ref{fig:mobile_data_growth_device} are reprinted from the Cisco VNI white paper\cite{Cisco_2014:whitepaper_mobile_data_growth_forcast}, with permission of Cisco.} the global mobile data traffic grew $81\%$ in 2013, up from 0.82 exabytes (\gls{EB}), i.e. $0.82 \times 10^{18}$ bytes per month at the end of 2012 to 1.5 EB per month at the end of 2013; furthermore, as predicted in Fig. \ref{fig:mobile_data_growth_overall}, it will increase nearly 11-fold between 2013 and 2018, which translates to a compound annual growth rate (\gls{CAGR}) of $61\%$ for the period spanning from 2013 to 2018, reaching 15.9 EB per month by 2018\cite{Cisco_2014:whitepaper_mobile_data_growth_forcast, Cisco_2014:VNI_project}. As seen from Fig. \ref{fig:mobile_data_growth_device}, this explosive growth is mainly fuelled by the prevalence of smartphones, laptops and tablets, as well as by the emergence of machine-to-machine (\gls{M2M})\index{M2M} communications\cite{Lawton_2004:M2M, Cha_2009:M2M, Bob_2010:M2M, Starsinic_2010:M2M, Chen_2010:M2M, Fadlullah_2011:M2M_smart_grid, Niyato_2011:M2M, Lien_2011:M2M, Lu_2011:M2M, Wu_2011:M2M, Zhang_2011:M2M, Lien_2011:M2M_CL}. Additionally, the design of wireless communication systems is highly constrained by the paucity of radio spectrum,  which is exemplified by the overcrowded frequency allocation chart of the United States\cite{US_aug_2011:frequency_allocation_chart}. As a consequence of the combined effect of the mobile data traffic growth trend and the scarcity of favorable radio spectrum in the low-loss frequency-range, the forthcoming fifth generation (\gls{5G}) communication systems have to resort to the employment of massive/large-scale multiple-input multiple-output (\gls{LS-MIMO})\index{LS-MIMO} transmission techniques, which invoke a large number of antenna elements at the transmitter and/or receiver for achieving a high spectral-efficiency\cite{Paulraj_2003:introduction_to_MIMO, Gesbert_2007:shifting_MIMO_paradigm, Mietzner_2009:multi_antenna_survey, Gesbert_2010:multicell_MIMO, Rusek_2013:massive_MIMO, Larsson_2014:massive_MIMO_overview, Boccardi_2014:five_tech_5G, Andrews_2014:5G_overview, Lu_2014_JSTSP_massive_MIMO_overview} and high energy-efficiency\cite{ Han_2011:green_radio, Rusek_2013:massive_MIMO, Larsson_2013:EE_SE_massive_MIMO, Renzo_2014:spatial_modulation}. 

\begin{figure}[t]
\centering
\includegraphics[width=0.9\linewidth]{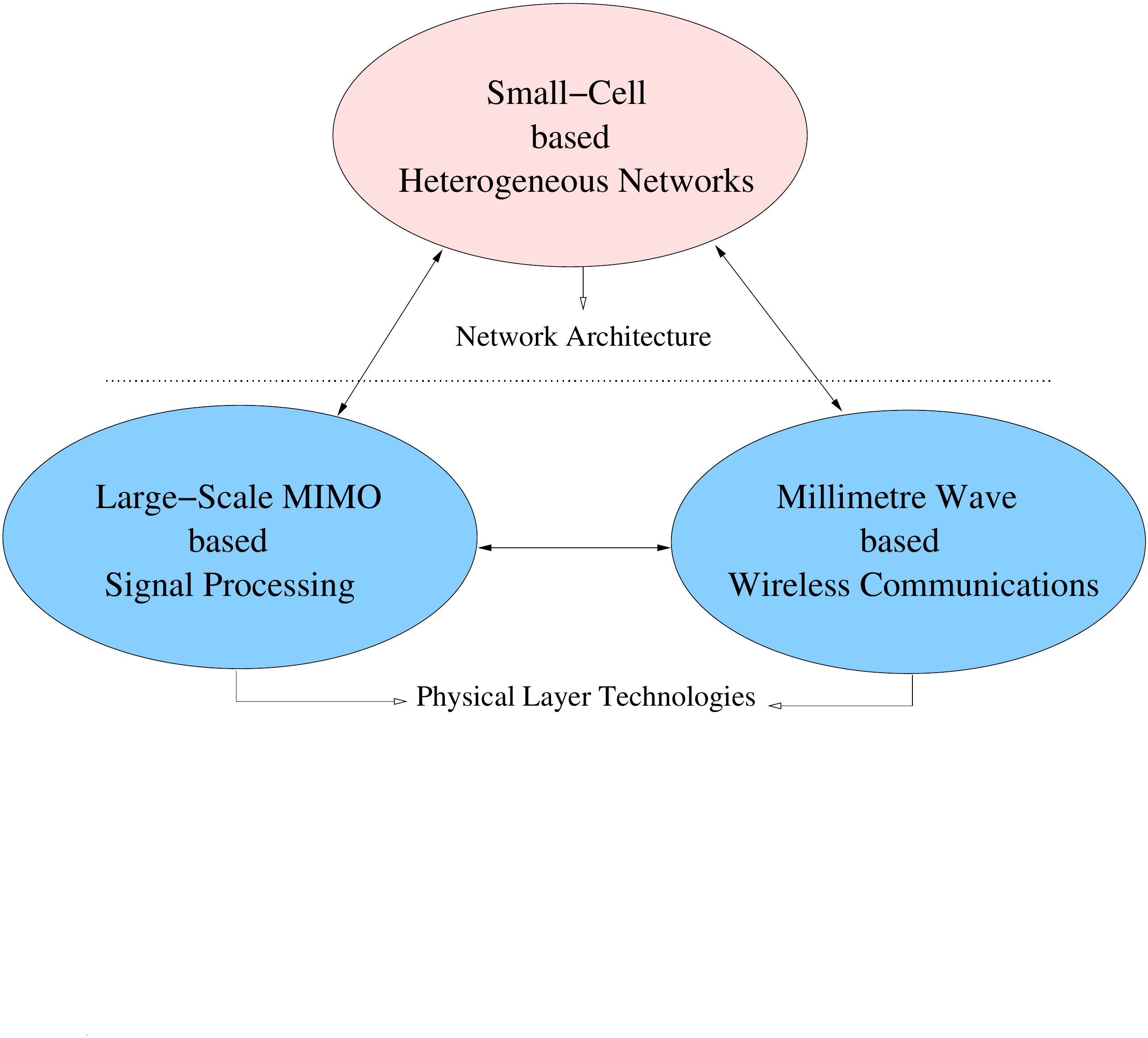}
\caption{There is a natural marriage amongst LS-MIMO, millimetre wave, and small-cell based HetNet, which constitute three fundamental technologies for 5G wireless communications. } \label{fig:3_tech_5G}
\end{figure}

A range of other fundamental technologies conceived for 5G communications are closely related to LS-MIMO. For example, both millimetre wave  communications\cite{Rappaport2013:MMW_5G} and LS-MIMOs may be regarded as enabling techniques facilitating \textit{high-dimensional} physical-layer communication technologies. Their difference is that LS-MIMOs achieve high dimensionality in the spatial domain, while millimetre wave communication systems achieve a high dimensionality in the frequency domain by operating at frequencies ranging from \textit{about} 30 GHz to 300 GHz, which is much higher than the operating frequencies of contemporary third generation (\gls{3G})/4G systems (from 450 MHz to 3.5 GHz). Furthermore, owing to the much shorter wavelength, millimetre wave technologies may facilitate compacting a large number of antenna elements in a relatively small space. Additionally, the coverage area of a single cell of millimetre wave communication systems may be significantly smaller than a single cell of 3G/4G systems. As a result, small-cell based heterogeneous network (HetNet) architecture is required. Therefore, as shown in Fig. \ref{fig:3_tech_5G}, there is a natural marriage amongst LS-MIMO, millimetre wave and small-cell technologies. 

\subsection{Why is MIMO Detection Important and Challenging?}
\begin{figure}[tbp]
\centering
\includegraphics[width=\linewidth]{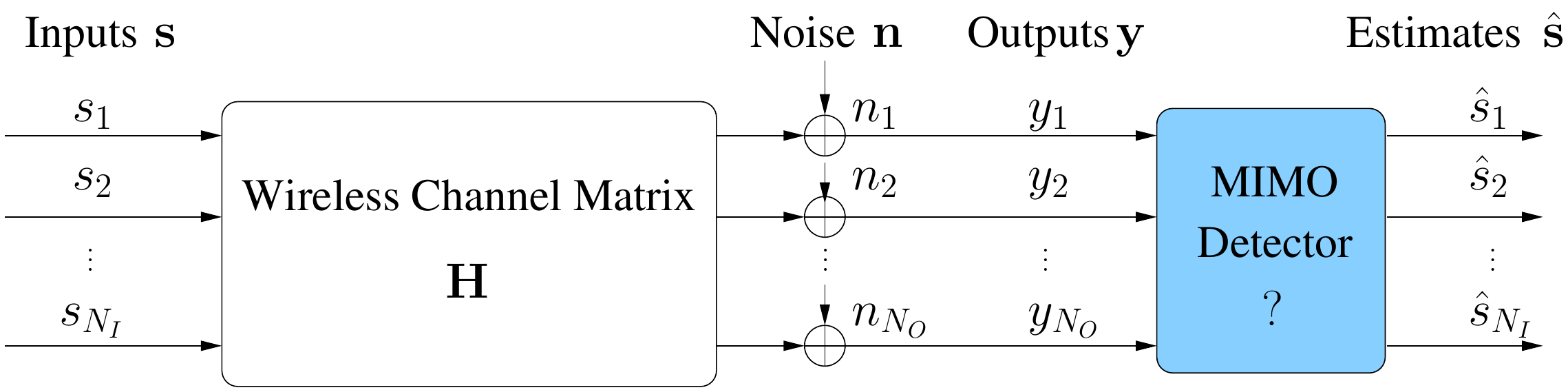}
\caption{Conceptual illustration of the MIMO detection problem. } \label{fig:concept_MIMO_detection_problem}
\end{figure}

As Claude Shannon pointed out, ``The fundamental problem of communication is that of reproducing at one point either exactly or approximately a message selected at another point''\cite{Shannon_1948:math_theory_comm}. Compared to conventional single-input single-output systems, e.g.  the single-antenna point-to-point system, in MIMO systems we have multiple interfering messages/symbols transmitted concurrently, and at the receiver these symbols are expected to be detected/decoded subject to the contamination of random noise or interference, as shown in Fig. \ref{fig:concept_MIMO_detection_problem}. The multiple symbols may be detected separately or jointly. As opposed to separate detection, in joint detection each symbol has to be detected taking into account the characteristics of the other symbols. As a beneficial result, typically joint detection is capable of achieving a significantly better performance than separate detection, although joint detection imposes a higher computational complexity.

The joint detection of multiple symbols in MIMO\index{MIMO} systems is of central importance for the sake of realizing the substantial benefits of various MIMO\index{MIMO} techniques. This is because the co-channel interference (\gls{CCI})\index{CCI} routinely encountered in MIMO-based\index{MIMO} communication systems constitutes a fundamental limiting characteristic\cite{Hanly:initial_interference_limited, Wyner:initial_interference_limited, Shamai:interference_limited_part_I, Shamai:interference_limited_part_II, Somekh:work_of_shamai's_student, Catreux:MIMO_cellular_capacity, Catreux:Throughput_MIMO_cellular, Blum:analysis_MIMO_cellular_capacity, Dai:downlink_capacity_MIMO_cellular, Gesbert_2010:multicell_MIMO, Tse:Fundamental}.  Unfortunately, the optimum MIMO\index{MIMO} detection problem was proven non-deterministic polynomial-time hard (\gls{NP-hard})\index{NP-hard} \cite{Boas_1981:prove_NP_hard_closest_point_search,Verdu_1989:complexity_optimal_MUD,Micciancio_2001:simpler_proof_NP_hard_closest_point_search}, thus all known algorithms conceived for solving the problem optimally have a complexity exponentially increasing with the number of decision variables. As a result, the computational complexity of the optimum maximum-likelihood (\gls{ML})\index{ML} criterion or the maximum \textit{a posteriori} (\gls{MAP}) criterion based MIMO\index{MIMO} detection algorithms quickly become excessive as the number of decision variables increases. Thanks to the rapid development of the semiconductor industry, the hardware computing power has been dramatically increasing over the years, and in some cases a ``not-so-extreme'' computational complexity is no longer regarded as a bottleneck of practical applications. However, it should be noted that while transistors get faster and smaller, supply voltages cannot be reduced significantly in modern complementary metal-oxide semiconductor (\gls{CMOS}) processes. Therefore, virtually all modern integrated circuits (\glspl{IC}) encounter an integration density limit owing to the maximum tolerable internal temperature imposed by the excessive power consumption or power density. In other words, this power bottleneck still limits today's IC development. As a consequence, one cannot simply rely on Moore's law, and even modest-complexity MIMO detection algorithms could be too power hungry for battery-powered devices.
Hence low-complexity, yet high-performance suboptimum MIMO detection algorithms are needed for practical MIMO applications.

\subsection{The Contributions of This Paper}
\begin{figure*}[t]
\centering
\includegraphics[width=6.3in]{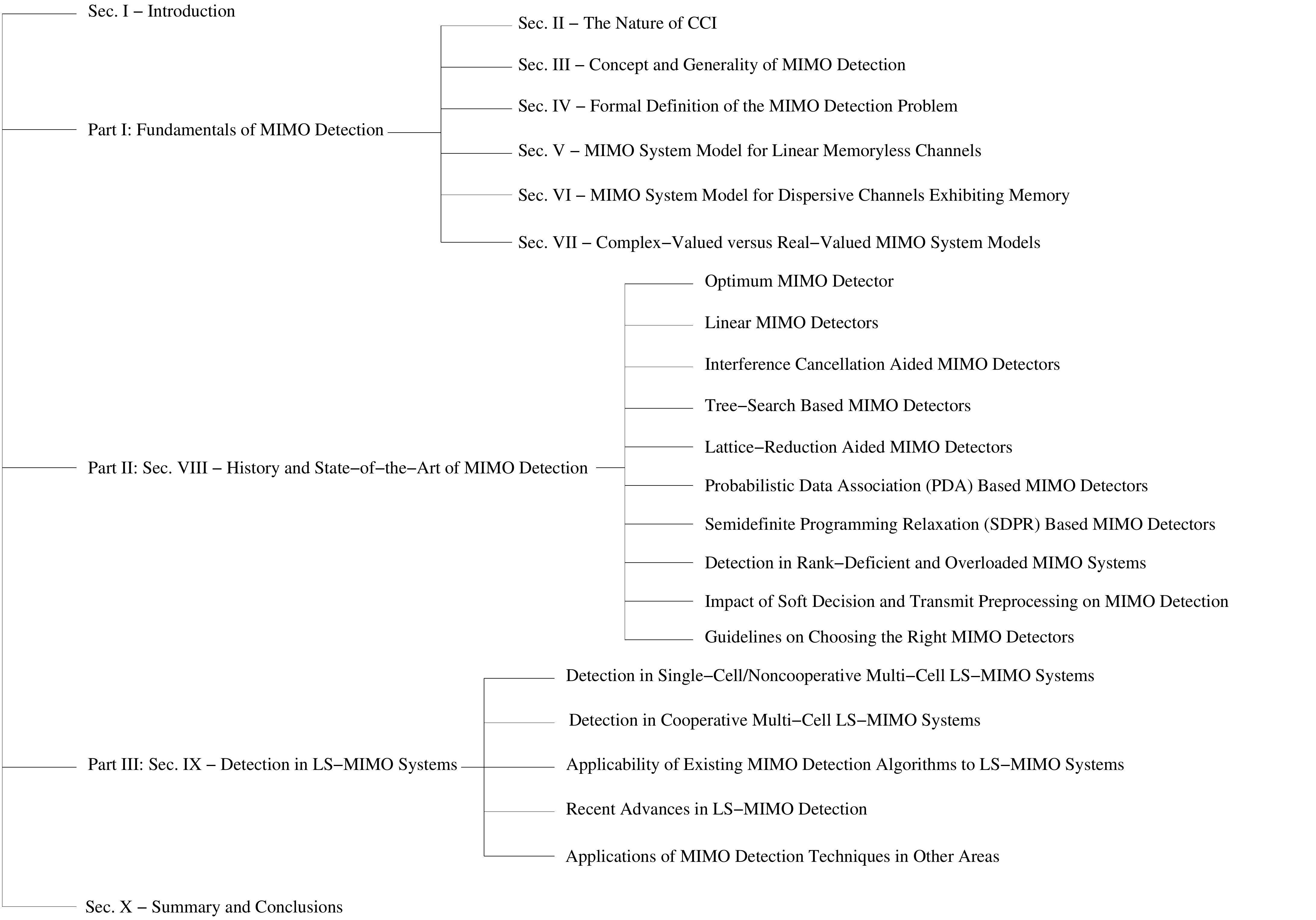}
\caption[]{The organization of this paper.} \label{fig:organization}
\end{figure*}
In this paper, an extensive review of the family of representative MIMO detection methods invented during the past fifty years is presented in a unified mathematical model\footnote{This means that the algorithms conceived for the equalization, multiuser detection and multi-antenna detection problems can be treated under the same umbrella of the MIMO detection model of \eqref{eq:general_MIMO_model_matrix_form}. More discussions on the similarities and differences amongst these three problems are provided in Section \ref{sec:generality_MIMO_detection:chap_intro} and Section \ref{sec:definition_MIMO_detection:chap_intro}, as well as are found in the first two paragraphs of Section \ref{sec:history_MIMO_detection:chap_intro} and the last but one paragraph of Section \ref{subsec:linear_MIMO_detector:chap_intro}.}, although practical MIMO schemes have various subtleties. Our particular focus is on complexity-scalable MIMO detection algorithms potentially applicable to LS-MIMO systems\cite{Rusek_2013:massive_MIMO}. The algorithms surveyed include the classic linear MIMO detection, the interference-cancellation based MIMO detection, the tree-search based MIMO detection, the lattice-reduction (\gls{LR}) aided MIMO detection, the probabilistic data association (\gls{PDA}) based MIMO detection, and the semidefinite programming relaxation (\gls{SDPR}) based MIMO detection. Several high-quality books or reviews were published on MIMO detection\cite{Hallen_1995:MUD_CDMA_overview, Moshavi_1996:MUD_overview_CDMA, Verdu:MUD_book, Honig:advances_MUD_edited, Bai_2012:low_complexity_MIMO_detection}. They were predominantly dedicated to CDMA systems in the 1990s\cite{Hallen_1995:MUD_CDMA_overview, Moshavi_1996:MUD_overview_CDMA, Verdu:MUD_book} or to conventional small-/medium-scale MIMO systems \cite{Honig:advances_MUD_edited, Bai_2012:low_complexity_MIMO_detection}, whereas LS-MIMOs just became a hot research topic at the time of writing\cite{Rusek_2013:massive_MIMO}. On the other hand, they mainly covered the most common suboptimum MIMO detection methods, such as the linear zero-forcing (\gls{ZF})\index{ZF} detector, the linear minimum mean-square error (\gls{MMSE})\index{MMSE} detector and various interference cancellation based detectors\cite{Hallen_1995:MUD_CDMA_overview, Moshavi_1996:MUD_overview_CDMA, Verdu:MUD_book, Honig:advances_MUD_edited}, or focused largely on a single type of MIMO detector\cite{Bai_2012:low_complexity_MIMO_detection}. In comparison, there is a paucity of reviews on more advanced MIMO detection methods, such as the tree-search based MIMO detectors (the sphere decoder (\gls{SD})\index{SD} constitutes an instance of the tree-search based MIMO detectors)\cite{Viterbo_1999:SD, Agrell:closest_point_search_in_lattice, Damen_2000:lattice_decoder_STC, Damen_2000:generalized_SD_STC, Damen_2003:MLD_closest_lattice_point_search, Hassibi_2005:SD_complexity_part_1, Vikalo_2005:SD_complexity_part_2, Jalden:SD_complexity_journal, Burg_2005:VLSI_depth_first_SD,
Wong_2002:K-best_SD_VLSI, Guo_2006:implementation_K_best_SD_MIMO, Chen_2007:K-best_VLSI, Studer_2008:soft_SD_implementation, Murugan_2006:unified_tree_search_sequential_decoder, Gowaikar_2007:statistical_pruning_SD_depth_first, Lee_2007:short_path_SD, Stojnic_2008:speed_up_SD_via_infinity_H_norm, Kim_2010:best_first_search, Chang_2012:best_first_search_A_algorithm, Chang_2012:best_first_search,Choi:sphere_decoder_look_ahead, Hochwald:SD_near_capacity, Boutros_2003:SISO_SD_MIMO, Vikalo_2004:IDD_SD_MIMO, Wang_2006:approaching_MIMO_capacity_hard_SD, Studer_2010:SISO_single_SD, Rachid_2010:best_first_search, Barbero:fixed_complexity_SD_journal, Barbero:soft_fixed_complexity_SD, Jalden:FCSD_error_prob}, the LR based MIMO detectors\cite{Yao_2002:LR_MIMO_detector, Fischer_2003:MIMO_precoding_detection, Wubben_2004:LR_reduction_MMSE, Windpassinger_2004:LR_precoding_MIMO, Taherzadeh_2007:LLL_LR_receiver_diversity, Ling_2007:LLL_LR, Jalden_2008:complexity_LLL, Seethaler_2007:SA_LR_MIMO, zhang_2008:analysis, Burg_2007:LR_Brun_algorithm, Shabany_2008:LR_VLSI_implementation, Gestner_2008:LR_VLSI_CLLL, Gan_2005:CLLL_conf, Gan_2009:CLLL_MIMO_detection, Ma_2008:performance_analysis_CLLL, Ma_2008:CLLL, Silvola_2006:soft_LR_MAP, Qi_2007:soft_LR, Ponnampalam_2007:LR_soft_outputs,Wei_2010:soft_LR_MIMO_detector, Zhou_2013:Lattice_reduction_LS_MIMO_detector}, as well as the PDA\cite{Luo:PDA_Sync_CDMA, Luo:PDA_thesis, Pham:PDA_Async_CDMA, Luo_2003:sliding_window_PDA, Penghui_2003:iterative_PDA_MUD, Yin_2004:turbo_equalization_PDA, Huang_2004:generalized_PDA, Pham:GPDA, Liu:CPDA-apx, Liu:Kalman_PDA_freq_selective, Latsoudas_2005:hybrid_PDA_SD, Jia_2005:Gaussian_approximation_mixture_reduction_MIMO, Fricke:Impact_of_Gaussian_approximation, Shaoqian2005:turbo_PDA, Penghui_2006:asymptotic_optimum_PDA, Cai2006:iterative_PDA, Jia:CPDA, Cao:Relation_of_PDA_and_MMSE-SIC, Bavarian_2007:distributed_BS_cooperation_uplink, Bavarian_2008:distributed_BS_cooperation_uplink_journal, Jia_2008:multilevel_SGA_PDA, Grossmann_2008:turbo_equalization_PDA, Kim_2008:noncoherent_PDA, Shaoshi_2008:PDA_JD, Mohammed_2009:PDA_STBC, Bavarian:SDE_PDA_freq_selective, Shaoshi2011:B_PDA, Shaoshi2011:DPDA, Shaoshi2013:Turbo_AB_Log_PDA, Shaoshi_2013:EB_Log_PDA_journal} and the SDPR\cite{Penghui:SDP_CDMA, Ma:SDR_CDMA_BPSK,
Ma:SDR_CDMA_QPSK, Luo:SDR_performance_analysis, Jalden:SDR_diversity, Luo:SDP_MPSK, Ma:SDR_MPSK,
Wiesel:PI_SDR_16QAM, Yijin:SDR_16QAM_tight, Sidiropoulos:SDR_HOM, Mobasher:SDR_QAM_journal, Mao:SDR_LMR, Ma:equivalence_SDR, Shaoshi_2013:DVA_SPDR_journal} based detectors etc., although a concise tutorial on some of these detectors was given in\cite{Larsson:MIMO_detection_overview}. Additionally, most of the existing research on LS-MIMO is focused on the precoding/beamforming based downlink of a special case of LS-MIMO, where one side of the communication link has a significantly higher number of antennas than the other. By contrast, only limited attention has been dedicated to the uplink of general LS-MIMOs. Hence, our goal is to fill these gaps in the open literature.
For the sake of clarity, the organization of this paper is shown in Fig. \ref{fig:organization}.

\section{The Nature of Co-Channel Interference}
\label{sec:essence_CCI:chap_intro}
\begin{figure}[t]
\centering
\includegraphics[width=3.5in]{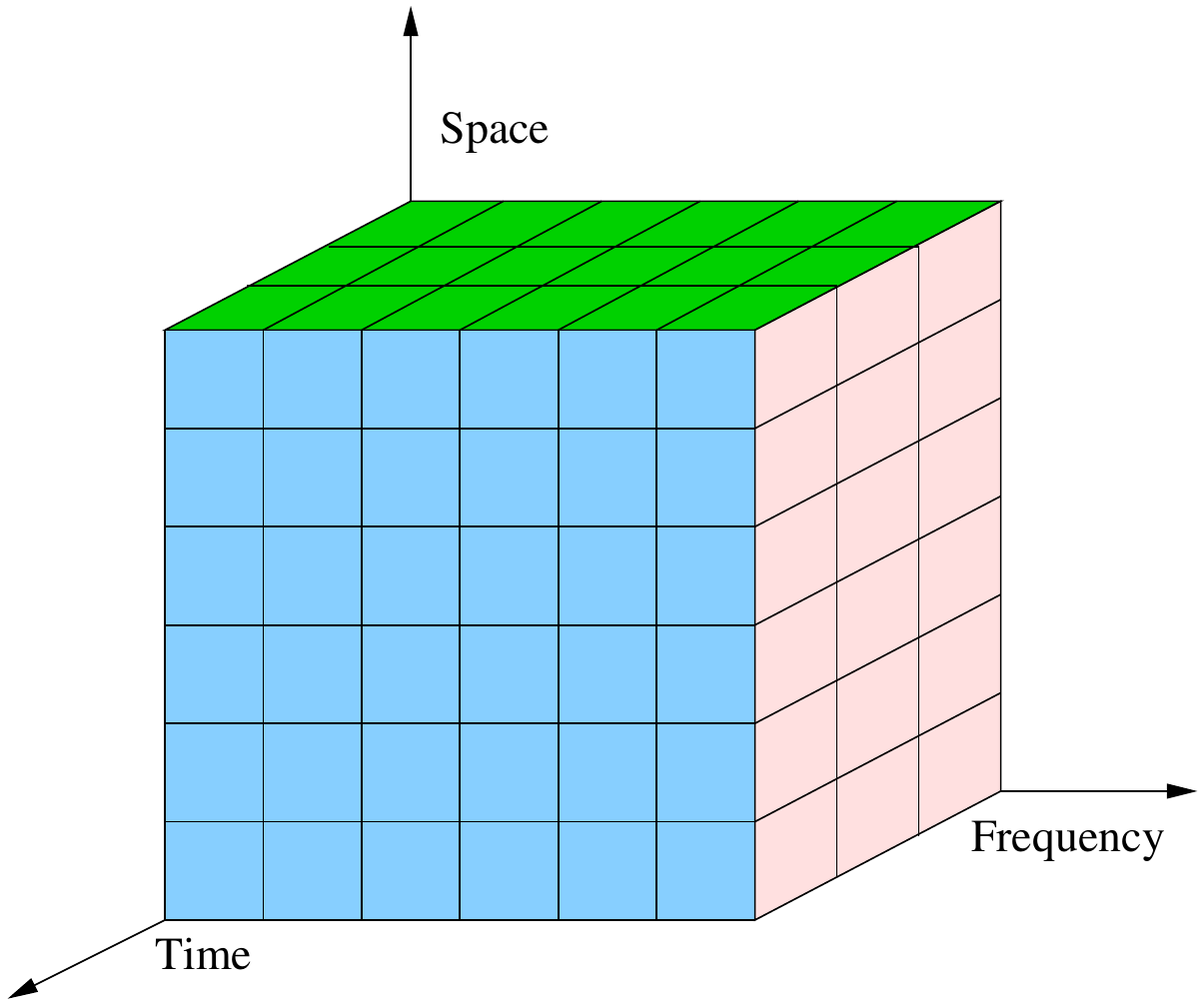}
\caption[Three fundamental domains of characterizing signals.]{The multiple signals have to be distinguishable in at least one of the three fundamental domains of time, frequency and space.} \label{fig:resource_space}
\end{figure}

To gain profound insights into the intricacies of the MIMO detection problem, let us briefly reflect on the nature of the CCI\index{CCI} in this section.
The nature of CCI\index{CCI} depends on the specific context. In this paper, it is defined in its most generic form as \textit{the interfering signal imposed by multiple transmissions taking place on channels which are mutually non-orthogonal.} Mathematically, CCI may also be interpreted as interfering signals that span a subspace having a ``non-empty'' intersection with the subspace spanned by the desired signals. The channel-induced non-orthogonality may be observed in the frequency, time and/or space domain, as shown in
Fig. \ref{fig:resource_space}. To recover the desired signal at the receiver, the desired signal has to be distinguishable from the interference in \textit{at least} one domain. In the extreme case, if the multiple transmissions are highly non-orthogonal in all domains, then it may become impossible to recover the desired signal by any means.

\begin{figure}[t]
% \begin{minipage}{0.501\linewidth}
\centering
\subfigure[Time-domain waveform]{
\label{subfig:rectangular_pulse}
\includegraphics[width = 1.0 \linewidth]{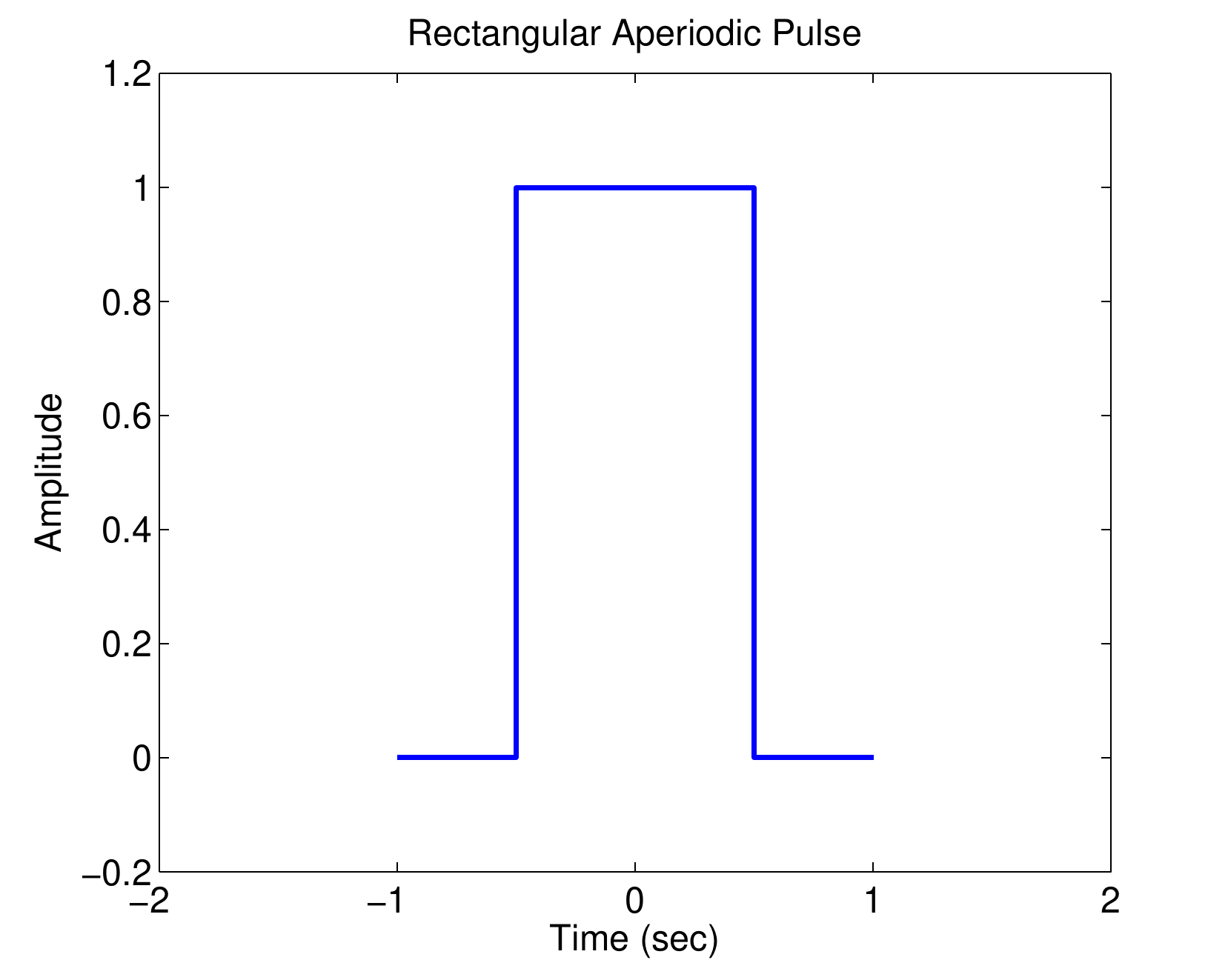}
}
% \end{minipage}
% \begin{minipage}{0.501\linewidth}
\centering
\subfigure[Frequency-domain response]{
\label{subfig:Sinc_Func}
\includegraphics[width = 1.0 \linewidth]{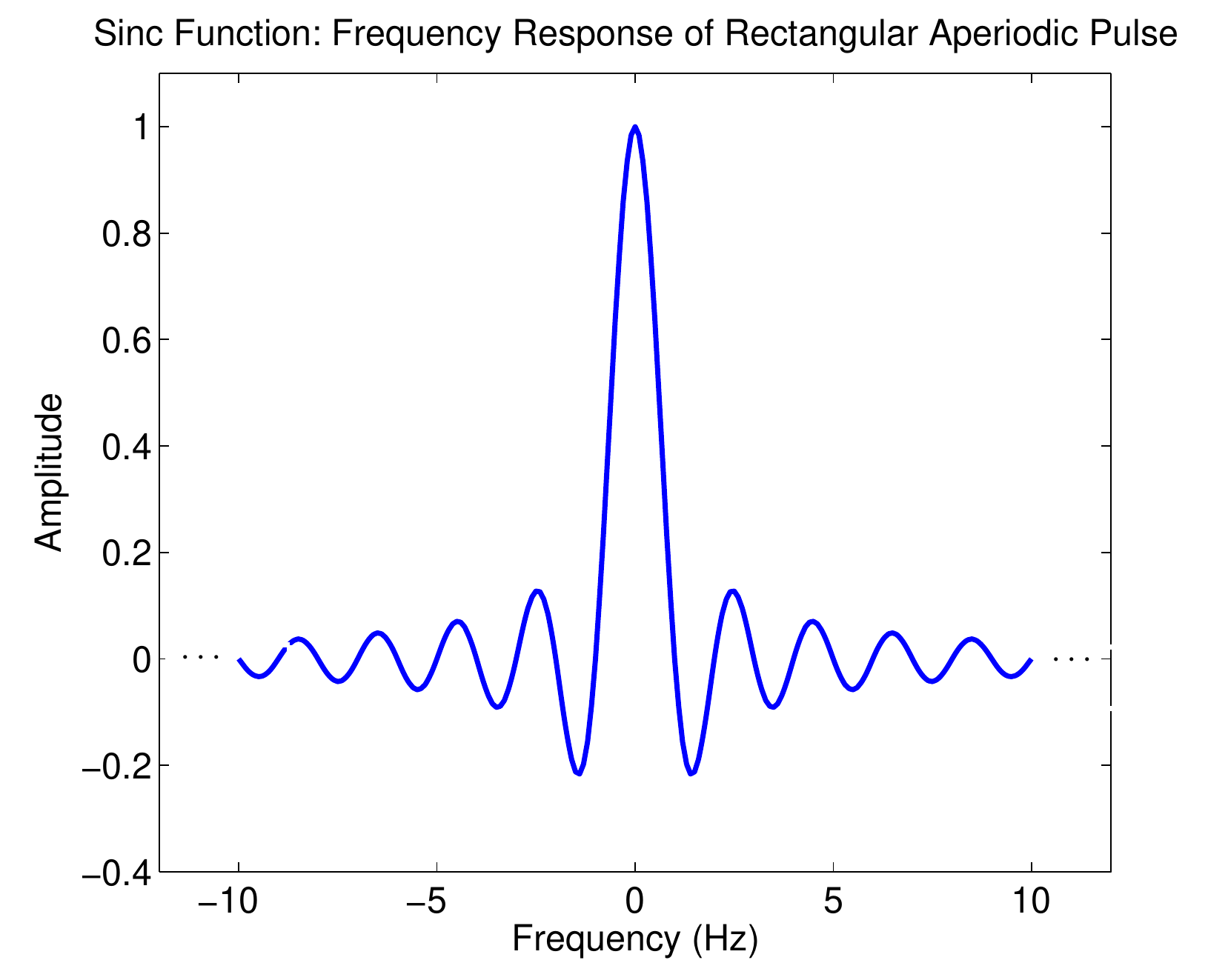}
}
% \end{minipage}
\caption[Limited duration-time vs infinite bandwidth: the reason why every active radio transmitter has an impact on every operating radio receiver.]{All realizable, duration-time limited waveforms has a infinite frequency band.}
\label{fig:time_fre_pair}
\end{figure}

In essence, the CCI\index{CCI} originates from \textit{signal-feature-overlapping} of multiple transmissions. For example,
in spectrum-efficient communication systems such as the code-division multiplexing /multiple-access (\gls{CDM}/\gls{CDMA})\index{CDM}\index{CDMA} systems\cite{Lee_1991:overview_CDMA, Jung_1993:advantages_of_CDMA_over_TDMA_FDMA, Viterbi_1995:CDMA_book} and the space-division multiplexing /multiple-access (\gls{SDM}/\gls{SDMA})\index{SDM}\index{SDMA} systems\cite{Roy1996:SDMA, Gerlach1995:SDMA_thesis, Ottersten1996:SDMA, Roy1997:SDMA, Lotter1998:SDMA, vandenameele2001:SDMA, Paulraj_2003:introduction_to_MIMO, Tse:Fundamental, Goldsmith_2005:Wireless_Comm}, multiple transmissions are often deliberately arranged to
take place simultaneously over the same frequency band. These ``frequency sharing'' and ``time sharing'' strategies result in a ``frequency-overlapping'' and a ``time-overlapping'' phenomenon, respectively.
It is worth pointing out that as far as radio waves are concerned, rigorously the CCI\index{CCI} always tends to exist in the frequency, time and space domains. For example, when no deliberate frequency-overlapping is arranged, the ``frequency-overlapping'' is due to the underlying fact that \textit{for all realizable, time-limited radio waveforms, their absolute bandwidth is infinite} \cite{Oppenheim:1996:signals_and_systems, Sklar:digital_communications}, as shown in Fig. \ref{fig:time_fre_pair}.
In other words, \textit{every active radio transmitter has an impact on every operating radio receiver}. Similarly, for a strictly bandwidth-limited signal, its time duration has to be infinite. With respect to the space domain, it is well known that the propagation of
electromagnetic energy in free space
is determined by the inverse square law\cite{Rappaport_2002:Wireless_textbook, Paulraj_2003:introduction_to_MIMO, Goldsmith_2005:Wireless_Comm}, i.e. we have $S = {P_t}/{4\pi d^2}$, where $S$ is the power per unit area or power spatial density (in Watts per metre-squared) at distance $d$, and $P_t$ is the total power transmitted (in Watts). Hence, theoretically, the radio signals cannot be stopped, they are only attenuated in the frequency, time and space domains.

In engineering practice, fortunately, by using well-designed filters\cite{Proakis_2007:DSP,Haykin_2003:adaptive_filtering}, typically the waveform of the time-limited signal can be shaped so that most energy of the signal can be kept within a given limited frequency-band, and thus the signal energy leakage outside the target frequency-band can be reduced to a sufficiently low level. Similarly, in the space domain, two transmissions taking place at a sufficiently far distance can also be regarded as non-interfering with each other. Therefore, despite the fact that the signal-feature-overlapping in frequency, time and space domains is inevitable from the theoretic point of view, in practical spectrum-efficient systems we can typically assume that the signal-feature-overlapping in these three domains is a result of a \textit{deliberate} design. In this context, \textit{the signals are made as much distinguishable as possible in one domain, and as much overlapping as possible in the remaining domains.} Our task is to recover the desired signal based on this deliberate arrangement.

Since the frequency, time and space domains represent the fundamental physical features of signal transmission, each of them corresponds to a distinct multiplexing/multiple-access scheme\cite{Jamalipour:TDMA}, namely the frequency-division multiplexing/multiple-access (\gls{FDM}/\gls{FDMA})\index{FDM}\index{FDMA}, time-division multiplexing/multiple-access (\gls{TDM}/\gls{TDMA})\index{TDM}\index{TDMA}, and SDM/SDMA, respectively. All of these multiple-access techniques aim for avoiding CCI by orthogonalizing the channel access in a certain domain.  It is worth noting that compared to these three fundamental domains, the \textit{spreading code sequences} used in CDM/CDMA\index{CDM}\index{CDMA} systems do \textit{not} constitute an independent domain. This is because the orthogonality of the spreading code sequences is essentially \textit{a special case of time-domain orthogonality}. In systems using spreading codes, in principle we pursue to transmit orthogonal code sequences to minimize the inter-code interference. Although it is mathematically possible to construct perfectly orthogonal code sequences, the orthogonality of these code sequences is typically degraded in practical transmissions\cite{Viterbi_1995:CDMA_book}. Moreover, since the number of theoretically orthogonal code sequences is rather limited, often quasi-orthogonal code sequences are adopted in practice\cite{Viterbi_1995:CDMA_book}. Therefore, typically substantial interference is imposed by the non-orthogonality of spreading code sequences in practical CDM/CDMA\index{CDM}\index{CDMA} systems\cite{Kandukuri_2002:power_control_CDMA}.

In analogy to CDM/CDMA\index{CDM}\index{CDMA}, the conventional frequency-division pattern, time-division pattern and space-division pattern can also be regarded as a special case of ``spreading codes'' in the frequency domain, time domain and space domain, respectively. Note, however, that there exist a certain degree of differences in terms of \textit{their multiplexing/multiple-access resolution} in these three domains. More specifically, in practice, by using guard intervals in the corresponding domain, a good resolution of frequency-division and time-division may be readily maintained -- in other words, the orthogonality of ``spreading codes'' in frequency- and time-division systems may be relatively easy to obtain\cite{Jamalipour:TDMA}. By contrast, the resolution of space-division tends to be undermined by the physical size of transmitters/receivers and by the random propagation channel, hence typically substantial interference is imposed by the non-orthogonality of ``spreading codes'' in practical space-division systems\cite{Roy1996:SDMA, Gerlach1995:SDMA_thesis, Ottersten1996:SDMA, Roy1997:SDMA, Lotter1998:SDMA, vandenameele2001:SDMA, Paulraj_2003:introduction_to_MIMO, Tse:Fundamental, Goldsmith_2005:Wireless_Comm}. This is similar to the case in CDM/CDMA\index{CDM}\index{CDMA} systems and explains why MIMO detection typically represents a more significant problem in CDM/CDMA\index{CDM}\index{CDMA} systems and SDM/SDMA systems than in FDM/FDMA\index{FDM}\index{FDMA} and TDM/TDMA\index{TDM}\index{TDMA} systems.

Additionally, there are more advanced \textit{multicarrier} based orthogonal multiple-access techniques, such as  orthogonal frequency-division multiple-access (\gls{OFDMA})\cite{Koffman:OFDMA, Morelli_2007:synchronization_OFDMA_tutorial, Necker:Interference_OFDMA}, single-carrier frequency-division multiple-access (\gls{SC-FDMA})\footnote{Note that the so-called SC-FDMA is actually a multicarrier multiple-access technique, although somewhat misleadingly it has ``single-carrier'' in its name.}\cite{Benvenuto:SC-FDMA}, and multicarrier CDMA \cite{Hara_1997:MC_CDMA, Lieliang_2002:MC_DS_CDMA_Nakagami, Lieliang_2003:MC_DS_CDMA, Adachi_2005:MC_CDMA, Hanzo:CDMA_book, Lieliang_2009:Multicarrier, Chockalingam_2008:HNN_LAS_based_LS_MIMO_detector, Kadir_2015:MIMO_multicarrier_STSK}. Despite their potential advantages in averaging interference over different subcarriers for different users, they usually suffer high sensitivity to frequency offset, which leads to \textit{intercarrier interference}. Therefore, judicious  frequency offset compensation scheme and frequency reuse scheme have to be designed to minimize the intercarrier interference.

In this paper, the CCI\index{CCI} considered mainly refers to the interference in SDM/SDMA or CDM/CDMA\index{CDM}\index{CDMA} systems, where multiple transmissions often take place \textit{simultaneously}, or \textit{partially simultaneously} over the same frequency. Depending on specific applications, CCI is often alternatively termed as intersymbol interference (\gls{ISI}), interchannel interference (\gls{ICI}), interantenna interference (\gls{IAI}), multiuser interference (\gls{MUI}), multiple-access interference (\gls{MAI}), and multiple-stream interference (\gls{MSI}) etc. 

\section{Concept and Generality of MIMO Detection}
\label{sec:generality_MIMO_detection:chap_intro}
\begin{figure}[t]
\centering
\includegraphics[width=3.5in]{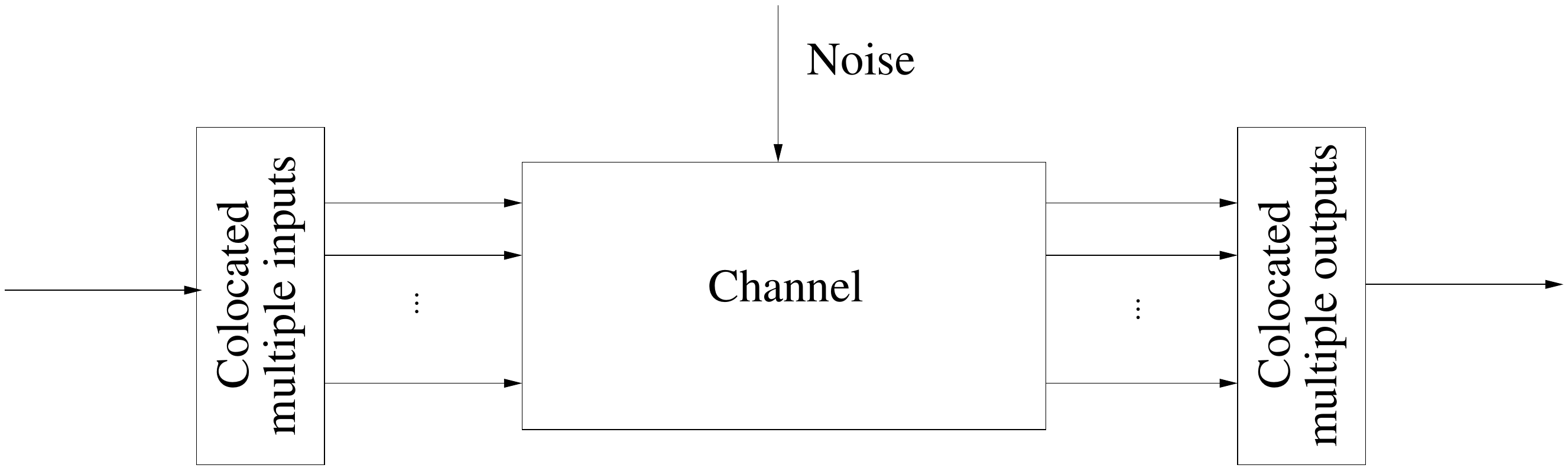}
\caption[Schematic of the point-to-point MIMO channel.]{Point-to-point MIMO channel.} \label{fig:SU_MIMO}
\end{figure}

\begin{figure}[tbp]
\centering
\includegraphics[width=3.5in]{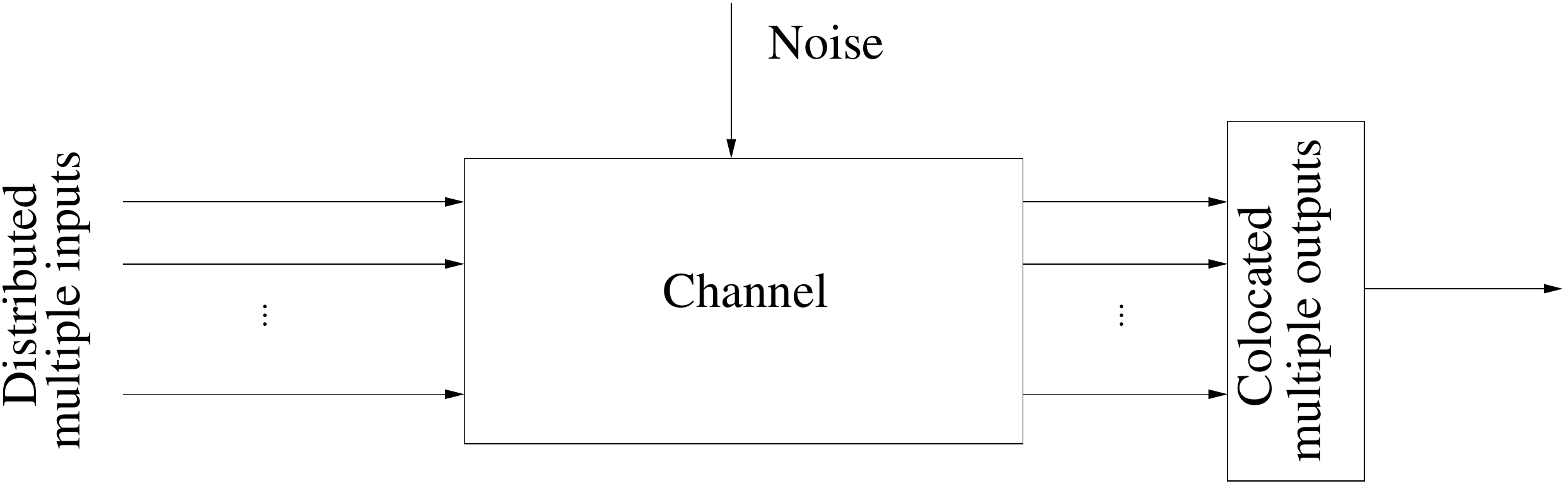}
\caption[Schematic of the MIMO multiple-access channel.]{MIMO multiple-access channel.} \label{fig:MA_MIMO}
\end{figure}

As a family of general techniques for physical-layer CCI\index{CCI} management, MIMO detection deals with the joint detection of several information-bearing symbols transmitted over a communication channel having multiple inputs and multiple outputs. This problem is of \textit{fundamental} importance for modern high-throughput digital communications. Rigorously, the MIMO detection problem arises \textit{if and only if} the respective \textit{subchannels} of the multiple inputs are \textit{not} orthogonal to each other, and hence there exists interference between the outputs. As a \textit{generic} mathematical model, the MIMO detection problem underpins numerous relevant applications, while the physical meaning of the inputs and outputs herein may vary in different contexts. For instance, in band-limited ISI channels requiring equalization, the inputs refer to a sequence of symbols interfering with each other, and the outputs are the signals received within a given observation window in the time/frequency domain\cite{Tuchler_2011:turbo_equalisation_overview}. In single-user SDM-MIMO systems equipped with multiple transmit and receive antennas\cite{Foschini:MIMO, Wolniansky:VBLAST, Golden_1999:VBLAST_first_journal}, the inputs refer to the vector of modulated symbols that are transmitted from multiple \textit{collocated} transmit antennas, while the outputs refer to the vector of received signals recorded at multiple \textit{collocated} receive antennas, as shown in Fig. \ref{fig:SU_MIMO}. This is indeed a canonical scenario of investigating MIMO detection algorithms\cite{Larsson:MIMO_detection_overview}.
A third example is the uplink of multiuser multiple-antenna systems\cite{Sfar_2003:uplink_MU_MIMO_MUD, Serbetli_2004:transceiver_design_MU_MIMO}, where the inputs may be multiple transmitted symbols belonging to a cluster
of geographically \textit{distributed} single-antenna mobile stations (\glspl{MS}), and the outputs may be the signals received at the serving base station (\gls{BS}) equipped with multiple \textit{collocated} antennas, as shown in Fig. \ref{fig:MA_MIMO}. This is actually the so-called SDMA system\cite{Roy1996:SDMA, Gerlach1995:SDMA_thesis, Ottersten1996:SDMA, Roy1997:SDMA, Lotter1998:SDMA, vandenameele2001:SDMA, Paulraj_2003:introduction_to_MIMO, Tse:Fundamental, Goldsmith_2005:Wireless_Comm}. Yet another important example represented by Fig. \ref{fig:MA_MIMO} is the uplink of CDMA\index{CDMA} systems\cite{Hallen_1995:MUD_CDMA_overview, Moshavi_1996:MUD_overview_CDMA, Verdu:MUD_book, Honig:advances_MUD_edited}, where the inputs are the transmitted symbols of distributed single-antenna MSs, and the outputs are typically generated by filtering the signal received at the single-antenna BS with a bank of matched filters (\glspl{MF}), whose impulse responses are matched to a set of \textit{a priori} known user-signature waveforms.

Here, it should be emphasized that whether the multiple inputs and/or the multiple outputs are ``collocated or not'' is extremely important in determining the signal processing techniques to be used. If multiple inputs/outputs are collocated, the \textit{cooperative joint encoding/decoding} of the inputs/outputs can be conducted\cite{Zhao_2003:cooperative_signal_processing, Hunter_2002:cooperative_diversity_coding, Laneman_2003:distributed_protocol_cooperative_diversity, Laneman_2004:cooperative_diversity, Nosratinia_2004:cooperative_comm, Janani_2004:coded_cooperation, Hunter_2006:diversity_coded_cooperation, Paulraj_2003:introduction_to_MIMO, Tse:Fundamental, Goldsmith_2005:Wireless_Comm}, which renders joint MIMO transmission/detection feasible. For example, the single-user MIMO system shown in Fig. \ref{fig:SU_MIMO} has both its transmit and receive antennas collocated, hence it enjoys the privilege of performing both joint encoding and joint decoding. As a benefit, both simultaneous transmission and simultaneous reception can be attained relatively simply. By contrast, the multiple-access MIMO system of Fig. \ref{fig:MA_MIMO} is typically not capable of joint encoding at the user side, hence the uplink transmissions of both CDMA\index{CDMA} and SDMA\index{SDMA} systems are \textit{asynchronous} by nature.

\begin{figure}[t]
\centering
\includegraphics[width=3.5in]{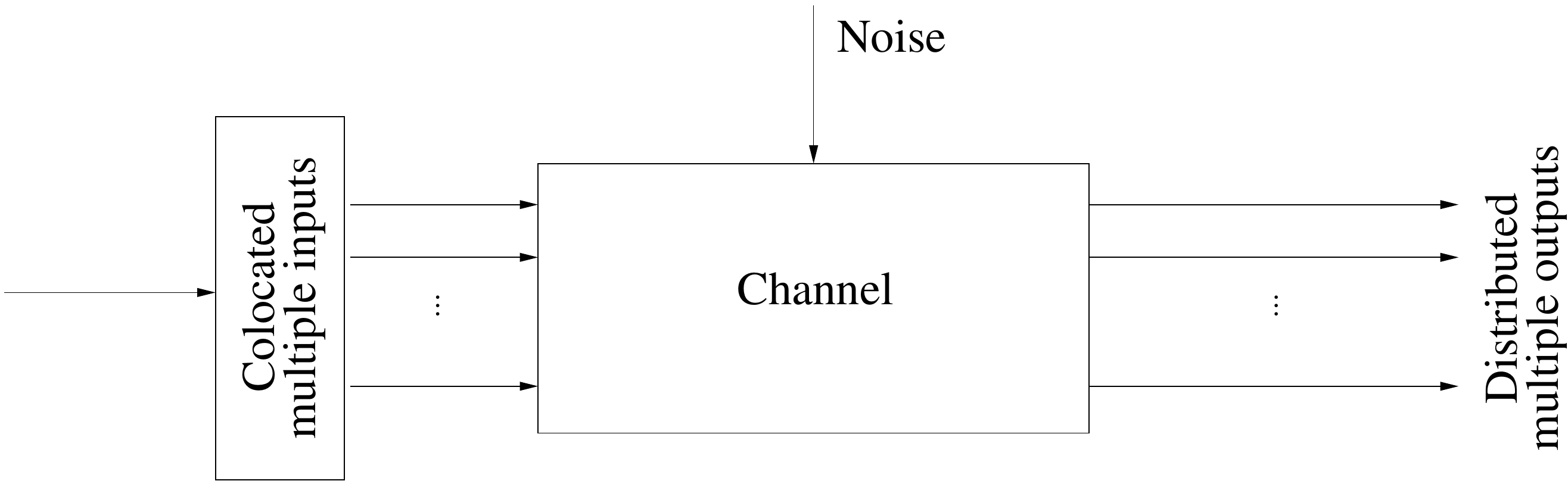}
\caption[Schematic of the MIMO broadcast channel.]{MIMO broadcast channel, which is not suitable for MIMO detection.} \label{fig:BC_MIMO}
\end{figure}

Additionally, as far as the downlink of multiuser MIMO systems, namely the multiuser MIMO broadcast channel of Fig. \ref{fig:BC_MIMO} is concerned, typically most of the sophisticated signal processing tasks are conducted in the form of transmit preprocessing (i.e. precoding) at the BS, where collocated inputs are available for cooperative joint encoding\cite{Spencer_2004:ZF_DL_MU_MIMO, Spencer:intro_multi-user_MIMO_downlink, Choi_2004:preprocessing_DL_MU_MIMO, Zhang:CCI_mitigation_downlink, Caire_2008:MU_MIMO_partial_cooperation_IA, Caire_2010:MU_MIMO_DL}. As a result, detection at the user becomes \textit{less} challenging. Since the investigation of MIMO transmit preprocessing techniques is beyond the scope of this paper, we will not elaborate on it in the sequel.

\begin{figure}[tbp]
\centering
\includegraphics[width=3.5in]{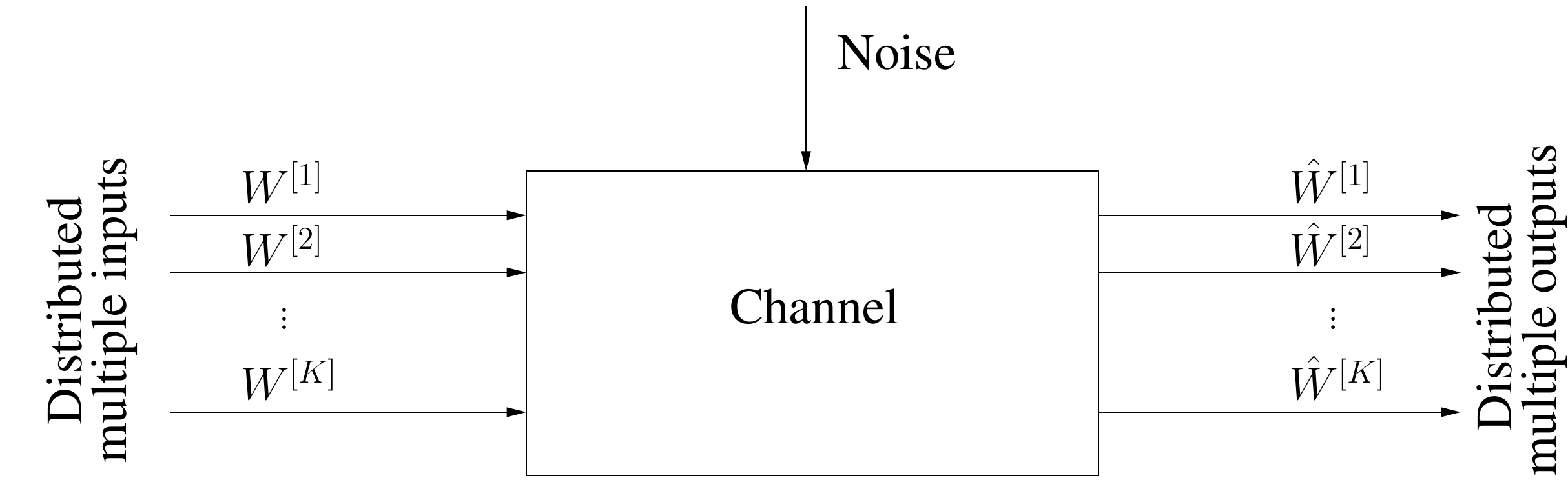}
\caption[Schematic of the MIMO interference channel.]{MIMO interference channel, where $W^{[k]}$ represents the message that originates at transmitter $i$ and is intended for its dedicated receiver $i$, while $\hat W^{[i]}$ denotes the recovered version of $W^{[k]}$, $k = 1, \cdots, K$. } \label{fig:IC_MIMO}
\end{figure}
\begin{figure}[t]
\centering
\includegraphics[width=3.5in]{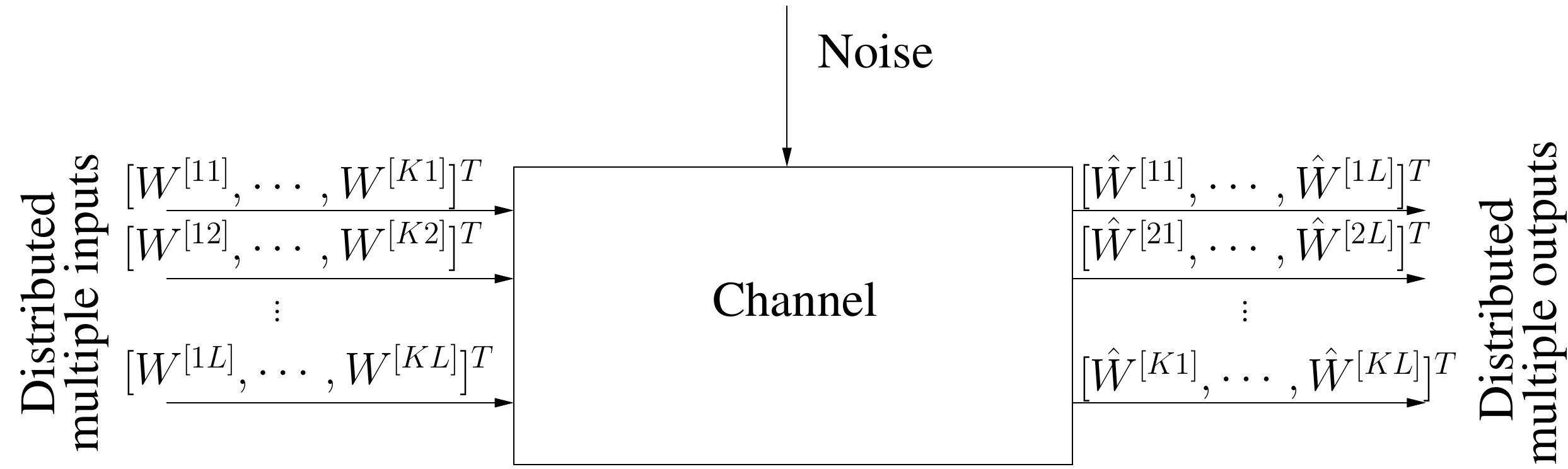}
\caption[Schematic of the MIMO X channel.]{MIMO \textit{X} channel, where $W^{[kl]}$ represents the message that originates at transmitter $l$ and is intended for receiver $k$, while $\hat W^{[kl]}$ denotes the recovered version of $W^{[kl]}$, $k = 1, \cdots, K$, $l = 1, \cdots, L$. } \label{fig:X_MIMO}
\end{figure}
Finally, when both the transmitters and the receivers are geographically distributed, the MIMO channel turns into either an \textit{interference channel}\cite{Carleial_1975:earliest_interference_cancellation_idea, Han_1981:interference_channel_rate_region, Sato_1981:interference_channel_strong_interference, Costa_1985:Gaussian_interference_channel, Costa_1987:capacity_interference_channel, Sason_2004:rate_regions_Gaussian_interference_channel, Kramer_2006:review_interference_channel_rate_regions, Jafar_2007:DoF_MIMO_interference_channel, Etkin_2008:interference_channel_with_one_bit} or an \textit{X channel}\cite{Vishwanath_2003:Z_channel, Ali_2008:interference_alignment_journal, Jafar_2008:DoF_MIMO_X_channel, Cadambe2009:IA_for_Wirelss_X_networks}, which are shown in Fig. \ref{fig:IC_MIMO} and Fig. \ref{fig:X_MIMO}, respectively.  An interference channel characterizes a situation where each transmitter, potentially equipped with multiple antennas, only wants to communicate with its dedicated receiver, and each receiver, possibly equipped with multiple antennas as well, only cares about the information arriving from the corresponding transmitter. There is a \textit{strict one-to-one correspondence} between the multiple transmitters and the multiple receivers. Therefore, each transmission link interferes with the others. By comparison, in the MIMO \textit{X} channel relying on $L$ transmitters and $K$ receivers, each transmitter has an independent message for each receiver. Hence, there are a total of $KL$ independent messages to deliver. The MIMO $X$ channel is a more generalized model, which encompasses the MIMO multiple access channel of Fig. \ref{fig:MA_MIMO}, the MIMO broadcast channel of Fig. \ref{fig:BC_MIMO} and the MIMO interference channel of Fig. \ref{fig:IC_MIMO} as its special cases. Despite their difference, the MIMO interference channel and $X$ channel share a key common feature, namely they both have distributed transmitters and receivers. The \textit{distributed nature} of transmitters and receivers makes the signal processing required for mitigating the detrimental effects of the MIMO interference channel and $X$ channel far more challenging compared to the single-user MIMO channel. In fact, the capacity analysis and the signal processing techniques for MIMO interference channel and $X$ channel still constitute a largely open field, and most of existing efforts have aimed for \textit{transforming} the MIMO interference channel and $X$ channel so that \textit{cooperation} at the transmitter/receiver side can be exploited to some degree, at least in some specific scenarios. For example, in multicell systems, \textit{BS cooperation}\cite{Grant:one_dimensional_Wyner_model, Zhang:JT_and_BS_selection, Zhang:CCI_mitigation_downlink, Zhang:CCI_Asynchronous, Hadisusanto:BD_and_dual_decomposition, Mayer:Turbo_BS_cooperation_interference_cancellation, Khattak:distributed_max_log_MAP, Aktas:Belief_Propagation_2D_Wyner_model, Ng:BS_cooperation_downlink_beamforming, Shaoshi_2011:BS_cooperation_DPDA_conf, Shaoshi2011:DPDA, Zakhour_2012:BS_cooperation_downlink_large_system}, also known as \textit{joint multicell processing}\cite{Zhang:CCI_mitigation_downlink, Gesbert_2010:multicell_MIMO}, has been advocated for the sake of transforming the MIMO interference channel and $X$ channel to a number of cooperative multiuser MIMO channels. Additionally, the recent advances in the capacity analysis of the MIMO interference channel and $X$ channel have stimulated significant interests in \textit{interference alignment}\cite{Ali_2006:first_interference_alignment, Ali_2006:interference_alignment_report_part_1, Ali_2006:interference_alignment_report_part_2, Ali_2008:interference_alignment_journal, Jafar_2007:DoF_MIMO_X_channel_conf, Jafar_2008:DoF_MIMO_X_channel, Cadambe_Jafar_2008:interference_alignment_journal, Jafar_2008:DoF_MIMO_X_channel, Cadambe2009:IA_for_Wirelss_X_networks,  Jafar_2010:interference_alignment_tutorial}, which is essentially constituted by a family of precoding/beamforming techniques for the MIMO interference channel and $X$ channel. The problems related to interference alignment are also beyond the scope of this paper and will not be discussed in detail.

\section{Formal Definition of the MIMO Detection Problem}
\label{sec:definition_MIMO_detection:chap_intro}
Despite the fact that similar problems have been known for a while\cite{Shnidman1967:earliest_ISI_cross_talk_equivalent, Kaye_1970:transmit_multiplexed_PAM_over_Multi_channel, Etten_1975:optimum_linear_receiver_for_multi_channel, Etten_1976:ML_receiver_for_multi_channel,Horwood_1975:signal_design_multiple_access, Schneider_1979:linear_ZF_CDMA, Schneider_1980:crosss_talk_resistant_receiver, Timor_1980:improved_decoding_CDMA, Timor_1981:multistage_decoding_CDMA, Verdu_1983:earliest_optimal_MUD_conf_ISIT, Verdu_1983:earliest_optimal_MUD_conf_milcom, Verdu_1986:optimal_MUD_asynchronous_CDMA, Verdu_1986:optimum_MUD_asymptotic_efficiency, Verdu_1989:complexity_optimal_MUD, Lupas_1989:linear_MUD_synchrohous_CDMA, Lupas_1990:near_far_MUD_asynchronous, Kohno_1983:PIC_CDMA, Kohno_1990:IC_CDMA, Kohno_1990:PIC_CDMA_journal, Varanasi_1990:multistage_detection_asynchronous_CDMA, Varanasi_1991:multistage_MUD_synchronous_CDMA, Varanasi_1991:noncoherent_MUD,  Yoon_1993:PIC_CDMA_journal, Divsalar_1998:PIC_MUD_CDMA, Buehrer_1996:adaptive_multistage_IC_CDMA, Masamura_1988:earliest_interference_cancellation_SSMA, Viterbi_1990:earliest_interference_cancellation_approach_capacity, Xie_1990:sequential_MUD_for_async_CDMA, Xie_1990:Linear_MMSE_WLS_MUD, Xie_1993:joint_signal_detection_estimation, Hallen_1993:decorrelating_DFD_synch_CDMA, Hallen:1995:DFD_asynchronous_CDMA, Patel_1994:SIC_CDMA, Varanasi:DFD, Hui_1998:SIC_asynch_CDMA, Verdu:MUD_book, Honig:advances_MUD_edited, Verdu_1999:spectral_efficiency_CDMA, Wang_2009:Wireless_advanced_reception}, the term ``MIMO detection'' became widespread mainly with the advent of multiple-antenna techniques during the mid-1990s\footnote{Note that the first multi-antenna based MIMO system was attributed to a patent granted to Paulraj and Kailath in 1994\cite{Paulraj1994:MIMO}. Gerlach\cite{Gerlach1995:SDMA_thesis}, Roy\cite{Roy1996:SDMA, Roy1997:SDMA} and Ottersten\cite{Roy1996:SDMA, Ottersten1996:SDMA} initiated the earliest research on SDMA systems. The earliest contribution demonstrating the huge capacity of multi-antenna based MIMO systems may be attributed to Telatar\cite{Telatar1995:MIMO_capacity_report,Telatar_1999:MIMO_capacity}, as well as Foschini\cite{Foschini:MIMO, Foschini:MIMO_capacity} and Gans\cite{Foschini:MIMO_capacity}, followed by other members of the team at Bell Labs\cite{Wolniansky:VBLAST, Golden_1999:VBLAST_first_journal}. On the other hand, Tarokh, Jafarkhani, Calderbank, Naguib \textit{et al.}\cite{Tarokh:Space-Time_code, Tarokh:STBC, Naguib_1998:STC_modem, Tarokh_1999:STBC_performance_results, Tarokh_1999:STC_criteria_CE, Jafarkhani_2001:quasi_orthogonal_STBC, Jafarkhani_2003:super_orthogonal_STTC} as well as Alamouti\cite{Alamouti_1998:Alamouti_code} are pioneers of space-time code design. }\cite{Paulraj1994:MIMO, Telatar1995:MIMO_capacity_report, Roy1996:SDMA, Gerlach1995:SDMA_thesis, Ottersten1996:SDMA, Roy1997:SDMA, Lotter1998:SDMA, Wolniansky:VBLAST, Golden_1999:VBLAST_first_journal, Foschini:MIMO, Foschini:MIMO_capacity, Telatar_1999:MIMO_capacity, Tarokh:Space-Time_code, Tarokh:STBC, Alamouti_1998:Alamouti_code, Naguib_1998:STC_modem, Tarokh_1999:STBC_performance_results, Tarokh_1999:STC_criteria_CE, Jafarkhani_2001:quasi_orthogonal_STBC, Jafarkhani_2003:super_orthogonal_STTC}. As a result, in the narrow sense, MIMO detection usually refers to the symbol detection problem encountered in narrow-band SDM based multiple-antenna systems, such as the vertical Bell Laboratories layered space-time (\gls{VBLAST}) system\cite{Foschini:MIMO, Wolniansky:VBLAST, Golden_1999:VBLAST_first_journal}. However, we emphasize that as a family of important signal processing techniques, MIMO detection should be interpreted based on a generic mathematical model, as detailed below.

In the generic sense, the MIMO detection problem can be defined for an $N_I$-input linear system whose transfer function is described by a matrix having \textit{non-orthogonal columns} and its $N_O$ outputs are contaminated by additive random noise. Note that the noise does not necessarily obey the Gaussian distribution. The multiple inputs can be denoted as a vector $\bf s$, which is randomly drawn from the set $\mathbb{A}^{N_I}$ composed by $N_I$-element vectors, whose components are from a finite set $\mathbb{A} = \{ a_m| m = 1, \cdots, M \}$ and the \textit{a priori} probability of selecting each vector from $\mathbb{A}^{N_I}$ is identical. The set $\mathbb{A}$ is usually referred to as the constellation alphabet, whose elements can take either real or complex values. Additionally, $\underline{\bf s}_n$, $n = 1, \cdots, M^{N_I}$, represents the realizations of $\bf s$, hence they are the elements of $\mathbb{A}^{N_I}$. Then the relationship between the inputs and the outputs of this linear system can be characterized by\footnote{In the additive random noise contaminated ISI channels, the received signal is given by $y_n = \sum_{i=0}^{M_h}h_is_{n-i} + w_n$, $n=0, 1, \cdots, L-1$, or by ${\bf y} = {\bf Hs} +  {\bf w}$, where ${\bf y} = [y_0, y_1, \cdots, y_{L-1}]^T$, ${\bf w} = [w_0, w_1, \cdots, w_{L-1}]^T$, $L$ is the number of symbols observed, and $M_h$ is the memory length of the ISI channel.}
\begin{equation}\label{eq:general_MIMO_model_matrix_form}
{\bf{y}} = {\bf{Hs}} + {\bf{n}},
\end{equation}
where ${\bf y} \in \mathbb{F}^{N_O}$ is the received signal vector, ${\bf H}\in \mathbb{F}^{N_O\times N_I}$ is the transfer function/channel matrix of the system, and ${\bf n}\in \mathbb{F}^{N_O}$ represents the additive random noise vector. Depending on the specific applications considered, $\mathbb{F}$ can be either the field of real numbers, $\mathbb{R}$, or the field of complex numbers, $\mathbb{C}$. Concisely speaking, any system having multiple inputs, multiple outputs and subject to additive random noise can be regarded as a MIMO system, but the MIMO detection problem considered herein is only confronted in MIMO systems whose channel matrix is non-orthogonal in columns. It is worth noting that the constellation alphabet $\mathbb{A}$, the number of inputs $N_I$ and the number of outputs $N_O$ are typically regarded as constant quantities\footnote{However, in an adaptive system both the constellation alphabet $\mathbb{A}$ and the number of inputs $N_I$ might be varying. But this adaptation is typically constrained by a discrete size-limited codebook.} for a given system. Hence, they are assumed to be known by default, although this will not be explicitly emphasized, unless necessary. As a further note, when the input symbol vectors of multiple consecutive time slots are associated with each other via space-time coding\cite{Tarokh:Space-Time_code, Naguib_1998:STC_modem, Alamouti_1998:Alamouti_code, Tarokh:STBC, Tarokh_1999:STBC_performance_results, Tarokh_1999:STC_criteria_CE, Jafarkhani_2001:quasi_orthogonal_STBC, Jafarkhani_2003:super_orthogonal_STTC, Jafarkhani2005:STC_book}, the MIMO system model is given by 
\begin{equation}\label{eq:MIMO_model_matrix_form}
{\bf{Y}} = {\bf{HC}} + {\bf{N}},
\end{equation}                                                                                                                                                                                                                                                                                                                                                                                                                                                                                                                                                                                                                                                                                                                                                                                                                                                                                                                                                                                                                                                                                                                                                                                                                                                         
where $\bf Y$ is a matrix denoting the signal received in multiple time slots, $\bf C$ is a matrix representing the space-time codeword, and $\bf N$ is the corresponding noise matrix. We can obtain (\ref{eq:general_MIMO_model_matrix_form}) from (\ref{eq:MIMO_model_matrix_form}) by setting the number of time slots considered to one. In this regard, (\ref{eq:MIMO_model_matrix_form}) is more general than (\ref{eq:general_MIMO_model_matrix_form}). However, (\ref{eq:MIMO_model_matrix_form}) is mainly used for characterizing space-time coding aided MIMO systems, where typically the MIMO detection problem defined in this paper does not exist. This is because the optimal ML decoding can be simply implemented using the separate symbol-by-symbol decoding strategy [for orthogonal space-time block codes (\glspl{STBC})] or the pairwise decoding strategy (for quasi-orthogonal STBCs)\cite{Jafarkhani2005:STC_book}. Therefore, in most cases associated with MIMO detection, we rely on the system model (\ref{eq:general_MIMO_model_matrix_form}).      
                                                                                                                                                                                                                                                                                                                                                                                                                                                                                                                                                                                                                                                                                                                                                                                                                                                                                                                                                                                                                                                                                                                                                                                                                 
Based on the generic mathematical model of (\ref{eq:general_MIMO_model_matrix_form}), \textit{the basic task of MIMO detection is to estimate the input vector $\bf s$ relying on the knowledge of the received signal vector $\bf y$ and the channel matrix $\bf H$.} Note that for $\bf y$, typically its exact value has to be known, while for $\bf H$, sometimes only the knowledge of its statistical parameters is available. To elaborate a little further, if the \textit{instantaneous} value of $\bf H$ is known from \textit{explicit} channel estimation, the detection of $\bf s$ is said to rely on \textit{coherent} detection. By contrast, if the explicit estimation of the instantaneous channel state is avoided, the detection of $\bf s$ belongs to the family of \textit{noncoherent} detection schemes. In the latter case, the channel estimation is either performed implicitly in signal detection, or it is completely avoided, whereas typically the statistical knowledge of the channel matrix $\bf H$ is invoked for supporting signal detection. Additionally, the noncoherent MIMO detection schemes usually require that the input symbols are subject to some form of \textit{differential encoding}, which imposes correlation on the input symbols, and as a result, typically a \textit{block-by-block based sequence detection} has to be employed. This is the so-called multiple-symbol differential detection\cite{Divsalar_1990:Multiple_symbol_differential_detection, Tarokh_2000:differential_decoding, Jafarkhani_2001:MSDD_STC, Hughes_2000:differential_ST, Warrier_2002:spectrally_efficient_noncoherent, Schober_2002:noncoherent_receiver_differential_STC, Hassibi_2003:how_much_training_needed, Lampe_2005:MSDSD, Jafarkhani2005:STC_book, hanzo2010:mimo_coherent_vs_noncoherent}, which usually leads to higher computational complexity than the symbol-by-symbol based detectors of coherent MIMO systems. Moreover, the noncoherent detectors typically exhibit degraded power efficiency, which results in an inherent performance loss compared to their coherent counterparts, unless the block size is sufficiently large. Therefore, we have to consider the performance-versus-complexity tradeoff in choosing the proper block size. However, similar to the coherent detection of STBCs, there exist simple symbol-by-symbol or pairwise noncoherent detection schemes\cite{Tarokh_2000:differential_decoding, Jafarkhani_2001:MSDD_STC} for differential space-time modulation. As a result, the decoding complexity of the differential space-time modulation increases linearly, instead of exponentially, with the number of antennas\cite{Jafarkhani2005:STC_book}. In this paper, we focus our attention on coherent MIMO detection. Then, from the perspective of mathematical mapping, a coherent MIMO detector is defined as:
\begin{equation}\label{eq:MIMO_detector_coherent}
\hat{{\bf s}} = \mathfrak{D}({\bf y}, {\bf H}): \mathbb{F}^{N_O} \times \mathbb{F}^{N_O\times N_I} \mapsto \mathbb{A}^{N_I},
\end{equation}
where $\hat{{\bf s}}$ is the estimate of $\bf s$.

\section{MIMO System Model for Linear Memoryless Channels}
\label{sec:MIMO_model_memoryless:chap_intro}
\begin{figure}[t]
\centering
\includegraphics[width=3.5in]{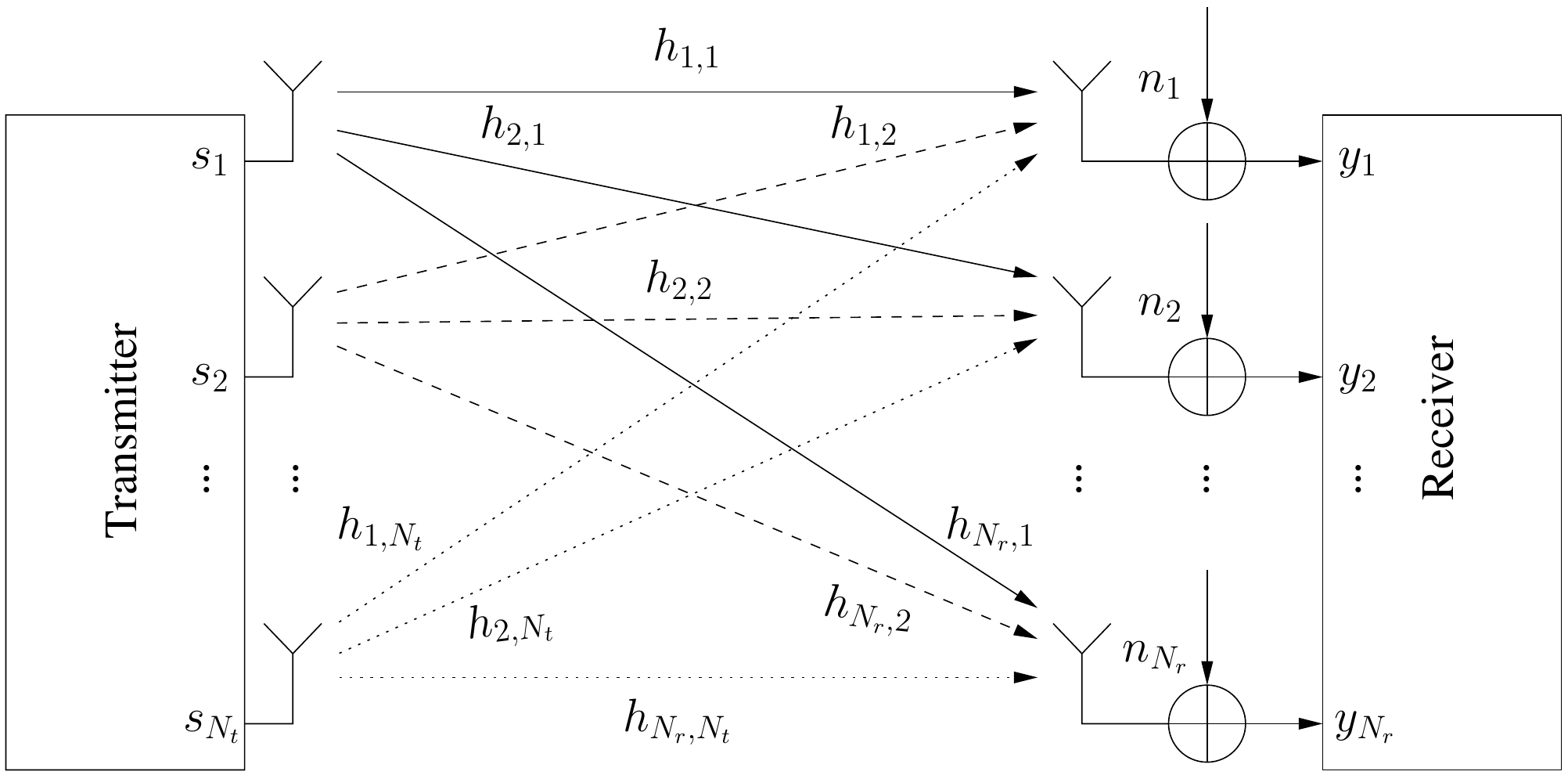}
\caption{Schematic of VBLAST-style SDM-MIMO systems communicating over flat fading channels. }
\label{fig:VBLAST}
\end{figure}
Bearing in mind specific applications, the system model of (\ref{eq:general_MIMO_model_matrix_form}) may be established either in the time domain or in the frequency domain, and may be applied to both memoryless channels and dispersive channels exhibiting memory\cite{Verdu:MUD_book, Wang_2009:Wireless_advanced_reception, Honig:advances_MUD_edited,  Paulraj_2003:introduction_to_MIMO, Tse:Fundamental, Goldsmith_2005:Wireless_Comm}. With respect to linear memoryless MIMO channel, a canonical example is the narrowband single-carrier synchronous VBLAST-style SDM-MIMO system\cite{Foschini:MIMO, Wolniansky:VBLAST, Golden_1999:VBLAST_first_journal} communicating over flat fading channels, as shown in Fig. \ref{fig:VBLAST}. Because the system's outputs at the current time interval are independent of the system's inputs at previous time intervals, its baseband equivalent discrete-time (i.e. sampled) system model, representing an instance of the generic model (\ref{eq:general_MIMO_model_matrix_form}), can be written as
\begin{align}\label{eq:memoriless_MIMO_model_scalar_form}
\left[ {\begin{array}{*{20}{c}}
{{y_1}}\\
{{y_2}}\\
 \vdots \\
{{y_{{N_r}}}}
\end{array}} \right] & =  \left[ {\begin{array}{*{20}{c}}
{{h_{1,1}}}&{{h_{1,2}}}& \cdots &{{h_{1,{N_t}}}}\\
{{h_{2,1}}}&{{h_{2,2}}}& \cdots &{{h_{2,{N_t}}}}\\
 \vdots & \vdots & \ddots & \vdots \\
{{h_{{N_r},1}}}&{{h_{{N_r},2}}}& \cdots &{{h_{{N_r},{N_t}}}}
\end{array}} \right]\left[ {\begin{array}{*{20}{c}}
{{s_1}}\\
{{s_2}}\\
 \vdots \\
{{s_{{N_t}}}}
\end{array}} \right] \\ \nonumber 
& + \left[ {\begin{array}{*{20}{c}}
{{n_1}}\\
{{n_2}}\\
 \vdots \\
{{n_{{N_r}}}}
\end{array}} \right]. 
\end{align}
In this specific application, we have $N_t = N_I$ and $N_r = N_O$, which represent the number of transmit and receive antennas, respectively. Furthermore,  $h_{j,i}$ denotes the (complex-valued) impulse response between the $i$th transmit antenna and the $j$th receive antenna, with $i = 1, 2, \cdots, N_t$ and $j = 1, 2, \cdots, N_r$. Another example is the multiple-antenna aided orthogonal frequency-division multiplexing (\gls{OFDM}) system\cite{hanzo2010:mimo_coherent_vs_noncoherent} communicating over frequency-selective channels, where each subcarrier subjected to a specific frequency-domain attenuation is narrowband and the model (\ref{eq:memoriless_MIMO_model_scalar_form}) applies to each subcarrier. Notably, for this linear memoryless MIMO channel, the \textit{one-shot detection} which relies only on a single received signal vector ${\bf y} = [y_1, y_2, \cdots, y_{N_r}]^T$ is adequate. Additionally, for the sake of fair comparison with the single-input single-output systems, typically the energy normalization of $\mathcal{E}(s_i) = 1$ or $\mathcal{E}({\bf s}) = 1$ is imposed on the transmitted symbols.

\section{MIMO System Model for Dispersive Channels Exhibiting Memory}
\label{sec:MIMO_model_memory:chap_intro}
On the other hand, when considering the stand-alone wideband VBLAST system\footnote{Multicarrier techniques, such as OFDM, are not used in this system. However, when MIMO-OFDM systems\cite{Stuber_2004:MIMO_OFDM} are considered, the MIMO detection is carried out on each subcarrier separately.} communicating over frequency-selective MIMO channels\cite{Gamal_2003:ST_SF_MIMO_fre_selec, Abe_2003:ST_Turbo_equalization_fre_selec_MIMO, Zhu_2004:SF_equalization_MIMO_fre_selet, Liu:CPDA-apx, Ma_2005:optimum_training_MIMO_fre_selec}, the link between each input-output pair may be modelled by a linear finite impulse response (\gls{FIR}) dispersive channel, whose sampled version can be denoted as the (possibly complex-valued) vector ${\bf h}_{j,i} = (h_{j,i}^0, h_{j,i}^1, \cdots, h_{j,i}^{L-1})^T$. Here $L$ is the maximum number of multipath components in each link, and it is also known as the channel memory length. In this case, the \textit{one-shot detection} which utilizes a single $N_r$-element received signal vector is not optimal. Instead, the \textit{sequence detection} using multiple $N_r$-element received vectors has to be used. 

We assume that a block-based transmission structure relying on zero-padding for eliminating the interblock interference is used, which is beneficial for alleviating the performance degradation imposed by noise enhancement or error propagation\cite{Wang_2000:multicarrier_fourier_meets_shannon}. Following zero-padding, a transmission block becomes a frame which occupies $K = N + P$ sampling intervals, where $N$ is the number of sampling intervals occupied by information-bearing symbol vectors in the frame, while $P \ge L -1$ represents the number of sampling intervals during which $P$ consecutive $N_t$-element zero vectors are inserted at the tail of the frame. Here we set $P = L-1$. Given the above-mentioned transmitted frame, the entire received signal vector may be generated by a concatenation of $K$ noise-contaminated sampled received signal vectors, namely, ${\bf y} = ({\bf y}^T[0], {\bf y}^T[1], \cdots, {\bf y}^T[K-1])^T$, where ${\bf y}[k] = (y_1[k], y_2[k], \cdots, y_{N_r}[k])^T$ represents the $N_r$ outputs at the $k$th sampling instant, $k =0, 1, \cdots, K-1$. Then, the baseband signal received by the $j$th receive antenna at the $k$th sampling instant is given by
\begin{equation}\label{eq:jth_receive_antenna}
{y_j}\left[ k \right] = \sum\limits_{i = 1}^{{N_t}} {\sum\limits_{l = 0}^{L-1} {h_{j,i}^l{s_i}\left[ {k - l} \right] + {n_j}\left[ k \right]} },
\end{equation}
where $h_{j,i}^l$, the $l$th element of ${\bf h}_{j,i} $, $l = 0, 1, \cdots, L-1$, denotes the channel gain of the $l$th path between the $i$th transmit antenna and the $j$th receive antenna. Furthermore, $s_i[k]$ is the symbol transmitted from the $i$th transmit antenna at the $k$th sampling instant, and $n_j[k]$ represents the noise imposed on the $j$th receive antenna at the $k$th sampling instant. Similar to ${\bf y}[k]$, we define ${\bf s}[k] = (s_1[k],s_2[k],\cdots,s_{N_t}[k])^T$ and ${\bf n}[k] = (n_1[k],n_2[k],\cdots,n_{N_r}[k])^T$. Additionally, similar to $\bf y$, we may construct ${\bf s} = ({\bf s}^T[0], {\bf s}^T[1], \cdots, {\bf s}^T[N-1])^T$ and ${\bf n} = ({\bf n}^T[0], {\bf n}^T[1], \cdots, {\bf n}^T[K-1])^T$. Then, the received signal corresponding to a transmitted frame can also be written following the matrix notation of (\ref{eq:general_MIMO_model_matrix_form}), where the size of $\bf y$ and $\bf n$ is $N_O = KN_r = (N+L-1)N_r$, while that of $\bf s$ is $N_I = NN_t$, and the MIMO channel matrix $\bf H$ exhibits the banded Toeplitz structure\cite{Wang_2000:multicarrier_fourier_meets_shannon} of:
\begin{equation}\label{eq:wideband_MIMO_channel_matrix}
{\bf H} =  \left[ {\begin{array}{*{20}{c}}
{{{\bf{H}}^0}}&{\bf{0}}& \cdots &{\bf{0}}\\
 \vdots &{{{\bf{H}}^0}}& \ddots & \vdots \\
{{{\bf{H}}^{L - 1}}}& \vdots & \ddots &{\bf{0}}\\
{\bf{0}}&{{{\bf{H}}^{L - 1}}}& \ddots &{{{\bf{H}}^0}}\\
 \vdots & \vdots & \ddots & \vdots \\
{\bf{0}}&{\bf{0}}& \cdots &{{{\bf{H}}^{L - 1}}}
\end{array}} \right],
\end{equation}
whose dimension is $(KN_r \times NN_t)$, and each entry ${\bf H}^l$ of (\ref{eq:wideband_MIMO_channel_matrix}) is an $(N_r \times N_t)$-element matrix containing the channel gains between all pairs of transmit and receive antennas for the $l$th path, i.e. we have
\begin{equation}\label{eq:single_path_channel_matrix}
{\bf H}^l =   \left[ {\begin{array}{*{20}{c}}
{{h^l_{1,1}}} & {{h^l_{1,2}}} & \cdots  & {{h^l_{1,{N_t}}}}\\
{{h^l_{2,1}}} & {{h^l_{2,2}}} & \cdots & {{h^l_{2,{N_t}}}}\\
 \vdots & \vdots & \cdots & \vdots \\
{{h^l_{{N_r},1}}} & {{h^l_{{N_r},2}}} & \cdots & {{h^l_{{N_r},{N_t}}}}
\end{array}} \right].
\end{equation}

It is worth noting that the linear memoryless MIMO model and the linear MIMO model exhibiting memory may also be used for characterizing the family of \textit{synchronous} and \textit{asynchronous} CDMA\index{CDMA} systems, respectively. Additionally, the asynchronous CDMA\index{CDMA} systems can also be characterized by (\ref{eq:general_MIMO_model_matrix_form}) in the $z$ domain\cite{Verdu:MUD_book}.

\section{Complex-Valued versus Real-Valued MIMO System Model}
\label{sec:complex_vs_real_model}
As we mentioned in Section \ref{sec:definition_MIMO_detection:chap_intro}, the generic MIMO system model of (\ref{eq:general_MIMO_model_matrix_form}) can be defined both in the field of real numbers, $\mathbb{R}$, and in the field of complex numbers, $\mathbb{C}$. Since the complex-valued modulation constellations, such as quadrature amplitude modulation (\gls{QAM}) and phase-shift keying (\gls{PSK}), are often employed in digital communications, the complex-valued MIMO system model is typically a natural and more concise choice for the formulation and performance analysis of the algorithms considered. 

The complex-valued and the real-valued MIMO system models are often mutually convertible. More specifically, if we assume that the generic MIMO system model of (\ref{eq:general_MIMO_model_matrix_form}) is defined in $\mathbb{C}$, and assume that the real part and the imaginary part of $\bf s$ are uncorrelated,\footnote{For rectangular QAM, this uncorrelatedness assumption is almost always adopted for ease of decoding, even in single-input single-output systems. In channel-coded systems, where the channel codes may introduce correlation between the coded bits. However, in such systems, typically an interleaver is invoked after the encoder, which mitigates the correlation.} then the complex-valued MIMO system model of (\ref{eq:general_MIMO_model_matrix_form}) can be transformed to an equivalent real-valued system model of
\begin{equation}\label{eq:real_valued_MIMO_model}
\tilde{\bf y} = \tilde {\bf H}\tilde{\bf s} + \tilde{\bf n}, 
\end{equation} 
where ${\tilde {\bf y}} = \left[ {\begin{array}{*{20}{c}}
\Re({\bf y})\\
\Im({\bf y})
\end{array}} \right] $, ${\tilde {\bf s}} = \left[ {\begin{array}{*{20}{c}}
\Re({\bf s})\\
\Im({\bf s})
\end{array}} \right] $, ${\tilde {\bf n}} = \left[ {\begin{array}{*{20}{c}}
\Re({\bf n})\\
\Im({\bf n})
\end{array}} \right] $, and ${\tilde {\bf H}} = \left[ {\begin{array}{*{20}{c}}
\Re({\bf H}) & -\Im({\bf H})\\
\Im({\bf H}) & \Re({\bf H}) 
\end{array}} \right] $.

However, the above real-valued decomposition is only applicable to MIMO systems employing real-valued constellations or rectangular QAM constellations (but not for PSK, star QAM\cite{Webb_1991:star_QAM, Webb_1995:star_QAM, Dong_1998:star_QAM_fre_sel, Dong_1999:star_QAM} and near-Gaussian constellations\cite{Forney_1998:Gaussian_constellation, Nevat_2010:MIMO_detect_Gaussian_constellation} etc.), which severely limits its applicability. Furthermore, in many applications, complex-valued operations are more efficient for hardware implementation. The reason for this fact is twofold. Firstly, decomposing an $(N_t \times N_r)$-element MIMO system into a real-valued system requires storage of the $(2N_r \times 2N_t)$-element real-valued channel matrix $\tilde {\bf H}$, which is twice larger than having $2N_tN_r$ real-valued elements that would be needed for a standard representation of the original complex-valued channel matrix $\bf H$. Secondly, implementing complex-valued arithmetics in hardware (e.g. very-large-scale integration (\gls{VLSI}) based circuits) is straightforward and does not result in more complex hardware. For example, a complex-valued multiplier can be built using 4 real multipliers and 2 real adders, because we have $(a+jb)(c+jd) = ac -bd + j(ad+bc)$, or using 3 real multiplications and 5 real additions, because alternatively we have $(a+jb)(c+jd) = ac -bd + j[(a+b)(c+d)-ac-bd]$, which is known as the Gaussian technique of multiplying complex numbers\cite{Knuth_1997:art_of_programming}. As a result, the complex-valued model imposes a lower silicon complexity than that required by the real-valued decomposition based model. Therefore, in many cases the real-valued decomposition can be detrimental and hence MIMO detector designers are typically in favor of the complex-valued system model, owing to its flexibility concerning the choice of constellations and its efficiency in VLSI implementations.

On the other hand, the real-valued MIMO system model may also enjoy some advantages, such as the increased \textit{freedom of manipulation} in signal processing. To elaborate a little further, as far as the achievable performance is concerned, in most cases signal processing algorithms based on the complex-valued model of (\ref{eq:general_MIMO_model_matrix_form}) and the real-valued MIMO system model of (\ref{eq:real_valued_MIMO_model}) deliver an equivalent performance. For example, \cite{Telatar_1999:MIMO_capacity} showed the equivalence between the complex-valued and the real-valued MIMO system models in the derivation of the optimal ML detector and the MIMO channel capacity.  However, this equivalence does not always hold. For example, it was shown in \cite{Fischer_2003:real_versus_complex_VBLAST} that the real-valued VBLAST detector outperforms its complex-valued counterpart, owing to its additional freedom in selecting the optimum detection ordering. Hence, a beneficial performance gain may be gleaned from transforming the complex-valued system model to the double-dimensional real-valued system model. This is also true for the tree-search based MIMO detectors, which will be introduced in Section \ref{subsec:tree_search_MUD:chap_intro}, when they invoke symbol detection ordering. More generally, the key insight inferred here is that for all MIMO detection algorithms whose performance is related to detection ordering, the real-valued system model based formulation is capable of providing a better performance than its complex-valued counterpart. This gain is achieved at the expense of extra redundancy in storage of the channel matrix $\tilde {\bf H}$, and if the symbol detection ordering technique is invoked, this redundancy cannot be avoided.

Additionally, it is worth noting that the real-valued formulation of the complex-valued MIMO system model is \textit{not} unique. For example, a pairwise real-valued MIMO system model was used in\cite{Siti_2006:pairwise_real_model, Azzam_2009:pairwise_real_model_SD}, which was shown to result in a reduced complexity compared to the conventional real-valued MIMO system model of (\ref{eq:real_valued_MIMO_model}). A more comprehensive investigation of the complex-valued versus real-valued MIMO detectors was presented in\cite{Liu_2012:real_versus_complex_MIMO_detection}.  

Finally, we emphasize that if the complex-valued random signals considered are \textit{improper} or \textit{noncircular}, the more advanced complex-valued signal processing techniques of\cite{Neeser:proper_Gaussian_process, Adali2011:complex_valued_SP, Mandic_2009:complex_valued_signal_processing} that rely on additional statistics and tools for fully characterizing the complex-valued random signals have to be used. More specifically, for a complex-valued random vector $\bf x$, in addition to the conventional covariance matrix of  \begin{equation}
\textrm{C}_{xx} = \mathcal{E}[({\bf x}-{\boldsymbol \mu}_x)({\bf x}-{\boldsymbol \mu}_x)^H],
\end{equation}
another second-order statistic, namely the pseudo-covariance matrix defined as 
\begin{equation}
\bar{\textrm{C}}_{xx} = \mathcal{E}[({\bf x}-{\boldsymbol \mu}_x)({\bf x}-{\boldsymbol \mu}_x)^T],
\end{equation} 
has to be introduced for fully describing the complex-valued random vector. For a proper complex-valued random vector, the pseudo-covariance matrix vanishes, which is formulated as $\bar{\textrm{C}}_{xx} = {\bf 0}$. This results in the fact that for a proper complex-valued random scalar, the real and imaginary parts must have the same variance and be uncorrelated. Additionally, a circularly (symmetric) complex-valued random variable has a probability distribution that is invariant under rotation in the complex plane, namely the distribution of $\bf x$ must be the same as the distribution of $e^{j\theta}{\bf x}$, where we have $\theta \in [0, 2\pi)$. 
The conventional signal processing techniques often assume, usually implicitly, that the complex-valued random signals are proper or circularly symmetric\footnote{For complex-valued Gaussian random vector $\bf Z$, circular symmetry implies that $\bf Z$ is zero mean and proper.}. However, these assumptions may not be justified in some applications, hence the complex-valued signal processing techniques may in certain circumstances achieve a better performance\cite{Jia:CPDA, Shaoshi2011:B_PDA, Shaoshi2011:DPDA, Shaoshi2013:Turbo_AB_Log_PDA, Shaoshi_2013:EB_Log_PDA_journal}. For more details on the complex-valued signal processing, please refer to \cite{Neeser:proper_Gaussian_process, Adali2011:complex_valued_SP, Mandic_2009:complex_valued_signal_processing}.

\section{History and State-of-the-Art of MIMO detection}
\label{sec:history_MIMO_detection:chap_intro}
\begin{figure}[t]
\centering
\includegraphics[width=3.3in]{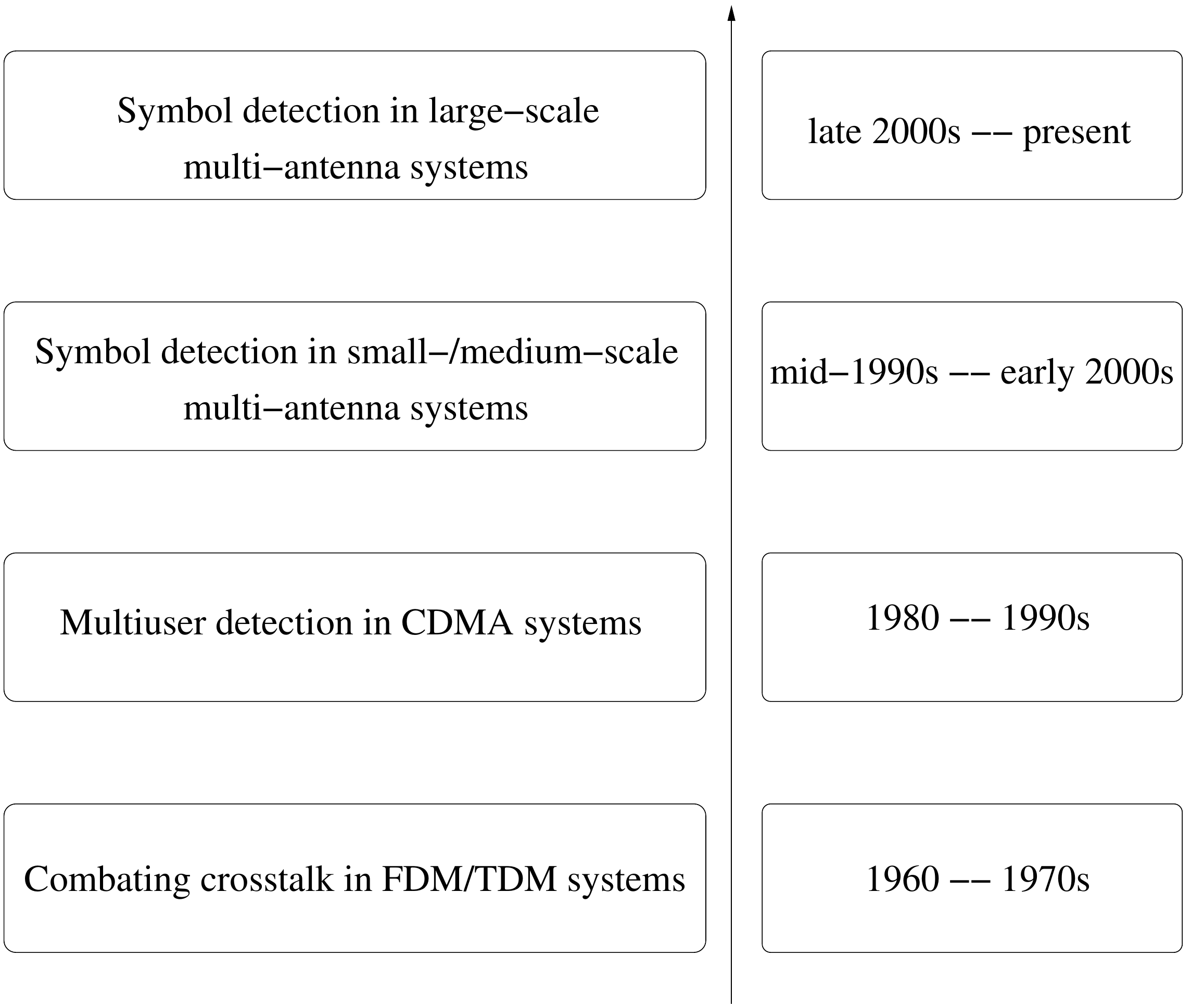}
\caption[The four historical periods in the development of MIMO detection.]{The four historical periods in the development of MIMO detection.} \label{fig:history_MIMO_detect}
\end{figure}
\begin{figure*}[tbp]
\centering
\includegraphics[width=5.1in]{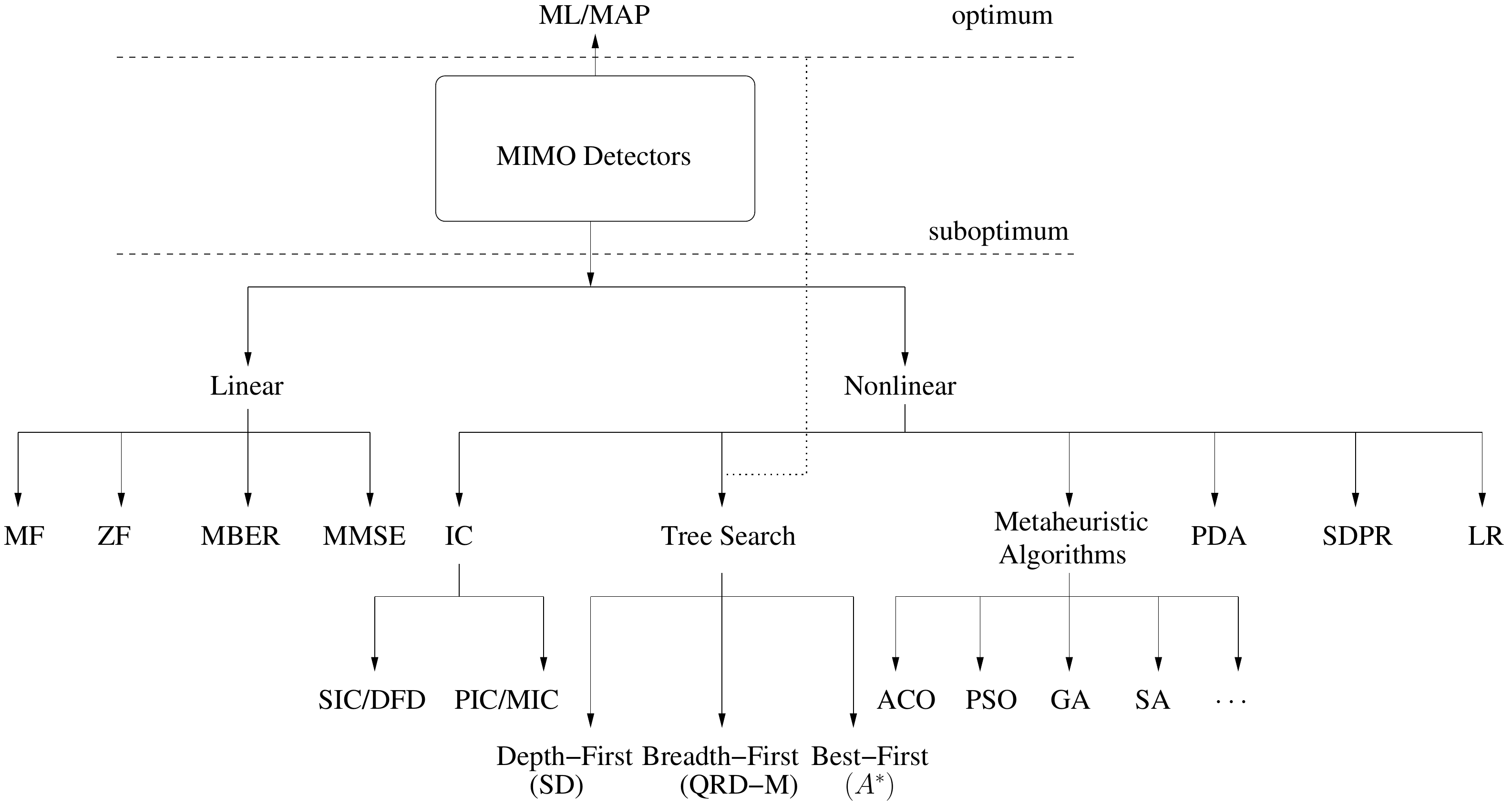}
\caption[Overview of representative MIMO detectors considered in this paper.]{Overview of representative MIMO detectors considered in this paper. All the acronyms used have been defined in our Glossary. Note that the tree-search based detector is in general suboptimum, but it has the flexibility to strike different tradeoffs between the achievable performance and the computational complexity.  Even the optimum ML performance may be achieved by tree-search based detectors in certain scenarios.} \label{fig:MIMO_Detector}
\end{figure*}

The research of MIMO detection is a broad and vibrant area. Its embryonic concept dates back to the 1960s. The earliest contribution on MIMO detection was sparked off in 1967\cite{Shnidman1967:earliest_ISI_cross_talk_equivalent}, when Shnidman considered the \textit{equalization} problem of a bandwidth-limited pulse modulation system. This system was modelled with the aid of $M$ waveforms, each of which is amplitude-scaled and simultaneously transmitted over a single physical channel, which has $M$ outputs corresponding to each signal waveform. In order to eliminate both the ISI between the pulse train and the interference between different waveforms (also known as crosstalk), Shnidman formulated a \textit{generalized Nyquist criterion} and proposed an optimum linear receiver. This landmark contribution was essentially inspired by the classic \textit{Nyquist's problem}\cite{Tufts_1965:Nyquist_problem_ISI}, which aims for the joint optimization of the transmitter and receiver for the sake of combating the ISI when communicating over a conventional single-input single-output channel. Since then, the MIMO detection problem has been studied in the context of diverse applications and under possibly different names. This half-century history can be roughly divided into four periods, as seen in Fig. \ref{fig:history_MIMO_detect}, namely the period of combating crosstalk in the context of the early single-user FDM/TDM\index{FDM}\index{TDM} systems (1960s -- 1970s)\cite{Shnidman1967:earliest_ISI_cross_talk_equivalent, Kaye_1970:transmit_multiplexed_PAM_over_Multi_channel, Etten_1975:optimum_linear_receiver_for_multi_channel, Etten_1976:ML_receiver_for_multi_channel, Schneider_1980:crosss_talk_resistant_receiver}, the period of multiuser detection (\gls{MUD}) during the prevalence of CDM/CDMA\index{CDM}\index{CDMA} systems (1980s -- 1990s)\cite{Horwood_1975:signal_design_multiple_access, Schneider_1979:linear_ZF_CDMA, Timor_1980:improved_decoding_CDMA, Timor_1981:multistage_decoding_CDMA, Verdu_1983:earliest_optimal_MUD_conf_ISIT, Verdu_1983:earliest_optimal_MUD_conf_milcom, Verdu_1986:optimal_MUD_asynchronous_CDMA, Verdu_1986:optimum_MUD_asymptotic_efficiency, Verdu_1989:complexity_optimal_MUD, Lupas_1989:linear_MUD_synchrohous_CDMA, Lupas_1990:near_far_MUD_asynchronous, Kohno_1983:PIC_CDMA, Kohno_1990:IC_CDMA, Kohno_1990:PIC_CDMA_journal, Varanasi_1990:multistage_detection_asynchronous_CDMA, Varanasi_1991:multistage_MUD_synchronous_CDMA, Varanasi_1991:noncoherent_MUD,  Yoon_1993:PIC_CDMA_journal, Divsalar_1998:PIC_MUD_CDMA, Buehrer_1996:adaptive_multistage_IC_CDMA, Masamura_1988:earliest_interference_cancellation_SSMA, Viterbi_1990:earliest_interference_cancellation_approach_capacity, Xie_1990:sequential_MUD_for_async_CDMA, Xie_1990:Linear_MMSE_WLS_MUD, Xie_1993:joint_signal_detection_estimation, Hallen_1993:decorrelating_DFD_synch_CDMA, Hallen:1995:DFD_asynchronous_CDMA, Patel_1994:SIC_CDMA, Varanasi:DFD, Hui_1998:SIC_asynch_CDMA, Verdu:MUD_book,  Verdu_1999:spectral_efficiency_CDMA, Wang_2009:Wireless_advanced_reception, Honig:advances_MUD_edited}, the period of joint symbol detection in the small-/medium-scale multiple-antenna systems (mid-1990s -- mid-2000s)\cite{Foschini:MIMO, Wolniansky:VBLAST, Golden_1999:VBLAST_first_journal, Li_2002:MIMO_OFDM_SIC_detection_CE, Lu_2002:MIMO_OFDM_LDPC_SIC, Zanella_2005:MMSE_SIC_MIMO, Chen_2006:MBER_MIMO_Detection, Palomar_2005:MBER_MIMO_tranceiver, Wubben_2003:MMSE_DSNR_ordering, Wai_2000:IMSE_DSNR_ordering_MMSE_SIC, Bohnke_2003:MMSE_SIC_GSNR, Hassibi_2000:sqrt_algorithm_BLAST_LMSE_ordering, Benesty_2003:MMSE_SIC_LMSE_ordering, Liu_2009:MMSE_SIC_LMSE, Chin_2000:PIC_BLAST, Luo_2008:PIC_MIMO, Studer_2011:PIC_MIMO_ASIC_implementation, Viterbo_1993:SD_conf, Viterbo_1999:SD, Agrell:closest_point_search_in_lattice,Damen_2003:MLD_closest_lattice_point_search,Hassibi_2005:SD_complexity_part_1, Vikalo_2005:SD_complexity_part_2,Burg_2005:VLSI_depth_first_SD, Jalden:SD_complexity_journal, Wong_2002:K-best_SD_VLSI, Guo_2006:implementation_K_best_SD_MIMO, Chen_2007:K-best_VLSI, Wu_2008:early_pruning_SD, Barbero:fixed_complexity_SD_journal,Jalden:FCSD_error_prob, Fukatani_2004:best_first_search, Lee_2007:short_path_SD, Okawado_2008:best_first_search, Stojnic_2008:speed_up_SD_via_infinity_H_norm, Kim_2010:best_first_search, Chang_2012:best_first_search, Chang_2012:best_first_search_A_algorithm, Wang_2009:Wireless_advanced_reception, Luo:PDA_Sync_CDMA, Luo:PDA_thesis, Pham:PDA_Async_CDMA, Luo_2003:sliding_window_PDA, Penghui_2003:iterative_PDA_MUD, Yin_2004:turbo_equalization_PDA, Huang_2004:generalized_PDA, Pham:GPDA, Liu:CPDA-apx, Liu:Kalman_PDA_freq_selective, Latsoudas_2005:hybrid_PDA_SD, Jia_2005:Gaussian_approximation_mixture_reduction_MIMO, Fricke:Impact_of_Gaussian_approximation, Shaoqian2005:turbo_PDA, Penghui_2006:asymptotic_optimum_PDA, Cai2006:iterative_PDA, Jia:CPDA, Cao:Relation_of_PDA_and_MMSE-SIC, Bavarian_2007:distributed_BS_cooperation_uplink, Bavarian_2008:distributed_BS_cooperation_uplink_journal, Jia_2008:multilevel_SGA_PDA, Grossmann_2008:turbo_equalization_PDA, Kim_2008:noncoherent_PDA, Mohammed_2009:PDA_STBC, Bavarian:SDE_PDA_freq_selective, Luo:SDR_simplest, Luo:Convex_Optimization_for_SP_Comm, Penghui:SDP_CDMA, Ma:SDR_CDMA_BPSK,
Ma:SDR_CDMA_QPSK, Luo:SDR_performance_analysis, Jalden:SDR_diversity, Luo:SDP_MPSK, Ma:SDR_MPSK,
Wiesel:PI_SDR_16QAM, Yijin:SDR_16QAM_tight, Sidiropoulos:SDR_HOM, Mobasher:SDR_QAM_journal, Mao:SDR_LMR, Ma:equivalence_SDR}, and the period of symbol detection in the large-scale multiple-antenna systems\cite{Penghui_2006:asymptotic_optimum_PDA, Liang_2006:block_iterative_DFEs_massive_MIMO, Liang_2007:MMSE_large_system_performance, Liang_2008:relation_between_MMSE_SIC_and_BI_GDFE, Chockalingam_2008:HNN_LAS_based_LS_MIMO_detector,  Mohammed_2009:PDA_STBC, Chockalingam_2009:RTS_LS_MIMO_detector, Chockalingam_2009:LAS_non_orthogonal_STBC, Chockalingam_2010:low_complexity_LS_MIMO_detection, Chockalingam_2011:hybrid_RTS_BP_LS_MIMO_detector, Chockalingam_2011:graphical_model_LS_MIMO_detection, Chockalingam_2011:randomized_MCMC_and_search_LS_MIMO_detection, Chockalingam_2013:MC_Sampling_receiver_LS_MIMO, Chockalingam_2013:LR_detection_LS_MIMO, Suthisopapan_2013:capacity_approaching_LDPC_MMSE_detection_LS_MIMO_journal, Chockalingam_2014:Channel_harderning_message_passsing, Larsson_2014:SUMIS_LS_MIMO_detector, Wu_2014:message_passing_soft_iterative_LS_MIMO_detection, Studer_2014:implementation_LS_MIMO_detector, Studer_2014:implementation_LS_MIMO_detector_journal}. Diverse MIMO detectors have been proposed for meeting the requirements imposed by a multiplicity of applications. These MIMO detectors can be categorised from various perspectives, such as optimum/suboptimum, linear/nonlinear, sequential/one-shot, adaptive/non-adaptive, hard-decision/soft-decision, blind/non-blind, iterative/non-iterative, synchronous/asynchronous, coded/uncoded etc. Note that a detailed discourse on the application of soft-decision MIMO detectors in near-capacity turbo/iterative receivers was provided in \cite{Shinya_2012:CST_near_capacity_transceiver}, while coherent and noncoherent MIMO detectors in the context of the emerging ``space-time shift keying (\gls{STSK})'' based multicarrier MIMO systems was presented in \cite{Kadir_2015:MIMO_multicarrier_STSK}. The representative MIMO detectors considered in this paper are summarized in Fig. \ref{fig:MIMO_Detector}.  

Owing to the similarities between the classic equalization problem encountered in channels imposing ISI and the \textit{generic} MIMO detection problem defined by (\ref{eq:general_MIMO_model_matrix_form}) and (\ref{eq:MIMO_detector_coherent}), it is not surprising that the techniques, which were found to be effective in combating ISI were also often extended to the context of MIMO detection problems\cite{Andrews:Interference_Cancellation_overview}. Some of the equalization algorithms which have been adapted for MIMO detection include, but not limited to, the ML sequence estimation (Viterbi algorithm)\cite{Viterbi_1967:viterbi_algorithm, Omura_1969:Viterbi_algorithm_new_interpretation, Forney_1972:MLSE_ISI, Forney_1973:Viterbi_algorithm_tutorial, Viterbi_2006:viterbi_algorithm_history} based equalization, linear ZF equalization\cite{Rappaport_2002:Wireless_textbook}, linear MMSE equalization\cite{Rappaport_2002:Wireless_textbook}, ZF/MMSE aided decision-feedback equalization\cite{Rappaport_2002:Wireless_textbook}, adaptive equalization\cite{Qureshi_1985:adaptive_equalization, honig1995:blind_adaptive_MUD}, blind equalization\cite{Tong_1994:blind_identification_equalization,Wang_1998:blind_equalization_and_MUD_CDMA} etc, as detailed below.

\subsection{Optimum MIMO Detector}
\label{subsec:ML_detector:chap_intro}
The earliest work on \textit{optimum} MIMO detectors dates back to 1976, when van Etten\cite{Etten_1976:ML_receiver_for_multi_channel} derived an ML sequence estimation based receiver for combating both ISI and interchannel interference (\gls{ICI}) in multiple-channel transmission systems. Explicitly, he demonstrated that under certain conditions, the performance of the ML receiver asymptotically approaches that of the optimum receiver of the idealized system which is free from both ISI and ICI. The significance of this work was not fully recognized until the research interests in commercial CDMA\index{CDMA} systems and multiple-antenna systems intensified.\footnote{In fact, van Etten's pioneering companion papers on the optimum MIMO detector\cite{Etten_1976:ML_receiver_for_multi_channel} and on the optimum linear MIMO detector\cite{Etten_1975:optimum_linear_receiver_for_multi_channel} were included in the book \textit{The best of the best: Fifty years of communications and networking research}\cite{Tranter_2007:best_of_best}, which was compiled by the IEEE Communications Society in 2007, and he is the only researcher who has two 
sole-author papers included in this selection.}

\subsubsection{Matched Filter (MF) versus Optimum MIMO Detector}\label{subsubsec:MF_vs_ML}
\begin{figure}[t]
\centering
\includegraphics[width=3.3in]{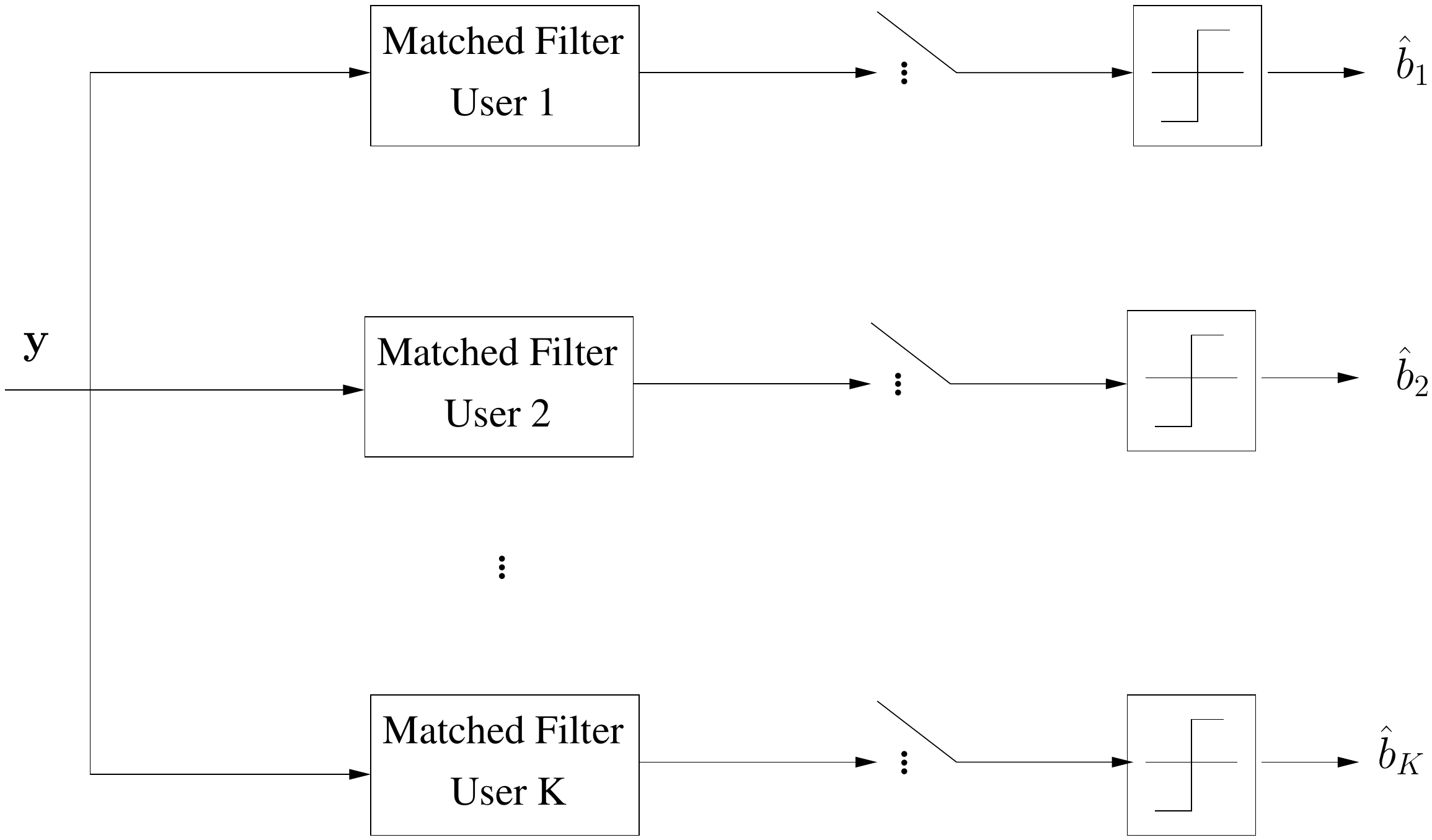}
\caption[]{An implementation example of the MF detector for synchronous CDMA systems employing BPSK modulation.} \label{fig:MF}
\end{figure}

Although it is widely recognized at the time of writing that MIMO detection provides significant performance gains compared to conventional single-stream detection, there was a widespread misconception until the early 1980s that the MUI can be accurately modelled as a white Gaussian random process, and thus the conventional single-user MF (\gls{SUMF}) based detector, as illustrated in Fig. \ref{fig:MF}, was believed to be essentially optimum. In 1983, this conventional wisdom was  explicitly proven wrong by Verd{\'u}\cite{Verdu_1983:earliest_optimal_MUD_conf_ISIT,Verdu_1983:earliest_optimal_MUD_conf_milcom} with the introduction of the optimal MUD in the context of asynchronous/synchronous Gaussian multiple-access channels shared by $K$ users. The full analysis and derivation of the optimum MUD was reported later in \cite{Verdu_1984:Phd_thesis, Verdu_1986:optimal_MUD_asynchronous_CDMA}, demonstrating that there is, in general, a substantial gap between the performance of the conventional SUMF and the optimal MUD performance. Additionally, upon identifying the non-Gaussian nature of the MUI, Poor and Verd{\'u}\cite{Poor_1988:SUD_for_MU} also designed nonlinear single-user detectors for CDMA\index{CDMA} systems operating in diverse scenarios such as weak interferers, high spreading gains and high signal-to-noise ratio (\gls{SNR}). The performance of the ML based optimum MIMO detector has been analyzed in\cite{Verdu_1984:Phd_thesis, Verdu_1986:optimal_MUD_asynchronous_CDMA,Grant_1998:ML_diversity_order,Grant_2000:further_results_ML_diversity_order,Awater_2000:ML_MMSE_diversity_order,Murch_2002:ML_performance_analysis,Lee_2006:ML_VER}. 

There does exist some situations where a \textit{bona fide} application of the central limit theorem is feasible and hence the MUI can be rigorously proven to be asymptotically Gaussian.\footnote{A specific example of such a situation is that an infinite-population multiuser signal model with the individual amplitudes going to zero at the appropriate speed -- in other words, when the overall interference power is fixed and the number of equal-power interferers tends to infinity\cite{Verdu_1997:MUD_progress_misconception}.} However, even if the MUI may be accurately modelled as a Gaussian variable, the SUMF is still not the optimal receiver. This is because the output of the MF for the desired user does not constitute a \textit{sufficient statistic} in the presence of  MUI\cite{Verdu_1997:MUD_progress_misconception}. In other words, the SUMF is optimal only in the context of the single-user channel contaminated by additive white Gaussian noise (\gls{AWGN}). By contrast, in multiple-access systems, unless the multiplexed signals (after passing through the channel) are orthogonal, the outputs of the MFs corresponding to the interfering users contain valuable information which may be exploited for the detection of the symbol of interest, and hence more intelligent joint detection strategies capable of exploiting all MFs' outputs are required for achieving better detection performance.

\subsubsection{Optimum Decision Criteria}\label{subsubsec:optimality_criteria}
When designing \textit{optimal} detectors/receivers for communication systems, it is usually necessary to clarify in what sense the word ``optimum'' is referred to. This is because the specific choice of an optimal detector/receiver is strongly dependent on the specific assumptions and criteria of ``goodness''. An optimal detector/receiver is the one that best satisfies the given criterion of goodness under a given set of assumptions. If either the criterion or the assumptions change, typically the choice of the optimal detector/receiver also changes. If the assumptions used in the theoretical analysis are inconsistent with the conditions of the realistic environment considered, then it is possible that the theoretically optimal detector/receiver obtained fails to provide valid insights and results for the practically achievable performance and designs. Special attention has to be paid to the definition of ``optimum'', since the theoretically optimal results obtained are mainly invoked as a benchmark or bound, against which any other results can be compared.    
There are many criteria of goodness. As far as the performance of the detectors/receivers used in communication systems is concerned, the \textit{minimum error probability criterion} is of primary interest, and hypothesis testing as well as likelihood ratios are of great importance. In Bayesian inference, the optimum decision criterion which minimizes the error probability based only on the observed signals and a given set of hypotheses is the maximum \textit{a posteriori} (\gls{MAP}) criterion. The error probability in communication systems can be measured in multiple scales, such as bit-error rate (\gls{BER}), symbol-error rate (\gls{SER}), and packet-/frame-/block-/vector-error rate (\gls{PER}/\gls{FER}/\gls{BLER}/\gls{VER}). When considering the MIMO system model of (\ref{eq:general_MIMO_model_matrix_form}), the MAP criterion based MIMO detector which is optimal in the sense of minimum \textit{VER} is formulated as
\begin{equation}\label{eq:MAP}
\mathfrak{D}_{\text{MAP}}: \hat{{\bf s}} = \arg \mathop {\max }\limits_{{\bf s} \in {\mathbb A}^{N_I}}  \Pr ({\bf s}|{\bf y}).
\end{equation}
Using Bayes' rule, the \textit{a posteriori} probability (\gls{APP}) in (\ref{eq:MAP}) may be expressed as
\begin{equation}\label{eq:MAP_to_ML}
\Pr ({\bf{s}}|{\bf{y}}) = \frac{{p({\bf{y}}|{\bf{s}})\Pr ({\bf{s}})}}{{p({\bf{y}})}} = \frac{{p({\bf{y}}|{\bf{s}})\Pr ({\bf{s}})}}{{\sum\limits_{{\bf s'} \in {\mathbb A}^{N_I} } {p({\bf{y}}|{\bf{s'}})\Pr ({\bf{s'}})} }},
\end{equation}
where $ \Pr ({\bf{s}})$ is the \textit{a priori} probability of $\bf s$, and $p({\bf y}|{\bf{s}})$ is the conditional probability density function (\gls{PDF}) of the observed signal vector $\bf y$ given $\bf s$. The MAP criterion can be simplified when each vector in ${\mathbb A}^{N_I}$ has an identical \textit{a priori} probability, i.e. we have $ \Pr ({\bf{s}}) = 1/|\mathbb{A}|^{N_I}$ for all realizations of $\bf s$, where $|\mathbb{A}|$ represents the number of elements, i.e. the cardinality of the constellation alphabet $\mathbb{A}$. Furthermore, considering the fact that $p({\bf{y}})$ is independent of which particular signal vector is transmitted, then the MAP detector of (\ref{eq:MAP}) becomes equivalent to the ML detector of
\begin{equation}\label{eq:ML}
\mathfrak{D}_{\text{ML}}: \hat{{\bf s}} = \arg \mathop {\max }\limits_{{\bf s} \in {\mathbb A}^{N_I}}  p ({\bf y}|{\bf s}).
\end{equation}
Therefore, the MAP criterion is usually used in the iterative detection and decoding (\gls{IDD}) aided receiver of forward-error-correction (\gls{FEC})-coded systems, where the \textit{a priori} probabilities of the transmitted symbols, $ \Pr ({\bf{s}})$, may be obtained with the aid of a backward-and-forward oriented iterative information exchange between the signal detector and the channel decoder. By contrast, the ML criterion is usually used in FEC-uncoded systems, where the \textit{a priori} probabilities of the transmitted symbols cannot be made available by the channel decoder. If $\bf n$ is AWGN, then we have
\begin{equation}\label{eq:likelihood_func}
p({\bf y}|{\bf s}) \propto \exp(-\left\| {{\bf{y}} - {\bf{Hs}}} \right\|_2^2),
\end{equation}
where the symbol $\propto$ represents the relationship ``is proportional to''. Consequently, we have
\begin{equation}\label{eq:equivalent_to_min_ED}
  \max\limits_{{\bf s} \in {\mathbb A}^{N_I}} p({\bf y}|{\bf s}) \Leftrightarrow \min\limits_{{\bf s} \in {\mathbb A}^{N_I}} \left\| {{\bf{y}} - {\bf{Hs}}} \right\|_2^2,
\end{equation}
where the symbol $\Leftrightarrow $ represents the relationship ``is equivalent to''.
Therefore, the ML detection problem for the system model of (\ref{eq:general_MIMO_model_matrix_form}) can be reformulated as the finite-set constrained least-squares (\gls{LS}) optimization problem of
\begin{equation} \label{eq:ML_minimum_Euclidean_distance}
{\hat {\bf{s}}}_{{\rm{ML}}} = \arg \mathop {\min }\limits_{{\bf{s}}
\in \mathbb{A}^{N_I}} {\rm{ }}  \left\| {{\bf{y}} - {\bf{Hs}}}
\right\|_2^2,
\end{equation}
which can also be interpreted as the \textit{minimum Euclidean distance (\gls{MED}) criterion}. 

Note, however, that the above-mentioned MAP, ML and MED criterion based MIMO detectors all aim for minimizing the VER, but do not guarantee achieving the minimum BER and minimum SER, which are two metrics of particular importance in many applications, such as in FEC-coded systems. There are other frequently used criteria in MIMO detector design. The linear MF criterion is optimal for maximizing the received SNR in the presence of additive stochastic noise. The linear ZF criterion is optimal for maximizing the received signal-to-interference ratio (\gls{SIR}). By contrast, the linear MMSE criterion based detector is optimal amongst all linear detectors\footnote{In general, the MMSE detector and the linear MMSE detector are not necessarily the same. The former only aims at minimizing mean-square error (\gls{MSE}) and does not impose any constraint on the form of the MMSE estimator. The latter assumes that the MMSE estimator is a linear function of the observed signal vector $\bf y$. If $\bf y$ and the transmitted signal vector $\bf s$ are jointly Gaussian, then the MMSE estimator is linear. In this case, for finding the MMSE estimator, it is sufficient to find the linear MMSE estimator.} in terms of achieving the MMSE, and in essence it is also optimal for maximizing the received signal-to-interference-plus-noise ratio (\gls{SINR}) amongst linear detectors\cite{Li_2006:MMSE_SINR_distribution, Mehana_2012:MMSE_diversity}. Additionally, the linear minimum bit-error rate (\gls{MBER}) criterion based detector achieves the lowest BER amongst all linear detectors, as detailed in Section \ref{susubsec:linear_MBER}.  

\subsubsection{Computational Complexity}
\begin{figure}[t]
\centering
\includegraphics[width=2.4in]{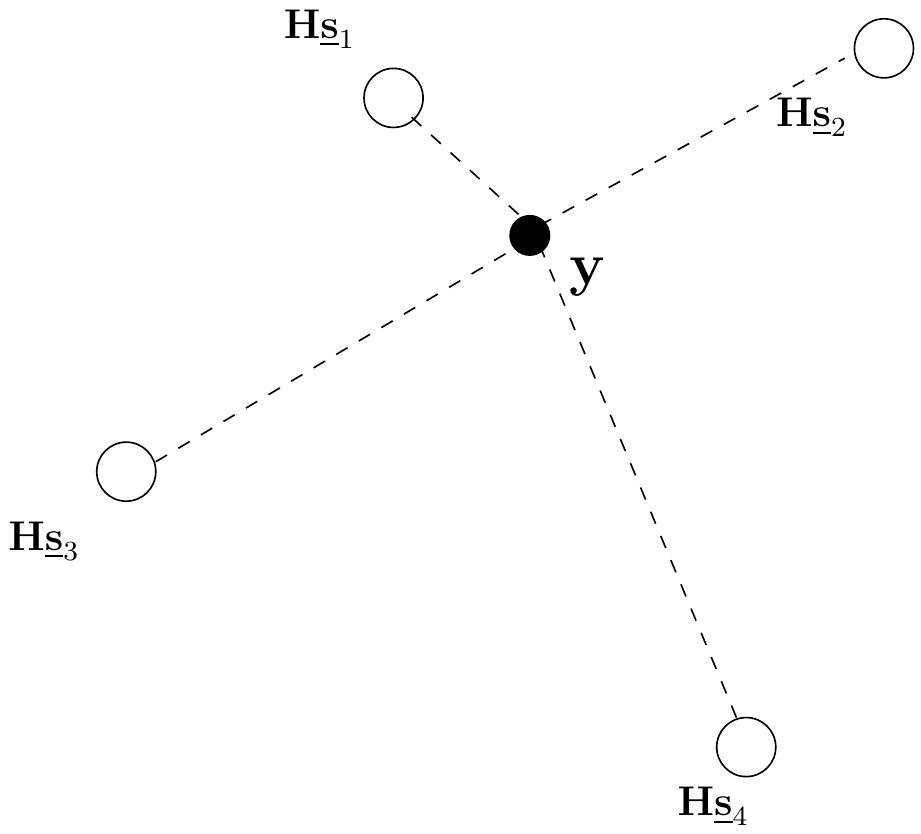}
\caption{Example of the optimal ML based MIMO detector in the context of $N_I = 2$ and BPSK modulation. } \label{fig:MIMO_D_geographical}
\end{figure}

The optimization problem of (\ref{eq:ML_minimum_Euclidean_distance}) can be solved by ``brute-force'' search over ${\mathbb A}^{N_I}$, resulting in an exponentially increasing computational complexity of $|\mathbb{A}|^{N_I}$. To elaborate a little further, let us consider the example shown in Fig. \ref{fig:MIMO_D_geographical}, where binary phase-shift keying (\gls{BPSK}) modulation ($M = 2$) and $N_I = 2$ are employed. Hence, there are a total of $M^{N_I} =4$ possible realizations for the transmitted symbol vector $\bf s$, and they are denoted as $\underline{\bf s}_1 = [1,1]^T$, $\underline{\bf s}_2 = [1, -1]^T$, $\underline{\bf s}_3 = [-1, -1]^T$, $\underline{\bf s}_4 = [-1,1]^T$.  

To gain deeper understanding of the computational complexity of the optimum MIMO detector formulated in (\ref{eq:ML}), let us examine its implementations in practical CDMA systems. More explicitly, the optimum MUD proposed in\cite{Verdu_1986:optimal_MUD_asynchronous_CDMA} for asynchronous CDMA\index{CDMA} systems consists of a bank of MFs followed by a dynamic programming algorithm of the forward (Viterbi) type\cite{Viterbi_1967:viterbi_algorithm, Omura_1969:Viterbi_algorithm_new_interpretation, Forney_1972:MLSE_ISI, Forney_1973:Viterbi_algorithm_tutorial, Viterbi_2006:viterbi_algorithm_history} (for ML criterion based detection) or of the backward-forward type \cite{Abend_1968:earliest_MAP_sequence_detector, Abend_1970:earliest_MAP_sequence_detection, Bahl_1974:BCJR_algorithm, Verdu_1984:decision_algorithms,Verdu_1987:decision_algorithms} (for minimum error probability criterion based detection). As mentioned in Section \ref{sec:MIMO_model_memory:chap_intro}, asynchronous CDMA systems can be modelled relying on the MIMO system model given in Section \ref{sec:MIMO_model_memory:chap_intro} for transmission over linear dispersive channels exhibiting memory. Therefore, the optimum MUD conceived for asynchronous CDMA\index{CDMA} constitutes a \textit{sequence} detector, while the optimum MUD of synchronous CDMA\index{CDMA} is a \textit{one-shot} detector, and as such it is a special case of the asynchronous optimum MUD. The optimum MUD relying on brute-force search\cite{Verdu_1986:optimal_MUD_asynchronous_CDMA} requires that the transmitted energies of each user were known to the receiver. More critically, the computational complexity of the optimum decision algorithms suggested in \cite{Verdu_1986:optimal_MUD_asynchronous_CDMA, Verdu_1984:decision_algorithms,Verdu_1987:decision_algorithms} increases exponentially with the number of active users, i.e. it is on the order of $\mathcal{O}(2^K)$ per bit for asynchronous transmission and $\mathcal{O}(2^K/K)$ per bit for synchronous transmission,  where $K$ is the number of active users. This is because the optimum MUD of both the synchronous and asynchronous CDMA scenarios was proven by Verd{\'u}\footnote{In fact, the optimum MIMO detection problem of (\ref{eq:ML_minimum_Euclidean_distance}) constitutes an instance of the general \textit{closest lattice-point search (\gls{CLPS})} problem, whose complexity had been analyzed earlier by Boas\cite{Boas_1981:prove_NP_hard_closest_point_search} in 1981, showing that this problem is NP-hard. Additionally, Micciancio \cite{Micciancio_2001:simpler_proof_NP_hard_closest_point_search} provided a simpler proof for the hardness of the CLPS problem in 2001.} to be an NP-hard and a non-deterministic polynomial-time complete (\gls{NP-complete}) problem\cite{Verdu_1989:complexity_optimal_MUD,Verdu_1984:Phd_thesis}. Thus, \textit{all known algorithms} designed for solving this problem optimally exhibit an exponentially increasing  computational complexity in the number of decision variables. Therefore, the optimum MUD becomes computationally intractable for a large number of active users. 

\begin{figure}[tbp]
\centering
\includegraphics[width=3.3in]{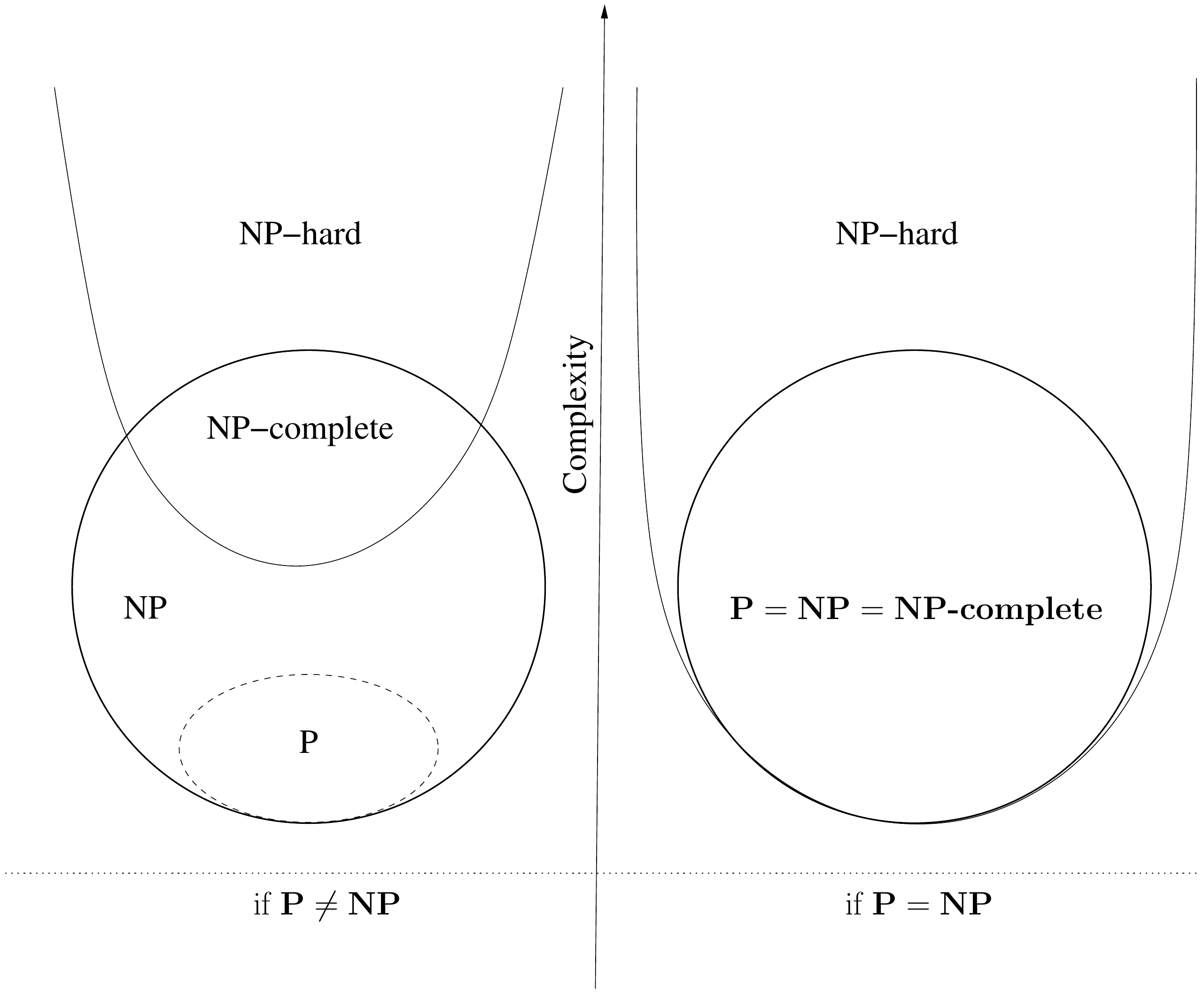}
\caption[Euler diagram for P, NP, NP-complete, NP-hard set of problems under both P$\neq $NP and P$ = $NP assumptions.]{Euler diagram for P, NP, NP-complete, NP-hard set of problems under both P$\neq $NP and P$ = $NP assumptions\cite{Garey_1979:complexity_theory, Cormen_2009:introduction_to_algorithm}. } \label{fig:complexity_class}
\end{figure}

It should be noted that the optimal MIMO detection problem would only have a polynomially increasing complexity \textit{if and only if} a polynomial-time solution could be found for any NP-complete problem, such as the famous \textit{travelling salesman problem} and the \textit{integer linear programming problem} which have been so far widely believed insolvable within polynomial time. However, the question of whether there exists a polynomial-time solution for NP-complete problems has not been answered by a rigorous proof to date. It is widely recognized that in computational complexity theory, the complexity class of ``P'' represents one of the most fundamental complexity classes, and it contains all decision problems that can be solved by a \textit{deterministic Turing machine} using a polynomially increasing amount of computation time (this is conventionally abbreviated to the parlance of ``polynomial-time'' for convenience). In fact, the most important open question in computational complexity theory\cite{Garey_1979:complexity_theory, Cormen_2009:introduction_to_algorithm} has been the formal proof of ``Is P $=$ NP?'', which explicitly poses the dilemma whether polynomial-time algorithms actually do exist for NP-complete problems, and by corollary, for all NP problems.
Fig. \ref{fig:complexity_class} concisely depicts the Euler diagram characterizing the relationships amongst the P, NP, NP-complete, and NP-hard set of problems under both the P$\neq $NP and P$ = $NP assumptions.

Additionally, it is worth mentioning that for some algorithms, such as the tree-search based MIMO detectors to be detailed in Section \ref{subsec:tree_search_MUD:chap_intro}, the computational complexity may vary in different scenarios. As such, the average computational complexity, the worst-case computational complexity and the distribution of computational complexity become important metrics to examine. Finally, in practical algorithm implementations, it is also important to consider the hardware complexity, which, in simplest form, can be measured by the silicon area and the number of NAND2 gates required in the IC implementation.

\subsubsection{Milestone Contributions}
For the sake of clarity, the main contributions to the development of the optimum MIMO detector are summarized in Table \ref{Table:optimal_MIMO_detector}.
\begin{table*}[tbp]
\setlength{\tabcolsep}{2pt}
\renewcommand{\arraystretch}{1.3}
\extrarowheight 3pt
\caption{Milestones in the development of the optimal MIMO detector}
\label{Table:optimal_MIMO_detector}
\centering
 \begin{footnotesize}
\begin{tabular}{|l|p{2.2cm}|p{11cm}|}
\hlinewd{0.9pt}
  Year & Authors & Contributions \\
\hline \hline
  1976 & van Etten\cite{Etten_1976:ML_receiver_for_multi_channel} & Derived an ML sequence estimation based receiver for combating both the ISI and ICI in multiple-channel transmission systems and demonstrated that under certain conditions, the performance of the ML receiver is asymptotically as good as if both the ISI and ICI were absent. \\
\hline
  1981 & Boas\cite{Boas_1981:prove_NP_hard_closest_point_search} & Analyzed the complexity of the generic problem of ``closest point search in an $N_I$-dimensional lattice'', which is identical to the optimum MIMO detection problem, as a function of the dimension $N_I$ of the decision-variable vector, and proved that this problem is NP-hard. Thus, all known algorithms conceived for solving the generic MIMO detection problem optimally have an exponentially increasing computational complexity.
 \\
\hline
  1983 - 1986 & Verd{\'u}\cite{Verdu_1983:earliest_optimal_MUD_conf_ISIT,Verdu_1983:earliest_optimal_MUD_conf_milcom, Verdu_1984:Phd_thesis,  Verdu_1986:optimal_MUD_asynchronous_CDMA} & First presented a full derivation and analysis of the ML based multiuser detector for asynchronous/synchronous CDMA\index{CDMA} systems; showed that there is, in general, a huge gap between the performance of the conventional SUMF and the optimal attainable performance; showed that the infamous near-far problem was not an inherent flaw of CDMA\index{CDMA} but a consequence of the inability of the SUMF to exploit the structure of the MUI; introduced the performance measure of multiuser asymptotic efficiency, which was later widely used in the asymptotic analysis of multiuser detectors at the high-SNR region. \\
\hline
  1984 - 1989 & Verd{\'u}\cite{Verdu_1984:Phd_thesis, Verdu_1989:complexity_optimal_MUD} & Independently proved that the optimum MUD problem in CDMA\index{CDMA} systems is NP-hard\index{NP-hard} and NP-complete\index{NP-complete}. \\
\hline
  2001 & Micciancio\cite{Micciancio_2001:simpler_proof_NP_hard_closest_point_search} & Presented a simpler proof of the NP-hardness of the problem of closest point search in an $N_I$-dimensional lattice.\\
\hline 
 2003 &  Garrett \textit{et al.} \cite{Garrett_2003:first_ML_VLSI_implementation}  & Proposed the first VLSI implementation of a soft-output ML detector having a 19.2 Mbps uncoded data rate supporting up to $4 \times 4$ QPSK MIMO. \\
\hline 
  2003 & Burg \textit{et al.} \cite{Burg_2003:VLSI_hard_ML} &  Presented an efficient VLSI implementation of hard-decision optimum ML detector for QPSK MIMO. The proposed method does not compromise optimality of the ML detector. Instead it uses the special properties of QPSK modulation, together with algebraic transformations and architectural optimizations, to achieve low hardware complexity and high speed up to 50 Mbps. \\
\hlinewd{0.9pt}
\end{tabular}
\end{footnotesize}
\end{table*}

The substantial performance and complexity differences between the optimum MIMO detector and the conventional SUMF detector stimulated a lot of interests in the development of suboptimum MIMO detection algorithms that are capable of achieving good performance at a low computational cost. Some representative classes of suboptimum MIMO detectors include the linear detectors, the interference cancellation aided detectors, the tree-search based detectors, the PDA based detectors, the SDPR based detectors and the LR based detectors etc., as seen in Fig. \ref{fig:MIMO_Detector} and detailed below.

\subsection{Linear MIMO Detectors}
\label{subsec:linear_MIMO_detector:chap_intro}
\begin{figure}[tbp]
\centering
\includegraphics[width=3.5in]{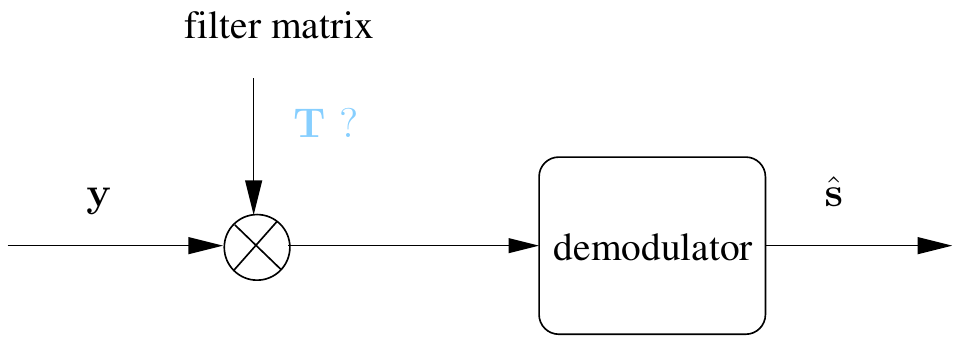}
\caption[]{Conceptual illustration of linear MIMO detectors.} \label{fig:linear_MUD}
\end{figure}

The linear MIMO detectors of Fig. \ref{fig:MIMO_Detector} are based on a linear transformation of the output signal vector $\bf y$. In general, they are known for their appealingly low complexity, but suffer from a considerable performance loss in comparison to the ML detector. More explicitly, the decision statistics of linear MIMO detectors may be expressed as
\begin{equation}\label{eq:linear_MIMO_detector}
{\bf d} = {\bf Ty},
\end{equation}
where $\bf T$ is the linear transformation (or filtering) matrix to be designed using various criteria. A conceptual illustration of the linear MIMO detectors is given in Fig. \ref{fig:linear_MUD}.

\subsubsection{MF Detector} 
For the sake of illuminating the philosophy of linear MIMO detectors, let us rely on (\ref{eq:memoriless_MIMO_model_scalar_form}) and continue by considering the MF detector, which has the lowest computational complexity among all MIMO detectors and its linear transformation matrix is given by 
 \begin{equation}\label{eq:linear_MF_detector}
{\bf T}_{\rm MF} = {\bf H}^{H}.
\end{equation}
Upon using the MF detector of (\ref{eq:linear_MF_detector}), we obtain  
\begin{equation}\label{eq:MF_detector_data}
{\bf d} = {\bf H}^H{\bf Hs} + {\bf H}^H{\bf n}.
\end{equation}
The MF detector is well known as the optimal linear filter designed for maximizing the output SNR in the presence of additive stochastic noise.   
In Section \ref{subsubsec:MF_vs_ML}, we have provided some discussions regarding the MF detector in order to justify the motivations of developing  \textit{joint detection} based MIMO detectors. To elaborate a little further, the MF detector had been widely used before the concept of MIMO detection was born, and it is essentially based on the single-user detection philosophy. Hence, strictly speaking, it does not belong to the \textit{joint detection} based MIMO detection family, and typically it exhibits a poor performance in CCI-limited MIMO systems. However, in \textit{certain} LS-MIMO contexts\cite{Rusek_2013:massive_MIMO, tenBrink2013:massive_MIMO}, the MF detector is capable of approaching the performance of the optimal ML detector, as it will be further discussed in Section \ref{Sec:detection_in_massive_MIMO}.

\subsubsection{Linear ZF Detector}
Assuming that the noise vector is zero, (\ref{eq:memoriless_MIMO_model_scalar_form}) becomes a system of linear equations, and the MIMO detection problem becomes equivalent to ``finding the solution for $N_t$ unknown variables subject to $N_r$ linear equations''. Therefore, if $\bf H$ is a square matrix (i.e. $N_r = N_t$) and of full rank, the solution of this system of linear equations is given by ${\bf s} = {\bf H}^{-1}{\bf y}$. To generalize this problem a little further, if the matrix $\bf H$ satisfies $N_r > N_t$ and has a full column rank of $N_t$, we have $\bf s = {\bf H}^{\dagger}{\bf y}$, where ${\bf H}^{\dagger} = ({\bf H}^H{\bf H})^{-1}{\bf H}^H$ is the left-multiplying Moore-Penrose pseudoinverse of $\bf H$. This example actually conveys the essential idea of the ZF  criterion based MIMO detector, for which the linear transformation matrix is given by \begin{equation}\label{eq:linear_ZF_detector}
{\bf T}_{\rm ZF} = {\bf H}^{\dagger},
\end{equation}
and if $\bf H$ is invertible, the left-multiplying pseudoinverse ${\bf H}^{\dagger}$ and the inverse coincides, i.e. we have ${\bf H}^{\dagger} = {\bf H}^{-1}$. Upon using the ZF detector, we have ${\bf d} = {\bf s} + {\bf H}^{\dagger}{\bf n}$, which indicates that the interference amongst the multiple inputs is completely eliminated, albeit the noise power is augmented.

Similar to the case of the optimum ML-based MIMO detector, the ZF criterion based linear MIMO detector of Fig. \ref{fig:MIMO_Detector} was also first proposed by van  Etten\cite{Etten_1975:optimum_linear_receiver_for_multi_channel} in 1975 for a multiple-channel multiplexing transmission system subjected to both ISI and ICI. As far as CDMA\index{CDMA} systems are concerned, this solution was first proposed by Schneider\cite{Schneider_1979:linear_ZF_CDMA} in 1979 for synchronous CDMA\index{CDMA} systems transmitting equal-energy multiuser signals, where he sought to minimize the probability of bit error, but erroneously arrived at the ZF detector. From 1986 to 1990, Lupas and Verd{\'u} systematically investigated this detector in the context of both synchronous\cite{Lupas_1986:linear_MUD_synchrohous_CDMA_conf, Lupas_1989:linear_MUD_synchrohous_CDMA} and asynchronous\cite{Lupas_1988:linear_MUD_asynchrohous_CDMA_conf, Lupas_1990:near_far_MUD_asynchronous} CDMA\index{CDMA} systems. They referred to it as the linear \textit{decorrelating} multiuser detector. It was shown that if the transmitted energies of each user are unknown to the receiver, then both the ML amplitude estimates and the ML decisions on the transmitted bits are obtained by the ZF detector, regardless of the values of the received energies of each user. As a beneficial result, the ZF detector achieves the \textit{same} degree of resistance to the infamous \textit{near-far problem} as the optimum ML detector, despite its significantly reduced computational complexity. The insight that the near-far problem was not an inherent flaw of CDMA\index{CDMA} but a consequence of the SUMF's inability to exploit the non-Gaussian structure of the MUI\cite{ Verdu_1986:optimal_MUD_asynchronous_CDMA}, and the fact that the joint detection based MUDs, including its linear versions, achieve a significantly better near-far resistance\cite{Lupas_1986:linear_MUD_synchrohous_CDMA_conf, Lupas_1989:linear_MUD_synchrohous_CDMA, Lupas_1988:linear_MUD_asynchrohous_CDMA_conf, Lupas_1990:near_far_MUD_asynchronous} became another major incentive for the subsequent research activities dedicated to MUD in CDMA\index{CDMA}. Additionally, with the advent of the multiple-antenna technologies conceived during the mid-1990s, the ZF detector was first studied in the SDM-based VBLAST systems by Foschini, Wolniansky, Golden and Valenzuela\cite{Foschini:MIMO, Wolniansky:VBLAST, Golden_1999:VBLAST_first_journal}.

\subsubsection{Linear MMSE Detector}
As seen in Fig. \ref{fig:MIMO_Detector}, the linear transformation matrix $\bf T$ of (\ref{eq:linear_MIMO_detector}) can also be designed according to the MMSE criterion, which minimizes the mean-square error between the actual transmitted data and the channel's output data after using the linear transformation matrix $\bf T$. To be more specific, $\bf T$ is obtained by solving the optimization problem of
\begin{equation} \label{eq:MMSE_criterion_metric}
{\bf T}_{\rm MMSE} = \arg \mathop {\min }\limits_{\bf{T}}
 {\rm{ }} {\mathcal E} \left( \left\| {{\bf{s}} - {\bf{Ty}}}
\right\|_2^2\right).
\end{equation}
Using the \textit{orthogonality principle}\cite{Kay_1993:Fundamentals_statistical_signal_processing_vol_1}, we have 
\begin{equation} \label{eq:MMSE_orthogonality_principle}
{\mathcal E} \lbrack {({\bf s} - {\bf Ty}){\bf y}^H} \rbrack = {\bf 0},
\end{equation}
then ${\bf T}_{\rm MMSE}$ may be derived as
\begin{equation}
{\bf T}_{\rm MMSE} = ({\bf H}^H{\bf H} + 2\sigma^2{\bf I})^{-1}{\bf H}^H,
\end{equation}
where $\sigma^2$ is the noise power per real dimension, and $\mathcal{E}({\bf s}) = 1$ is assumed. Compared to the linear ZF detector, the linear MMSE detector achieves a better balance between the MUI elimination and noise enhancement by jointly minimizing the total error imposed by both the MUI and the noise. Hence, the linear MMSE detector achieves a better performance at low SNRs than the ZF detector.

The MMSE criterion based linear MIMO detector was first proposed by Shnidman\cite{Shnidman1967:earliest_ISI_cross_talk_equivalent} in 1967, and hence it is the oldest MIMO detector found in the literature. The \textit{generalized Nyquist criterion} formulated by Shnidman first indicates that the ISI and crosstalk\footnote{Crosstalk may be interpreted as a special case of ICI. For example, as mentioned before, in \cite{Shnidman1967:earliest_ISI_cross_talk_equivalent}, crosstalk means the interference between the multiplexed different waveforms.} between multiplexed signals essentially represent identical phenomena. Then,  relying on this insight, he proposed a linear receiver that is optimal in the sense of the MMSE criterion for combating both the ISI and crosstalk in single-channel multiple-waveform-multiplexed pulse-amplitude modulation (\gls{PAM}) systems. In 1970, Kaye and George \cite{Kaye_1970:transmit_multiplexed_PAM_over_Multi_channel} explicitly extended the MMSE receiver of \cite{Shnidman1967:earliest_ISI_cross_talk_equivalent} to the family of general multiple-channel systems transmitting multiplexed PAM signals and/or providing diversity.
The MMSE criterion based linear detector for CDMA\index{CDMA} systems was proposed by Xie, Rushforth and Short in 1989\cite{Xie_1989:Linear_MMSE_WLS_MUD_conf, Xie_1990:Linear_MMSE_WLS_MUD}. A decade later, it was also revisited by Foschini, Wolniansky, Golden and Valenzuela in the context of SDM-based multi-antenna systems\cite{Foschini:MIMO, Wolniansky:VBLAST, Golden_1999:VBLAST_first_journal}. The performance of linear ZF/MMSE based MIMO detectors depends on the SINR experienced at the output of these detectors, which was first analyzed by Poor and Verd{\'u}\cite{Poor_1997:probability_of_error_MMSE_MUD} in 1997, and investigated in more depth later from various other perspectives, such as the error probability, outage probability, diversity-multiplexing tradeoff (\gls{DMT})\cite{Lizhong_2003:DMT, Tse_2004:DMT}, as well as asymptotic distribution of the SINR (in terms of antenna number and high/low SNR regimes)\cite{Li_2006:MMSE_SINR_distribution, Hedayat_2007:outage_diversity_ZF_MMSE, Jorswieck_2007:outage_prob_MIMO,Moustakas_2009:MMSE_massive_MIMO,Kumar_2009:asymptotic_antenna_SNR_linear_MIMO_receiver,Jiang_2011:performance_analysis_ZF_MMSE,Mehana_2012:MMSE_diversity}. 
\begin{table*}[tbp]
\setlength{\tabcolsep}{2pt}
\renewcommand{\arraystretch}{1.3}
\extrarowheight 3pt
\caption{Milestones in the development of linear MIMO detectors}
\label{Table:linear_MIMO_detector}
\centering
 \begin{footnotesize}
\begin{tabular}{|l|p{2.5cm}|p{10.7cm}|}
\hlinewd{0.9pt}
  Year & Authors & Contributions \\
\hline \hline
  1967 & Shnidman\cite{Shnidman1967:earliest_ISI_cross_talk_equivalent}	 & First formulated a generalized Nyquist criterion, which pointed out that the ISI and crosstalk between multiplexed signals are essentially identical phenomena; he then proposed a linear MMSE receiver for combating both ISI and crosstalk in single-channel multiple-waveform-multiplexed PAM systems.  \\
\hline
  1970 & Kaye \textit{et al.} \cite{Kaye_1970:transmit_multiplexed_PAM_over_Multi_channel} & Extended the MMSE receiver of \cite{Shnidman1967:earliest_ISI_cross_talk_equivalent} to the general multiple-channel systems transmitting multiplexed PAM signals and/or providing diversity.     \\
\hline
  1975 & van Etten\cite{Etten_1975:optimum_linear_receiver_for_multi_channel} & Developed linear receivers based on both the ZF criterion and the minimum error probability criterion for a multiple-channel transmission system similar to that of \cite{Kaye_1970:transmit_multiplexed_PAM_over_Multi_channel}; these two detectors heralded the linear ZF and the linear MBER multiuser detectors of CDMA\index{CDMA} systems.      \\
\hline
  1975 & Horwood \textit{et al.} \cite{Horwood_1975:signal_design_multiple_access} & Proposed two linear signal-correlation based detectors for synchronous digital multiple-access systems; one of them assumes that each user only knows its own signature, while the other assumes that each user knows all users' signatures; this is the first attempt in multiple-access systems to exploit the structure of the signals simultaneously sent, which is the key idea of MUD in CDMA\index{CDMA} systems.   \\
\hline
  1979 & Schneider\cite{Schneider_1979:linear_ZF_CDMA} & First made an attempt to conceive MUD for CDMA\index{CDMA} systems; he proposed the linear decorrelating detector, namely the linear ZF detector, which represents one of the mainstream MUD approaches conceived for CDMA\index{CDMA} systems; this detector was also extended to the scenario of combating crosstalk in $M$-ary multiplexed transmission systems in 1980 \cite{Schneider_1980:crosss_talk_resistant_receiver}. \\
\hline
  1986-1990 & Lupas \textit{et al.} \cite{Lupas_1986:linear_MUD_synchrohous_CDMA_conf, Lupas_1989:linear_MUD_synchrohous_CDMA, Lupas_1988:linear_MUD_asynchrohous_CDMA_conf, Lupas_1990:near_far_MUD_asynchronous} & Systematically investigated the linear ZF MUD in the context of both synchronous \cite{Lupas_1986:linear_MUD_synchrohous_CDMA_conf, Lupas_1989:linear_MUD_synchrohous_CDMA} and asynchronous \cite{Lupas_1988:linear_MUD_asynchrohous_CDMA_conf, Lupas_1990:near_far_MUD_asynchronous} CDMA\index{CDMA} systems; they showed that the ZF detector achieves exactly the same degree of resistance to the infamous near-far problem as the optimum ML detector, despite its much lower computational and implementation complexity; they also first proposed a linear MAME MUD, which is capable of equivalently minimizing the probability of bit error in the limit as the noise approaches zero.\\
\hline
  1989 - 1990 & Xie \textit{et al.} \cite{Xie_1989:Linear_MMSE_WLS_MUD_conf, Xie_1990:Linear_MMSE_WLS_MUD} & First proposed the MMSE criterion based linear MUD, the modified linear equalizer based MUD, and the WLS linear MUD for CDMA\index{CDMA} systems. In contrast to the linear ZF detector, to the linear MMSE detector, and to the modified linear equalizer based detector, the linear WLS detector is capable of providing an unbiased estimate of the transmitted symbols. \\
\hline
  1993 - 1997 & Mandayam \textit{et al.} \cite{Mandayam_1993:MBER_optical_CDMA_MUD_conf, Mandayam_1997:MBER_optical_CDMA_MUD_part_1, Mandayam_1997:MBER_optical_CDMA_MUD_part_2, Mandayam_1997:MBER_CDMA_MUD} & First proposed the MBER criterion based linear MIMO detectors for CDMA\index{CDMA} systems; the linear MBER detector is capable of outperforming the linear MMSE detector when either the signature cross-correlation is high or the background noise is non-Gaussian. \\
\hline
  1996 - 1999 &  Foschini \textit{et al.} \cite{Foschini:MIMO, Wolniansky:VBLAST, Golden_1999:VBLAST_first_journal} &  First discussed the application of linear ZF/MMSE detectors in multiple-antenna aided SDM-MIMO systems. \\
\hline
  2006 & Chen \textit{et al.} \cite{Chen_2006:MBER_MIMO_Detection} & Proposed the MBER criterion based linear detector for multi-antenna aided MIMO systems. \\
\hline 
  2006 & Burg \text{et al.} \cite{Burg_2006:linear_MMSE_VLSI} &  Presented an algorithm and a corresponding VLSI architecture for the implementation of linear MMSE detection in packet-based MIMO-OFDM communication systems. The algorithm also supports the extraction of soft information for channel decoding. \\
\hline 
  2009 & Yoshizawa \textit{et al.} \cite{Yoshizawa_2009:VLSI_implementation_MMSE_MIMO_OFDM} &  Reported a VLSI implementation for a  $4 \times 4$ MIMO-OFDM transceiver relying on linear MMSE, which achieves a target data rate of 1 Gbps. \\ 
\hline 
  2014 & Yin \textit{et al.} \cite{Studer_2014:implementation_LS_MIMO_detector_journal} &  Presented the first application-specific integrated circuit (\gls{ASIC}) implementation for the soft-output linear MMSE detector based large-scale MIMO system which uses 128 BS antennas to support 8 users, and a sum-rate of 3.8 Gbps was achieved. \\
\hline
\hlinewd{0.9pt}
\end{tabular}
\end{footnotesize}
\end{table*}

\subsubsection{Other Linear Detectors}\label{susubsec:linear_MBER}
As observed in the family-tree of Fig. \ref{fig:MIMO_Detector}, there are a range of other criteria for designing the linear transformation matrix $\bf T$. 
\begin{itemize}
\item
For example, in\cite{Lupas_1986:linear_MUD_synchrohous_CDMA_conf, Lupas_1989:linear_MUD_synchrohous_CDMA}, Lupas and Verd{\'u} also proposed a maximum asymptotic-multiuser-efficiency (\gls{MAME}) based linear detector, which is capable of minimizing the probability of bit errors in the limit as the noise approaches zero. The asymptotic-multiuser-efficiency (\gls{AME}) is a metric which characterizes the performance of the MUD in the high-SNR region. It implies the performance loss of the desired user in the high-SNR region due to the interference imposed by other active users. To be more specific, it is defined as the limit of the ratio between the effective SNR (that is required by a single-user system to achieve the same asymptotic error probability) and the actual SNR of the desired user, when the noise power tends to zero. Furthermore, the linear MAME detector was designed by exploiting the assumption that the individual transmitted energies of all the users are fixed and known to the receiver. By contrast, the ZF detector does \textit{not} require the knowledge of the transmitted energies of the users. 
\item 
Additionally, since a common disadvantage of the linear ZF and MMSE detectors is that their estimates of the transmitted symbols are \textit{biased}, Xie, Rushforth and Short\cite{Xie_1989:Linear_MMSE_WLS_MUD_conf, Xie_1990:Linear_MMSE_WLS_MUD} proposed the so-called weighted least-squares (\gls{WLS}) linear detector, which is capable of providing an \textit{unbiased} estimate of the transmitted symbols. It is worth pointing out that except for the linear MF and ZF detectors, other linear MIMO detectors -- including the linear MMSE detector, the linear MAME detector and the linear WLS detector -- were typically derived under the assumption that the system parameters such as the signal's phase, power and delay are known. As a result, in practice these parameters must be estimated and the receiver's structure has to be regularly modified to reflect the updated estimates. 
\item 
Another important class of linear MIMO detectors are based on the MBER criterion. The linear MBER detector is capable of outperforming the linear MMSE detector when either the signature cross-correlation is high or the background noise is non-Gaussian\cite{Psaromiligkos_1999:MBER_CDMA}. Again, the MBER based MIMO detector was first considered by van Etten \cite{Etten_1975:optimum_linear_receiver_for_multi_channel} in 1975 in the context of a multiple-channel multiplexing transmission system subjected to both ISI and ICI. This MBER criterion was later studied in the context of CDMA\index{CDMA} systems \cite{Mandayam_1993:MBER_optical_CDMA_MUD_conf, Mandayam_1997:MBER_optical_CDMA_MUD_part_1, Mandayam_1997:MBER_optical_CDMA_MUD_part_2, Mandayam_1997:MBER_CDMA_MUD, Wang_1999:MBER_MUD_conf, Wang_2000_MBER_MUD, Psaromiligkos_1999:MBER_CDMA, Yeh_1998:MBER_conf,Yeh_2000:MBER_MUD, Chen_2001:adpative_MBER_linear_MUD} and multi-antenna systems\cite{Chen_2006:MBER_MIMO_Detection, Palomar_2005:MBER_MIMO_tranceiver}. 
\item 
Finally, we would like to mention that the linear MIMO detector can also be designed from the perspective of a linear equalizer\cite{Xie_1989:Linear_MMSE_WLS_MUD_conf, Xie_1990:Linear_MMSE_WLS_MUD}, since the mathematical models of the MIMO detection problem and of the equalization problem are similar\cite{Andrews:Interference_Cancellation_overview}. To elaborate a little further, in MIMO systems each symbol's interference is imposed by other simultaneous transmissions, while in the band-limited ISI channels requiring equalization, the interference of a particular symbol is due to other symbols that are transmitted sequentially in the time domain.
\end{itemize}

The main contributions to the development of linear MIMO detectors are summarized in Table \ref{Table:linear_MIMO_detector}.

\subsection{Interference Cancellation Aided MIMO Detectors}
\label{subsec:IC_MIMO_detector:chap_intro}
Another important class of suboptimum MIMO detectors portrayed in Fig. \ref{fig:MIMO_Detector} are constituted by the \textit{interference cancellation}  based MIMO detectors, which are \textit{nonlinear} and generally achieve a better performance than linear MIMO detectors. The concept of interference cancellation was first studied in 1974 by Bergmans and Cover\cite{Bergmans_1974:interference_cancellation_earliest, Cover_1975:earliest_interference_cancellation_idea}, as well as by Carleial\cite{Carleial_1975:earliest_interference_cancellation_idea} in 1975, in their information-theoretic studies of broadcast channels and of interference channels, respectively. In the context of CDMA\index{CDMA} and multi-antenna systems, this class of MIMO detectors have numerous variants due to the associated design flexibility, including the successive interference cancellation (\gls{SIC})\index{SIC} detector\cite{Viterbi_1990:earliest_interference_cancellation_approach_capacity, Dent_1992:SIC_CDMA, Foschini:MIMO, Wolniansky:VBLAST, Golden_1999:VBLAST_first_journal}, the parallel interference cancellation (\gls{PIC}) detector\cite{Kohno_1983:PIC_CDMA, Divsalar_1998:PIC_MUD_CDMA, Studer_2011:PIC_MIMO_ASIC_implementation}, the multistage interference cancellation (\gls{MIC}) detector\cite{Timor_1981:multistage_decoding_CDMA, Varanasi_1990:multistage_detection_asynchronous_CDMA, Varanasi_1991:multistage_MUD_synchronous_CDMA, Moshavi_1996_multistage_linear_receivers_CDMA}, and the decision-feedback detector (\gls{DFD})\cite{Xie_1990:Linear_MMSE_WLS_MUD, Hallen_1993:decorrelating_DFD_synch_CDMA, Hallen:1995:DFD_asynchronous_CDMA, Varanasi:DFD} etc. The interference cancellation based MIMO detectors are typically capable of providing a significantly better performance than their linear counterparts at the expense of a higher complexity, especially in the absence of channel coding\cite{Verdu_1999:spectral_efficiency_CDMA}, albeit this is not necessarily always the case. In practice, a common drawback of the interference cancellation based MIMO detectors is that they often suffer  from error propagation. Hence their performance only approaches that of the optimum ML based MUD when the interfering users have a much stronger signal strength than the desired user. From this perspective, the weakest user benefits most from the employment of the interference cancellation detector.
\begin{figure}[t]
\centering
\includegraphics[width=\linewidth]{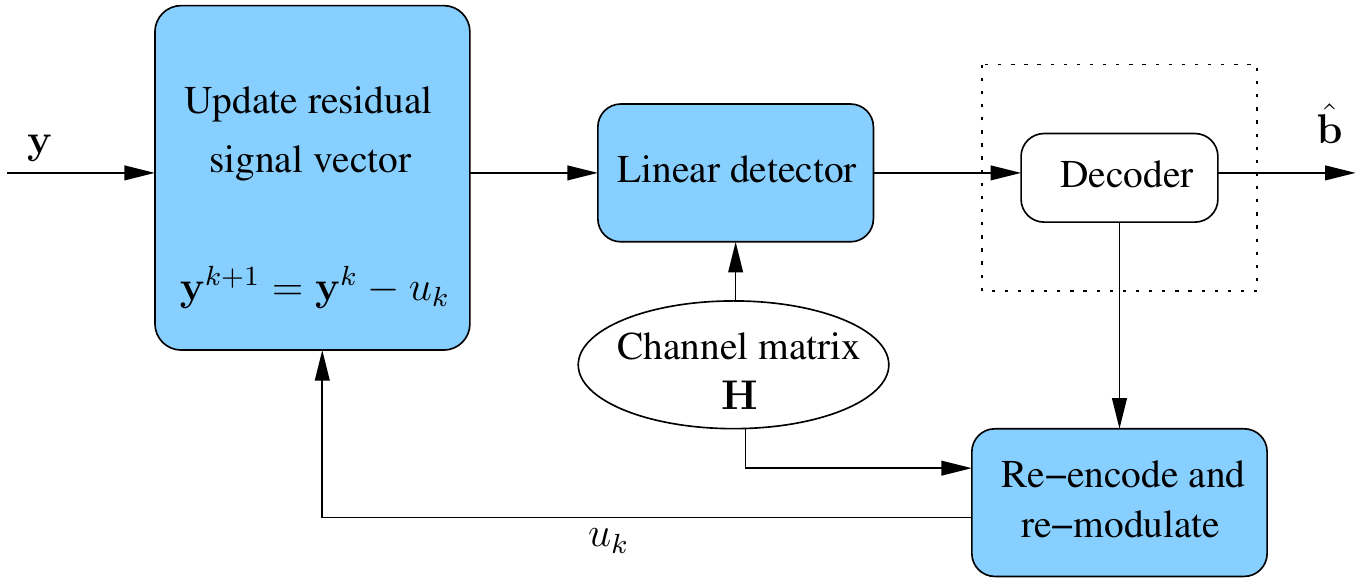}
\caption[]{The basic principle of the SIC/DFD based MIMO detectors.} \label{fig:SIC}
\end{figure}
\begin{figure}[tbp]
\centering
\includegraphics[width=\linewidth]{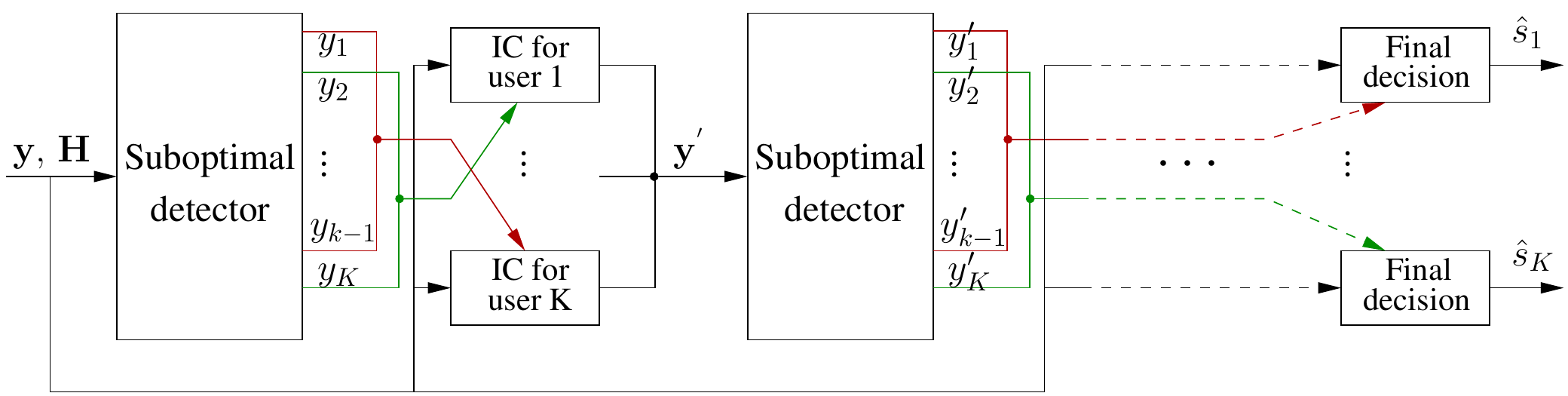}
\caption[]{The basic principle of the PIC/MIC based MIMO detectors.} \label{fig:PIC}
\end{figure}

\begin{itemize}
 \item 
\textit{SIC:} In the most popular SIC based MIMO detector, a single symbol $s_i$ is detected at a time. Then the interference imposed by this particular symbol on the other symbols ${\{s_k\}}_{k \ne i}$ yet to be detected is subtracted after recreating the interference upon generating the modulated signal corresponding to this symbol. In this scheme, it is most important to cancel the effect of the strongest interfering signal before detecting the weaker signals. Therefore, the specific symbol detection ordering, which can be designed according to various criteria, is quite critical for the SIC detector's performance. Some of the typical ordering criteria for ordered SIC (\gls{OSIC}) include the decreasing signal-to-noise ratio (\gls{DSNR}) criterion\cite{Wubben_2003:MMSE_DSNR_ordering, Wai_2000:IMSE_DSNR_ordering_MMSE_SIC}, the greatest signal-to-noise ratio (\gls{GSNR}) criterion\cite{Bohnke_2003:MMSE_SIC_GSNR}, the increasing mean-square error (\gls{IMSE}) criterion\cite{Wai_2000:IMSE_DSNR_ordering_MMSE_SIC}, and the least mean-square error (\gls{LMSE}) criterion\cite{Hassibi_2000:sqrt_algorithm_BLAST_LMSE_ordering, Benesty_2003:MMSE_SIC_LMSE_ordering, Liu_2009:MMSE_SIC_LMSE}. The SIC method performs well when there is a substantial difference in the received signal strength of the multiple simultaneously transmitted symbols. However, this condition is not always satisfied in practical applications, which renders the SIC detector potentially sensitive to decision error propagation. Therefore, the SIC detector is well-suited for multiple-access systems suffering from the near-far problem, such as the family of CDMA\index{CDMA} or SDMA\index{SDMA} systems. In the SIC detector, there is a need for detection reordering at each iteration of the SIC detector, and the number of detection iterations increases linearly with the number of symbols in $\bf s$. Therefore, for a system which has a high-dimensional transmitted symbol vector $\bf s$, the SIC technique imposes a substantial complexity, which ultimately increases the processing delay. The SIC detector designed for CDMA\index{CDMA} systems was first proposed by Viterbi\cite{Viterbi_1990:earliest_interference_cancellation_approach_capacity}. Later it was studied extensively in \cite{Dent_1992:SIC_CDMA, Kubota_1992:SIC_CDMA, Kubota_1992:SIC_CDMA_base_station, Patel_1993:SIC_CDMA, Patel_1994:SIC_CDMA, Holtzman_1994:SIC_CDMA, Holtzman_1994:SIC_CDMA_milcom, Hui_1998:SIC_asynch_CDMA, Gupta_2007:SIC_constellation_structure}. In the context of multi-antenna based SDM systems, the SIC scheme was first studied by Foschini, Wolniansky, Golden and Valenzuela in \cite{Foschini:MIMO, Wolniansky:VBLAST, Golden_1999:VBLAST_first_journal}, and it was later studied more comprehensively by numerous other researchers in\cite{Li_2002:MIMO_OFDM_SIC_detection_CE, Lu_2002:MIMO_OFDM_LDPC_SIC,Zanella_2005:MMSE_SIC_MIMO}. Among these schemes,  Viterbi\cite{Viterbi_1990:earliest_interference_cancellation_approach_capacity} proposed an SIC scheme for a convolutionally coded direct-sequence CDMA (\gls{DS-CDMA}) system and revealed that with the aid of the SIC based receiver, the aggregate data rate of all simultaneous users may approach the Shannon capacity of the Gaussian channel. It should be emphasized that although theoretically the SIC method achieves the Shannon capacity in the multiple-access channel by assuming perfectly error-free detection (hence avoiding decision error propagation), this is not necessarily true in practice,  because the SIC method is sensitive to decision error propagation, and hence MIMO detectors that are more robust to decision error propagation might outperform the SIC detector in practice. Another fact worth mentioning is that the performance degradation imposed by error propagation in the SIC detector can be mitigated by accurate power control\cite{Andrews:power_control_SIC}.
\item
\textit{PIC:} Alternatively, in the PIC based MIMO detector, all symbols are detected simultaneously. For each symbol, the coarse initial estimate of the interfering symbols can be used for regenerating the interference and then for deducting it from each of the composite received signals. Then this PIC detection process may be repeated for several iterations. Therefore, sometimes the PIC detection is also regarded as a MIC technique, or vice versa. Compared to the SIC detector, the PIC detector has lower processing delay, and is more robust to inter-stream error propagation. However, its near-far resistance is inferior to that of the SIC detector, because some users might have much weaker received signal strength than others. Hence, the PIC is suitable for similar-power signals, while the SIC performs better for different-power streams. In the context of CDMA\index{CDMA} systems, the earliest contribution to PIC may be attributed to Kohno \textit{et al.}\cite{Kohno_1983:PIC_CDMA, Kohno_1990:IC_CDMA, Kohno_1990:PIC_CDMA_journal}. Later significant contributions to PIC were also attributed to Yoon\cite{Yoon_1992:PIC_CDMA, Yoon_1993:PIC_CDMA_journal}, Divsalar\cite{Divsalar_1998:PIC_MUD_CDMA}, Buehrer\cite{Buehrer_1999:linear_versus_nonlinear_IC} and Guo\cite{Guo_2000:PIC_CDMA} \textit{et al.}. In the context of multi-antenna MIMO systems, the PIC detector was studied mainly in\cite{Chin_2000:PIC_BLAST, Luo_2008:PIC_MIMO, Studer_2011:PIC_MIMO_ASIC_implementation}.
\item
\textit{MIC:} In the MIC based MIMO detector, the first stage can be the conventional SUMF detector, the linear ZF/MMSE detector, the SIC detector or any other suboptimum detector. \textit{The decisions made for all symbols $\bf s$} by the $(n-1)$th stage are employed as the input of the $n$th stage for the sake of cancelling the MUI. Note that historically, the MIC detector was developed independently of the PIC, although they share similar concepts. The MIC detector was first proposed by Timor for frequency-hopped CDMA (\gls{FH-CDMA})\index{FH-CDMA}\index{CDMA} systems\cite{Timor_1980:improved_decoding_CDMA, Timor_1981:multistage_decoding_CDMA}. Then, it was extensively studied in the context of both \textit{asynchronous DS-CDMA} systems \cite{Varansi_1988:multistage_detector_asynchronous_CDMA, Varanasi_1990:multistage_detection_asynchronous_CDMA} and \textit{synchronous DS-CDMA} systems\cite{Varanasi_1988:multistage_MUD_synchronous_CDMA_conf,Varanasi_1991:multistage_MUD_synchronous_CDMA, Moshavi_1996_multistage_linear_receivers_CDMA}. An analytical framework was proposed for adaptive MIC in\cite{Buehrer_1996:adaptive_multistage_IC_CDMA}.
\begin{table*}[tbp]
\setlength{\tabcolsep}{2pt}
\renewcommand{\arraystretch}{1.3}
\extrarowheight 3pt
\caption{Milestones in the development of interference cancellation MIMO detectors}
\label{Table:IC_MUD}
\centering
 \begin{footnotesize}
\begin{tabular}{|l|p{3.2cm}|p{10.1cm}|}
\hlinewd{0.9pt}
  Year & Authors & Contributions \\
\hline \hline
1974 - 1975 & Bergmans and Cover\cite{Bergmans_1974:interference_cancellation_earliest, Cover_1975:earliest_interference_cancellation_idea} & First demonstrated the effectiveness of the SIC concept from an information-theoretic perspective for broadcast channels.\\
\hline
1975  &  Carleial\cite{Carleial_1975:earliest_interference_cancellation_idea} & First characterized the effectiveness of the SIC principle from an information-theoretic perspective for interference channels.\\
\hline
1980-1981  & Timor\cite{Timor_1980:improved_decoding_CDMA, Timor_1981:multistage_decoding_CDMA} & First proposed a two-stage\cite{Timor_1980:improved_decoding_CDMA} MUD and a multistage\cite{Timor_1981:multistage_decoding_CDMA} MUD for FH-CDMA \index{FH-CDMA}\index{CDMA}systems employing multiple frequency-shift keying (\gls{MFSK}) modulation; showed that the mutual interference between the users of a FH-CDMA\index{FH-CDMA}\index{CDMA} system may be significantly reduced by making use of the well-defined algebraic structure of the users' signature waveforms, and that introducing an extra stage of interference cancellation may further improve the detector's performance.  \\
\hline
1990 & Viterbi\cite{Viterbi_1990:earliest_interference_cancellation_approach_capacity} & First conceived an SIC scheme for a convolutionally coded DS-CDMA system, and revealed that with the aid of the SIC based receiver the aggregate data rate of all simultaneous users can approach the Shannon capacity of the Gaussian channel.\\
\hline
1983 - 1990 & Kohno \textit{et al.} \cite{Kohno_1983:PIC_CDMA, Kohno_1990:IC_CDMA, Kohno_1990:PIC_CDMA_journal} & First proposed a PIC based MUD for CDMA\index{CDMA} systems. \\
\hline
1988 - 1991 & Varanasi \textit{et al.} \cite{Varansi_1988:multistage_detector_asynchronous_CDMA, Varanasi_1990:multistage_detection_asynchronous_CDMA, Varanasi_1988:multistage_MUD_synchronous_CDMA_conf,Varanasi_1991:multistage_MUD_synchronous_CDMA} & Designed and systematically characterized the MIC MUDs for both asynchronous and synchronous CDMA\index{CDMA} systems. \\
\hline
1989 - 1990 & Xie \textit{et al.}\cite{Xie_1989:Linear_MMSE_WLS_MUD_conf, Xie_1990:Linear_MMSE_WLS_MUD} & First proposed a DFD based MUD for \textit{asynchronous DS-CDMA systems}. \\
\hline
1991 - 1995 & Duel-Hallen \textit{et al.}\cite{Hallen_1991:DFD_conf, Hallen_1993:decorrelating_DFD_synch_CDMA, Hallen_1993:DFE_asynchronous_CDMA, Hallen:1995:DFD_asynchronous_CDMA} & Systematically investigated DFD MUDs conceived for both synchronous\cite{Hallen_1991:DFD_conf, Hallen_1993:decorrelating_DFD_synch_CDMA} and asynchronous\cite{Hallen_1993:DFE_asynchronous_CDMA, Hallen:1995:DFD_asynchronous_CDMA} CDMA\index{CDMA} systems from a receiver filter optimization perspective.\\
\hline
1996 - 1999 & Foschini \textit{et al.} \cite{Foschini:MIMO, Wolniansky:VBLAST, Golden_1999:VBLAST_first_journal} &  First discussed the ZF based SIC detector conceived for multiple-antenna aided SDM-MIMO systems. \\
\hline
2002 & Chin \textit{et al.} \cite{Chin_2000:PIC_BLAST} &  Extended the PIC detector to the multiple-antenna aided SDM-MIMO systems.\\
\hline
2003 & W{\"u}bben \textit{et al.} \cite{Wubben_2003:MMSE_DSNR_ordering} & Proposed a QR-decomposition (\gls{QRD}) based MMSE-SIC detector for multiple-antenna aided SDM-MIMO systems.\\
\hline
2003 & Guo \textit{et al.} \cite{Guo_2003:VLSI_VBLAST_detector} & Presented a VLSI implementation of the V-BLAST detector for a $4 \times 4$ MIMO system employing QPSK, and a detecting throughput of 128 Mbps was achieved. \\
\hline
2011 & Studer \textit{et al.} \cite{Studer_2011:PIC_MIMO_ASIC_implementation} & Reported an ASIC implementation of a soft-input soft-output MMSE based PIC detector for multiple-antenna aided SDM-MIMO systems.\\
\hline
\hlinewd{0.9pt}
\end{tabular}
\end{footnotesize}
\end{table*}
\item
\textit{DFD:} The concept of DFD is based on the same premise as that of the family of decision-feedback equalizers\cite{Hallen_1989:delayed_DFSE, Hallen_1992:Equalizers_for_MIMO_PAM}. Although DFD also relies on the SIC idea, its emphasis is on the receiver filter's optimization, which consists of a feedforward filter and a feedback filter optimization. The first DFD scheme was proposed by Xie \textit{et al.}\cite{Xie_1989:Linear_MMSE_WLS_MUD_conf, Xie_1990:Linear_MMSE_WLS_MUD} for \textit{asynchronous DS-CDMA systems}. Other major contributors of the subject of DFD include Duel-Hallen\cite{Hallen_1991:DFD_conf, Hallen_1993:decorrelating_DFD_synch_CDMA, Hallen_1993:DFE_asynchronous_CDMA, Hallen:1995:DFD_asynchronous_CDMA}, who comprehensively investigated decision-feedback MUDs designed for both synchronous\cite{Hallen_1991:DFD_conf, Hallen_1993:decorrelating_DFD_synch_CDMA} and asynchronous\cite{Hallen_1993:DFE_asynchronous_CDMA, Hallen:1995:DFD_asynchronous_CDMA} CDMA\index{CDMA} systems. Furthermore, Varanasi\cite{Varanasi:DFD}  analyzed the performance of a general class of DFDs using a new performance metric constituted by the probability that at least one user is detected erroneously, and also proposed algorithms for determining the most beneficial detection ordering.
\end{itemize}

The basic principles of the SIC/DFD detectors and the PIC/MIC detectors are illustrated in Fig. \ref{fig:SIC} and Fig. \ref{fig:PIC}, respectively. The main contributions to the development of the interference cancellation based MIMO detectors are summarized in Table \ref{Table:IC_MUD}.
A more comprehensive exposition of the above-mentioned MIMO detectors developed in the context of CDMA\index{CDMA} systems can be found in\cite{Verdu:MUD_book,Verdu_1989:recent_progress_in_MUD,Hallen_1995:MUD_CDMA_overview, Moshavi_1996:MUD_overview_CDMA, Wang_2009:Wireless_advanced_reception, Honig:advances_MUD_edited}.

\subsection{Tree-Search Based MIMO Detectors}
\label{subsec:tree_search_MUD:chap_intro}
The tree-search based MIMO detectors are arguably the most popular detectors investigated in the era of multi-antenna MIMO systems. This is because 1) the introduction of the powerful SD algorithm happened to coincide with the development of multi-antenna MIMO techniques; 2) some profound research results on the CLPS problem showed that the tree-search MIMO detectors enjoy significant design flexibility in terms of striking an attractive tradeoff between approaching the optimum ML performance and reducing the computational complexity.

Indeed, some tree-search based MUDs had been reported earlier in the context of CDMA\index{CDMA} systems\cite{Xie_1988:sequential_decoding_MUD_async_CDMA, Xie_1990:sequential_MUD_for_async_CDMA, Xie_1990:tree_search_MUD_conf, Xie_1993:joint_signal_detection_estimation, Wei_1997:tree-search_MUD_CDMA}. For example, the so-called (depth-first) stack sequential detection was proposed by Xie in \cite{Xie_1988:sequential_decoding_MUD_async_CDMA, Xie_1990:sequential_MUD_for_async_CDMA}, while the (breadth-first) $K$-best tree-search detection was proposed, again, by Xie in \cite{Xie_1990:tree_search_MUD_conf, Xie_1993:joint_signal_detection_estimation}, which was then further studied by Wei \textit{et al.} in \cite{Wei_1997:tree-search_MUD_CDMA}. Looking back to the earlier history, because of the convertibility between the trellis structure and the tree structure, the tree-search detection methods proposed for CDMA\index{CDMA} systems, including the classic $M$-algorithm\cite{Anderson_1969:M-algorithm_MSc_thesis, Anderson_1971:M-algorithm} and $T$-algorithm\cite{Waltmann_1965:T-algorithm, Simmons_1986:T-algorithm_trellis_decoder, Simmons_1990:T-algorithm_conf, Simmons_1990:T-algorithm_trellis_decoder}, were actually extensions of their counterparts used in trellis decoding\cite{Eyuboglu_1988:reduced_state_SE_DF, Waltmann_1965:T-algorithm, Fano_1963:fano_algorithm_sequential_decoding, Zigangirov_1966:stack_algorithm_sequential_decoding, Jelinek_1969:stack_algorithm_sequential_decoding, Massey_1972:variable_length_codes_Fano_metric, Anderson_1969:M-algorithm_MSc_thesis, Anderson_1971:M-algorithm, Simmons_1986:T-algorithm_trellis_decoder, Simmons_1990:T-algorithm_trellis_decoder, Simmons_1990:T-algorithm_conf, Anderson_1984:sequential_decoding_survey, Anderson_1989:sequential_decoding_survey, Pottie_1989:sequential_decoding_survey}.
However, these tree-search based detectors did not attract as much attention as the linear detectors and the interference cancellation aided detectors in the era of CDMA\index{CDMA} systems.
\begin{figure}[tbp]
\centering
\includegraphics[width=\linewidth]{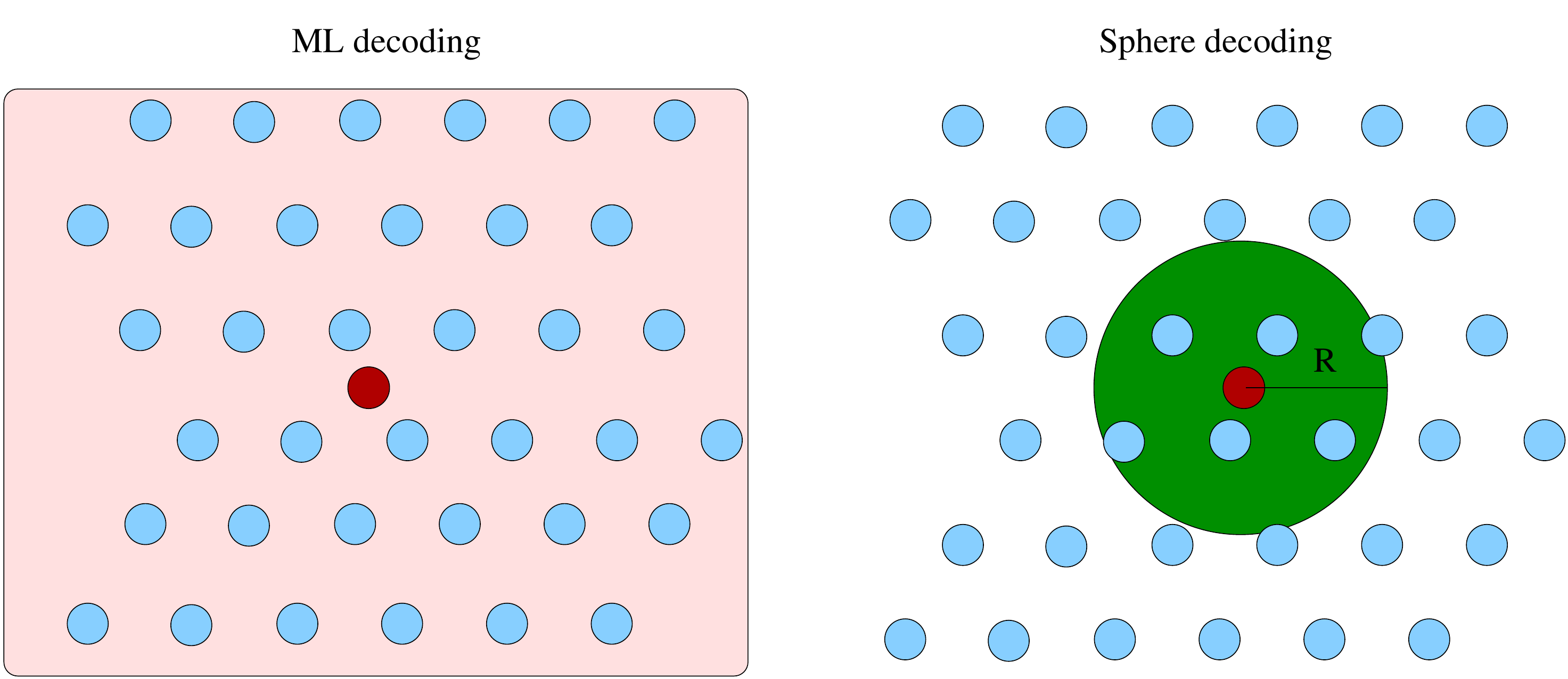}
\caption[]{The basic principle of the SD based MIMO detectors. In short, it solves the problem of finding the ``closest'' lattice point to a given vector $\bf y$ in the skewed and rotated (caused by the MIMO channel) lattice ${\bf Hs}$.} \label{fig:SD_MIMO}
\end{figure}

The research interests related to tree-search based MIMO detectors were largely stimulated by the seminal work of Viterbo \textit{et al.}\cite{Viterbo_1993:SD_conf, Viterbo_1999:SD}, who first applied the depth-first SD algorithm to the ML detection of multidimensional constellations transmitted over single-antenna fading channels. The basic principle of the SD algorithm is illustrated in Fig. \ref{fig:SD_MIMO}. Compared to the optimal brute-force search based ML decoding, the SD algorithm aims for reducing the computational complexity by only searching over the noiseless received signals that lie within a hypersphere of radius $R$ around the received signal. Note that before it was applied in digital communications, the SD algorithm, also known as the Fincke-Pohst algorithm, had been reported in\cite{Pohst_1981:computation_lattice_vectors_minimum, Fincke_1985:improved_lattice_vectors_calculation}. Later, Agrell \textit{et al.}\cite{Agrell:closest_point_search_in_lattice} proposed to employ the Schnorr-Euchner (\gls{SE}) refinement\cite{Schnorr_1994:lattice_basis_reduction} of the Fincke-Pohst algorithm\cite{Pohst_1981:computation_lattice_vectors_minimum, Fincke_1985:improved_lattice_vectors_calculation} for solving the \textit{CLPS problem}, and they concluded that the SE enumeration is more efficient than the Viterbo-Boutros (\gls{VB}) implementation\cite{Viterbo_1999:SD} of the SD algorithm. More recently, Damen \textit{et al.}\cite{Damen_2003:MLD_closest_lattice_point_search} proposed two improved SD algorithms for finding the closest lattice point, both of which were shown to offer a significant complexity reduction compared to the VB-SD\cite{Viterbo_1999:SD} and to the
SE-SD\cite{Agrell:closest_point_search_in_lattice}, respectively. There exist a number of other variants of the tree-search based MIMO detectors, which mainly fall into three categories: the depth-first tree-search detector\cite{Viterbo_1993:SD_conf, Viterbo_1999:SD, Agrell:closest_point_search_in_lattice,Damen_2003:MLD_closest_lattice_point_search,Hassibi_2005:SD_complexity_part_1, Vikalo_2005:SD_complexity_part_2,Burg_2005:VLSI_depth_first_SD, Jalden:SD_complexity_journal, Barbero:fixed_complexity_SD_journal,Jalden:FCSD_error_prob}, the breadth-first tree-search detector\cite{Wong_2002:K-best_SD_VLSI, Guo_2006:implementation_K_best_SD_MIMO, Chen_2007:K-best_VLSI, Wu_2008:early_pruning_SD} and the best-first tree-search detector\cite{Fukatani_2004:best_first_search, Lee_2007:short_path_SD, Okawado_2008:best_first_search, Stojnic_2008:speed_up_SD_via_infinity_H_norm, Kim_2010:best_first_search, Chang_2012:best_first_search, Chang_2012:best_first_search_A_algorithm, Shen_2012:best_first_tree_search_VLSI}.

The major momentum which propels the enormous research activities on tree-search based MIMO detectors is that they were shown to be capable of achieving near-ML performance, or even exact ML performance at the expense of significantly reduced complexity\cite{Hassibi_2001:SD_complexity_first, Hassibi_2002:SD_complexity, Hassibi_2003:ML_ILS_complexity, Hassibi_2005:SD_complexity_part_1, Vikalo_2005:SD_complexity_part_2}. However, we would like to emphasize that this claim is shown not true in general\cite{Jalden:SD_complexity_conf, Jalden:SD_complexity_journal}. More specifically, Hassibi and Vikalo\cite{Hassibi_2001:SD_complexity_first, Hassibi_2002:SD_complexity, Hassibi_2003:ML_ILS_complexity, Hassibi_2005:SD_complexity_part_1, Vikalo_2005:SD_complexity_part_2} first studied the expected complexity, averaged over the noise and over the lattice, of the Fincke-Pohst SD based MIMO detectors. It was claimed that although the worst-case complexity of the SD algorithm is exponential, the expected complexity of the SD algorithm is polynomial, in fact, often roughly cubic, for a wide range of SNRs and number of antennas. Contrary to this claim, Jald\'{e}n and Ottersten\cite{Jalden:SD_complexity_conf, Jalden:SD_complexity_journal} demonstrated that the expected complexity of SD based MIMO detectors is given by $\mathcal{O}(M^{\beta N_t})$, where $ \beta \in \left( {0,1} \right]$ is a small factor depending on the value of SNR. In other words, the expected complexity of the SD algorithm is still exponential for fixed SNR values. Therefore, in general the tree-search based MIMO detectors are not efficient for MIMO systems which operate under low-SNR condition and/or have a large number of inputs. Notably, in order to avoid the varying-complexity characteristic of tree-search based MIMO detectors, recently a suboptimal fixed-complexity SD (\gls{FCSD}) was proposed for MIMO systems\cite{Barbero:fixed_complexity_SD_journal}. It was shown that the FCSD achieves a near-ML performance with a complexity of $\mathcal{O}(M^{\sqrt {N_t}})$ \cite{Jalden:FCSD_error_prob} regardless of the specific SNR encountered, which represents an attractive solution of facilitating an efficient hardware implementation compared to the conventional exponential-complexity SD. The main contributions in the development of the depth-first tree-search MIMO detectors, the breadth-first tree-search MIMO detectors and the best-first tree-search MIMO detectors are summarized in Table \ref{Table:depth_first_tree_search}, Table \ref{Table:breadth_first_tree_search} and Table \ref{Table:best_first_tree_search}, respectively.
\begin{table*}[tbp]
\setlength{\tabcolsep}{2pt}
\renewcommand{\arraystretch}{1.3}
\extrarowheight 3pt
\caption{Milestones in the development of the tree-search MIMO detectors: Depth-first type}
\label{Table:depth_first_tree_search}
\centering
 \begin{footnotesize}
\begin{tabular}{|l|p{3.2cm}|p{10cm}|}
\hlinewd{0.9pt}
  Year & Authors & Contributions \\
\hline \hline
1981 - 1985 & Pohst and Fincke \cite{Pohst_1981:computation_lattice_vectors_minimum, Fincke_1985:improved_lattice_vectors_calculation} & First proposed the SD algorithm, which is hence known as the Fincke-Pohst algorithm, for calculating vectors of short length in a lattice at a reduced complexity; this work laid the mathematical foundation of applying the SD algorithm to the MIMO detection problem. \\
\hline
1988 - 1990 & Xie \textit{et al.} \cite{Xie_1988:sequential_decoding_MUD_async_CDMA, Xie_1990:sequential_MUD_for_async_CDMA} & First proposed a stack sequential decoding based MUD for asynchronous CDMA\index{CDMA} systems; this detector is essentially a depth-first tree-search MIMO detector.\\
\hline
1993 - 1999 & Viterbo \textit{et al.}  \cite{Viterbo_1993:SD_conf, Viterbo_1999:SD} & Applied the depth-first SD algorithm to the ML detection of multidimensional constellations transmitted over single-antenna fading channels, which largely stimulated the research interests of tree-search based MIMO detectors. \\
\hline
1994 & Schnorr and Euchner \cite{Schnorr_1994:lattice_basis_reduction} & Proposed a more efficient variation, known as SE refinement, of the Fincke-Pohst SD algorithm; the SE-SD algorithm was based on the lattice basis reduction philosophy and represents a popular solution to the MIMO detection problem. \\
\hline
2001 - 2003 & Hochwald \textit{et al.}\cite{Hochwald_2001:IDD_LSD_near_capacity_conf, Hochwald:SD_near_capacity} & Proposed a complex-valued SD and the list-SD (\gls{LSD}) for a FEC-coded MIMO using IDD receiver, showing that a near-capacity performance can be achieved with the aid of a soft-SD based IDD receiver. \\
\hline
2002 & Agrell \textit{et al.} \cite{Agrell:closest_point_search_in_lattice} & First proposed to use the SE refinement \cite{Schnorr_1994:lattice_basis_reduction} of the Fincke-Pohst SD algorithm\cite{Pohst_1981:computation_lattice_vectors_minimum, Fincke_1985:improved_lattice_vectors_calculation} to the \textit{CLPS problem}, and concluded that the SE enumeration technique is more efficient than the VB implementation\cite{Viterbo_1999:SD} of the SD algorithm designed for MIMO detection. \\
\hline
2003 & Damen \textit{et al.}\cite{Damen_2003:MLD_closest_lattice_point_search} & Proposed a pair of improved SD algorithms for finding the closest lattice point, both of which were shown to offer a significant complexity reduction compared to the VB-SD of\cite{Viterbo_1999:SD} and to the SE-SD of \cite{Agrell:closest_point_search_in_lattice}.\\
\hline
2001 - 2005 & Hassibi and Vikalo\cite{Hassibi_2001:SD_complexity_first, Hassibi_2002:SD_complexity, Hassibi_2003:ML_ILS_complexity, Hassibi_2005:SD_complexity_part_1, Vikalo_2005:SD_complexity_part_2} & Analyzed the expected complexity of the SD algorithm, and concluded that the expected complexity of SD algorithm is dependent on both the problem size and the SNR; showed that when the SNR is high, the expected complexity of SD can be approximated by a polynomial function for a small problem size. \\
\hline
2004 - 2005 & Jald\'{e}n and Ottersten\cite{Jalden:SD_complexity_conf, Jalden:SD_complexity_journal} & Further analyzed the expected complexity of the SD algorithm, and demonstrated that the expected complexity of the SD algorithm increases exponentially for a fixed SNR with a search-space, which contradicts previous claims; therefore, strictly speaking, the SD algorithm has an exponential lower bound in terms of both the expected complexity as well as the worst-case complexity, although it can be efficient at high SNRs and for problems of moderate size.  \\
\hline 
2004 & Garrett \textit{et al.}\cite{Garrett_2004:VLSI_ML_SD_silicon_complexity}  & First reported the VLSI implementation of a soft-output depth-first SD based detector for a $4\times 4$ 16QAM MIMO system, achieving 38.8 Mbps over a 5-MHz channel. \\ 
\hline
2005 & Burg \textit{et al.}\cite{Burg_2005:VLSI_depth_first_SD} & Proposed two ASIC implementations of depth-first MIMO SD. The first ASIC attains the ML performance with an average throughput of 73 Mb/s at an SNR of 20 dB; the second ASIC achieves a throughput of 170 Mb/s at the same SNR with only a negligible BER degradation. The proposed implementations rely on four key contributing factors to achieve high throughput and low complexity: depth-first tree traversal with radius reduction, implemented in a one-node-per-cycle architecture, the use of the $\it{l}^\infty$-instead of $\it{l}^2$-norm, and an efficient implementation of the enumeration approach. \\
\hline
2006 - 2008 & Barbero \textit{et al.} \cite{Barbero_2006:fixed_complexity_SD, Barbero:fixed_complexity_SD_journal} & Proposed a noise-level independent fixed-complexity tree-search MIMO detector, which overcomes the two main limitations of the SD from an implementation point of view: its variable complexity and its sequential nature. \\
\hline 
2008 & Studer \textit{et al.} \cite{Studer_2008:soft_SD_implementation} &  Presented the VLSI implementation of a soft-output depth-first SD based MIMO detector, which demonstrated that single tree-search, sorted QR-decomposition, channel matrix regularization, log-likelihood ratio clipping, and imposing runtime constraints are the key ingredients for realizing soft-output MIMO detectors with near max-log performance.\\
\hline
2009 & Jald\'{e}n \textit{et al.} \cite{Jalden:FCSD_error_prob} & Presented analytical study of the error probability of the fixed-complexity SD in MIMO systems having an arbitrary number of antennas, proving that it achieves the same diversity order as the ML detector, regardless of the constellation size used. \\
\hline
\hlinewd{0.9pt}
\end{tabular}
\end{footnotesize}
\end{table*}
\begin{table*}[tbp]
\setlength{\tabcolsep}{2pt}
\renewcommand{\arraystretch}{1.3}
\extrarowheight 3pt
\caption{Milestones in the development of the tree-search MIMO detectors: Breadth-first type}
\label{Table:breadth_first_tree_search}
\centering
 \begin{footnotesize}
\begin{tabular}{|l|p{3.3cm}|p{10cm}|}
\hlinewd{0.9pt}
  Year & Authors & Contributions \\
\hline \hline
1990 - 1993 & Xie \textit{et al.} \cite{Xie_1990:tree_search_MUD_conf, Xie_1993:joint_signal_detection_estimation, Wei_1997:tree-search_MUD_CDMA} & First conceived a breadth-first $K$-best tree search MUD for asynchronous CDMA\index{CDMA} systems; proposed a joint signal detection and parameter estimation scheme based on their breadth-first tree search MUD. \\
\hline
1997 & Wei \textit{et al.} \cite{Wei_1997:tree-search_MUD_CDMA} & Studied both the $M$-algorithm and the $T$-algorithm based breath-first tree-search MUD in the context of CDMA\index{CDMA} systems operating in fading channels. \\
\hline
2002 & Wong \textit{et al.} \cite{Wong_2002:K-best_SD_VLSI} & Proposed and implemented a breadth-first $K$-best tree-search MIMO detector using a VLSI architecture, which is capable of achieving a decoding throughput of 10 Mb/s at 100 MHz clock frequency in a 16-QAM aided $(4 \times 4)$-element SDM-MIMO system. \\
\hline
2004 - 2006 & Guo \textit{et al.} \cite{Guo_2004:K_best_implementation, Guo_2006:implementation_K_best_SD_MIMO} & Proposed and implemented both hard and soft SE-strategy based $K$-best tree-search MIMO detectors, which are capable of supporting up to 53.3 Mb/s throughput at 100 MHz clock frequency for a 16-QAM aided $(4 \times 4)$-element SDM-MIMO system.\\
\hline
2006 & Wenk \textit{et al.} \cite{Wenk_2006:K_best_SD_VLSI} & Presented a new VLSI architecture for the implementation of the K-best algorithm, which relies on a more parallel approach and the ASIC design achieves up to 424 Mbps throughput. \\
\hline
2007 & Chen \textit{et al.} \cite{Chen_2007:K-best_VLSI} & Reported a VLSI implementation of a soft-output breadth-first tree search aided MIMO detector for a $(4 \times 4)$-element MIMO system employing 64-QAM, which is capable of achieving a throughput of above 100 Mb/s. \\
\hline 
2010 & Patel \textit{et al.} \cite{Patel_2010:K_best_LTE_WiMAX_soft} &  Presented a VLSI architecture of a novel soft-output K-Best MIMO detector. This implementation attains a peak throughput of 655 Mbps for a $4\times 4$ 64-QAM MIMO system with 0.13um CMOS. Synthesis results in 65nm CMOS show the potential to support a sustained throughput up to 2 Gbps, which may meet the requirements of for mobile WiMAX and LTE-A standards. \\
\hline
\hlinewd{0.9pt}
\end{tabular}
\end{footnotesize}
\end{table*}
\begin{table*}[tbp]
\setlength{\tabcolsep}{2pt}
\renewcommand{\arraystretch}{1.3}
\extrarowheight 3pt
\caption{Milestones in the development of the tree-search MIMO detectors: best-first type}
\label{Table:best_first_tree_search}
\centering
 \begin{footnotesize}
\begin{tabular}{|l|p{2.8cm}|p{11.5cm}|}
\hlinewd{0.9pt}
  Year & Authors & Contributions \\
\hline \hline
  2004 & Fukatani \textit{et al.}\cite{Fukatani_2004:best_first_search} &  Applied Dijkstra's algorithm\cite{Dijkstra_1959:Dijkstra_algorithm} for reducing the complexity of the SD based MIMO detector at the expense of an increased storage size.  \\
\hline
  2004 & Xu \textit{et al.} \cite{Xu_2004:best_first_stack_algorithm} & Applied the stack  algorithm\cite{Jelinek_1969:stack_algorithm_sequential_decoding} to the best-first tree search based MIMO detector. \\
\hline
 2006 & Murugan \textit{et al.} \cite{Murugan_2006:unified_tree_search_sequential_decoder} & Proposed a unified framework for tree search decoding, which  encompasses all existing SDs as special cases, hence unifying the depth-first search, the breadth-first search and the best-first search based on the proposed framework. \\
\hline
  2012 & Chang \textit{et al.} \cite{Chang_2012:best_first_search} & A generalization of Dijkstra's algorithm was developed as a unified tree-search detection framework; the proposed framework incorporates a parameter triplet that allows the configuration of the memory usage, detection complexity and  the sorting dynamic associated with the tree-search algorithm; by tuning the different parameters, beneficial performance-complexity tradeoffs are attained and a fixed-complexity version can be conceived. \\
\hline
  2012 & Chang \textit{et al.} \cite{Chang_2012:best_first_search_A_algorithm} & First applied the A* algorithm to the best-first tree-search based MIMO detection problem. \\
\hline 
  2012 & Shen \textit{et al.} \cite{Shen_2012:best_first_tree_search_VLSI} & Proposed the algorithms and VLSI architectures for both the best-first soft- and hard-decision tree-search based MIMO decoders in the context of a $4 \times 4$ 64-QAM system using 65-nm CMOS technology at 333 MHz clock frequency.\\
\hline
\hlinewd{0.9pt}
\end{tabular}
\end{footnotesize}
\end{table*}

\subsection{Lattice-Reduction Aided Detectors}\label{subsec:LR_MIMO_detectord}
Lattice-reduction (LR) aided detectors constitute another important class of MIMO detectors, which rely on the algebraic concept of ``lattice'' originating from classic geometry and group theory\footnote{Recall that the lattice perspective -- many detection problems can be interpreted as the problem of finding the closest lattice point\cite{Mow_1994:lattice_MLSE, Agrell:closest_point_search_in_lattice, Damen_2003:MLD_closest_lattice_point_search} -- is also the foundation for the tree-search based MIMO detectors described in Section \ref{subsec:tree_search_MUD:chap_intro}.}. A lattice in $\mathbb{R}^n$ is a discrete subgroup of $\mathbb{R}^n$, which spans the real-valued vector space $\mathbb{R}^n$. Each lattice in $\mathbb{R}^n$ can be generated from a \textit{basis} of the vector space by forming all linear combinations with integer coefficients. In the MIMO transmission model of \eqref{eq:general_MIMO_model_matrix_form}, the received signal vector $\bf y$ is the noisy observation of the vector $\bf Hs$, which is in the lattice spanned by the column vectors $\bf H$, since both the real and imaginary parts of all the elements of $\bf s$ may be transformed to integers by shifting and scaling.

A lattice typically has multiple sets of basis vectors. Some bases spanning the same lattice as $\bf H$ might be ``closer'' to orthogonality than $\bf H$ itself. The process of finding a basis closer to orthogonality is referred to as LR. Theoretically, finding the optimal (closest to orthogonality) set of basis vectors is computationally expensive. Therefore, in practice LR algorithms typically aim for finding a ``better'' channel matrix ${\boldsymbol {\mathcal H}} = \bf HL$, where the real and imaginary parts of all the entries of the unimodular matrix $\bf L$ are integers and the determinant of $\bf L$ is $\pm 1$ or $\pm j$. As a result, the LR preprocessing technique is capable of improving the ``quality'' of the MIMO channel matrices. 

There is a variety of LR algorithms developed by mathematicians\cite{Wubben_2011:LR_magazine}. Some of them, such as Gaussian reduction\cite{Daude_1994:Gaussian_reduction}, Minkowski reduction and Korkine-Zolotareff (KZ) reduction\cite{KZ_1873:LR_reduction}, are capable of finding the optimal basis for a lattice, but they are computationally prohibitive for communications systems\cite{Yao_2003:LR_MIMO_detector, Zhang_2009:wireless_MIMO_receiver_design, KZ_1873:LR_reduction, SE_1994:LR_reduction}. Other well-known LR algorithms include the Lenstra-Lenstra-Lov\'{a}sz (\gls{LLL}) algorithm\cite{LLL_1982:LLL_LR_algorithm}, Seysen's algorithm\cite{Seysen_1993:SA_LLR_algorithm, Seethaler_2007:SA_LR_MIMO, zhang_2008:analysis} and Brun's algorithm\cite{Burg_2007:LR_Brun_algorithm, Seethaler_2006:Brun_LR_algorithm, Clarkson_1997:Brun_LR_algorithm}, which are all suboptimal. The most popular LR algorithm is the LLL algorithm, which does not guarantee to find the optimal basis, but it guarantees to find a basis within a factor to the optimal one in polynomial time\cite{Yao_2003:LR_MIMO_detector,Ling_2006:approximate_lattice_decoding, Ling_2007:LLL_LR, Jalden_2008:complexity_LLL}. For example, it was formally proved in \cite{Ling_2007:LLL_LR} that an upper bound on the average computational complexity of LLL is $\mathcal{O}(N_t^3\log N_t)$, where $N_t$ is the size of $\bf s$. Furthermore, a tighter upper bound of $\mathcal{O}(N_t^2\log \frac{N_t}{N_r-N_t+1})$ was found in \cite{Jalden_2008:complexity_LLL}, where $N_r$ is the size of $\bf y$. Note that the worst-case computational complexity of the LLL algorithm can be infinite\cite{Yao_2003:LR_MIMO_detector, Jalden_2008:complexity_LLL}. However, in practice the probability of the scenario which leads to this worst-case complexity is zero. There are mainly two variants for the LLL algorithm, namely the real-valued LLL\cite{Fischer_2003:MIMO_precoding_detection, Wubben_2004:LR_reduction_MMSE} and the complex-valued LLL\cite{Gan_2005:CLLL_conf, Gan_2009:CLLL_MIMO_detection, Ma_2008:performance_analysis_CLLL, Ma_2008:CLLL}. The real-valued LLL is applied to the real-valued MIMO system model of Section \ref{sec:complex_vs_real_model}, while the complex-valued LLL is designed for directly using it in the complex-valued MIMO system model. Additionally, the authors of \cite{Napias_1996:generalization_CLLL} proposed an LLL algorithm which is not only applicable to the complex-valued model, but also applicable to the Euclidean ring in general.    

In principle, LR can be combined with virtually all the other MIMO detectors to further improve their performance. For example, the LR technique was used in conjunction with traditional linear ZF and nonlinear ZF-SIC detectors in \cite{Yao_2002:LR_MIMO_detector}, as well as with linear MMSE and nonlinear MMSE-SIC detectors in\cite{Wubben_2004:LR_reduction_MMSE, Wubben_2004:LR_reduction_MMSE_WSA}, both achieving a substantial performance gain with little additional computational complexity. As a further advance over \cite{Yao_2002:LR_MIMO_detector}, a real-valued LLL-based LR algorithm was used in\cite{Fischer_2003:MIMO_precoding_detection}, which enables the application of the algorithm in MIMO systems with arbitrary numbers of dimensions. In addition, it was shown in \cite{Fischer_2003:MIMO_precoding_detection, Windpassinger_2004:LR_precoding_MIMO} that LR can also be favorably applied in MIMO systems that use precoding. The LLL based LR algorithm was shown to be capable of achieving the maximum attainable/full receive diversity in MIMO decoding\cite{Taherzadeh_2007:LLL_LR_receiver_diversity}. The VLSI implementation of the LR technique aided precoder and of the $K$-best MIMO detector was reported in \cite{Burg_2007:LR_Brun_algorithm} and \cite{Shabany_2008:LR_VLSI_implementation}, respectively. LR-aided soft-decision MIMO detectors are studied in \cite{Silvola_2006:soft_LR_MAP, Qi_2007:soft_LR, Ponnampalam_2007:LR_soft_outputs,Wei_2010:soft_LR_MIMO_detector}. Recently, element-based LR algorithms, which reduce the diagonal elements of the noise covariance matrix of linear detectors and thus enhance the asymptotic performance of linear detectors, were proposed for large-scale MIMO systems\cite{Zhou_2013:Lattice_reduction_LS_MIMO_detector}. The main contributions in the development of LR-aided MIMO detection are summarized in Table \ref{Table:LR_MIMO}. 
\begin{table*}[tbp]
\setlength{\tabcolsep}{2pt}
\renewcommand{\arraystretch}{1.3}
\extrarowheight 3pt
\caption{Milestones in the development of the LR-based MIMO detectors}
\label{Table:LR_MIMO}
\centering
 \begin{footnotesize}
\begin{tabular}{|l|p{2.8cm}|p{11.5cm}|}
\hlinewd{0.9pt}
  Year & Authors & Contributions \\
\hline \hline
 1982 & Lenstra \textit{et al.} \cite{LLL_1982:LLL_LR_algorithm} & First proposed the LLL algorithm for LR, which becomes the most popular LR algorithm used in practice. \\
\hline
  2002 & Yao \textit{et al.}\cite{Yao_2002:LR_MIMO_detector} &  First applied the LR technique in conjunction with the traditional linear ZF and nonlinear ZF-SIC detector, showing a substantial performance gain at a modest additional computational complexity.   \\
\hline 
  2003-2004 & Windpassinger and W{\"u}bben \textit{et al.} \cite{Fischer_2003:MIMO_precoding_detection, Wubben_2004:LR_reduction_MMSE} & Presented LR-aided MIMO detectors relying on real-valued LLL algorithms. \\ 
\hline
 2003-2004 & Windpassinger\textit{et al.} \cite{Fischer_2003:MIMO_precoding_detection, Windpassinger_2004:LR_precoding_MIMO} &  Proposed a real-valued LLL-based LR algorithm, which enables the application of the algorithm in MIMO systems having arbitrary numbers of dimensions. It was also shown that LR can be favorably applied in MIMO systems that use precoding. \\
\hline
  2004 & W{\"u}bben \textit{et al.} \cite{Wubben_2004:LR_reduction_MMSE, Wubben_2004:LR_reduction_MMSE_WSA} & Extended the LR-aided linear ZF and nonlinear ZF-SIC MIMO detectors to their MMSE based counterparts. \\
\hline
  2007 & Taherzadeh \textit{et al.} \cite{Taherzadeh_2007:LLL_LR_receiver_diversity} & Demonstrated that the LLL based LR algorithm is capable of achieving full receive diversity of MIMO decoding. \\
\hline
  2007-2008 & Ling and Jald{\'e}n \textit{et al.} \cite{Ling_2007:LLL_LR, Jalden_2008:complexity_LLL} & Provided upper bounds for the average computational complexity of the LLL algorithm, namely $\mathcal{O}(N_t^3\log N_t)$ and $\mathcal{O}(N_t^2\log \frac{N_t}{N_r-N_t+1})$, respectively.    \\
\hline 
  2007-2008 & Seethaler and Zhang \textit{et al.}\cite{Seethaler_2007:SA_LR_MIMO, zhang_2008:analysis} & Studied the performance of the Seysen's algorithm based LR techniques in MIMO detection problems. \\ 
\hline 
  2007 & Burg \textit{et al.} \cite{Burg_2007:LR_Brun_algorithm} & The first VLSI implementation of the LR technique relying on Brun's algorithm was reported. \\
\hline 
  2008 & Shabany \textit{et al.} \cite{Shabany_2008:LR_VLSI_implementation} &  Presented a VLSI implementation of the LR-aided $K$-best MIMO detector. \\
\hline 
  2008 & Gestner \textit{et al.} \cite{Gestner_2008:LR_VLSI_CLLL} & The first VLSI implementation of the LR technique relying on the complex-valued LLL algorithm was reported. \\
\hline 
  2005-2009 & Gan and Ma \textit{et al.} \cite{Gan_2005:CLLL_conf, Gan_2009:CLLL_MIMO_detection, Ma_2008:performance_analysis_CLLL, Ma_2008:CLLL} &  Proposed a number of complex-valued LLL algorithms which can be directly used in the complex-valued MIMO system model.\\
\hline 
  2006 - 2010 & Silvola, Qi, Ponnampalam and Zhang \textit{et al.} \cite{Silvola_2006:soft_LR_MAP, Qi_2007:soft_LR, Ponnampalam_2007:LR_soft_outputs,Wei_2010:soft_LR_MIMO_detector} & Studied a range of LR-aided soft-output MIMO detectors, including LR aided K-best\cite{Qi_2007:soft_LR}, LR-aided MAP\cite{Silvola_2006:soft_LR_MAP}, LR-aided fixed radius algorithm, fixed candidates algorithm, fixed memory-usage algorithm etc.\cite{Wei_2010:soft_LR_MIMO_detector}. \\ 
\hline 
  2013 & Zhou \textit{et al.} \cite{Zhou_2013:Lattice_reduction_LS_MIMO_detector} & Proposed a class of element-based LR algorithms, which reduce the diagonal elements of the noise covariance matrix of linear detectors and thus enhance the asymptotic performance of linear detectors, in large-scale MIMO systems having hundreds of BS antennas.  \\
\hline 

\hline
\hlinewd{0.9pt}
\end{tabular}
\end{footnotesize}
\end{table*}

\subsection{Probabilistic Data Association Based Detectors}
\label{subsec:PDA_MUD:chap_intro}
The PDA filter technique is a statistical approach originally invented by Bar-Shalom\cite{Shalom_1975:invention_PDA} in 1975 for the problem of target tracking and surveillance in a cluttered environment, where measurements are unlabelled and may be spurious. To elaborate a little further, it was developed for solving the problem of \textit{plot-to-track association} in a radar tracker. In this context, all of the potential candidates for association to a specific track are combined into a single statistically most likely update, taking account of the statistical distributions of both the tracking errors and the clutter, while assuming that only one of the candidates is the desired target with the rest of them representing false alarms. A major extension of the PDA filter is the joint probabilistic data association (\gls{JPDA}) filter\cite{Fortmann_1980:JPDA_invention_conf, Fortmann_1983:JPDA_sonar_tracking_journal}, which takes account of the situation that multiple targets are present out of all the potential candidates, and hence seeks to compute the joint decision probabilities for the multiple targets. In addition to their wide applications in radar, sonar, electro-optical sensor networks and navigation systems\cite{Shalom:PDA_original, Shalom_2004:estimation_applications_to_tracking_navigation, Shalom_2005:PDA_target_tracking_sonar_radar, Shalom2009:PDA_filter, Shalom_1975:invention_PDA, Fortmann_1980:JPDA_invention_conf, Fortmann_1983:JPDA_sonar_tracking_journal, Chang_1984:JPDA_multi_target_tracking, Chang_1986:JPDA_distributed_sensor_networks, Roecker_1993:suboptimal_JPDA, Musicki_1994:IPDA, Kershaw_1997:WSPDA, Blom_2000:PDA_avoid_track_coalescence, Kirubarajan_2004:PDA_tracking_clutter, Musicki_2004:JIPDA}, the PDA techniques have also been applied in the field of computer vision for solving the visual target tracking problem\cite{Cox_1993:review_statistical_data_association, Schulz_2001:tracking_robots, Schulz_2003:people_tracking_robots, Rasmussen_2001:PDA_in_computer_vision}.
\begin{figure*}[tbp]
   \centering
   \subfigure[Initial distribution]{
   \label{four_modal:subfig:a}
   \includegraphics[width=2.4in]{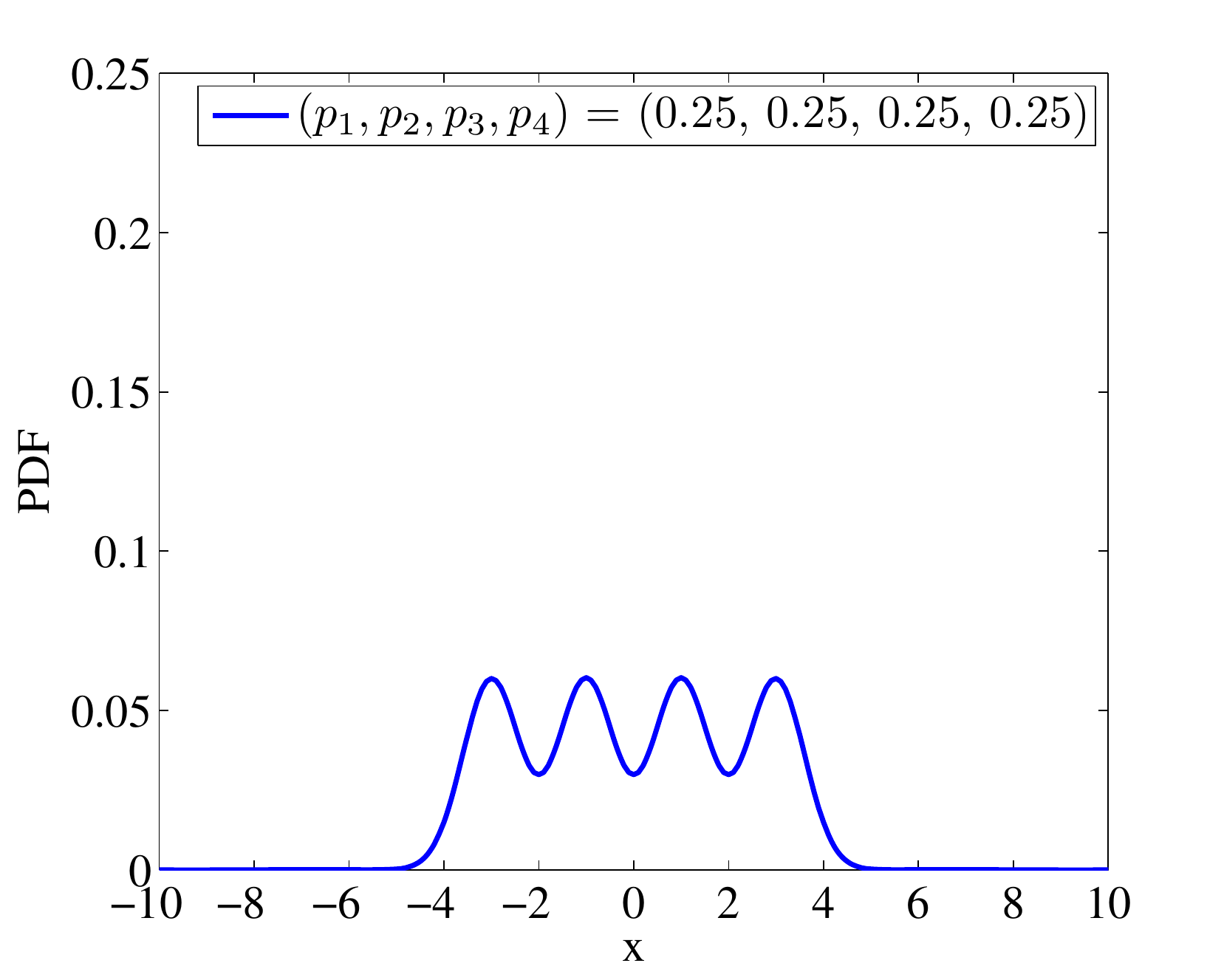}}
   \subfigure[Updated distribution: 1st stage]{
   \label{four_modal:subfig:b}
   \includegraphics[width=2.4in]{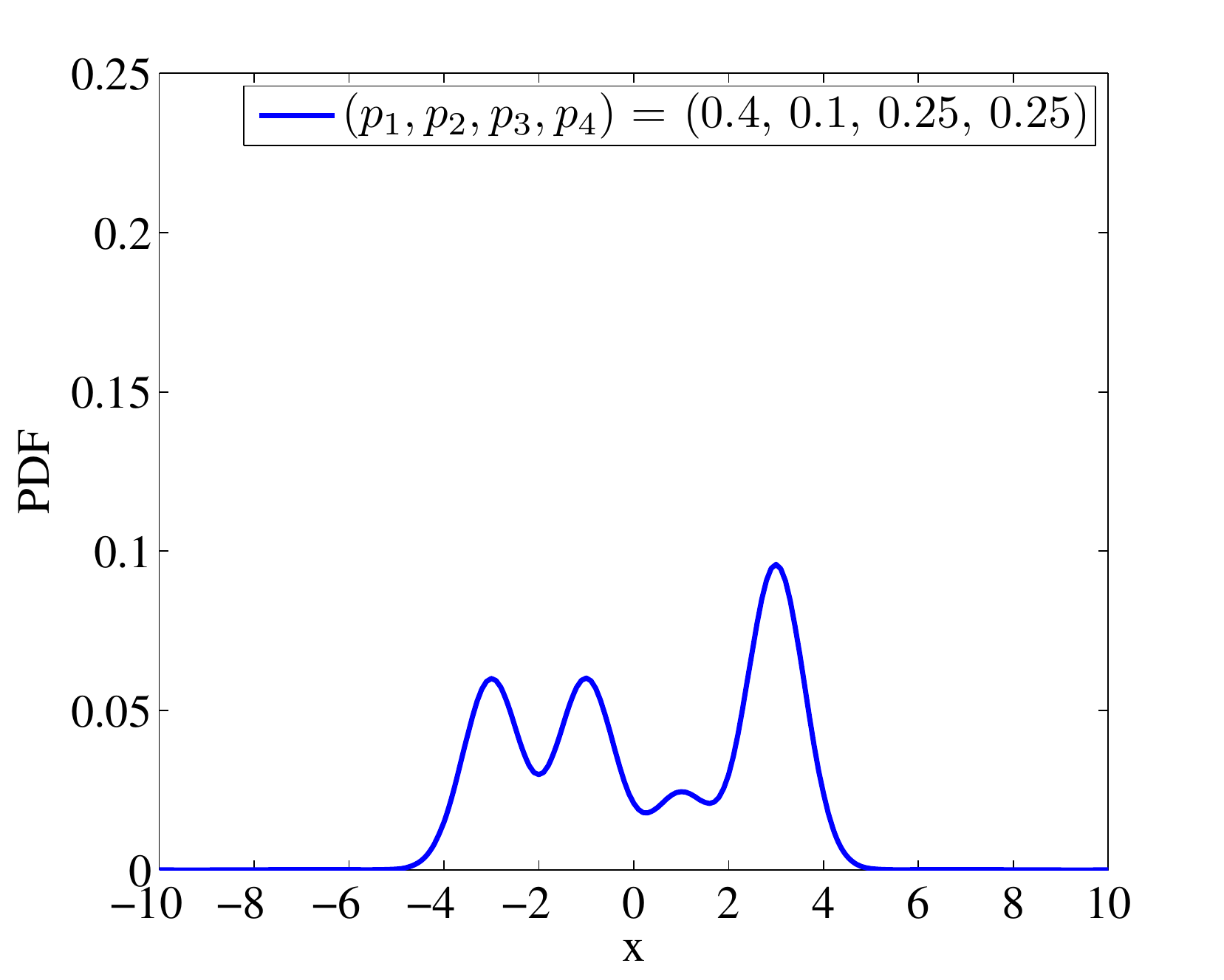}}
   \subfigure[Updated distribution: 2nd stage]{
   \label{four_modal:subfig:c}
   \includegraphics[width=2.4in]{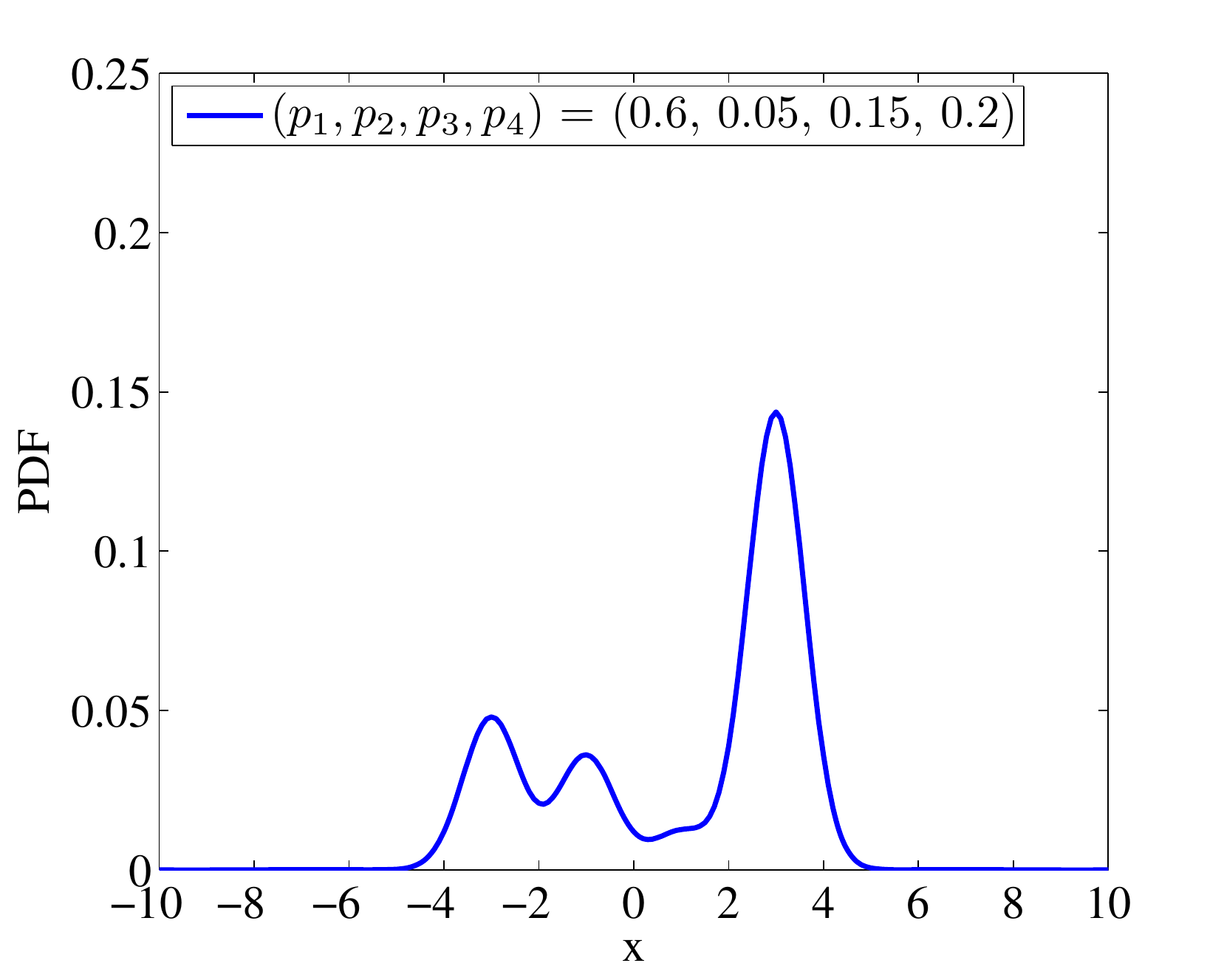}}
%    \subfigure[Updated distribution: 3rd stage]{
%    \label{four_modal:subfig:d}
%    \includegraphics[width=2.4in]{figures/fourmodal_apri_dot_8.eps}}
   \subfigure[Updated distribution: 3rd stage]{
   \label{four_modal:subfig:e}
   \includegraphics[width=2.4in]{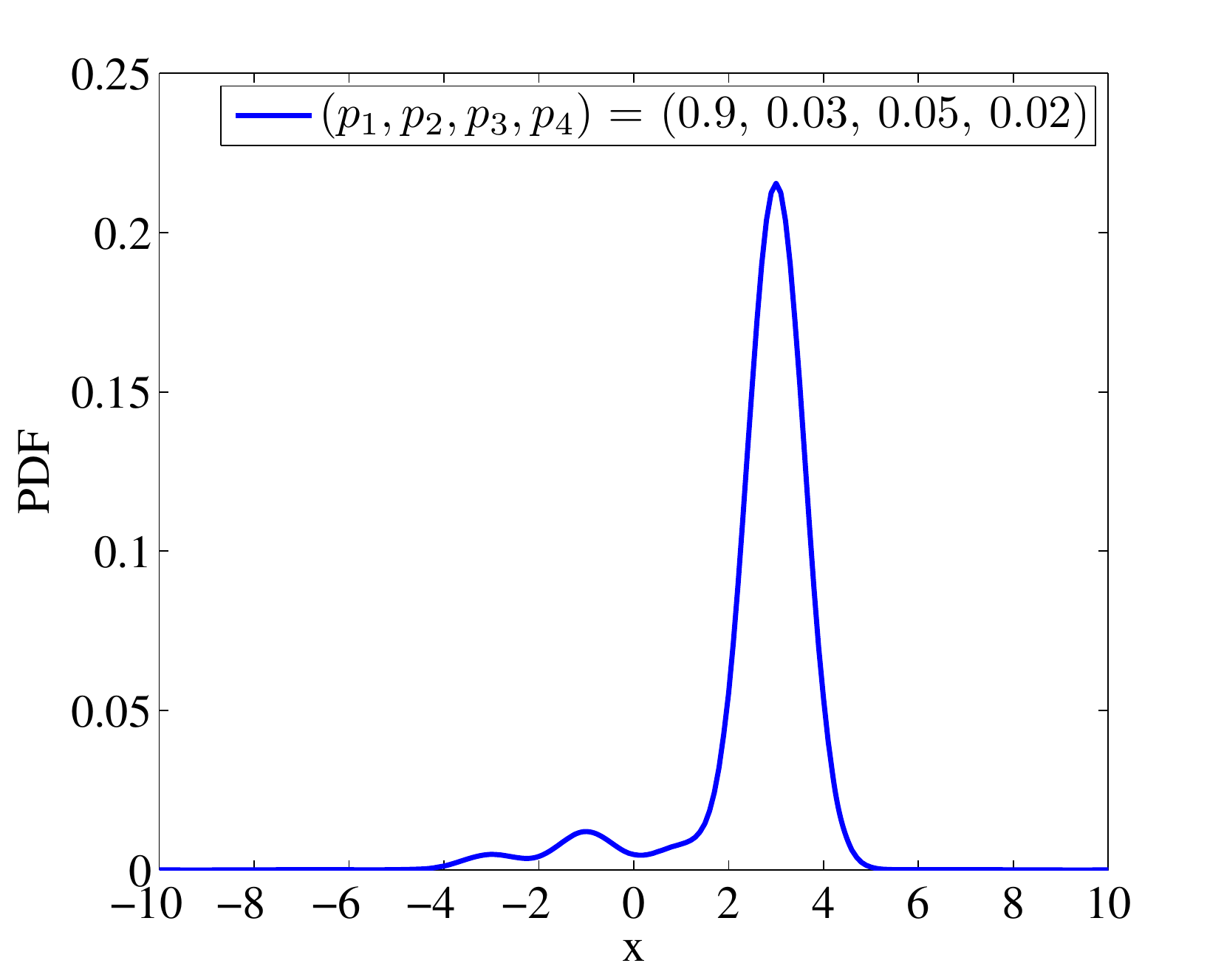}}
 \caption{\small{The basic principle of the PDA based MIMO detectors: An example process of approximating a single-variate four-modal Gaussian mixture distribution by a single Gaussian distribution.}} \label{four_modal}
\end{figure*}

The PDA approach may also be applied for solving challenging problems in digital communications. For example, it may be developed as a reduced-complexity design alternative to the optimal soft-decision aided MAP detectors/equalizers of MIMO channels\cite{Luo:PDA_Sync_CDMA, Luo:PDA_thesis, Pham:PDA_Async_CDMA, Luo_2003:sliding_window_PDA, Penghui_2003:iterative_PDA_MUD, Yin_2004:turbo_equalization_PDA, Huang_2004:generalized_PDA, Pham:GPDA, Liu:CPDA-apx, Liu:Kalman_PDA_freq_selective, Latsoudas_2005:hybrid_PDA_SD, Jia_2005:Gaussian_approximation_mixture_reduction_MIMO, Fricke:Impact_of_Gaussian_approximation, Shaoqian2005:turbo_PDA, Penghui_2006:asymptotic_optimum_PDA, Cai2006:iterative_PDA, Jia:CPDA, Cao:Relation_of_PDA_and_MMSE-SIC, Bavarian_2007:distributed_BS_cooperation_uplink, Bavarian_2008:distributed_BS_cooperation_uplink_journal, Jia_2008:multilevel_SGA_PDA, Grossmann_2008:turbo_equalization_PDA, Kim_2008:noncoherent_PDA, Shaoshi_2008:PDA_JD, Mohammed_2009:PDA_STBC, Bavarian:SDE_PDA_freq_selective, Shaoshi2011:B_PDA, Shaoshi2011:DPDA, Shaoshi2013:Turbo_AB_Log_PDA, Shaoshi_2013:EB_Log_PDA_journal}, and it is also applicable to channel estimation of MIMO systems\cite{Wang_2003:PDA_CE, Wang_2005:PDA_CE_journal}. Since we mainly focus on MIMO detection in this paper, a more detailed discussion of the PDA-based MIMO channel estimation will not be included in the sequel. As far as the PDA-based MIMO detection is concerned, it is Luo \textit{et al.}\cite{Luo:PDA_Sync_CDMA} who first applied the PDA approach to the MUD problem of BPSK-modulated synchronous CDMA\index{CDMA} systems in 2001, showing a near-optimum performance at a significantly lower computational complexity than the ML detector. Thereafter, the PDA-based detector was naturally extended to the scenario of BPSK-modulated asynchronous CDMA\index{CDMA} systems\cite{Pham:PDA_Async_CDMA, Luo_2003:sliding_window_PDA}. Recently, it was also extended to the symbol detection of QAM-aided SDM-MIMO systems \cite{Pham:GPDA, Liu:CPDA-apx, Liu:Kalman_PDA_freq_selective, Fricke:Impact_of_Gaussian_approximation, Jia:CPDA, Shaoshi2011:B_PDA}, striking an attractive tradeoff between the attainable performance and the complexity imposed. More specifically, in\cite{Pham:GPDA} a real-valued PDA (\gls{RPDA}) was formulated for $M$-QAM constellations, which is based on the equivalent real-valued MIMO signal model previously discussed in Section \ref{sec:complex_vs_real_model}. Additionally, in\cite{Liu:CPDA-apx} an approximate complex-valued PDA (\gls{A-CPDA}) detector was proposed, in which the complex-valued Gaussian distribution is approximately characterized by a matched mean and a matched covariance only. Furthermore, the pseudo-covariance, as defined by Neeser and Massey in \cite{Neeser:proper_Gaussian_process}, was employed in\cite{Jia:CPDA} to fully characterize the complex-valued Gaussian distribution, and the resultant formulation of complex-valued PDA (\gls{CPDA})\cite{Jia:CPDA} was shown to outperform both the RPDA\cite{Pham:GPDA} and the A-CPDA\cite{Liu:CPDA-apx}.

\begin{table*}[tbp]
\setlength{\tabcolsep}{2pt}
\renewcommand{\arraystretch}{1.3}
\extrarowheight 3pt
\caption{Milestones in the development of the PDA-based MIMO detectors}
\label{Table:PDA_detectors}
\centering
 \begin{small}
\begin{tabular}{|l|p{2.7cm}|p{11.4cm}|}
\hlinewd{0.9pt}
  Year & Authors & Contributions \\
\hline \hline
  2001 & Luo \textit{et al.}\cite{Luo:PDA_Sync_CDMA} & First applied the PDA filter technique to the MUD problem of synchronous CDMA\index{CDMA} systems, showing a near-optimum performance at a significantly reduced complexity. \\
\hline
  2002 & Pham \textit{et al.} \cite{Pham:PDA_Async_CDMA} & Proposed a PDA-Kalman MUD approach for asynchronous CDMA\index{CDMA} systems. \\
\hline
  2003 & Luo \textit{et al.}\cite{Luo_2003:sliding_window_PDA} & Conceived a sliding-window PDA based MUD approach for asynchronous CDMA\index{CDMA} systems. \\
\hline
  2003 & Tan \textit{et al.} \cite{Penghui_2003:iterative_PDA_MUD} & Designed a PDA-based IDD receiver for a coded CDMA\index{CDMA} system using BPSK modulation. \\
\hline
 2004 & Pham \textit{et al.} \cite{Pham:GPDA} & Extended the PDA detector to SDM-MIMO systems based on a real-valued signal model. \\
\hline
 2004 & Liu \textit{et al.} \cite{Liu:CPDA-apx} & Proposed a PDA-based soft equalization scheme for frequency-selective MIMO channels. \\
\hline
 2005 & Liu \textit{et al.} \cite{Liu:Kalman_PDA_freq_selective} & Extended the PDA-Kalman MUD approach of\cite{Pham:PDA_Async_CDMA} to the soft equalization of frequency-selective MIMO channels. \\
\hline
 2005 & Latsoudas \textit{et al.}\cite{Latsoudas_2005:hybrid_PDA_SD} & Proposed a hybrid MIMO detector that combined the SD and the PDA detectors. \\
\hline
2005 & Fricke \textit{et al.}\cite{Fricke:Impact_of_Gaussian_approximation} & Studied the impact of Gaussian approximation on the performance of the PDA based MIMO detector. \\
\hline
2006 & Jia \textit{et al.} \cite{Jia:CPDA} &  Proposed a complex-valued PDA (CPDA) detector which takes the pseudo-covariance into account during the derivation of the complex-valued PDA detector.\\
\hline
2008 & Kim \textit{et al.} \cite{Kim_2008:noncoherent_PDA} & Applied the PDA method as a component of an iterative receiver designed for non-coherent
MIMO systems.\\
\hline 
2008 & Yang \textit{et al.} \cite{Shaoshi_2008:PDA_JD} & Proposed a PDA detector for correlated source bits using joint detection of multiple consecutive symbol vectors.\\
\hline
2009 & Mohammed \textit{et al.} \cite{Mohammed_2009:PDA_STBC} & Applied the PDA algorithm to the problem of decoding large non-orthogonal space-time block codes (\glspl{STBC}). \\
\hline
2011 & Yang \textit{et al.} \cite{Shaoshi2011:B_PDA} & Proposed a unified direct bit-based PDA approach for detecting linear mapping based high-order rectangular QAM symbols, achieving a better performance at a lower computational complexity than the CPDA detector of\cite{Jia:CPDA}.\\
\hline
2011 & Yang \textit{et al.} \cite{Shaoshi2011:DPDA} & Proposed a distributed soft combining based PDA receiver for BS cooperation aided multi-cell multiuser MIMO systems.\\
\hline
2013 & Yang \textit{et al.} \cite{Shaoshi2013:Turbo_AB_Log_PDA, Shaoshi_2013:EB_Log_PDA_journal} & Investigated the PDA based iterative receiver design for FEC-coded MIMO systems: revealed that the outputs of the conventional PDA detectors are indeed the normalized symbol likelihoods rather than the true APPs; proposed a pair of PDA based MIMO iterative receivers, namely the approximate and the exact Bayesian theorem based iterative PDA receivers. \\
\hline
\hlinewd{0.9pt}
\end{tabular}
\end{small}
\end{table*}
In these PDA-based MIMO detectors/equalizers, the probabilities of the potential candidate symbols serve as the soft input/output information and are typically estimated relying on a self-iterative process. The key operation in this process is the iterative approximation of the interference-plus-noise term obeying a \textit{multimodal Gaussian mixture} distribution by an \textit{ever-updated} multivariate Gaussian distribution \cite{Luo:PDA_Sync_CDMA, Pham:GPDA, Liu:CPDA-apx, Liu:Kalman_PDA_freq_selective, Bavarian:SDE_PDA_freq_selective}. Therefore, the performance of the PDA based MIMO detectors is largely determined by the accuracy of the iterative Gaussian approximation, whose impact on the performance of the PDA based detectors was investigated in \cite{Fricke:Impact_of_Gaussian_approximation}. In order to further improve the accuracy of the Gaussian approximation, the authors of \cite{Shaoshi_2008:PDA_JD} proposed a PDA detector for correlated source bits using the joint detection of multiple consecutive symbol vectors. Additionally, in \cite{Shaoshi_2011:BPDA_conf, Shaoshi2011:B_PDA} a unified direct bit-based PDA approach was proposed for detecting linear mapping based high-order rectangular QAM symbols, which achieves a better performance at a lower computational complexity than the CPDA detector of\cite{Jia:CPDA}. Furthermore, the PDA based iterative receiver design of FEC-coded MIMO systems was investigated in \cite{Shaoshi2012gc:Turbo_AB_Log_PDA, Shaoshi2013:Turbo_AB_Log_PDA, Shaoshi2013gc:Turbo_EB_Log_PDA, Shaoshi_2013:EB_Log_PDA_journal}, where it was revealed that the outputs of the conventional PDA detectors in\cite{Luo:PDA_Sync_CDMA, Pham:GPDA, Liu:CPDA-apx, Jia:CPDA, Shaoshi_2008:PDA_JD, Shaoshi2011:B_PDA} are indeed the normalized symbol likelihoods, rather than the true APPs. Based on this insight, a pair of PDA based MIMO iterative receivers, namely the approximate and the exact Bayesian theorem based iterative PDA receivers were proposed in \cite{Shaoshi2012gc:Turbo_AB_Log_PDA, Shaoshi2013:Turbo_AB_Log_PDA} and \cite{Shaoshi2013gc:Turbo_EB_Log_PDA, Shaoshi_2013:EB_Log_PDA_journal}, respectively. Additionally, a distributed soft combining based PDA receiver was conceived in \cite{Shaoshi_2011:BS_cooperation_DPDA_conf, Shaoshi2011:DPDA} for BS cooperation aided multi-cell multiuser MIMO systems.

The advantages of the PDA based detectors are summarized as follows. 
\begin{itemize}
\item 
First, it may achieve a near-optimal detection performance in certain circumstances, for example in the context of FEC-uncoded CDMA\index{CDMA} systems\cite{Luo:PDA_Sync_CDMA, Pham:PDA_Async_CDMA, Luo_2003:sliding_window_PDA, Luo:PDA_thesis}. 
\item 
Second, it has a low complexity that increases no faster than $\mathcal{O}\left( M_i{{N_I}^4 } \right)$ per symbol vector, where $M_i$ is the number of PDA iterations, while $N_I$ represents either the number of users in CDMA\index{CDMA}\cite{Luo:PDA_Sync_CDMA, Pham:PDA_Async_CDMA, Luo_2003:sliding_window_PDA, Luo:PDA_thesis}, or the number of transmit antennas in  multi-antenna aided MIMO systems\cite{Pham:GPDA, Liu:CPDA-apx, Jia:CPDA}. 
\item 
Third, it is inherently an soft-input soft-output algorithm, which is eminently applicable in combination with FEC codes such as convolutional codes, turbo codes\cite{Berrou:Turbo_coding_conference, Berrou:Turbo_coding_journal} or low-density parity-check (\gls{LDPC}) codes\cite{Gallager:LDPC_code, MacKay:LDPC_code}. 
\item 
Furthermore, the higher the number of transmit antennas or users, the better its performance, provided that the channel is not overloaded ($N_I > N_O$) or rank-deficient \cite{Fricke:Impact_of_Gaussian_approximation}. However, due to its nature of approximation and iteration, the PDA based MIMO detector has not been well-understood compared to other mature MIMO detectors. 
\end{itemize}

For the sake of more explicitly clarifying the fundamental principle of the PDA based MIMO detector, its Gaussian approximation process is conceptually illustrated in Fig. \ref{four_modal}, which is based on the assumption that the interference-plus-noise term to be processed by the PDA detector obeys a single-variate multimodal (four-modal) Gaussian mixture distribution of $p_{\text{M}}(x) = p_1 \times f_1(x) +p_2 \times f_2(x) + p_3 \times
f_3(x) + p_4 \times f_4(x)$. Here, the ``single variate'' assumption indicates that only a single interfering symbol, say $s_i$, exists for the other symbol to be detected. In other words, a $(2\times 2)$-element VBLAST system is assumed.  
More specifically, the four-modal distribution observed in Fig. \ref{four_modal} stands for the case of a
4PAM-like scenario, which is a simplified real-valued example for $M$-QAM. More specifically, $p_{\text{M}}(x)$ is constructed by a mixture of four
constituent Gaussian distributions $f_1(x)$, $f_2(x)$, $f_3(x)$,
$f_4(x)$ having the same variance, but different means of $m_1 = -3$,
$m_2 = -1$, $m_3 = +1$, $m_4 = +3$ and different constituent
probabilities of $p_1$, $p_2$, $p_3$, $p_4$. The four constituent probabilities
correspond to the different probabilities that an interfering symbol
has the value $s_i = -3$, $s_i = -1$, $s_i = +1$ and $s_i = +3$,
respectively. 

The main contributions to the development of the PDA based MIMO detectors are summarized in Table \ref{Table:PDA_detectors}.

\subsection{Semidefinite Programming Relaxation Based Detectors}
\label{subsec:SDP_MUD:chap_intro}
In contrast to other MIMO detectors, the SDPR approach relies on a relaxation of the optimum MIMO detection problem to the mathematical model of semidefinite programming (\gls{SDP}) \cite{Vandenberg:semidefinite_programming,
Luo:SDR_simplest, Luo:Convex_Optimization_for_SP_Comm}, which is a subfield of convex optimization theory
\cite{Boyd:Convex_Optimization}. 

\begin{figure}[t]
\centering
\includegraphics[width=\linewidth]{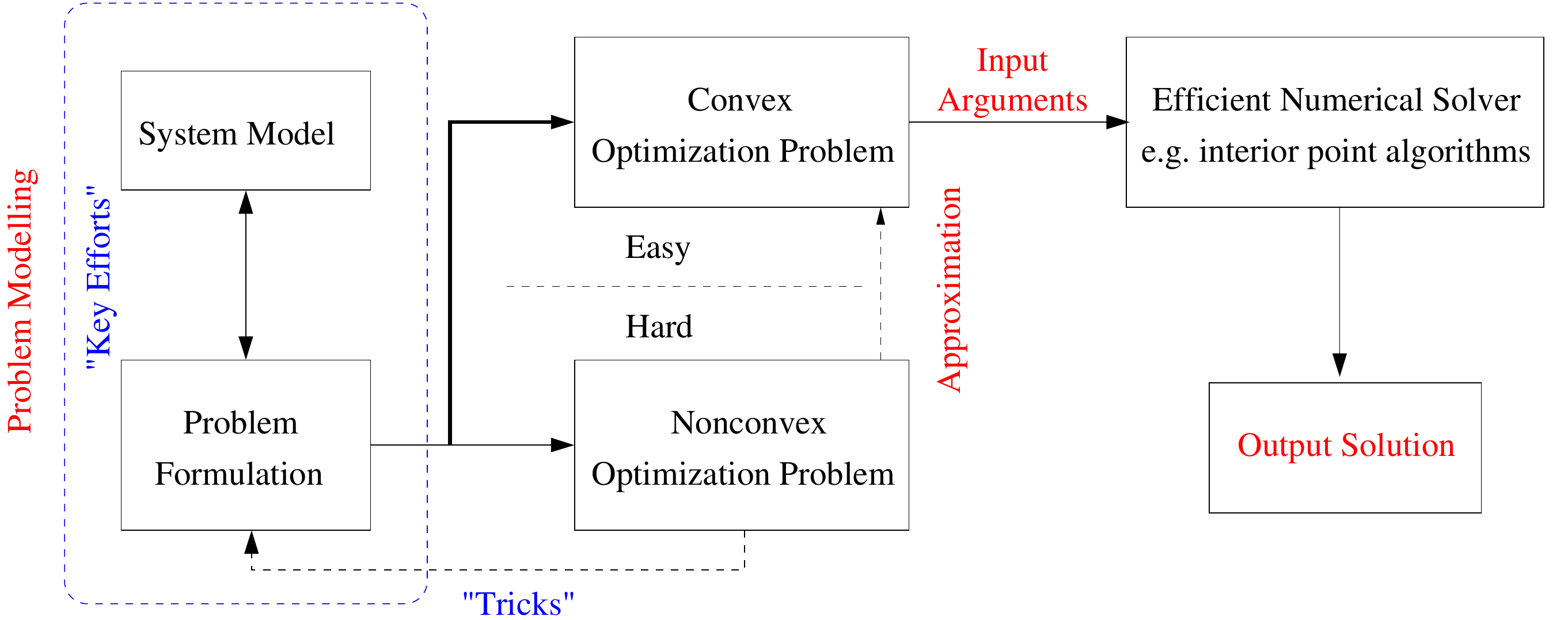}
\caption[]{ Framework of solving problems using convex optimization.} \label{fig:framework_convex_optimization}
\end{figure}

Convex optimization constitutes a subfield of the generic mathematical optimization problem. It studies the minimization of a convex objective function over convex sets. Fig. \ref{fig:framework_convex_optimization} illustrates the basic framework of solving mathematical optimization problems using convex optimization. If a mathematical optimization problem is identified as a convex
optimization problem, it is mathematically regarded as an “easy” problem, because powerful
numerical algorithms, such as the interior-point methods\cite{Helmberg:interior_point_algorithm}, exist for efficiently finding
the optimal solution of convex problems. Therefore, in mathematical optimization theory,
the dividing line between the family of “easy” and “difficult” problems is “convex versus
nonconvex”, rather than “linear versus nonlinear”. In other words, convex optimization problems are efficiently solvable, whereas nonconvex optimization problems are generally difficult
to solve. Convex optimization has a range of other important properties. For example, in convex optimization problems, every locally optimal solution constitutes the globally optimal solution, hence there is no risk of being trapped in a local optimum. Additionally, a
rigorous optimality condition and a duality theory exist for verifying the optimal nature of
a solution in convex optimization problems. For more details of convex optimization, please
refer to\cite{Luo:Convex_Optimization_for_SP_Comm, Boyd:Convex_Optimization}.

\begin{figure}[tbp]Fig
\centering
\includegraphics[width=\linewidth]{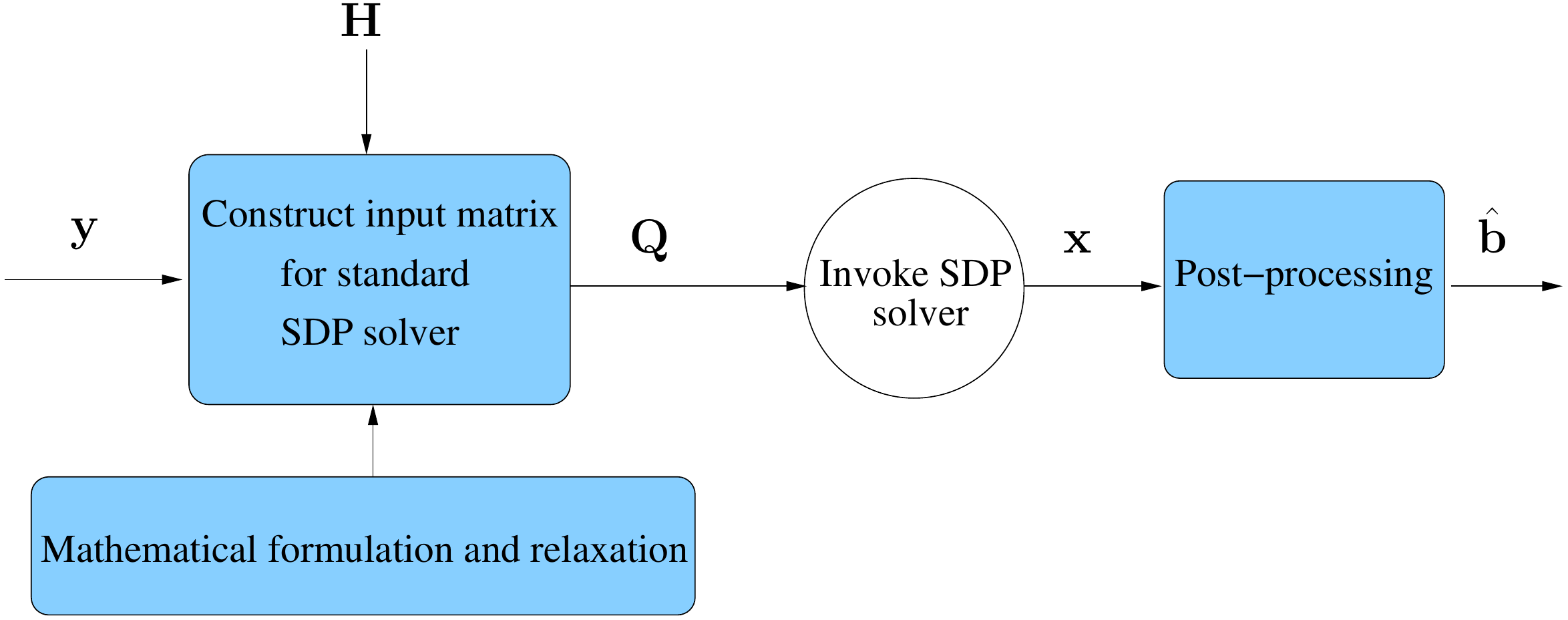}
\caption[]{The basic principle of the SDPR based MIMO detectors.} \label{fig:flow_chart_SDPR}
\end{figure}

The SDPR based MIMO detectors have recently received substantial research attention \cite{Penghui:SDP_CDMA, Ma:SDR_CDMA_BPSK,
Ma:SDR_CDMA_QPSK, Luo:SDR_performance_analysis, Jalden:SDR_diversity, Luo:SDP_MPSK, Ma:SDR_MPSK,
Wiesel:PI_SDR_16QAM, Yijin:SDR_16QAM_tight, Sidiropoulos:SDR_HOM, Mobasher:SDR_QAM_journal, Mao:SDR_LMR, Ma:equivalence_SDR}. The most
attractive characteristic of the SDPR-aided detectors is that they guarantee a so-called polynomial-time\footnote{The computational
complexity increases as a polynomial function of $N_I$.} worst-case computational complexity, while achieving a high performance in certain circumstances.   
\begin{table*}[t]
\setlength{\tabcolsep}{2pt}
\renewcommand{\arraystretch}{1.3}
\extrarowheight 3pt
\caption{Milestones in the development of the SDPR-based MIMO detectors}
\label{Table:SDPR_detector}
\centering
 \begin{small}
\begin{tabular}{|l|p{3.2cm}|p{10cm}|}
\hlinewd{0.9pt}
  Year & Authors & Contributions \\
\hline \hline
 2001 -2003 & Tan \textit{et al.} \cite{Penghui_2001:SDP_CDMA_conf, Penghui:SDP_CDMA}, Ma \textit{et al.} \cite{Ma_2001:SDP_CDMA_conf, Ma:SDR_CDMA_BPSK}, and  Wang \textit{et al.}  \cite{Wang_2001:SDP_CDMA_conf, Wang_2003:SDP_CDMA_journal} & These authors independently proposed a SDPR based MUD for BPSK-modulated synchronous CDMA\index{CDMA} systems; the eigen-decomposition based method of\cite{Penghui_2001:SDP_CDMA_conf, Penghui:SDP_CDMA, Wang_2001:SDP_CDMA_conf, Wang_2003:SDP_CDMA_journal} and the randomization method of\cite{Ma_2001:SDP_CDMA_conf, Ma:SDR_CDMA_BPSK} were proposed for converting the continuous-valued solution of the SDP problem into the binary decision output. Additionally, a cutting plane method was introduced for further improving the performance of the SDPR detector for systems supporting a large number of users\cite{Penghui_2001:SDP_CDMA_conf, Penghui:SDP_CDMA}; it was shown that the classic MUDs, such as the linear ZF/MMSE detector, can be interpreted as degenerate forms of the SDPR based MUD\cite{Ma_2001:SDP_CDMA_conf, Ma:SDR_CDMA_BPSK}. \\
\hline
2003 & Steingrimsson \textit{et al.} \cite{luo:soft_SDR} & Proposed a soft SDPR detector for an IDD receiver of QPSK-aided MIMO systems employing LDPC codes.\\
\hline
 2004 & Ma \textit{et al.} \cite{Ma:SDR_CDMA_QPSK} & Conceived a SDPR based MUD for BPSK/QPSK aided asynchronous CDMA\index{CDMA} systems with multiple receive antennas in frequency-selective fading environments; based on a flexible block alternating likelihood maximization (\gls{BALM}) principle, the large-scale ML detection problem was decomposed into smaller subproblems, and each subproblem was solved by the SDPR detector. \\
\hline
 2003 -2004 & Luo \textit{et al.}\cite{Luo:SDP_MPSK} and Ma \textit{et al.}\cite{Ma:SDR_MPSK} & Proposed SDPR detectors for general $M$-PSK aided synchronous CDMA\index{CDMA} systems.  \\
\hline
 2005  & Kisialiou \textit{et al.}\cite{Luo:SDR_performance_analysis} &  Provided the first analytical study of the SDPR detector for BPSK-aided MIMO systems; it was shown that the SDPR detector is capable of achieving the same BER performance as that of the ML detector in high-SNR scenarios, while at the low SNR region, the SDPR detector serves as a constant factor approximation to the ML detector in large systems. \\
\hline
2005 & Wiesel \textit{et al.} \cite{Wiesel:PI_SDR_16QAM} & Designed a PI-SDPR detector for 16-QAM aided MIMO systems, which can be extended to high-order $M$-QAM scenarios. \\
\hline
2006 & Sidiropoulos \textit{et al.} \cite{Sidiropoulos:SDR_HOM} & Advocated a BC-SDPR detector for employment in high-order $M$-QAM aided MIMO systems. \\
\hline
2007 & Mao \textit{et al.} \cite{Mao:SDR_LMR} & Proposed a VA-SDPR detector for $M$-QAM aided multicarrier CDMA(\gls{MC-CDMA})\index{MC-CDMA}\index{CDMA} systems; the method can directly operate at the bit-level in the context of linear mapping based $M$-QAM. \\
\hline
2007 & Mobasher \textit{et al.} \cite{Mobasher:SDR_QAM_journal} & Studied several variants of the SDPR detectors, and showed that it is possible to further improve the SDPR detector's performance by increasing their complexity. \\
\hline
2008 & Jald\'{e}n \textit{et al.} \cite{Jalden:SDR_diversity} & Analytically demonstrated that the SDPR based detector is capable of achieving full receive diversity order in BPSK-aided real-valued MIMO channels. \\
\hline
2009 & Ma \textit{et al.} \cite{Ma:equivalence_SDR} & Demonstrated that the PI-SDPR of \cite{Wiesel:PI_SDR_16QAM}, the BC-SDPR of \cite{Sidiropoulos:SDR_HOM}, and the VA-SDPR of \cite{Mao:SDR_LMR} are actually equivalent in the sense that they obtain the same symbol decisions, and hence exhibit an identical SER performance. \\
\hline 
2013 & Yang \textit{et al.}\cite{Shaoshi_2013:DVA_SPDR_journal} & Proposed a bit-based SDPR detector capable of directly detecting the nonlinear Gray mapping aided rectangular high-order QAM symbols, where the unequal error protection property (UEP) of QAM bits was exploited and the resultant SDPR detector outperforms that of \cite{Mao:SDR_LMR}.\\
\hline
\hlinewd{0.9pt}
\end{tabular}
\end{small}
\end{table*}
Most of the existing SDPR detectors are dependent on the specific modulation constellation. To elaborate a little further, SDPR was
first proposed for a BPSK-modulated CDMA\index{CDMA} system \cite{Penghui_2001:SDP_CDMA_conf, Penghui:SDP_CDMA, Ma_2001:SDP_CDMA_conf, Ma:SDR_CDMA_BPSK, Wang_2001:SDP_CDMA_conf, Wang_2003:SDP_CDMA_journal}, and then it was extended to quadrature phase-shift keying (\gls{QPSK})\cite{Ma:SDR_CDMA_QPSK}. Simulation results showed that the SDPR
detector is capable of achieving a near-ML BER performance, when using BPSK \cite{Penghui:SDP_CDMA} and QPSK
\cite{Ma:SDR_CDMA_QPSK}. The numerical and analytical results of \cite{Luo:SDR_performance_analysis, Jalden:SDR_diversity} confirmed
that the SDPR detector achieves the maximum possible diversity order, when using BPSK for transmission over a real-valued fading MIMO channel. The SDPR approach was also further developed for high-order modulation schemes, such as for $M$-ary PSK scenario in \cite{Luo:SDP_MPSK,
Ma:SDR_MPSK}, and for high-order rectangular QAM in \cite{Wiesel:PI_SDR_16QAM, Yijin:SDR_16QAM_tight, Sidiropoulos:SDR_HOM, Mobasher:SDR_QAM_journal, Mao:SDR_LMR}. As for the high-order QAM scenario, it was recently shown in \cite{Ma:equivalence_SDR} that the so-called polynomial-inspired
SDPR (\gls{PI-SDPR}) \cite{Wiesel:PI_SDR_16QAM}, the bound-constrained SDPR (\gls{BC-SDPR}) \cite{Sidiropoulos:SDR_HOM} and the virtually
antipodal SDPR (\gls{VA-SDPR}) \cite{Mao:SDR_LMR} are actually equivalent in the sense that they arrive at the same symbol decisions, and hence
they exhibit an identical SER performance.\footnote{More specifically, the solution equivalence
of the PI-SDPR and BC-SDPR schemes holds for $16$-QAM and $64$-QAM, while that between the BC-SDPR and VA-SDPR techniques holds for any $4^q$-QAM scheme, where $q$ is a
positive integer. The SDPR QAM detector of \cite{Mobasher:SDR_QAM_journal} exhibits a better performance than
that of \cite{Wiesel:PI_SDR_16QAM, Sidiropoulos:SDR_HOM, Mao:SDR_LMR}, but has a much higher complexity.} Furthermore, a bit-based SDPR detector capable of directly detecting the nonlinear Gray mapping aided rectangular high-order QAM symbols was proposed in\cite{Shaoshi_2013:DVA_SPDR_journal}, where the unequal error protection property (\gls{UEP}) of QAM bits was exploited and the resultant SDPR detector was shown to outperform that of \cite{Mao:SDR_LMR}. It should be noted, however, that for high-order modulation scenarios, the performance of the SDPR detectors is not as promising as that of the BPSK/QPSK scenario.  Therefore, there is a need to further improve the performance of the SDPR based MIMO detector designed for high-order QAM constellations, while maintaining its appealingly low computational complexity. The basic principle of SDPR based detectors is illustrated in Fig. \ref{fig:flow_chart_SDPR}, where the blue boxes represent the technical challenges. Furthermore, the main contributions to the development of the SDPR based MIMO detectors are summarized in Table \ref{Table:SDPR_detector}.

\subsection{Detection in Rank-Deficient and Overloaded MIMO Systems}
For MIMO detection, typically it is preferable to have a full-rank channel matrix, namely rank$({\bf H}) = N_I$ or $N_O$, In CDMA systems, this requirement may be satisfied by using well-designed spreading codes. In multi-antenna SDM systems, when an ideal rich scattering multipath environment is assumed, typically independently fading communications channels are encountered between each transmit/receive antenna pair. Then, the full-rank requirement may also be satisfied. However, in some propagation scenarios, the channel matrix $\bf H$ may not be of full-rank. For example, if the spatial separation between the antenna elements of the transmitter and/or the receiver is not large enough and hence the angular spread is small, the strong correlation between the antenna elements results in a rank-deficient channel matrix, i.e. we have rank$({\bf H})< \min (N_t, N_r)$. Hence, the spatial degrees of freedom available are reduced, which translates into a decreased MIMO capacity\cite{Foschini_1999:antenna_correlation_MIMO_capacity, Shiu_2000:fading_correlation_MIMO, Chizhik_2000:antenna_correlation_MIMO, Tse_2002:MIMO_capacity_correlation, Molisch_2002:correlated_MIMO_capacity_measured_channel, Shin_2003:keyhole_correlation_double_scattering_IT}. Furthermore, even if the spatial separation between antenna elements is sufficiently large, it is still possible that $\bf H$ is rank-deficient. This is due to the so-called ``keyhole/pinhole effects''\cite{Chizhik_2000:keyhole_letter, Chizhik_2002:keyhole_effect, Gesbert_2000:keyhole, Shin_2003:keyhole_correlation_double_scattering_IT, Molisch_2003:keyhole_measurement_letter, Molisch_2006:keyhole_MIMO_capacity_measurement}, which may be simply understood as a diffraction phenomenon, where a large obstacle with a small keyhole punched through it is placed between the MIMO transmitter and receiver, hence the only channel the radio wave can propagate through to the receiver is the keyhole. Due to this effect, the channel matrix $\bf H$ is degenerate and has only a single degree of freedom, i.e. we have rank$({\bf H}) = 1$, even though the entries of $\bf H$ are uncorrelated.   

Another preferable condition for the detection in CDMA and SDM-MIMO systems is that the system is \textit{not} overloaded. Then, the channel matrix is ``fat'' and does not have full column-rank (but it may still have full row-rank.). In the multi-antenna scenario, this means that $N_t \le N_r$, while in CDMA systems, it means that the number of users is higher than the dimension of the signal space/the processing gain of the system. As far as MIMO detection is considered, both the rank-deficient scenario discussed above and the overloaded scenario face the common challenge that the standard versions of most of the representative MIMO detectors, such as the linear ZF/MMSE detector, the ZF/MMSE-SIC detector and the original linear-decorrelation based PDA detector\cite{Luo:PDA_Sync_CDMA, Pham:GPDA} that invoke the inverse of $\bf H$, and the standard SD detector that invokes standard QRD or Cholesky's factorization\cite{Viterbo_1999:SD}, usually provide an unacceptably poor performance, because they are invoked for finding the solution of an under-determined linear system subject to random noise.  

Several strategies have been proposed to circumvent this predicament, such as the ``pseudo-inverse'' based linear detection\cite{Lupas_1989:linear_MUD_synchrohous_CDMA, Wolniansky:VBLAST}, the group detection\cite{Varanasi_1995:group_detection, Schlegel_1996:MUD_projection, Varanasi_2000:diversity_order_group_detection, Varanasi_2003:overloaded_MUD_CDMA, Zarikoff_2007:group_detection_overloaded_MIMO, Krause_2011:Group_detection_MIMO_overloaded}, the generalized SD detector\cite{Damen_2000:generalized_SD_STC, Damen_2003:MLD_closest_lattice_point_search, Dayal_2003:afast_SD_rank_deficient, Cui_2005:SD_rank_deficient_letter, Yang_2005:SD_rank_deficient, Hanzo_2006:SD_rank_deficient, Paulraj_2007:SD_rank_deficient, Wang2008:center_shifting_SD, Tian_2009:overloaded_MIMO_detection, Kanaras_2010:SD_ill_conditioned}, the modified non-decorrelated PDA detection\cite{Fricke:Impact_of_Gaussian_approximation, Liu:CPDA-apx, Cao:Relation_of_PDA_and_MMSE-SIC, Shaoshi2011:B_PDA}, the modified SDPR detection\cite{Cui_2006:SDR_overloaded,Tian_2009:overloaded_MIMO_detection, Shaoshi_2013:DVA_SPDR_journal}, the metaheuristics based detection\cite{Juntti_1997:genetic_algorithm_MUD_CDMA, Ergun_1998:genetic_algorithm_MUD, Ergun_2000:genetic_algorithm_MUD_journal, Wang_1998:genetic_algorithm_MUD, Yen_2001:genetic_algorithm_joint_MUD_CE,Yen_2003:genetic_algorithm_MUD_synch,Yen_2004:genetic_algorithm_MUD_asynch,Colman_2008:GA_overloaded, Dorigo_2006:ACO, Sun_2004:PSO, Hijazi_2004:ACO_MUD, Xu_2008:ACO_MUD, Xu_2009:ACO_MUD_letter, Lain_2010:ACO_overloaded,Tasneem_2012:overloaded_ACO_MIMO, Haris_2012:heuristics_MIMO_overloaded, Yang_2004:PSO, Zhao_2004:PSO_detection, Chen_2010:PSO_MIMO_detection, Taufik_2010:SA_MUD, Xia_2008:SA_MUD,    Datta_2012:rank_deficient_LS_MIMO_detection} etc. It is possible to design various scenario-dependent MIMO detectors for the rank-deficient and overloaded MIMO systems. However, it seems that the group detection strategy\cite{Varanasi_1995:group_detection, Schlegel_1996:MUD_projection, Varanasi_2000:diversity_order_group_detection, Varanasi_2003:overloaded_MUD_CDMA, Zarikoff_2007:group_detection_overloaded_MIMO, Krause_2011:Group_detection_MIMO_overloaded} and the search-based detection, regardless of the ML detector, the generalized SD detector\cite{Damen_2000:generalized_SD_STC, Damen_2003:MLD_closest_lattice_point_search, Dayal_2003:afast_SD_rank_deficient, Cui_2005:SD_rank_deficient_letter, Yang_2005:SD_rank_deficient, Hanzo_2006:SD_rank_deficient, Paulraj_2007:SD_rank_deficient, Wang2008:center_shifting_SD, Tian_2009:overloaded_MIMO_detection, Kanaras_2010:SD_ill_conditioned} and the metaheuristics based detector\cite{Juntti_1997:genetic_algorithm_MUD_CDMA, Ergun_1998:genetic_algorithm_MUD, Ergun_2000:genetic_algorithm_MUD_journal, Wang_1998:genetic_algorithm_MUD, Yen_2001:genetic_algorithm_joint_MUD_CE,Yen_2003:genetic_algorithm_MUD_synch,Yen_2004:genetic_algorithm_MUD_asynch,Colman_2008:GA_overloaded, Dorigo_2006:ACO, Sun_2004:PSO, Hijazi_2004:ACO_MUD, Xu_2008:ACO_MUD, Xu_2009:ACO_MUD_letter, Lain_2010:ACO_overloaded,Tasneem_2012:overloaded_ACO_MIMO, Haris_2012:heuristics_MIMO_overloaded, Yang_2004:PSO, Zhao_2004:PSO_detection, Chen_2010:PSO_MIMO_detection, Taufik_2010:SA_MUD, Xia_2008:SA_MUD,    Datta_2012:rank_deficient_LS_MIMO_detection}, are particularly suitable for rank-deficient and overloaded MIMO scenarios.

\subsection{Impact of Soft-Decision and Transmit Preprocessing on MIMO Detection}
In previous sections we aim for understanding the fundamental properties of MIMO detection algorithms. However, if we look at the entire process of communication, the assumption that only the receiver is responsible for signal recovery represents a passive and incomplete strategy. In fact, almost all practical systems invoke some form of encoding or transmit preprocessing, such as FEC, space-time coding and precoding/beamforming, to actively improve the performance of signal recovery or to reduce the detector's computational complexity from the transmitter side. 

To elaborate a little further, when FEC is used, tremendous efforts have been devoted to designing soft-input soft-output MIMO detectors that can fit into the powerful ``turbo processing principle''\cite{Berrou:Turbo_coding_conference, Berrou:Turbo_coding_journal, Hagenauer:iterative_decoding_concatenated_codes, Hagenauer1997:turbo_principle, Hagenauer1996:iterative_decoding_block_conv_codes} based IDD receiver architecture conceived for achieving near-optimum performance. All the MIMO detectors reviewed in Section \ref{sec:history_MIMO_detection:chap_intro} have their soft-decision versions to fit into IDD receivers. The representative contributions to iterative MIMO detection and decoding include: the optimal MAP detector based iterative receiver\cite{Moher_1997:turbo_MUD, Moher_1998:earliest_iterative_MUD_MAP, Reed1997:IDD_conf_PIMRC, Reed_1998:iterative_MUD_coded_CDMA, Tarkoy_1997:IDD_MUD}, the expectation-maximization (\gls{EM}) algorithm based soft-decision MUD \cite{Nelson_1996:EM_MUD}, the soft-decision MMSE detector assisted iterative receiver\cite{Wang_Xiaodong_1999:iterative_detection, Gamal_2000:iterative_MUD_coded_CDMA, Lee_2006:IDD_vblast}, the soft-decision SD based MIMO iterative receiver\cite{Hochwald:SD_near_capacity, DeJong_2005:iterative_tree_search, Studer_2008:soft_SD_implementation, Wang2008:center_shifting_SD, Barbero:soft_fixed_complexity_SD, Choi:sphere_decoder_look_ahead}, the PDA detector based iterative receiver\cite{Penghui_2003:iterative_PDA_MUD, Grossmann_2008:turbo_equalization_PDA, Shaoshi2013:Turbo_AB_Log_PDA, Shaoshi_2013:EB_Log_PDA_journal}, the soft-decision SDPR detector aided MIMO iterative receiver\cite{luo:soft_SDR, Nekuii:without_list_soft_SDR_QPSK_journal}, the soft-decision multiple symbol differential SD (\gls{MSDSD}) detector based non-coherent iterative receiver\cite{Pauli_2006:turbo_DPSK_MSDSD},  and the soft-decision iterative receiver for LS-MIMO systems\cite{Wu_2014:message_passing_soft_iterative_LS_MIMO_detection, Shen_2012:best_first_tree_search_VLSI} discussed in Section \ref{Sec:detection_in_massive_MIMO}. Yet another important contribution to IDD design is the extrinsic information transfer (\gls{EXIT}) chart invented by ten Brink\cite{ten_brink2001:EXIT_chart, hagenauer2004:EXIT_chart}, which is a powerful tool conceived not only for analyzing the convergence behavior of iterative receivers, but also for assisting near-capacity wireless system design\cite{Ariyavisitakul_2000:turbo_ST_processing, Hochwald:SD_near_capacity, Baro_2003:LISS_iterative_detection, Haykin_2004:turbo_MIMO, Hagenauer_2007:LISS_algorithm,  Peel_2005:vector_perturbation_near_capacity_part_1, Hochwald_2005:vector_perturbation_near_capacity_part_2, Wang_2006:approaching_MIMO_capacity_hard_SD, Alamri_2009:near_capacity_three_stage_iterative_detection, Hanzo_2011:near_capacity_proceeding_IEEE, Shinya_2012:CST_near_capacity_transceiver, Suthisopapan_2012:capacity_approaching_LDPC_MMSE_detection_LS_MIMO}. For more details on designing iterative MIMO receivers, please refer to\cite{Shinya_2012:CST_near_capacity_transceiver, Moher_1998:earliest_iterative_MUD_MAP, Reed_1998:iterative_MUD_coded_CDMA, Wang_Xiaodong_1999:iterative_detection, Hochwald:SD_near_capacity, Hanzo2009:near_capacity_multi_func_MIMO, Hanzo_2011:near_capacity_book, Hanzo:2011:near_capacity_variable_length_coding, Wymeersch2007:iterative_receiver_design}.    

Additionally, when space-time coding is employed, as we discussed in Section \ref{sec:definition_MIMO_detection:chap_intro}, the optimal ML decoding can be implemented with a simple separate symbol-by-symbol decoding strategy for orthogonal STBCs and with a linear-complexity pairwise decoding strategy for quasi-orthogonal STBCs\cite{Jafarkhani2005:STC_book}. As a result, the MIMO detection problem does not constitute a grave challenge for STBC aided MIMO systems. Similarly, when precoding/beamforming techniques\cite{Sampath_2001:MIMO_precoding, Fischer_2003:MIMO_precoding_detection, Palomar_2003:joint_TX_RX_beamforming, Love_2005:limited_feedback_precoding, Wiesel_2006:linear_precoder_MIMO,Sadek_2007:precoding_MU_MIMO, Paulraj_2007:MIMO_precoding_magazine, Heath_2009:network_MIMO_precoding} are employed in SDM-MIMO systems, the interference between the transmit antennas may be significantly mitigated or even completely removed (when using ZF-based linear precoding). As a result, the signal detection task of a precoded MIMO system becomes less challenging compared to that of SDM-MIMO systems invoking no preprocessing. The key insight gained here is that we can design an encoder or precoder to improve the performance or to reduce the computational complexity of decoders/detectors. 

\subsection{Guidelines on Choosing the Right MIMO Detectors}
As we mentioned in Section \ref{subsubsec:optimality_criteria}, the optimality of MIMO detectors strongly depends both on the criteria of ``goodness'' and on the assumptions made for specific application scenarios. Each type of MIMO detector has a different performance-and-complexity profile\footnote{Generally, ``performance'' and ``complexity'' may be interpreted in various ways. For example, the ``performance'' can be error probability, robustness to system imperfections, configuration flexibility, application generality etc., while the ``complexity'' could be computational complexity, hardware/silicon complexity and so on.}, and each of them has its own pros and cons. Therefore, in general there is no simple answer as to which algorithm is the best. In what follows, we first provide a qualitative comparison of the performance and complexity characteristics of the MIMO detectors reviewed, and then summarize their analytical performance and complexity results in Table \ref{table_performance_complexity_compare}. 
\begin{table*}[tbp]
\setlength{\tabcolsep}{3pt}
\renewcommand{\arraystretch}{1.3}
\extrarowheight 4pt \caption{Performance and complexity comparison of various hard-decision MIMO detectors in uncoded SDM-MIMO systems, where $N_r \ge N_t$.} \label{table_performance_complexity_compare} \centering
\begin{scriptsize}
\begin{tabular}{l|r|r|r|r}
\hlinewd{0.9pt}
\textbf{Detector} & \textbf{Receive diversity order}  & \textbf{Error probability/DMT/asymptotic}  &  \textbf{SNR penalty} & 			\textbf{Worst-case computational }  \\
                  & \textbf{at high SNR}              &  \textbf{analysis}    &              & \textbf{complexity order}    \\
& & & & \textbf{per symbol vector} \\
\hline
\hline
MAP/ML & $ N_r$\cite{Awater_2000:ML_MMSE_diversity_order, Grant_1998:ML_diversity_order} &  See\cite{Verdu_1984:Phd_thesis, Verdu_1986:optimal_MUD_asynchronous_CDMA, Awater_2000:ML_MMSE_diversity_order, Grant_1998:ML_diversity_order, Grant_2000:further_results_ML_diversity_order, Murch_2002:ML_performance_analysis,Lee_2006:ML_VER}  & Zero & $\mathcal{O}(M^{N_t})$ \\
\hline
Linear ZF & $N_r - N_t +1$\cite{Winters_1994:ZF_MMSE_diversity_order} & See\cite{Winters_1994:ZF_MMSE_diversity_order, Jiang_2011:performance_analysis_ZF_MMSE,Hedayat_2007:outage_diversity_ZF_MMSE, Kumar_2009:asymptotic_antenna_SNR_linear_MIMO_receiver}  & High & Between $\mathcal{O}(N_t^2)$ and $ \mathcal{O}(N_t^3)$  \\
\hline
Linear MMSE & $ N_r - N_t +1$\cite{Winters_1994:ZF_MMSE_diversity_order, Li_2006:MMSE_SINR_distribution, Mehana_2012:MMSE_diversity} & See\cite{Poor_1997:probability_of_error_MMSE_MUD,Winters_1994:ZF_MMSE_diversity_order},  & Lower than that  & Between $\mathcal{O}(N_t^2)$ and $\mathcal{O}(N_t^3)$  \\
& & \cite{Li_2006:MMSE_SINR_distribution,Jorswieck_2007:outage_prob_MIMO,Hedayat_2007:outage_diversity_ZF_MMSE, Kumar_2009:asymptotic_antenna_SNR_linear_MIMO_receiver,Moustakas_2009:MMSE_massive_MIMO, Mehana_2012:MMSE_diversity}  & of linear ZF & \\ 
\hline
ZF/MMSE-SIC & $ N_r - N_t +1$\cite{Varanasi_2004:DFD_performance_analysis,Tse:Fundamental, Loyka_2006:VBLAST_without_ordering_performance, Jiang_2005:Asymptotic_VBLAST, Jiang_2011:performance_analysis_ZF_MMSE} & See\cite{Varanasi_2004:DFD_performance_analysis,Tse:Fundamental, Loyka_2006:VBLAST_without_ordering_performance, Jiang_2005:Asymptotic_VBLAST, Jiang_2011:performance_analysis_ZF_MMSE} & Lower than that of  & Between $\mathcal{O}(N_t^3) $ and $ \mathcal{O}(N_t^4)$  \\ 
& & & linear ZF/MMSE & \\
\hline
ZF/MMSE-OSIC & $N_r - N_t +1 $\cite{Loyka_2004:VBLAST_performance_ordering, Varanasi_2004:DFD_performance_analysis, Jiang_2005:Asymptotic_VBLAST, Jiang_2011:performance_analysis_ZF_MMSE} & See\cite{Varanasi_2004:DFD_performance_analysis,Loyka_2004:VBLAST_performance_ordering, Jiang_2005:Asymptotic_VBLAST, Jiang_2011:performance_analysis_ZF_MMSE} & Lower than that of & Between $ \mathcal{O}(N_t^3) $ and $\mathcal{O}(N_t^4)$     \\
& & & ZF/MMSE-SIC & \\
\hline
SD & $N_r$\cite{Seethaler_2010:SD_infinity_norm}  & The same as that of ML,  & (Can be) zero & $\mathcal{O}(M^{\beta{N_t}})$, $\beta \in (0,1]$ \\
&   &  if it is used for obtaining & in general  &\cite{Hassibi_2005:SD_complexity_part_1,Vikalo_2005:SD_complexity_part_2, Jalden:SD_complexity_journal, Seethaler_2011:SD_complexity_distribution, Jalden_2012:SD_complexity_exponent}\\
& &  the exact ML solution. & & \\
\hline
FCSD & $\min[N_r, (N_r- N_t)(p+1)+(p+1)^2]$, & See\cite{Jalden:fixed_complexity_SD, Jalden:FCSD_error_prob} & Approach zero & $\mathcal{O}(M^{\sqrt{N_t}})$\cite{Jalden:fixed_complexity_SD, Jalden:FCSD_error_prob} \\ 
& where the first $p$ levels  & & at high-SNR & \\
& experience full search\cite{Jalden:fixed_complexity_SD, Jalden:FCSD_error_prob}. & &   & \\
\hline
$K$-best SD & $N_r-N_t+1$ to $N_r$, & unknown for arbitrary $K$, flexible & Between that of SIC and ML,  &  Between that of SIC and ML, \\
& depending on the value of $K$ & and suitable for VLSI implementation & depending on $K$\cite{Wenk_2006:K_best_SD_VLSI, Guo_2006:implementation_K_best_SD_MIMO} & depending on $K$ \\
\hline
LLL-LR-ZF/MMSE/SIC & $N_r$\cite{Taherzadeh_2007:LLL_LR_receiver_diversity,Ling_2006:LR_performance_analysis, Ma_2008:performance_analysis_CLLL, Gan_2009:CLLL_MIMO_detection, Ma_2008:CLLL} 
& See\cite{Ling_2006:LR_performance_analysis,Taherzadeh_2007:LLL_LR_receiver_diversity,Ma_2008:performance_analysis_CLLL,Ma_2008:CLLL, Gan_2009:CLLL_MIMO_detection, Taherzadeh_2010:LLL_LR_limitation, Ling_2011:LR_performance_analysis, Jalden_2010:DMT_optimality_LRA, Singh_2012:vanishing_gap_LR} & Can approach zero,   & Infinite in general (there  \\
& & & but the actual gap & is no universal upper bound   \\
& &　& depends on how well   & on the number of LLL \\
& & &  the particular channel & iterations)\cite{Yao_2003:LR_MIMO_detector, Jalden_2008:complexity_LLL}, but \\
& & &  can be reduced.  & can be substantially reduced  \\ 
& & &  & in many cases\cite{Jalden_2010:DMT_optimality_LRA,  Singh_2012:vanishing_gap_LR}\\ 
\hline
PDA & unknown & See\cite{Buehrer_1996:adaptive_multistage_IC_CDMA, Penghui_2006:asymptotic_optimum_PDA, Fricke:Impact_of_Gaussian_approximation} & Approach zero for large & $\mathcal{O}(M_iN_t^3)$ to $\mathcal{O}(M_iN_t^4)$,   \\
& & \cite{Cao:Relation_of_PDA_and_MMSE-SIC, Orlik_2013:PDA_performance_analysis,Shaoshi2013:Turbo_AB_Log_PDA} &  $N_t$\cite{Penghui_2006:asymptotic_optimum_PDA, Fricke:Impact_of_Gaussian_approximation, Luo:PDA_Sync_CDMA} & where $M_i$ is the number  \\
& & & & of PDA iterations\\
\hline
SDPR & $N_r$ only for BPSK transmission over & See\cite{Ma:SDR_CDMA_BPSK, Luo:SDR_performance_analysis, Jalden:SDR_optimality_conditions_BPSK} & Typically near-ML  & Constellation-dependent:   \\
& real-valued Gaussian fading channels \cite{Jalden:SDR_diversity} & \cite{Jalden:SDR_diversity, So:SDR_performance_low_SNR_QAM, Ma:equivalence_SDR}  & for BPSK/QPSK, but not  & $\mathcal{O}[(1+N_t\log_2M)^{3.5}]$  \\
& & &for high-order constellations & \cite{luo:soft_SDR, Ma:SDR_MPSK, Luo_2009:efficient_implementation_SDR}\\
& &  & unless complexity is & \cite{, Luo:SDR_simplest, Shaoshi_2013:DVA_SPDR_journal, Sidiropoulos:SDR_HOM} to\\
& &  & substantially increased\cite{Mobasher:SDR_QAM_journal}. & $\mathcal{O}[(\sqrt{M}(2N_t+1))^{6.5}]$ \\
& & & & \cite{Sidiropoulos:SDR_HOM, Mobasher:SDR_QAM_journal, Ma:equivalence_SDR} \\
\hlinewd{0.9pt}
\end{tabular}
\end{scriptsize}
\end{table*}
\begin{itemize}
 \item The MAP/ML based MIMO detectors relying on brute-force search have the optimal VER performance (not necessarily optimal BER or SER) and a computational complexity which increases exponentially with the system's dimension (e.g. the number of transmit antennas or users). Naturally, their computational complexity order $\mathcal{O}(M^{N_t})$ is the highest amongst all the MIMO detectors. Additionally, the MAP/ML algorithms have to be aware of the amplitudes of the transmitted symbols for calculating the decision statistics. However, the MAP/ML detector is insensitive to channel imperfections and rank-deficiency/overloading, and it has the best possible error probability performance across the entire SNR region. When the system's dimension is not too large, it remains possible to implement the exact MAP/ML algorithm in practical systems with the aid of state-of-the-art VLSI technologies.  
\item The linear MIMO detectors typically have the lowest computational complexity between $\mathcal{O}(N_t^2)$ and $\mathcal{O}(N_t^3)$,  although there exist subtle differences amongst the computational complexities of different linear detectors. Naturally, in general they have the least attractive error probability performance. However, in some scenarios, such as the large-scale MIMO systems to be detailed in Section \ref{Sec:detection_in_massive_MIMO}, where the receiver side has a significantly higher number of antennas than the transmitter side, the linear MF, ZF, MMSE, MBER etc. based MIMO detectors may achieve a near-ML error probability. Additionally, the linear MF and ZF detectors only have to know the channel matrix $\bf H$, but the linear MMSE detector additionally has to estimate the noise variance. Furthermore, as indicated in Section \ref{subsubsec:optimality_criteria}, the linear ZF detector is preferable in interference-dominated scenarios, the linear MF detector is preferable in noise-dominated scenarios, while the linear MMSE detector provides the highest SINR amongst all linear detectors, which makes it preferable in scenarios where the noise and the interference have a comparable level. Finally, the linear ZF and MMSE detectors exhibit an inadequate performance in rank-deficient/overloaded systems, where the number of independent inputs is higher than the dimension of the received signals, while the linear MF detector remains applicable. 
\item The interference cancellation based MIMO detectors have a computational complexity between $\mathcal{O}(N_t^3)$ and $\mathcal{O}(N_t^4)$, and typically they have a much more attractive error probability performance than the linear detectors.  Theoretically, the SIC/DFD based detectors are capable of approaching the Shannon capacity, provided that there is no error propagation at any of the decision stages. By contrast, the PIC/MIC based detectors do not have this property. Compared to PIC/MIC, the SIC/DFD detectors are more sensitive to error propagation. However, this makes them preferable in the ``near-far'' scenario, where the powers of different users are significantly different, such as those of the cell-center user and cell-edge user. Furthermore, the SIC/DFD detectors may have a higher processing delay than the PIC detectors. Additionally, similar to the linear ZF and MMSE detectors, the interference cancellation based detectors are not generally applicable to the rank-deficient/overloaded scenarios.    
\item The tree-search based MIMO detectors, especially the $K$-best detectors, have the flexibility to achieve different error probability versus computational complexity tradeoffs. They are even capable of attaining the optimum ML performance at a reduced complexity. In contrast to other types of MIMO detectors, the tree-search based detectors typically have a non-deterministic complexity, which is a challenge for hardware implementation, albeit it is possible to design fixed-complexity tree-search detectors. Therefore, the average computational complexity,  worst-case computational complexity and even the computational complexity distribution become important complexity metrics to consider. Note, however, that theoretically the tree-search based MIMO detectors still have an exponentially increasing worst-case/average computational complexity, in which case the exponent depends on different system parameters, such as the noise variance. As a result, the tree-search based detectors may not be suitable for low-SNR scenarios. Additionally, it may be possible to design tree-search based detectors for rank-deficient/overloaded scenarios. Furthermore, the tree-search based detectors rely on specific enumeration strategies, which by nature are not suitable for large-scale MIMO systems that have a high number of inputs. 
\item The LR algorithms constitute a family of powerful preprocessing techniques conceived for improving the ``quality'' of the effective channel matrix. They can be used in conjunction with all the other MIMO detectors. Since practically usable LR algorithms, such as the LLL algorithm, have a polynomially increasing computational complexity, the LR-aided MIMO detectors do not have a significantly increased total computational complexity. Hence, LR may be particularly useful for designing high-performance MIMO detectors maintaining a low complexity, which is critical in numerous practical implementations. However, the LR techniques do not fundamentally change the pros and cons of their baseline detectors. 
\item Compared to the other MIMO detectors mentioned above, the SDPR and PDA based MIMO detectors are not well-understood at the time of writing and they have not achieved the same degree of practical success, which is partially indicated by the lack of VLSI implementations of these two types of detectors. Although SDPR detectors have a favorable worst-case polynomial complexity, which is roughly between $\mathcal{O}[(1+N_t\log_2M)^{3.5}]$ and $\mathcal{O}[(\sqrt{M}(2N_t+1))^{6.5}]$, their achievable error probability performance becomes less attractive for high-order modulations (but they may achieve near-ML performance for BPSK and QPSK constellations). The Gaussian-mixture approximation based PDA detectors operate in a way similar to the classic soft interference cancellation, hence their computational complexity is similar to that of the soft SIC detectors, i.e. typically on the order between $\mathcal{O}(M_iN_t^3)$ and $\mathcal{O}(M_iN_t^4)$. As a result, the PDA detectors are also sensitive to error propagation, whilst exhibiting the nice property of preferring a large number of inputs, provided that the receive dimensions are no less than that of the inputs. Hence, for certain large-scale MIMO scenarios, both SDPR and PDA based detectors may be attractive. Finally, for large-scale MIMO systems which have a similarly large number of transmit and receive antennas, it might be valuable to resort to metaheuristics based algorithms, since all the other MIMO detectors might either be excessively complex  or fail to provide a high performance. Some of the metaheuristics based large-scale MIMO detectors are described in Section \ref{subsec:recent_LS_MIMO_detection}.          
\end{itemize}

\section{Detection in LS-MIMO Systems}\label{Sec:detection_in_massive_MIMO}
Having reviewed the representative families of MIMO detection algorithms in Section \ref{sec:history_MIMO_detection:chap_intro}, let us now shift our attention to the detection problem encountered in the emerging massive/LS-MIMO systems\cite{Marzetta2010:massive_MIMO,Rusek_2013:massive_MIMO, tenBrink2013:massive_MIMO}, where dozens or even hundreds of antennas may be invoked and an unprecedented spectral efficiency/diversity order may be achieved. The major benefits of LS-MIMOs can be deduced from the following well-known results. For transmission over a quasi-static channel where a codeword occupies only a single coherence-time and coherence-bandwidth interval, the outage probability of a point-to-point MIMO link scales according to 
\begin{equation}
{\Pr}_{outage} \propto \text{SNR}^{-N_tN_r},
\end{equation}
which indicates that potentially a diversity order of $(N_t\times N_r)$ may be achieved. In other words, the MIMO link's reliability quantified in terms of its error rate falls exponentially with $N_t$ and/or $N_r$ when $\text{SNR}$ increases. Additionally, on a fast-fading MIMO channel, the achievable rate scales as 
\begin{equation}
\min\{N_t, N_r\}\log_2(1+\text{SNR}), 
\end{equation}
which indicates that the achievable rate of a MIMO system scales linearly with $\min\{N_t, N_r\}$, and hence it is possible to attain a high data rate using a large $N_t$ and $N_r$. In conclusion, fundamentally, using more antennas grants us higher degrees 
of freedom in the spatial domain without increasing the bandwidth occupied.    

\begin{figure}[tbp]
\centering
\includegraphics[width=0.9\linewidth]{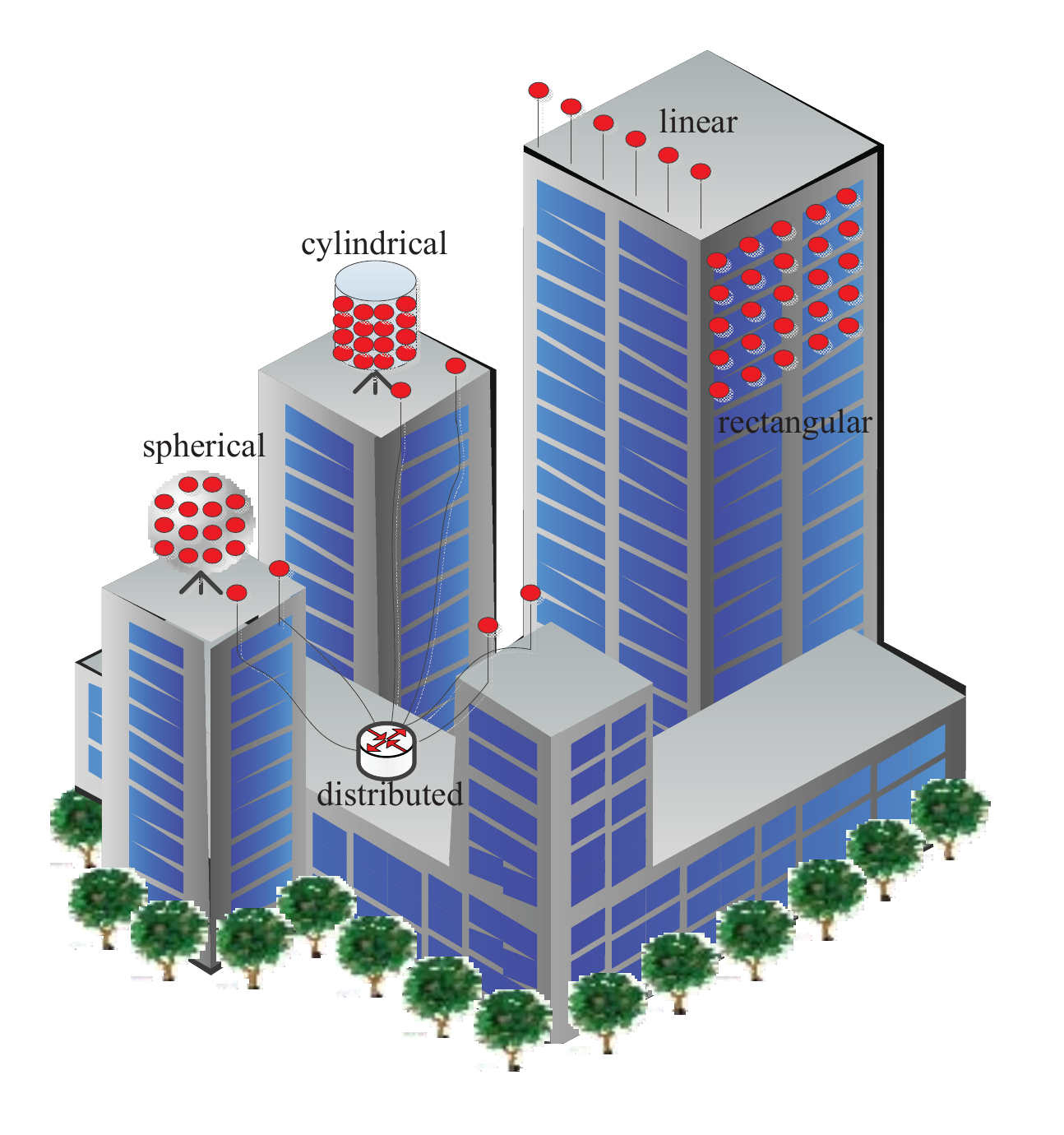}
\caption[]{Typical antenna array configurations and deployment scenarios of LS-MIMO systems\cite{Kan_2014:massive_MIMO_CM_survey}.} \label{fig:LS_MIMO_antenna_config}
\end{figure}

The LS-MIMO systems can be implemented in a variety of ways. For example, in the operational 3G/4G wireless communication systems, a point-to-point LS-MIMO system might be constructed to provide high-throughput wireless backhaul connectivity between the BSs by using a large number of antennas at each BS. However, apart from this particular application, it is typically quite challenging to construct a point-to-point LS-MIMO system where the antenna elements can have a sufficiently high spatial separation to guarantee a well-conditioned channel matrix. Furthermore, achieving the attractive multiplexing gains promised by point-to-point LS-MIMO schemes requires a high SNR. On the other hand, a multiuser LS-MIMO system\cite{Anzhong_2014:massive_MIMO_source_localization, Jiankang_2014:massive_MIMO_pilot_decontamination} can be envisaged, where the BS may be equipped with hundreds of antenna elements and serves dozens of MSs each having only a few antennas. Additionally, the LS-MIMO may be implemented in the extremely high frequency (EHF) band (i.e. at millimeter wave (MMW) frequencies ranging from 30 to 300 GHz and having wavelengths spanning from ten to one millimeter\cite{Rappaport2013:MMW_5G}). They may also be considered in the optical band for frequencies ranging from 300 GHz to 300 PHz and including the infrared, the visible and the ultraviolet band\cite{Haas2011:indoor_optical_wireless, Hanzo2012:myth_wireless}. Due to the adverse propagation properties, the coverage of the LS-MIMO systems operating in these high-frequency bands might be significantly limited, hence they are more applicable to indoor environments\cite{Haas2011:indoor_optical_wireless} or small-cell scenarios\cite{Hoydis2013:small_cell_massive_MIMO}. For the sake of more explicit clarity, several typical antenna configurations and deployment scenarios of LS-MIMOs are illustrated in Fig. \ref{fig:LS_MIMO_antenna_config}\cite{Kan_2014:massive_MIMO_CM_survey}. To elaborate a little further, the simplest linear array propagates signals on the two-dimensional plane and it typically occupies a large physical area. By contrast, the rectangular, cylindrical and spherical arrays are capable of radiating signals to any directions in the three-dimensional space. These antenna arrays are more complex, but also more compact, hence occupying a smaller physical area. Additionally, a virtual LS-MIMO may be constructed relying on distributed antenna arrays, which may be exploited to enhance the indoor coverage or outdoor cooperation\cite{Kan_2014:massive_MIMO_CM_survey}.      

As pointed out in Section  \ref{subsec:ML_detector:chap_intro}, the key motivation of studying the fundamental MIMO detection problem is that the computational complexity of the optimum ML/MAP MIMO detection increases exponentially with the problem size. Therefore, in principle the MIMO detection problem has intrinsically embedded the ``large-scale'' concept. In this regard, people may argue that the detection in LS-MIMO systems is not a novel problem, and consequently the detectors conceived for LS-MIMO systems might have no significant difference with respect to the existing MIMO detectors, except for the associated larger problem size. At first glance, this seems to be true. However, due to the limitations of practical applications, in the past large-scale MIMO systems have been regarded as being impractical and most of the research focused on small-scale MIMO systems. 
Nonetheless, in addition to their significant link reliability and throughput benefits, the LS-MIMO systems have been shown to enjoy some distinct advantages that are not available in small-scale MIMO systems. These benefits are mainly attributed to a range of relevant results in random matrix theory\cite{Tulino2004:random_matrix_wireless_comm, mehta2004:random_matrices}, and might be capable of circumventing signal processing problems in LS-MIMO systems. As a result, insights drawn from the detection in small-scale MIMO systems might have to be carefully adapted for the large-scale MIMO environments. Depending on the application scenarios considered, the detection problem of large-scale MIMO systems may be categorized as follows.
   
\subsection{Detection in Single-Cell/Noncooperative Multi-Cell LS-MIMO Systems}
\begin{figure}[t]
\centering
\includegraphics[width=0.9\linewidth]{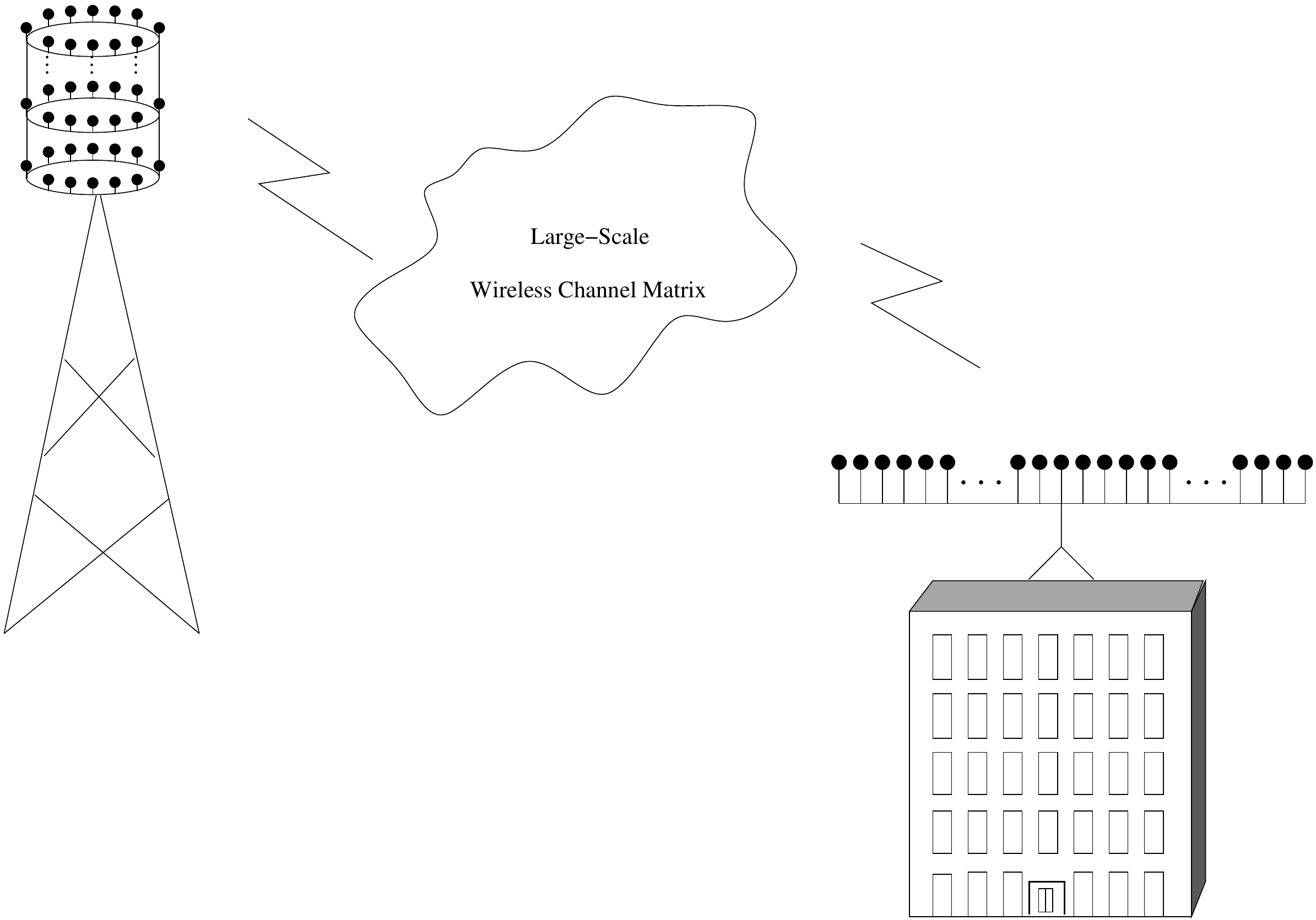}
\caption[]{An example of Type-I (point-to-point) LS-MIMO systems.} \label{fig:type_I_LS_MIMO}
\end{figure}

In a single-cell/noncooperative multi-cell MIMO system the BS is not concerned about the CCI imposed by the transmissions of other cells. In this scenario, as pointed out in Section \ref{sec:generality_MIMO_detection:chap_intro}, the detection problems encountered in both the point-to-point MIMOs (see Fig. \ref{fig:SU_MIMO}) and the multiple-access MIMOs (see Fig. \ref{fig:MA_MIMO}) can be characterized using the same received signal model of (\ref{eq:general_MIMO_model_matrix_form}). From the antenna configuration point of view, there are two types of single-cell/noncooperative multi-cell LS-MIMO systems. As shown in Fig. \ref{fig:type_I_LS_MIMO}, in the Type-I system, a large number of \textit{collocated} antennas may be mounted on the receiver, and \textit{also} a large number of \textit{collocated or distributed} antennas are used at the transmitter. Mathematically, the antenna configuration of the Type-I system may be characterized by 
\begin{equation}\label{eq:type_I_massive_MIMO}
\lim_{N_t, N_r \rightarrow \infty} \frac{N_t}{N_r} = c
\end{equation}
with $c$ being a positive constant. (\ref{eq:type_I_massive_MIMO}) indicates that both $N_t$ and $N_r$ tend to infinity at the same rate.  By comparison, in the Type-II system, \textit{only} the receiver is equipped with a large number of collocated antennas, while the total number of active antennas at the transmit side is significantly smaller. Hence, the antenna configuration of this system may be characterized as
\begin{equation}\label{eq:type_II_massive_mimo}
\lim_{N_r \rightarrow \infty} \frac{N_t}{N_r} = 0.
\end{equation}
% \begin{equation}\label{eq:type_II_massive_mimo}
% N_r \rightarrow \infty, \text{while}~N_t~\text{is fixed}.
% \end{equation}

For the Type-I system, it has been shown that the empirical distribution of the singular values of the random channel matrix $\bf H$ converges to a deterministic limiting distribution\footnote{This limiting distribution is the so-called \textit{quarter circle law}\cite[Chapter 8.2]{Tse:Fundamental}. } for almost all realizations of $\bf H$, which is a result of the Mar\v{c}enko and Pastur law\cite{Marvcenko1967:distribution_eigenvalues_random_matrix}. In other words, as $\bf H$ becomes larger (in terms of both $N_t$ and $N_r$), its singular values become less sensitive to the actual distributions of the i.i.d. entries of $\bf H$, and the channel becomes more and more deterministic. The Mar\v{c}enko and Pastur law also shows that as the size of $\bf H$ increases, the diagonal entries of ${\bf H}^H{\bf H}$ become increasingly larger in magnitude than the off-diagonal entries. This is the so-called ``channel-hardening'' behavior, which may be exploited for large-scale MIMO detection. To be more specific, the matrix inversion invoked by many MIMO detectors such as the ZF-aided detector, the MMSE-aided detector and the PDA-aided detector etc.,  may be conveniently approximated using the series expansion technique for large-dimensional random matrices\cite{Moshavi_1996_multistage_linear_receivers_CDMA}. Additionally, the channel-hardening phenomenon may allow low-complexity detection algorithms to achieve a good performance for large-scale MIMO systems\cite{chockalingam2014:large_MIMO_systems}. 

\begin{figure}[t]
\centering
\includegraphics[width=0.9\linewidth]{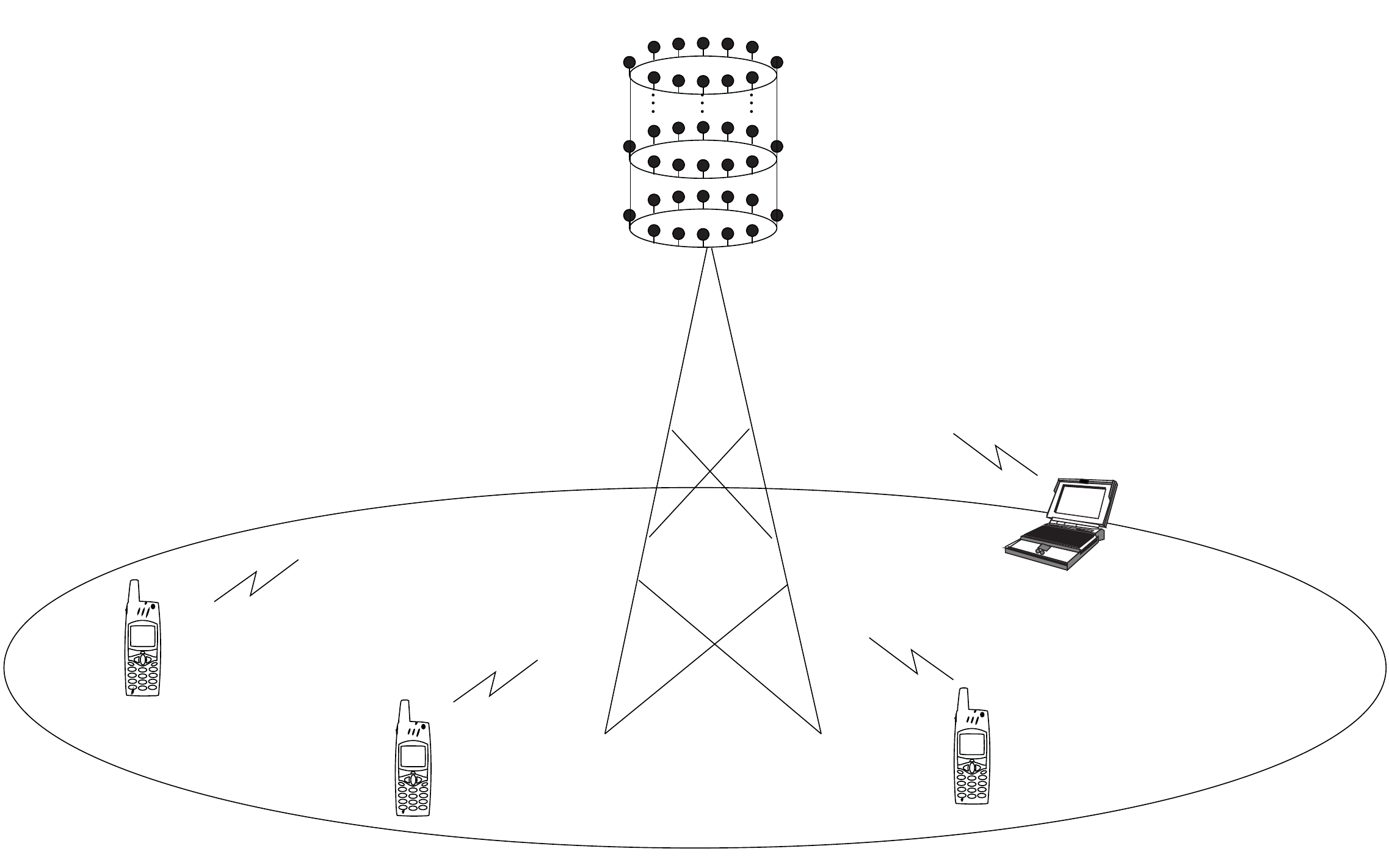}
\caption[]{An example of Type-II (multiuser) LS-MIMO systems.} \label{fig:type_II_LS_MIMO}
\end{figure}
The Type-II system essentially deals with the MIMO detection problem encountered on an \textit{underloaded} uplink,\footnote{On the downlink, large-scale MIMO precoding techniques may be employed, which facilitates the employment of simple receivers at each MS, because the precoder is capable of eliminating the IAI at the transmitter with the aid of accurate channel knowledge. } as shown in Fig. \ref{fig:type_II_LS_MIMO}. On the one hand, since the number of BS antennas may be significantly higher than the total number of active MS antennas, a very unbalanced antenna configuration is encountered, which results in a high receive diversity order. In the extreme case shown by (\ref{eq:type_II_massive_mimo}), the receive diversity gain obtained is so high that the impact of both the MUI and the noise diminishes. Additionally, the channel vectors associated with distinct MSs may become asymptotically orthogonal. Furthermore, another beneficial result of the Mar\v{c}enko and Pastur law\cite{Marvcenko1967:distribution_eigenvalues_random_matrix} is that very tall (with large $N_r$) and very wide (with large $N_t$) channel matrices $\bf H$ are very \textit{well conditioned}. Therefore, in the Type-II system, even the simplest MF detector is capable of achieving a near-optimum performance\cite{Marzetta2010:massive_MIMO, tenBrink2013:massive_MIMO}.  Similarly, when considering the precoding based downlink of the single-cell/noncooperative multi-cell TDD system, it was also revealed that increasing the number of BS antennas is always beneficial, even when the SNR is low and the channel estimate is poor. Furthermore, when the number of BS antennas tends to infinity, the effects of both the small-scale fast fading and uncorrelated noise are mitigated. In other words, a large number of BS antennas, regardless of whether the uplink or the downlink is considered, may be exploited to \textit{trade for} relevant performance improvements, such as compensating for the low SNR and/or poor channel estimates\cite{Marzetta2006:massive_MIMO_single_cell, Marzetta2010:massive_MIMO, tenBrink2013:massive_MIMO, Rusek_2013:massive_MIMO}.     

However, in the noncooperative multi-cell scenario, due to the so-called ``pilot contamination'' problem\footnote{This is essentially the  interference caused by reusing pilot sequences in adjacent cells.} \cite{Marzetta2010:massive_MIMO, Rusek_2013:massive_MIMO}, the interference emanating from other cells does exist and becomes the major limiting factor of the achievable performance\cite{Marzetta2010:massive_MIMO, Rusek_2013:massive_MIMO}. Therefore, in order to further enhance the achievable performance, the BS cooperation based multi-cell joint processing philosophy has to be adopted\cite{Shaoshi2011:DPDA}, as detailed below.       

\subsection{Detection in Cooperative Multi-cell Multiuser LS-MIMO Systems}
\begin{figure}[t]
% \begin{minipage}{0.501\linewidth}
\centering
\subfigure[Centralized BS cooperation based LS-MIMO cellular network. ]{
\label{subfig:cent_BS_coop_LS_MIMO}
\includegraphics[width = 1.0 \linewidth]{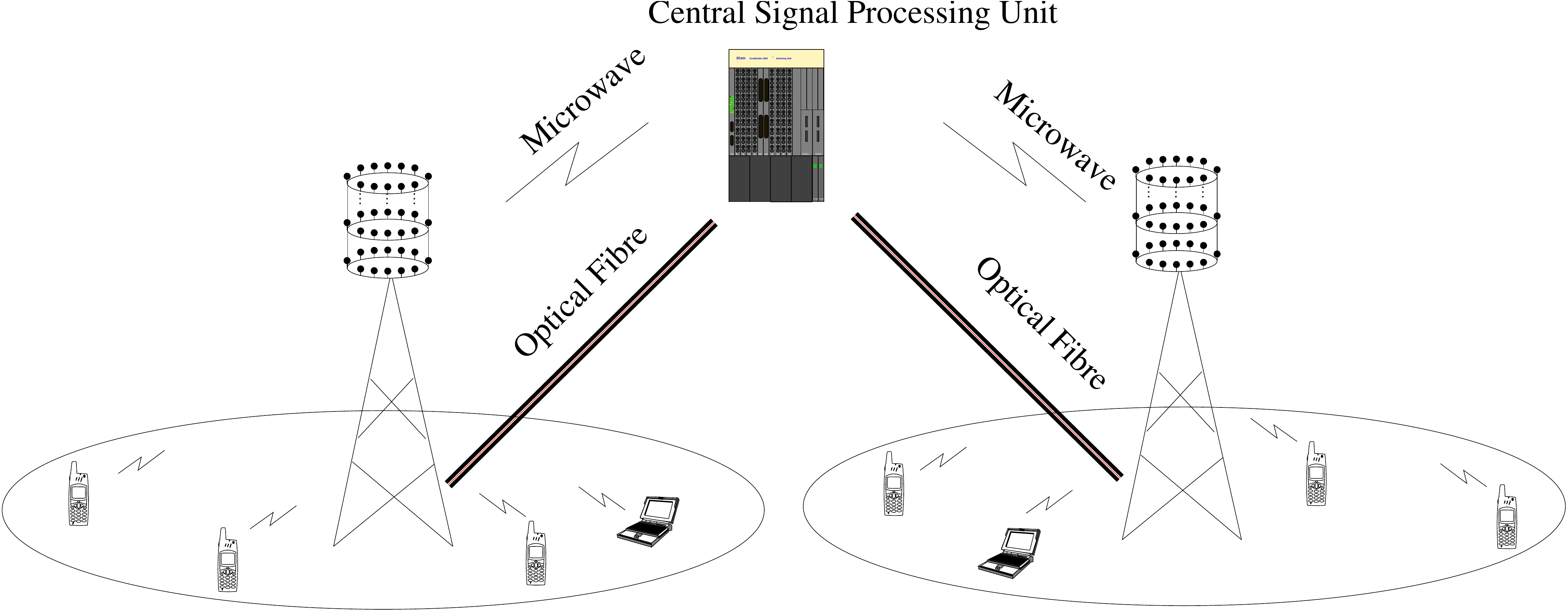}
}
% \end{minipage}
% \begin{minipage}{0.501\linewidth}
\centering
\subfigure[Distributed BS cooperation based LS-MIMO cellular network. ]{
\label{subfig:dist_BS_coop_LS_MIMO}
\includegraphics[width = 1.0 \linewidth]{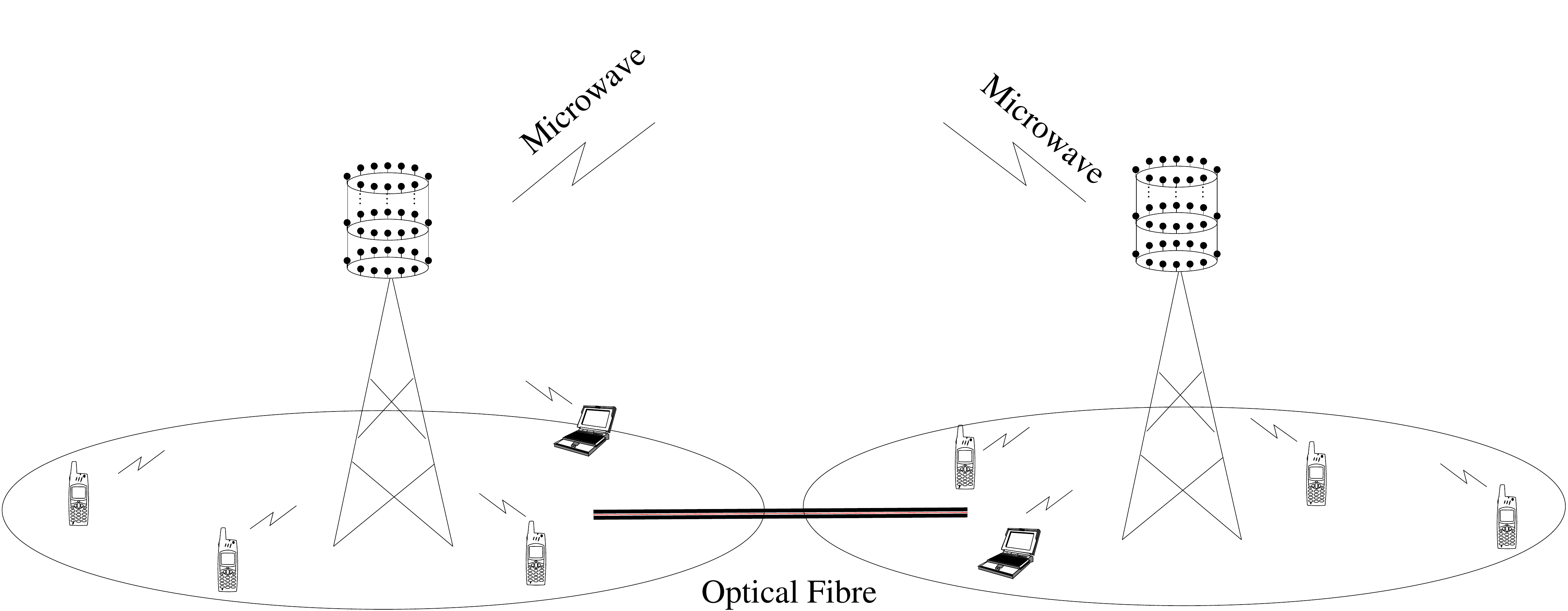}
}
% \end{minipage}
\caption{Centralized and distributed BS cooperation based multi-cell multiuser based LS-MIMO systems, which is also known as network-MIMO.}
\label{fig:BS_coop_LS_MIMO}
\end{figure}

As pointed out in Section \ref{sec:generality_MIMO_detection:chap_intro}, the multi-cell transmission scenario is characterized by the so-called ``MIMO interference channels'' of Fig. \ref{fig:IC_MIMO}. Fundamentally, in order to cope with the interference, it may be desirable to \textit{transform} the distributed model (such as the BSs of multiple cells) to a centralized model. This may be achieved by centralized/distributed BS cooperation\cite{Shaoshi2011:DPDA, Gesbert_2010:multicell_MIMO}, where multiple BSs of adjacent cells may be connected via high-capacity optic fiber or microwave links, as shown in Fig. \ref{fig:BS_coop_LS_MIMO}. As a result, effectively a physical/virtual super-BS is constructed to serve the cluster of collaborative cells, and this physical/virtual centralized model provides the performance \textit{upper bound} of the original distributed system model. As far as detection is concerned, in principle most of the detection algorithms developed for the single-cell/noncooperative multi-cell scenarios may be adapted to the uplink of the cooperative multi-cell LS-MIMO system. The BS cooperation aided network MIMO detectors may be designed based on two distinct philosophies, namely using either interference cancellation\cite{Khattak:distributed_max_log_MAP} or data fusion\cite{Shaoshi2011:DPDA}. However, the employment of BS cooperation might result in substantially increased backhaul traffic, which represents one of the major challenges facing the BS cooperation aided \textit{network MIMO}. 
         
\subsection{Applicability of Existing MIMO Detection Algorithms to LS-MIMO}
An inherent characteristic of LS-MIMO systems is their large dimension. Before investigating the applicability of existing MIMO detection algorithms in the LS-MIMO context, we have to identify which specific type of LS-MIMO systems is considered. On the one hand, in general most of the existing MIMO detectors would be applicable to a Type-II LS-MIMO system, where it is possible that low-complexity linear MIMO detectors might be capable of achieving near-optimum performance. Hence, the employment of more sophisticated MIMO detectors, such as the SD detector, may become unnecessary. On the other hand, some existing MIMO detection algorithms that have been specifically tailored for conventional small-/medium-scale MIMO systems might not be applicable to the Type-I LS-MIMO systems. To elaborate a little further, the family of tree-search based MIMO detectors, such as the popular SD detector that has a worst-case computational complexity increasing exponentially with the number of transmit antennas (see Section \ref{subsec:tree_search_MUD:chap_intro} for more details), will become less feasible in the Type-I LS-MIMO systems. Nonetheless, it might still be invoked in the Type-II LS-MIMO systems. By contrast, the PDA algorithm\cite{Shaoshi2011:B_PDA,Shaoshi2011:DPDA,Shaoshi2013:Turbo_AB_Log_PDA, Shaoshi_2013:EB_Log_PDA_journal}, which invokes the central limit theorem to perform stochastic interference modelling and imposes a polynomial-time worst-case computational complexity, will achieve an attractive performance versus complexity tradeoff in the Type-I LS-MIMO systems. Similarly, the convex optimization\footnote{Note that it is quite common to solve hundreds of unknown variables in a convex optimization problem.} based SDPR detectors, which also exhibit a polynomial-time worst-case complexity as a function of the number of transmit antennas, might potentially be applicable to the Type-I LS-MIMO systems\cite{Shaoshi_2013:DVA_SPDR_journal}.

\subsection{Recent Advances in LS-MIMO Detection}   \label{subsec:recent_LS_MIMO_detection}
The LS-MIMO systems have become a hot research topic following Marzetta's seminal work\cite{Marzetta2010:massive_MIMO}. However, in terms of detection, several earlier works had touched upon this topic from either a large system analysis or an asymptotic performance analysis perspective. To elaborate a little further, in 2006 Tan and Rasmussen\cite{Penghui_2006:asymptotic_optimum_PDA} derived a class of asymptotically optimal nonlinear MMSE MUDs based on a multivariate Gaussian approximation of the MUI for large-scale CDMA systems. This approach provided an alternative analytical justification for the structure of the PDA based detectors. The associated performance analysis showed that the BER performance of the PDA detectors can be accurately predicted and is close to the optimal detector's performance for large CDMA systems. Also in 2006, Liang \textit{et al.}\cite{Liang_2006:block_iterative_DFEs_massive_MIMO} proposed a block-iterative generalized decision feedback equalizer (\gls{BI-GDFE}) for LS-MIMO systems using PSK constellations. Their asymptotic performance analysis demonstrated that the BI-GDFE closely approaches the single-user matched-filter bound (\gls{MFB}) for large random MIMO channels, provided that the received SNR is sufficiently high\cite{Liang_2007:MMSE_large_system_performance}. Furthermore, in 2007 Liang \textit{et al.} derived both the limit and the asymptotic distribution of the SINR for a class of MMSE receivers invoked in large-scale CDMA systems supporting unequal-power users. Their solution relied on the random matrix theory. They also proved that the limiting SINR converges to a deterministic value when ${\lim_{K, N \rightarrow \infty}} \frac{K}{N} = c$, where $K$ is the number of users, $N$ is the number of degrees of freedom and $c$ is a positive constant. Recall that this insight is the same as that discussed in the context of (\ref{eq:type_I_massive_MIMO}). Additionally, they proved that the SINR of each particular user is asymptotically Gaussian for large $N$ and derived the closed-form expressions of the variance for the SINR variable under both real-valued spreading and complex-valued spreading. As a further advance, in 2008 Liang \textit{et al.}\cite{Liang_2008:relation_between_MMSE_SIC_and_BI_GDFE} investigated the relationship between the MMSE-SIC receiver and the BI-GDFE receiver. The asymptotic performance of the two receivers was compared for large random MIMO channel matrices, and it was shown that the two receivers have a similar convergence behavior, and that both of them are capable of achieving a BER performance approaching the single-user MFB for sufficiently high SNRs. 

Chockalingam \textit{et al.} also made significant contributions to the LS-MIMO detection problem, mainly using a variety of metaheuristics based local search algorithms invoked from machine learning/artificial intelligence\cite{Chockalingam_2010:low_complexity_LS_MIMO_detection}. More specifically, in 2008 they extended the low-complexity likelihood ascent search (\gls{LAS}) based MUD\cite{Sun_1998:HNN_MUD, Sun_2000:HNN_MUD_CDMA, Sun_2009:LAS_MUD_journal} to the Type-I LS-MIMO system having up to 600 transmit and receive antennas\cite{Chockalingam_2008:HNN_LAS_based_LS_MIMO_detector}. This detector relies on the local neighborhood search and has its roots in the family of Hopfield neural network (\gls{HNN})\footnote{The HNN algorithms were also proposed for the restoration of large image\cite{Sun_2000:HNN_image_restoration_part_I, Sun_2000:HNN_image_restoration_part_II}.} based MUD algorithms\cite{Paris_1988:neural_network_MUD_conf, Aazhang_1992:neural_networks_MUD_CDMA, Mitra_1992:neural_networks_MUD_CDMA, Mitra_1994:neural_network_adaptive_MUD,Mitra_1995:neutal_networks_near_far_CDMA, Kechriotis_1996:hopfield_neural_network_MUD}. It was shown that the LAS detector\footnote{A multiple-output selection based LAS detector, namely the list LAS detector, was also proposed in \cite{Murch_2010:LAS_LS_MIMO_detection} for LS-MIMO systems.} is capable of achieving near single-input single-output AWGN performance in a fading LS-MIMO environment at an average per-bit complexity of $O(N_tN_r)$\cite{Chockalingam_2008:HNN_LAS_based_LS_MIMO_detector}. Subsequently, they applied another local neighborhood search based algorithm, namely the reactive tabu search (\gls{RTS}) algorithm, to the detection of LS-MIMO systems. The RTS detector was shown to perform better than the LAS detector, because it relied on an efficient local minima exit strategy\cite{Mohammed_2009:PDA_STBC}. Additionally, a class of belief propagation (\gls{BP}) LS-MIMO detectors relying on graphical models were proposed in\cite{Chockalingam_2011:hybrid_RTS_BP_LS_MIMO_detector, Chockalingam_2011:graphical_model_LS_MIMO_detection, Chockalingam_2014:Channel_harderning_message_passsing}\footnote{Very recently, Wu \textit{et al.} also proposed an approximate message passing algorithm based iterative detector for FEC-coded large-scale MIMO-OFDM systems\cite{Wu_2014:message_passing_soft_iterative_LS_MIMO_detection}.}. A range of other detectors were studied by Chockalingam and his team in the context of LS-MIMO systems, including the randomized Markov chain Monte Carlo (\gls{R-MCMC}) detector\cite{Chockalingam_2011:randomized_MCMC_and_search_LS_MIMO_detection}, the randomized search (\gls{RS}) detector\cite{Chockalingam_2011:randomized_MCMC_and_search_LS_MIMO_detection}, the Monte-Carlo-Sampling based detector which jointly relies on a mixed Gibbs sampling (\gls{MGS}) strategy combined with a multiple restart (\gls{MR}) strategy\cite{Chockalingam_2013:MC_Sampling_receiver_LS_MIMO}, and the LR based\footnote{The application of LR detectors in LS-MIMO systems was also investigated by Zhou \textit{et al.} in \cite{Zhou_2013:Lattice_reduction_LS_MIMO_detector}.} detector\cite{Chockalingam_2013:LR_detection_LS_MIMO}. Additionally, they applied various detectors, including the MMSE detector, the PDA detector, the LAS detector and the RTS detector, in high-rate non-orthogonal STBC aided LS-MIMO systems\cite{Mohammed_2009:PDA_STBC, Chockalingam_2009:RTS_LS_MIMO_detector, Chockalingam_2009:LAS_non_orthogonal_STBC}. Furthermore, it was shown that non-binary LDPC coded LS-MIMO systems are capable of achieving a near-capacity performance with MMSE detection\cite{Suthisopapan_2013:capacity_approaching_LDPC_MMSE_detection_LS_MIMO_journal}. It should be noted that in principle a variety of other metaheuristics based MUDs, such as the genetic algorithm (\gls{GA}) based MUD\cite{Juntti_1997:genetic_algorithm_MUD_CDMA, Ergun_1998:genetic_algorithm_MUD, Ergun_2000:genetic_algorithm_MUD_journal, Wang_1998:genetic_algorithm_MUD, Yen_2001:genetic_algorithm_joint_MUD_CE,Yen_2003:genetic_algorithm_MUD_synch,Yen_2004:genetic_algorithm_MUD_asynch}, the ant colony optimization (\gls{ACO}) based MUD\cite{Dorigo_2006:ACO, Sun_2004:PSO, Hijazi_2004:ACO_MUD, Xu_2008:ACO_MUD, Xu_2009:ACO_MUD_letter}, the particle swarm optimization (\gls{PSO}) aided MUD\cite{Yang_2004:PSO, Zhao_2004:PSO_detection, Chen_2010:PSO_MIMO_detection}, and the simulated annealing (\gls{SA}) assisted MUD\cite{Taufik_2010:SA_MUD, Xia_2008:SA_MUD}, may also be extended to the LS-MIMO context. 

Finally, some soft-input soft-output LS-MIMO detectors having a relatively low complexity were proposed in\cite{Larsson_2014:SUMIS_LS_MIMO_detector, Wu_2014:message_passing_soft_iterative_LS_MIMO_detection}, which rely on the subspace marginalization aided interference suppression (\gls{SUMIS}) technique and an approximate message passing algorithm, respectively. The first ASIC design of an LS-MIMO detector invoking the  truncated Neumann series expansion technique was reported in \cite{Studer_2014:implementation_LS_MIMO_detector, Studer_2014:implementation_LS_MIMO_detector_journal}, which achieves a data rate of 3.8 Gb/s for a 3GPP Long Term Evolution-Advanced (\gls{LTE-A}) based LS-MIMO system having 128 BS antennas and supporting 8 users.

\subsection{Applications of MIMO Detection Techniques in Other Areas}
\begin{figure}[tbp]
\centering
\includegraphics[width=\linewidth]{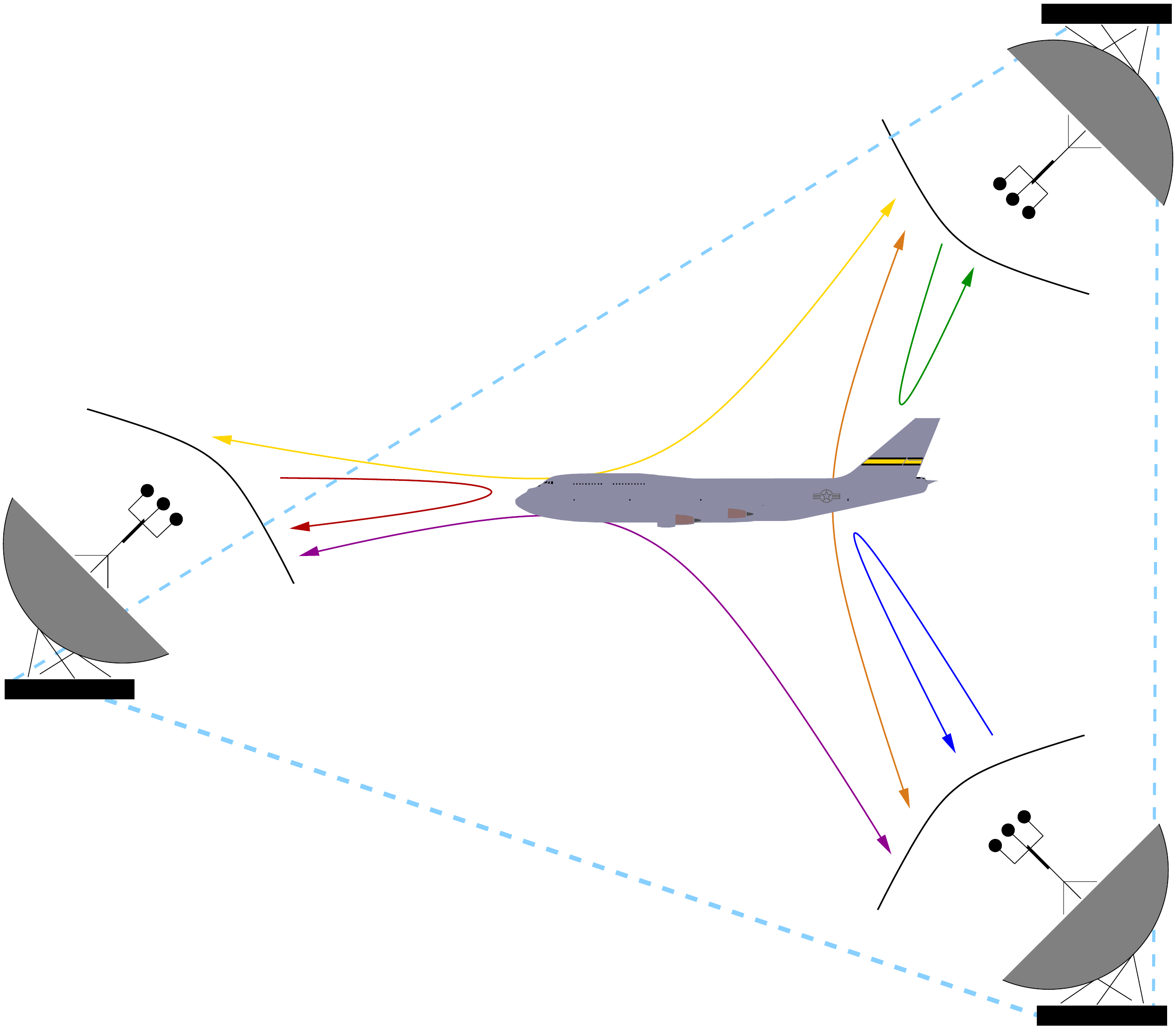}
\caption[]{Conceptual illustration of MIMO radar systems.} \label{fig:MIMO_radar}
\end{figure}
\begin{figure}[t]
\centering
\includegraphics[width=\linewidth]{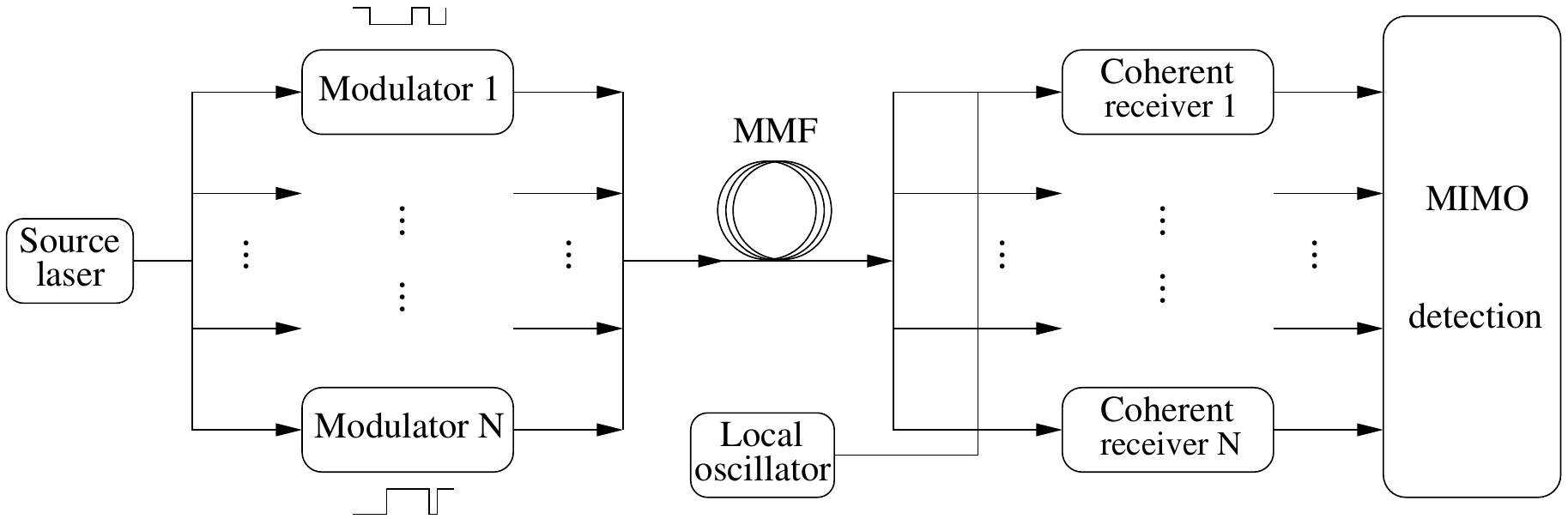}
\caption[]{An example of MIMO optical fiber system.} \label{fig:MIMO_optic_fibre}
\end{figure}
MIMO detection techniques may also be utilized in more advanced scenarios. For example, as a promising technique to utilize the precious radio spectrum more efficiently and flexibly, cognitive radio (\gls{CR})\cite{Mitola_1999:CR, Mitola_2000:Phd_thesis, Haykin_2005:cognitive_radio} has stimulated substantial research interests over the past decade. Relying on the software-defined radio (\gls{SDR}) concept, CR is defined as an intelligent wireless communication system that is capable of learning from the environment and adapting to statistical variations of the environment. Aiming for gleaning the benefits of both the CR and MIMO techniques, MIMO cognitive radio has also been studied from various perspectives\cite{Rui_2008:MIMO_cognitive_radio, Vishwanath_2008:MIMO_cognitive_radio_capacity, Palomar_2008:MIMO_cognitive_radio, Letaief_2009:MIMO_cognitive_radio, Gan_2009:MIMO_cognitive_radio_beamforming, Palomar_2010:MIMO_cognitive_radio_game_theory, Debbah_2010:MIMO_cognitive_radio, GBG_2011:MIMO_cognitive_radio, Zhang_2011:MIMO_cognitive_radio_SDP, Palomar_2011:robust_MIMO_cogntive_radio_game_theory, Varanasi_2012:MIMO_cognitive_radio_capacity}. Furthermore, MIMO techniques may also be integrated with the SDR or software-defined network (\gls{SDN}) for 5G wireless communication systems, where a network function virtualization (\gls{NFV}) based novel network architecture is envisaged\cite{Sun_2015:MIMO_SDR_SDN}. 
 
Apart from their dominant applications in wireless communications, MIMO detection techniques also significantly benefit a range of other research areas. For example, the idea of MIMO signal processing was extended to radar design, and the so-called ``MIMO radar'', as illustrated in Fig. \ref{fig:MIMO_radar}, has been a hot research topic since the 2000s \cite{fishler2004_MIMO_radar, fishler2006_MIMO_radar, Li_2007:MIMO_radar, haimovich_2008:MIMO_radar, Li_2009:MIMO_radar_signal_processing, Maio_2007:MIMO_radar_detector, Li2008:signal_detection_MIMO_radar}. Additionally, MIMO signal processing techniques are also instrumental in mode-division multiplexing (MDM) based multimode fiber (\gls{MMF}) optical communication systems, as shown in Fig. \ref{fig:MIMO_optic_fibre}.  For more details on MIMO aided high-speed optical communications, please refer to\cite{Shah_2005:coherent_MIMO, Tarighat2007:fundamentals_challenges_MIMO_fibre, Randel_2011_optical_fibre_MIMO, Ryf_2011:optical_fibre_MIMO_conf, Ryf_2011:optical_fibre_MIMO, Essiambre_2012:MIMO_optical_capacity, Arik_2014:MIMO_optic_overview}.

\section{Summary and Conclusions}
The concept of LS-MIMO systems may be regarded as a paradigm shift in the wireless communication and signal processing community. In this large dimensional context, the MIMO detection problem becomes even more challenging and important. To facilitate a better understanding of MIMO detection techniques,  in this survey, we provided a detailed clarification of the MIMO detection fundamentals, and recited the half-a-century history of MIMO detection. We also provided concise discussions on the distinct detection strategies for different types of LS-MIMO systems and concluded with the recent advances in LS-MIMO detection. Relevant insights and lessons were extracted from the rich heritage of small-/medium-scale MIMO detection. We note that when considering the design of LS-MIMO detectors, it is necessary to first identify which type of LS-MIMO system is considered. Specifically, the employment of several popular MIMO detectors, such as the SD based MIMO detectors, may become infeasible in Type-I LS-MIMO systems, while some low-complexity linear MIMO detectors may achieve near-optimum performance in Type-II LS-MIMO systems. Additionally, it was reported that in the LS-MIMO context, local neighborhood search based metaheuristics, Bayesian based message passing methods as well as convex optimization based methods may strike a promising performance versus complexity tradeoff.          

\section*{Acknowledgment}
The authors would like to thank Prof. Hamid Jafarkhani (University of California, Irvine), Prof. Arogyaswami Paulraj (Stanford University) and anonymous reviewers for their insightful comments and suggestions.   

\ifCLASSOPTIONcaptionsoff
  \newpage
\fi

% \footnotesize{\printglossaries}

\bibliography{../../../shaoshi_bib/IEEEfull,../../../shaoshi_bib/shaoshi_reference}
\begin{IEEEbiography}[{\includegraphics[width=1in,height=1.25in,clip,keepaspectratio]{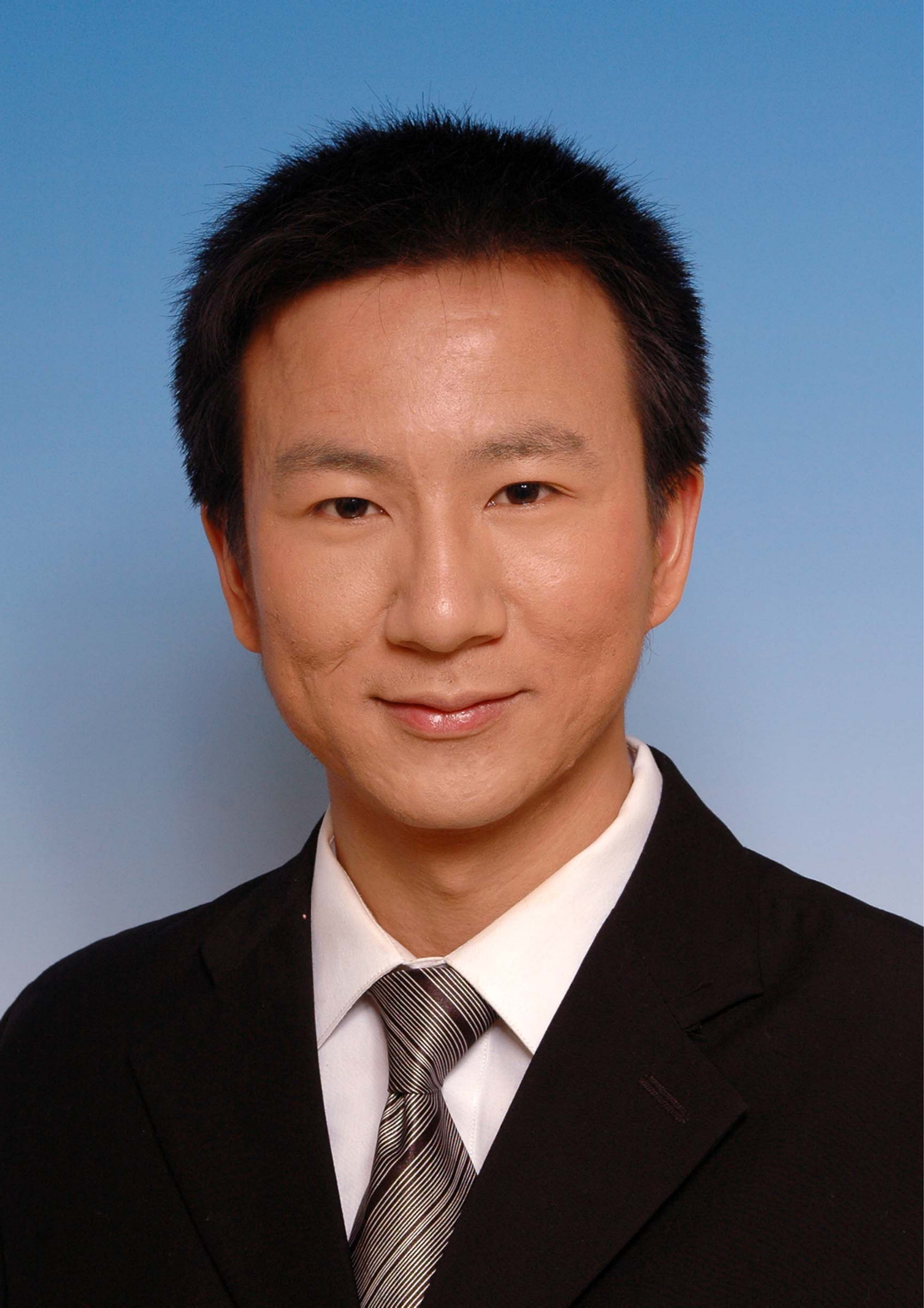}}] {Shaoshi Yang}
(SIEEE'09-MIEEE'13) received his B.Eng. degree in Information Engineering from Beijing University of Posts and Telecommunications (BUPT), Beijing, China in Jul. 2006, his first Ph.D. degree in Electronics and Electrical Engineering from University of Southampton, U.K. in Dec. 2013, and his second Ph.D. degree in Signal and Information Processing from BUPT in Mar. 2014. He is now working as the IU-ATC Senior Research Fellow in University of Southampton, U.K. From November 2008 to February 2009, he was an Intern Research Fellow with the Communications Technology Lab (CTL), Intel Labs, Beijing, China, where he focused on Channel Quality Indicator Channel (CQICH) design for mobile WiMAX (802.16m) standard. His research interests include MIMO signal processing, green radio, heterogeneous networks, cross-layer interference management, convex optimization and its applications. He has published in excess of 30 research papers on IEEE journals and conferences. 

Shaoshi has received a number of academic and research awards, including the PMC-Sierra Telecommunications Technology Scholarship at BUPT, the Electronics and Computer Science (ECS) Scholarship of University of Southampton, the Best PhD Thesis Award of BUPT, and the Dean's Award for Early Career Research Excellence of University of Southampton. He is a member of IEEE/IET, and a junior member of Isaac Newton Institute for Mathematical Sciences, Cambridge University, U.K. He also serves as a TPC member of several major IEEE conferences, including \textit{IEEE ICC, PIMRC, ICCVE, HPCC}, and as a Guest Associate Editor of \textit{IEEE Journal on Selected Areas in Communications.} (https://sites.google.com/site/shaoshiyang/) 
\end{IEEEbiography}

\begin{IEEEbiography}[{\includegraphics[width=1in,height=1.25in,clip,keepaspectratio]{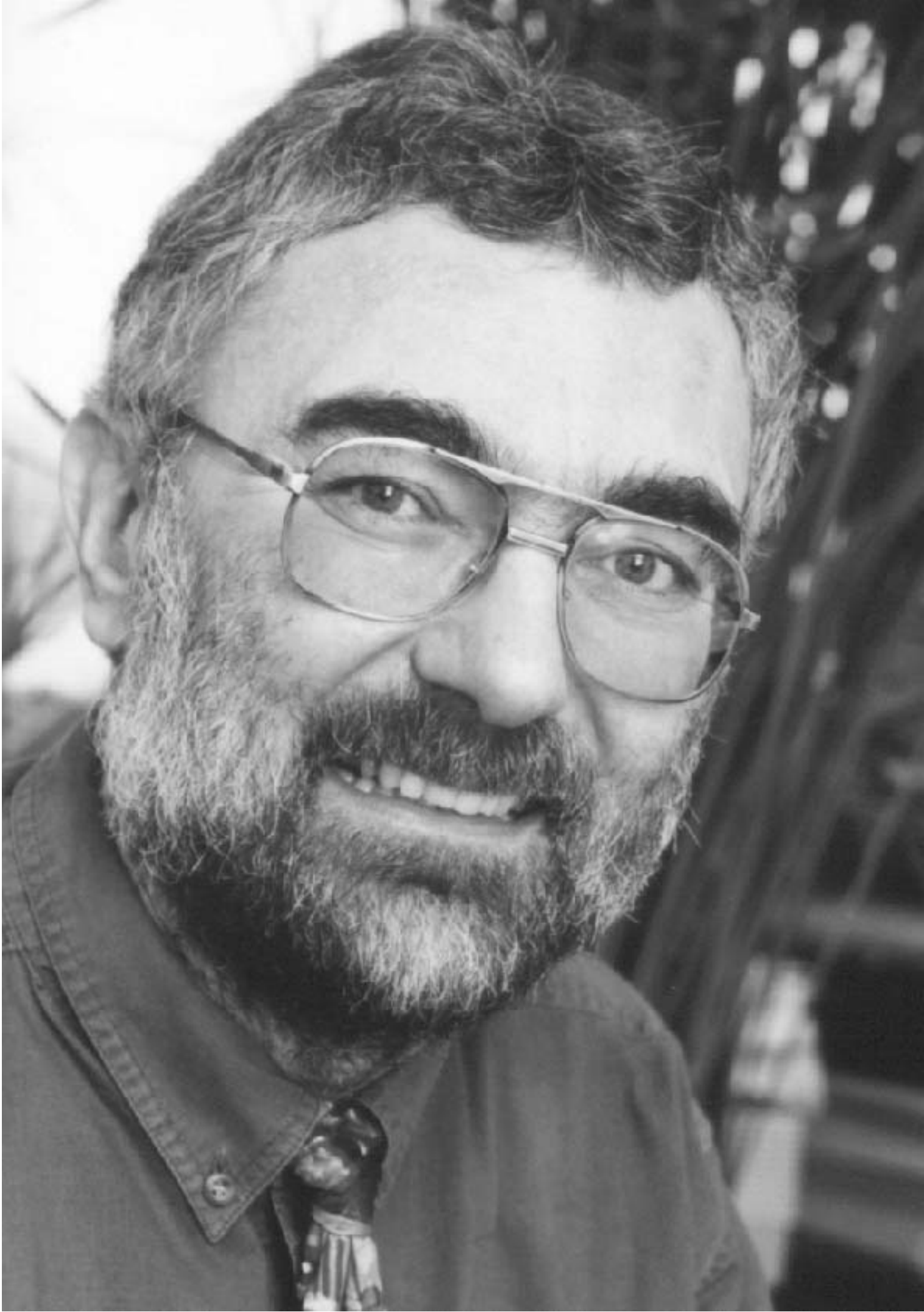}}] {Lajos Hanzo}
(FREng, FIEEE'04,
FIET, FEURASIP, DSc) received his degree in electronics in
1976 and his doctorate in 1983.  In 2009 he was awarded the honorary
doctorate ``Doctor Honoris Causa'' by the Technical University of
Budapest.  During his 37-year career in telecommunications he has held
various research and academic posts in Hungary, Germany and the
UK. Since 1986 he has been with the School of Electronics and Computer
Science, University of Southampton, UK, where he holds the chair in
telecommunications.  He has successfully supervised 80+ PhD students,
co-authored 20 John Wiley/IEEE Press books on mobile radio
communications totalling in excess of 10 000 pages, published 1400+
research entries at IEEE Xplore, acted both as TPC and General Chair
of IEEE conferences, presented keynote lectures and has been awarded a
number of distinctions. Currently he is directing a 100-strong
academic research team, working on a range of research projects in the
field of wireless multimedia communications sponsored by industry, the
Engineering and Physical Sciences Research Council (EPSRC) UK, the
European Research Council's Advanced Fellow Grant and the Royal
Society's Wolfson Research Merit Award.  He is an enthusiastic
supporter of industrial and academic liaison and he offers a range of
industrial courses.  He is also a Governor of the IEEE VTS.  During
2008 - 2012 he was the Editor-in-Chief of the IEEE Press and a Chaired
Professor also at Tsinghua University, Beijing. Lajos
has 20 000+ citations. For further information on research in progress and associated
publications please refer to http://www-mobile.ecs.soton.ac.uk
\end{IEEEbiography}

% \begin{IEEEbiography}[{\includegraphics[width=1in,height=1.25in,clip,keepaspectratio]{./Tiejun_Lv.eps}}] {Tiejun Lv}
% (MIEEE'08) received his M.S. and Ph.D. degrees in Electronic
% Engineering from University of Electronic Science and Technology
% of China (UESTC), in 1997 and 2000, respectively. From January 2001 to
% December 2002, he was a postdoctoral fellow in Tsinghua University,
% China. From September 2008 to March 2009, he was a visiting
% professor at the Department of Electrical Engineering, Stanford
% University, USA. He is currently a professor of the Key Lab of
% Universal Wireless Communications of Beijing University of Posts and
% Telecommunications (BUPT), China, and has published a number of
% technical papers on physical layer of wireless mobile
% communications. In 2006, he was awarded with the ``Program for New
% Century Excellent Talents in University (NCET)'' by the Ministry of
% Education, China. He is also a Member of the IEEE and a Senior
% Member of the Chinese Electronics Association.
% \end{IEEEbiography}
% 

\end{document}

%% file: glossaries.tex
\newglossaryentry{2G}{name=2G, description={second generation}}
\newglossaryentry{3G}{name=3G, description={third generation}}
\newglossaryentry{3GPP}{name=3GPP, description={Third Generation Partnership Project}}
\newglossaryentry{4G}{name=4G, description={fourth generation}}
\newglossaryentry{5G}{name=5G, description={fifth generation}}

\newglossaryentry{ACO}{name=ACO, description={ant colony optimization}}
\newglossaryentry{AF}{name=AF, description={amplify-and-forward}}
\newglossaryentry{AP}{name=AP, description={access point}}
\newglossaryentry{APP}{name=APP, description={\textit{a posteriori} probability}}
\newglossaryentry{AWGN}{name=AWGN, description={additive white Gaussian noise}}
\newglossaryentry{AME}{name=AME, description={asymptotic-multiuser-efficiency}}
\newglossaryentry{ASIC}{name=ASIC, description={application-specific integrated circuit}}
\newglossaryentry{A-CPDA}{name=A-CPDA, description={approximate complex-valued probabilistic data association}}
\newglossaryentry{AB-Log-PDA}{name=AB-Log-PDA, description={approximate Bayes' theorem based logarithmic-domain probabilistic data association}}

\newglossaryentry{BER}{name=BER, description={bit-error rate}}
\newglossaryentry{BC-SDPR}{name=BC-SDPR, description={bound-constrained semidefinite programming relaxation}}
\newglossaryentry{BCJR}{name=BCJR, description={Bahl-Cocke-Jelinek-Raviv}}
\newglossaryentry{BICM}{name=BICM, description={bit-interleaved coded modulation}}
\newglossaryentry{BLER}{name=BLER, description={block-error rate}}
\newglossaryentry{B-PDA}{name=B-PDA, description={bit-based probabilistic data association }}
\newglossaryentry{BPSK}{name=BPSK, description={binary phase-shift keying}}
\newglossaryentry{BS}{name=BS, description={base station}}
\newglossaryentry{BSC}{name=BSC, description={base station controller}}
\newglossaryentry{BALM}{name=BALM, description={block alternating likelihood maximization}}
\newglossaryentry{BQP}{name=BQP, description={Boolean quadratic programming}}
\newglossaryentry{BP}{name=BP, description={belief propagation}}
\newglossaryentry{BI-GDFE}{name=BI-GDFE, description={block-iterative generalized decision feedback equalizer}}

\newglossaryentry{CCI}{name=CCI, description={co-channel interference}}
\newglossaryentry{CCMC}{name=CCMC, description={continuous-input continuous-output memoryless channel}}
\newglossaryentry{CDF}{name=CDF, description={cumulative density function}}
\newglossaryentry{CDM}{name=CDM, description={code-division multiplexing}}
\newglossaryentry{CDMA}{name=CDMA, description={code-division multiple-access}}
\newglossaryentry{CPDA}{name=CPDA, description={complex-valued probabilistic data association}}
\newglossaryentry{CAGR}{name=CAGR, description={compound annual growth rate}}
\newglossaryentry{CMOS}{name=CMOS, description={complementary metal-oxide semiconductor}}

\newglossaryentry{CIR}{name=CIR, description={channel impulse response}}
\newglossaryentry{CP}{name=CP, description={cyclic prefixing}}
\newglossaryentry{CQI}{name=CQI, description={channel quality information}}
\newglossaryentry{CSI}{name=CSI, description={channel state information}}
\newglossaryentry{CSIR}{name=CSIR, description={channel state information at the receiver}}
\newglossaryentry{CSIT}{name=CSIT, description={channel state information at the transmitter}}
\newglossaryentry{CLPS}{name=CLPS, description={closest lattice-point search}}
\newglossaryentry{CSPU}{name=CSPU, description={central signal processing unit}}
\newglossaryentry{CR}{name=CR, description={cognitive radio}}

\newglossaryentry{DAS}{name=DAS, description={distributed antenna system}}
\newglossaryentry{DF}{name=DF, description={decode-and-forward}}
\newglossaryentry{DFE}{name=DFE, description={decision-feedback equalisation}}
\newglossaryentry{DFD}{name=DFD, description={decision-feedback detector}}
\newglossaryentry{DFT}{name=DFT, description={discrete Fourier transform}}
\newglossaryentry{DID}{name=DID, description={distributed iterative detection}}
\newglossaryentry{DPDA}{name=DPDA, description={distributed probabilistic data association}}
\newglossaryentry{DS-CDMA}{name=DS-CDMA, description={direct-sequence code-division multiple-access}}
\newglossaryentry{DSA}{name=DSA, description={dynamic subband/subcarrier allocation}}
\newglossaryentry{DT}{name=DT, description={direct transmission}}
\newglossaryentry{DSNR}{name=DSNR, description={decreasing signal-to-noise ratio}}
\newglossaryentry{DVA-SDPR}{name=DVA-SDPR, description={direct-bit-based virtually antipodal semidefinite programming relaxation}}
\newglossaryentry{DR}{name=DR, description={detectable range}}
\newglossaryentry{DSL}{name=DSL, description={digital subscriber line}}
\newglossaryentry{DQPSK}{name=DQPSK, description={differential quadrature phase-shift keying}}
\newglossaryentry{DMT}{name=DMT, description={diversity-multiplexing tradeoff}}

\newglossaryentry{EGC}{name=EGC, description={equal-gain combining}}
\newglossaryentry{EXIT}{name=EXIT, description={extrinsic information transfer}}
\newglossaryentry{EB-Log-PDA}{name=EB-Log-PDA, description={exact Bayes' theorem based logarithmic-domain probabilistic data association}}
\newglossaryentry{EB}{name=EB, description={exabytes}}
\newglossaryentry{EHF}{name=EHF, description={extremely high frequency}}
\newglossaryentry{EM}{name=EM, description={expectation-maximization}}

\newglossaryentry{FLOP}{name=FLOP, description={floating point operation}}
\newglossaryentry{FD}{name=FD, description={frequency-domain}}
\newglossaryentry{FDE}{name=FDE, description={frequency-domain equalisation}}
\newglossaryentry{FD-LE}{name=FD-LE, description={frequency-domain linear equalisation}}
\newglossaryentry{FD-DFE}{name=FD-DFE, description={frequency-domain decision-feedback equalisation}}
\newglossaryentry{FDM}{name=FDM, description={frequency-division multiplexing}}
\newglossaryentry{FDMA}{name=FDMA, description={frequency-division multiple-access}}
\newglossaryentry{FEC}{name=FEC, description={forward-error-correction}}
\newglossaryentry{FR}{name=FR, description={frequency reuse}}
\newglossaryentry{FH-CDMA}{name=FH-CDMA, description={frequency-hopped code-division multiple-access}}
\newglossaryentry{FIR}{name=FIR, description={finite impulse response}}
\newglossaryentry{FCSD}{name=FCSD, description={fixed-complexity sphere decoding/decoder}}
\newglossaryentry{FER}{name=FER, description={frame-error rate}}

\newglossaryentry{GSNR}{name=GSNR, description={greatest signal-to-noise ratio}}
\newglossaryentry{GA}{name=GA, description={genetic algorithm}}

\newglossaryentry{HSPA}{name=HSPA, description={high speed packet access}}
\newglossaryentry{HNN}{name=HNN, description={Hopfield neural network}}

\newglossaryentry{IC}{name=IC, description={integrated circuit}}
\newglossaryentry{ICI}{name=ICI, description={interchannel interference}}
\newglossaryentry{ICT}{name=ICT, description={information and communication technology}}
\newglossaryentry{ID}{name=ID, description={iterative decoding}}
\newglossaryentry{IDD}{name=IDD, description={iterative detection and decoding}}
\newglossaryentry{IDFT}{name=IDFT, description={inverse discrete Fourier transform}}
\newglossaryentry{ISI}{name=ISI, description={intersymbol interference}}
\newglossaryentry{IMSE}{name=IMSE, description={increasing mean-square error}}
\newglossaryentry{IEEE}{name=IEEE, description={Institute of Electrical and Electronics Engineers}}
\newglossaryentry{IVA-SDPR}{name=IVA-SDPR, description={index-bit-based virtually antipodal semidefinite programming relaxation}}
\newglossaryentry{IAI}{name=IAI, description={interantenna interference}}
\newglossaryentry{IPA}{name=IPA, description={interior point algorithm}}

\newglossaryentry{JPDA}{name=JPDA, description={joint probabilistic data association}}
\newglossaryentry{JMLD}{name=JMLD, description={joint maximum likelihood detection}}
\newglossaryentry{JD}{name=JD, description={joint detection}}

\newglossaryentry{LAS}{name=LAS, description={likelihood ascent search}}
\newglossaryentry{LE}{name=LE, description={linear equalisation}}
\newglossaryentry{LLR}{name=LLR, description={logarithmic likelihood ratio}}
\newglossaryentry{Log-MAP}{name=Log-MAP, description={logarithmic maximum a-posteriori probability}}
\newglossaryentry{LTE}{name=LTE, description={long-term evolution}}
\newglossaryentry{LTE-A}{name=LTE-A, description={Long Term Evolution-Advanced}}
\newglossaryentry{LTI}{name=LTI, description={linear time-invariant}}
\newglossaryentry{LS}{name=LS, description={least-squares}}
\newglossaryentry{LS-MIMO}{name=LS-MIMO, description={large-scale multiple-input multiple-output}}
\newglossaryentry{LMSE}{name=LMSE, description={least mean-square error}}
\newglossaryentry{LDPC}{name=LDPC, description={low-density parity-check}}
\newglossaryentry{LMR}{name=LMR, description={linear matrix representation}}
\newglossaryentry{LSD}{name=LSD, description={list sphere decoding}}
\newglossaryentry{LMI}{name=LMI, description={linear matrix inequality}}
\newglossaryentry{LZF}{name=LZF, description={linear zero-forcing}}
\newglossaryentry{LR}{name=LR, description={lattice-reduction}}
\newglossaryentry{LLL}{name=LLL, description={Lenstra-Lenstra-Lov\'{a}sz}}

\newglossaryentry{MA}{name=MA, description={multiple-access}}
\newglossaryentry{MAP}{name=MAP, description={maximum \textit{a posteriori}}}
\newglossaryentry{MBER}{name=MBER, description={minimum bit error rate}} 
\newglossaryentry{MC-CDMA}{name=MC-CDMA, description={multicarrier code-division multiple-access}}
\newglossaryentry{MCP}{name=MCP, description={multi-cell processing}}
\newglossaryentry{MF}{name=MF, description={matched filter}}
\newglossaryentry{MFB}{name=MFB, description={matched filter bound}}
\newglossaryentry{MIMO}{name=MIMO, description={multiple-input multiple-output}}
\newglossaryentry{MISO}{name=MISO, description={multiple-input single-output}}
\newglossaryentry{ML}{name=ML, description={maximum likelihood}}
\newglossaryentry{MLSE}{name=MLSE, description={maximum likelihood sequence estimator/estimation}}
\newglossaryentry{MIC}{name=MIC, description={multistage interference cancellation}}
\newglossaryentry{MMSE}{name=MMSE, description={minimum mean-square error}}
\newglossaryentry{MMF}{name=MMF, description={multimode fibre}}
\newglossaryentry{MAME}{name=MAME, description={maximum asymptotic-multiuser-efficiency}}
\newglossaryentry{MRC}{name=MRC, description={maximum ratio combining}}
\newglossaryentry{MSE}{name=MSE, description={mean-square error}}
\newglossaryentry{MS}{name=MS, description={mobile station}}
\newglossaryentry{MUD}{name=MUD, description={multiuser detection/detector}}
\newglossaryentry{MUI}{name=MUI, description={multiuser interference}}
\newglossaryentry{M2M}{name=M2M, description={machine-to-machine}}
\newglossaryentry{MFSK}{name=MFSK, description={multiple frequency-shift keying}}
\newglossaryentry{MAI}{name=MAI, description={multiple-access interference}}
\newglossaryentry{MSI}{name=MSI, description={multiple-stream interference}}
\newglossaryentry{MSDD}{name=MSDD, description={multi-symbol differential detection/detector}}
\newglossaryentry{MMW}{name=MMW, description={millimetre wave}}
\newglossaryentry{MGS}{name=MGS, description={mixed Gibbs sampling}}
\newglossaryentry{MR}{name=MR, description={multiple restart}}
\newglossaryentry{MSDSD}{name=MSDSD, description={multiple symbol differential sphere decoder}}
\newglossaryentry{MED}{name=MED, description={minimum Euclidean distance}}

\newglossaryentry{NSC}{name=NSC, description={non-systematic convolutional}}
\newglossaryentry{NP}{name=NP-hard, description={nondeterministic polynomial-time}}
\newglossaryentry{NP-hard}{name=NP-hard, description={nondeterministic polynomial-time hard}}
\newglossaryentry{NP-complete}{name=NP-complete, description={nondeterministic polynomial-time complete}}
\newglossaryentry{NFV}{name=NFV, description={network function virtualization}}

\newglossaryentry{OFDM}{name=OFDM, description={orthogonal frequency-division multiplexing}}
\newglossaryentry{OFDMA}{name=OFDMA, description={orthogonal frequency-division multiple-access}}
\newglossaryentry{OSIC}{name=OSIC, description={ordered successive interference cancellation}}

\newglossaryentry{PDA}{name=PDA, description={probabilistic data association}}
\newglossaryentry{PDF}{name=PDF, description={probability density function}}
\newglossaryentry{PE}{name=PE, description={partial equalisation}}
\newglossaryentry{PIC}{name=PIC, description={parallel interference cancellation}}
\newglossaryentry{P/S}{name=P/S, description={parallel-to-serial}}
\newglossaryentry{PAM}{name=PAM, description={pulse-amplitude modulation}}
\newglossaryentry{PI-SDPR}{name=PI-SDPR, description={polynomial-inspired semidefinite programming relaxation}}
\newglossaryentry{PSD}{name=PSD, description={positive semidefinite}}
\newglossaryentry{PSO}{name=PSO, description={particle swarm optimization}}
\newglossaryentry{PD-IPA}{name=PD-IPA, description={primal-dual interior-point algorithm}}
\newglossaryentry{PSK}{name=PSK, description={phase-shift keying}}
\newglossaryentry{PER}{name=PER, description={packet-error rate}}

\newglossaryentry{QoS}{name=QoS, description={quality-of-service}}
\newglossaryentry{QAM}{name=QAM, description={quadrature amplitude modulation}}
\newglossaryentry{QPSK}{name=QPSK, description={quadrature phase-shift keying}}
\newglossaryentry{QRD}{name=QRD, description={QR-decomposition}}

\newglossaryentry{RF}{name=RF, description={radio frequency}}
\newglossaryentry{RSC}{name=RSC, description={recursive systematic convolutional}}
\newglossaryentry{RPDA}{name=RPDA, description={real-valued probabilistic data association}}
\newglossaryentry{RTS}{name=RTS, description={reactive tabu search}}
\newglossaryentry{R-MCMC}{name=R-MCMC, description={randomized Markov chain Monte Carlo}}
\newglossaryentry{RS}{name=RS, description={randomized search}}

\newglossaryentry{SA}{name=SA, description={simulated annealing}}
\newglossaryentry{SC}{name=SC, description={soft combining}}
\newglossaryentry{SD}{name=SD, description={sphere decoding/decoder}}
\newglossaryentry{S/P}{name=S/P, description={serial-to-parallel}}
\newglossaryentry{SC-FDE}{name=SC-FDE, description={single-carrier frequency-domain equalisation}}
\newglossaryentry{SC-FDMA}{name=SC-FDMA, description={single-carrier frequency-division multiple-access}}
\newglossaryentry{SDM}{name=SDM, description={space-division multiplexing}}
\newglossaryentry{SDMA}{name=SDMA, description={space-division multiple-access}}
\newglossaryentry{SDP}{name=SDP, description={semidefinite programming}}
\newglossaryentry{SDPR}{name=SDPR, description={semidefinite programming relaxation}}
\newglossaryentry{SDR}{name=SDR, description={software-defined radio}}
\newglossaryentry{SDN}{name=SDN, description={software-defined network}}

\newglossaryentry{SIC}{name=SIC, description={successive interference cancellation}}
\newglossaryentry{SISO}{name=SISO, description={soft-input soft-output}}
\newglossaryentry{SIMO}{name=SIMO, description={single-input multiple-output}}
\newglossaryentry{SINR}{name=SINR, description={signal-to-interference-plus-noise ratio}}
\newglossaryentry{SIR}{name=SIR, description={signal-to-interference ratio}}
\newglossaryentry{SM}{name=SM, description={spatial multiplexing}}
\newglossaryentry{SNR}{name=SNR, description={signal-to-noise ratio}}
\newglossaryentry{SP}{name=SP, description={set partitioning}}
\newglossaryentry{SUMF}{name=SUMF, description={single-user matched filter}}
\newglossaryentry{SE}{name=SE, description={Schnorr-Euchner}}
\newglossaryentry{STBC}{name=STBC, description={space-time block code/coded}}
\newglossaryentry{SER}{name=SER, description={symbol-error rate}}
\newglossaryentry{SUD}{name=SUD, description={single-user detection}}
\newglossaryentry{SAIC}{name=SAIC, description={single-antenna interference cancellation}}
\newglossaryentry{SUMIS}{name=SUMIS, description={subspace marginalization aided interference suppression}}
\newglossaryentry{STSK}{name=STSK, description={space-time shift keying}}

\newglossaryentry{TCM}{name=TCM, description={trellis-coded modulation}}
\newglossaryentry{TTCM}{name=TTCM, description={turbo trellis-coded modulation}}
\newglossaryentry{TDD}{name=TDD, description={time-division-duplex}}
\newglossaryentry{TDM}{name=TDM, description={time-division multiplexing}}
\newglossaryentry{TDMA}{name=TDMA, description={time-division multiple-access}}

\newglossaryentry{UMR}{name=UMR, description={unified matrix representation}}
\newglossaryentry{UEP}{name=UEP, description={unequal error protection}}
\newglossaryentry{UE}{name=UE, description={user equipment}}
\newglossaryentry{UCS}{name=UCS, description={unified-client-server}}

\newglossaryentry{VA}{name=VA, description={virtually antipodal}}
\newglossaryentry{VBLAST}{name=VBLAST, description={vertical Bell Laboratories layered space-time}}
\newglossaryentry{VB}{name=VB, description={Viterbo-Boutros}}
\newglossaryentry{VLSI}{name=VLSI, description={very-large-scale integration}}
\newglossaryentry{VA-SDPR}{name=VA-SDPR, description={virtually antipodal semidefinite programming relaxation}}
\newglossaryentry{VNI}{name=VNI, description={visual network index}}
\newglossaryentry{VER}{name=VER, description={vector-error rate}}

\newglossaryentry{WLS}{name=WLS, description={weighted least-squares}}
\newglossaryentry{WiMAX}{name=WiMAX, description={Worldwide Interoperability for Microwave Access}}

\newglossaryentry{XOR}{name=XOR, description={exclusive or}}

\newglossaryentry{ZP}{name=ZP, description={zero-padding}}
\newglossaryentry{ZF}{name=ZF, description={zero-forcing}}

%% file: Fifty_Years_MIMO_Detection.bbl
\begin{thebibliography}{100}
\providecommand{\url}[1]{#1}
\csname url@samestyle\endcsname
\providecommand{\newblock}{\relax}
\providecommand{\bibinfo}[2]{#2}
\providecommand{\BIBentrySTDinterwordspacing}{\spaceskip=0pt\relax}
\providecommand{\BIBentryALTinterwordstretchfactor}{4}
\providecommand{\BIBentryALTinterwordspacing}{\spaceskip=\fontdimen2\font plus
\BIBentryALTinterwordstretchfactor\fontdimen3\font minus
  \fontdimen4\font\relax}
\providecommand{\BIBforeignlanguage}[2]{{%
\expandafter\ifx\csname l@#1\endcsname\relax
\typeout{** WARNING: IEEEtran.bst: No hyphenation pattern has been}%
\typeout{** loaded for the language `#1'. Using the pattern for}%
\typeout{** the default language instead.}%
\else
\language=\csname l@#1\endcsname
\fi
#2}}
\providecommand{\BIBdecl}{\relax}
\BIBdecl

\bibitem{Mcqueen_2009:momentum_LTE}
D.~Mcqueen, ``The momentum behind {LTE} adoption,'' \emph{{IEEE} Communications
  Magazine}, vol.~47, no.~2, pp. 44--45, Feb. 2009.

\bibitem{Lee_2010:Mobile_data_offloading_wifi}
\BIBentryALTinterwordspacing
K.~Lee, J.~Lee, Y.~Yi, I.~Rhee, and S.~Chong, ``Mobile data offloading: how
  much can {WiFi} deliver?'' in \emph{Proc. ACM 6th International Conference on
  emerging Networking EXperiments and Technologies (CoNEXT'10)}, Philadelphia,
  Pennsylvania, USA, Dec. 2010, pp. 26:1--26:12. [Online]. Available:
  \url{http://doi.acm.org/10.1145/1921168.1921203}
\BIBentrySTDinterwordspacing

\bibitem{Sayed_2011:mobile_data_explosion_monetizing}
\BIBentryALTinterwordspacing
M.~El-Sayed, A.~Mukhopadhyay, C.~Urrutia-Vald{\'e}s, and Z.~J. Zhao, ``Mobile
  data explosion: monetizing the opportunity through dynamic policies and {QoS}
  pipes,'' \emph{Bell Labs Technical Journal}, vol.~16, no.~2, pp. 79--99, Sep.
  2011. [Online]. Available: \url{http://dx.doi.org/10.1002/bltj.20504}
\BIBentrySTDinterwordspacing

\bibitem{Ranganathan_2011:data_explosion}
P.~Ranganathan, ``From microprocessors to nanostores: rethinking data-centric
  systems,'' \emph{IEEE Computer Magazine}, vol.~44, no.~1, pp. 39--48, Jan.
  2011.

\bibitem{Han_2012:mobile_data_offloading_opportunistic_social}
B.~Han, P.~Hui, V.~S.~A. Kumar, M.~V. Marathe, J.~Shao, and A.~Srinivasan,
  ``Mobile data offloading through opportunistic communications and social
  participation,'' \emph{{IEEE} Transactions on Mobile Computing}, vol.~11,
  no.~5, pp. 821--834, May 2012.

\bibitem{Cisco_2014:whitepaper_mobile_data_growth_forcast}
\BIBentryALTinterwordspacing
``{Cisco Visual Networking Index: Global Mobile Data Traffic Forecast Update,
  2013-2018},'' White Paper, Cisco, Feb. 2014. [Online]. Available:
  \url{http://www.cisco.com/c/en/us/solutions/collateral/service-provider/visual-networking-index-vni/white_paper_c11-520862.pdf}
\BIBentrySTDinterwordspacing

\bibitem{Cisco_2014:VNI_project}
\BIBentryALTinterwordspacing
Cisco {V}isual {N}etworking {I}ndex ({VNI}). Cisco. [Online]. Available:
  \url{http://www.cisco.com/c/en/us/solutions/service-provider/visual-networking-index-vni/index.html}
\BIBentrySTDinterwordspacing

\bibitem{Lawton_2004:M2M}
G.~Lawton, ``Machine-to-machine technology gears up for growth,'' \emph{IEEE
  Computer Magazine}, vol.~37, no.~9, pp. 12--15, Sep. 2004.

\bibitem{Cha_2009:M2M}
I.~Cha, Y.~Shah, A.~U. Schmidt, A.~Leicher, and M.~V. Meyerstein, ``Trust in
  {M2M} communication,'' \emph{{IEEE} Vehicular Technology Magazine}, vol.~4,
  no.~3, pp. 69--75, Sep. 2009.

\bibitem{Bob_2010:M2M}
\BIBentryALTinterwordspacing
B.~Emmerson, ``{M2M}: the internet of 50 billion devices,'' \emph{Huawei WinWin
  Magazine}, no.~4, pp. 19--22, Jan. 2010. [Online]. Available:
  \url{http://www-cnc.huawei.com/de/static/HW-072296.pdf}
\BIBentrySTDinterwordspacing

\bibitem{Starsinic_2010:M2M}
M.~Starsinic, ``System architecture challenges in the home {M2M} network,'' in
  \emph{Proc. IEEE 6th Annual Conference on Long Island Systems Applications
  and Technology (LISAT'10)}, Farmingdale, NY, USA, May 2010, pp. 1--7.

\bibitem{Chen_2010:M2M}
Y.~Chen and W.~Wang, ``Machine-to-machine communication in {LTE-A},'' in
  \emph{Proc. IEEE 72nd Vehicular Technology Conference (VTC'10-Fall)}, Ottawa,
  ON, Canada, Sep. 2010, pp. 1--4.

\bibitem{Fadlullah_2011:M2M_smart_grid}
Z.~M. Fadlullah, M.~M. Fouda, N.~Kato, A.~Takeuchi, N.~Iwasaki, and Y.~Nozaki,
  ``Toward intelligent machine-to-machine communications in smart grid,''
  \emph{{IEEE} Communications Magazine}, vol.~49, no.~4, pp. 60--65, Apr. 2011.

\bibitem{Niyato_2011:M2M}
D.~Niyato, L.~Xiao, and P.~Wang, ``Machine-to-machine communications for home
  energy management system in smart grid,'' \emph{{IEEE} Communications
  Magazine}, vol.~49, no.~4, pp. 53--59, Apr. 2011.

\bibitem{Lien_2011:M2M}
S.-Y. Lien, K.-C. Chen, and Y.~Lin, ``Toward ubiquitous massive accesses in
  {3GPP} machine-to-machine communications,'' \emph{{IEEE} Communications
  Magazine}, vol.~49, no.~4, pp. 66--74, Apr. 2011.

\bibitem{Lu_2011:M2M}
R.~Lu, X.~Li, X.~Liang, X.~Shen, and X.~Lin, ``{GRS}: the green, reliability,
  and security of emerging machine to machine communications,'' \emph{{IEEE}
  Communications Magazine}, vol.~49, no.~4, pp. 28--35, Apr. 2011.

\bibitem{Wu_2011:M2M}
G.~Wu, S.~Talwar, K.~Johnsson, N.~Himayat, and K.~D. Johnson, ``{M2M}: from
  mobile to embedded internet,'' \emph{{IEEE} Communications Magazine},
  vol.~49, no.~4, pp. 36--43, Apr. 2011.

\bibitem{Zhang_2011:M2M}
Y.~Zhang, R.~Yu, S.~Xie, W.~Yao, Y.~Xiao, and M.~Guizani, ``Home {M2M}
  networks: architectures, standards, and {QoS} improvement,'' \emph{{IEEE}
  Communications Magazine}, vol.~49, no.~4, pp. 44--52, Apr. 2011.

\bibitem{Lien_2011:M2M_CL}
S.-Y. Lien and K.-C. Chen, ``Massive access management for {QoS} guarantees in
  {3GPP} machine-to-machine communications,'' \emph{{IEEE} Communications
  Letters}, vol.~15, no.~3, pp. 311--313, Mar. 2011.

\bibitem{US_aug_2011:frequency_allocation_chart}
\BIBentryALTinterwordspacing
{United States Frequency Allocation Chart: August 2011 Edition}. National
  Telecommunications and Information Administration, United States Department
  of Commerce. [Online]. Available:
  \url{http://www.ntia.doc.gov/files/ntia/publications/spectrum_wall_chart_aug2011.pdf}
\BIBentrySTDinterwordspacing

\bibitem{Paulraj_2003:introduction_to_MIMO}
A.~Paulraj, R.~Nabar, and D.~Gore, \emph{Introduction to Space-Time Wireless
  Communications}.\hskip 1em plus 0.5em minus 0.4em\relax Cambridge University
  Press, 2003.

\bibitem{Gesbert_2007:shifting_MIMO_paradigm}
D.~Gesbert, M.~Kountouris, R.~W. Heath, C.-B. Chae, and T.~Salzer, ``Shifting
  the {MIMO} paradigm: from single-user to multiuser communications,''
  \emph{{IEEE} Signal Processing Magazine}, vol.~24, no.~5, pp. 36--46, Oct.
  2007.

\bibitem{Mietzner_2009:multi_antenna_survey}
J.~Mietzner, R.~Schober, L.~Lampe, W.~H. Gerstacker, and P.~A. Hoeher,
  ``Multiple-antenna techniques for wireless communications -- a comprehensive
  literature survey,'' \emph{IEEE Communications Surveys {\&} Tutorials},
  vol.~11, no.~2, pp. 87--105, Second Quarter 2009.

\bibitem{Gesbert_2010:multicell_MIMO}
D.~Gesbert, S.~V. Hanly, H.~Huang, S.~Shamai~(Shitz), O.~Simeone, and W.~Yu,
  ``Multi-cell {MIMO} cooperative networks: a new look at interference,''
  \emph{{IEEE} Journal on Selected Areas in Communications}, vol.~28, no.~9,
  pp. 1380--1408, Dec. 2010.

\bibitem{Rusek_2013:massive_MIMO}
F.~Rusek, D.~Persson, B.~K. Lau, E.~G. Larsson, T.~L. Marzetta, O.~Edfors, and
  F.~Tufvesson, ``Scaling up {MIMO}: opportunities and challenges with very
  large arrays,'' \emph{{IEEE} Signal Processing Magazine}, vol.~30, no.~1, pp.
  40--60, Jan. 2013.

\bibitem{Larsson_2014:massive_MIMO_overview}
E.~Larsson, O.~Edfors, F.~Tufvesson, and T.~Marzetta, ``Massive {MIMO} for next
  generation wireless systems,'' \emph{{IEEE} Communications Magazine},
  vol.~52, no.~2, pp. 186--195, Feb. 2014.

\bibitem{Boccardi_2014:five_tech_5G}
F.~Boccardi, R.~W. Heath, A.~Lozano, T.~L. Marzetta, and P.~Popovski, ``Five
  disruptive technology directions for {5G},'' \emph{{IEEE} Communications
  Magazine}, vol.~52, no.~2, pp. 74--80, Feb. 2014.

\bibitem{Andrews_2014:5G_overview}
J.~G. Andrews, S.~Buzzi, W.~Choi, S.~V. Hanly, A.~Lozano, A.~C.~K. Soong, and
  J.~C. Zhang, ``What will {5G} be?'' \emph{{IEEE} Journal on Selected Areas in
  Communications}, vol.~32, no.~6, pp. 1065--1082, Jun. 2014.

\bibitem{Lu_2014_JSTSP_massive_MIMO_overview}
L.~Lu, G.~Li, A.~Swindlehurst, A.~Ashikhmin, and R.~Zhang, ``An overview of
  massive {MIMO}: Benefits and challenges,'' \emph{IEEE Journal of Selected
  Topics in Signal Processing}, vol.~8, no.~5, Oct. 2014.

\bibitem{Han_2011:green_radio}
C.~Han, T.~Harrold, S.~Armour, I.~Krikidis, S.~Videv, P.~M. Grant, H.~Haas,
  J.~S. Thompson, I.~Ku, C.-X. Wang, T.~A. Le, M.~R. Nakhai, J.~Zhang, and
  L.~Hanzo, ``Green radio: radio techniques to enable energy-efficient wireless
  networks,'' \emph{{IEEE} Communications Magazine}, vol.~49, no.~6, pp.
  46--54, Jun. 2011.

\bibitem{Larsson_2013:EE_SE_massive_MIMO}
H.~Q. Ngo, E.~G. Larsson, and T.~L. Marzetta, ``Energy and spectral efficiency
  of very large multiuser {MIMO} systems,'' \emph{{IEEE} Transactions on
  Communications}, vol.~61, no.~4, pp. 1436--1449, Apr. 2013.

\bibitem{Renzo_2014:spatial_modulation}
M.~Di~Renzo, H.~Haas, A.~Ghrayeb, S.~Sugiura, and L.~Hanzo, ``Spatial
  modulation for generalized {MIMO}: Challenges, opportunities, and
  implementation,'' \emph{Proceedings of the IEEE}, vol. 102, no.~1, pp.
  56--103, Jan. 2014.

\bibitem{Rappaport2013:MMW_5G}
T.~S. Rappaport, S.~Sun, R.~Mayzus, H.~Zhao, Y.~Azar, K.~Wang, G.~N. Wong,
  J.~K. Schulz, M.~Samimi, and F.~Gutierrez, ``Millimeter wave mobile
  communications for 5{G} cellular: it will work!'' \emph{IEEE Access}, vol.~1,
  pp. 335--349, May 2013.

\bibitem{Shannon_1948:math_theory_comm}
C.~E. Shannon, ``A mathematical theory of communication,'' \emph{The Bell
  System Technical Journal}, vol.~27, no.~3, pp. 379--423, Jul. 1948.

\bibitem{Hanly:initial_interference_limited}
S.~V. Hanly and P.~Whiting, ``Information-theoretic capacity of multi-receiver
  networks,'' \emph{Telecommunication Systems}, vol.~1, no.~1, pp. 1--42, 1993.

\bibitem{Wyner:initial_interference_limited}
A.~D. Wyner, ``Shannon-theoretic approach to a {Gaussian} cellular
  multiple-access channel,'' \emph{{IEEE} Transactions on Information Theory},
  vol.~40, no.~6, pp. 1713--1727, Nov. 1994.

\bibitem{Shamai:interference_limited_part_I}
S.~Shamai~(Shitz) and A.~D. Wyner, ``Information-theoretic considerations for
  symmetric, cellular, multiple-access fading channels -- {Part I},''
  \emph{{IEEE} Transactions on Information Theory}, vol.~43, no.~6, pp.
  1877--1894, Nov. 1997.

\bibitem{Shamai:interference_limited_part_II}
------, ``Information-theoretic considerations for symmetric, cellular,
  multiple-access fading channels -- {Part II},'' \emph{{IEEE} Transactions on
  Information Theory}, vol.~43, no.~6, pp. 1895--1911, Nov. 1997.

\bibitem{Somekh:work_of_shamai's_student}
O.~Somekh and S.~Shamai~(Shitz), ``Shannon-theoretic approach to a {G}aussian
  cellular multiple-access channel with fading,'' \emph{{IEEE} Transactions on
  Information Theory}, vol.~46, no.~4, pp. 1401--1425, Jul. 2000.

\bibitem{Catreux:MIMO_cellular_capacity}
S.~Catreux, P.~F. Driessen, and L.~J. Greenstein, ``Simulation results for an
  interference-limited multiple-input multipleoutput cellular system,''
  \emph{{IEEE} Communications Letters}, vol.~4, no.~11, pp. 334--336, Nov.
  2000.

\bibitem{Catreux:Throughput_MIMO_cellular}
------, ``Attainable throughput of an interference-limited multiple-input
  multiple-output ({MIMO}) cellular system,,'' \emph{{IEEE} Transactions on
  Communications}, vol.~49, no.~8, pp. 1307--1311, Aug. 2001.

\bibitem{Blum:analysis_MIMO_cellular_capacity}
R.~S. Blum, J.~H. Winters, and N.~R. Sollenberger, ``On the capacity of
  cellular systems with {MIMO},'' \emph{{IEEE} Communications Letters}, vol.~6,
  no.~6, pp. 242--244, Jun. 2002.

\bibitem{Dai:downlink_capacity_MIMO_cellular}
H.~Dai, A.~F. Molisch, and H.~V. Poor, ``Downlink capacity of
  interference-limited {MIMO} system with joint detection,'' \emph{{IEEE}
  Transactions on Wireless Communications}, vol.~3, no.~2, pp. 442--453, Mar.
  2004.

\bibitem{Tse:Fundamental}
D.~N.~C. Tse and P.~Viswanath, \emph{Fundamentals of Wireless
  Communication}.\hskip 1em plus 0.5em minus 0.4em\relax Cambridge, UK:
  Cambridge University Press, 2005.

\bibitem{Boas_1981:prove_NP_hard_closest_point_search}
P.~van Emde~Boas, ``Another {NP}-complete partition problem and the complexity
  of computing short vectors in a lattice,'' Department of Mathematics,
  University of Amsterdam, The Netherlands, Tech. Rep. 81-04, Apr. 1981.

\bibitem{Verdu_1989:complexity_optimal_MUD}
\BIBentryALTinterwordspacing
S.~Verd{\'u}, ``Computational complexity of optimum multiuser detection,''
  \emph{Algorithmica}, vol.~4, no. 1-4, pp. 303--312, Jun. 1989. [Online].
  Available: \url{http://dx.doi.org/10.1007/BF01553893}
\BIBentrySTDinterwordspacing

\bibitem{Micciancio_2001:simpler_proof_NP_hard_closest_point_search}
D.~Micciancio, ``The hardness of the closest vector problem with
  preprocessing,'' \emph{{IEEE} Transactions on Information Theory}, vol.~47,
  no.~3, pp. 1212--1215, Mar. 2001.

\bibitem{Hallen_1995:MUD_CDMA_overview}
A.~Duel-Hallen, J.~M. Holtzman, and Z.~Zvonar, ``Multiuser detection for {CDMA}
  systems,'' \emph{{IEEE} Personal Communications Magazine}, vol.~2, no.~2, pp.
  46--58, Apr. 1995.

\bibitem{Moshavi_1996:MUD_overview_CDMA}
S.~Moshavi, ``Multi-user detection for {DS-CDMA} communications,'' \emph{{IEEE}
  Communications Magazine}, vol.~34, no.~10, pp. 124--136, Oct. 1996.

\bibitem{Verdu:MUD_book}
S.~Verd{\'u}, \emph{Multiuser Detection}.\hskip 1em plus 0.5em minus
  0.4em\relax Cambridge, UK: Cambridge University Press, 1998.

\bibitem{Honig:advances_MUD_edited}
M.~L. Honig, Ed., \emph{Advances in Multiuser Detection}.\hskip 1em plus 0.5em
  minus 0.4em\relax John Wiley \& Sons, Inc., 2009.

\bibitem{Bai_2012:low_complexity_MIMO_detection}
L.~Bai and J.~Choi, \emph{Low Complexity MIMO Detection}.\hskip 1em plus 0.5em
  minus 0.4em\relax Springer, 2012.

\bibitem{Viterbo_1999:SD}
E.~Viterbo and J.~Boutros, ``A universal lattice code decoder for fading
  channels,'' \emph{{IEEE} Transactions on Information Theory}, vol.~45, no.~7,
  pp. 1639--1642, Jul. 1999.

\bibitem{Agrell:closest_point_search_in_lattice}
E.~Agrell, T.~Eriksson, A.~Vardy, and K.~Zeger, ``Closest point search in
  lattices,'' \emph{{IEEE} Transactions on Information Theory}, vol.~48, no.~8,
  pp. 2201--2214, Aug. 2002.

\bibitem{Damen_2000:lattice_decoder_STC}
M.~O. Damen, A.~Chkeif, and J.-C. Belfiore, ``Lattice code decoder for
  space-time codes,'' \emph{{IEEE} Communications Letters}, vol.~4, no.~5, pp.
  161--163, May 2000.

\bibitem{Damen_2000:generalized_SD_STC}
M.~O. Damen, K.~Abed-Meraim, and J.-C. Belfiore, ``Generalised sphere decoder
  for asymmetrical space-time communication architecture,'' \emph{Electronics
  Letters}, vol.~36, no.~2, pp. 166--167, Jan. 2000.

\bibitem{Damen_2003:MLD_closest_lattice_point_search}
M.~O. Damen, H.~El~Gamal, and G.~Caire, ``On maximum-likelihood detection and
  the search for the closest lattice point,'' \emph{{IEEE} Transactions on
  Information Theory}, vol.~49, no.~10, pp. 2389--2402, Oct. 2003.

\bibitem{Hassibi_2005:SD_complexity_part_1}
B.~Hassibi and H.~Vikalo, ``On the sphere-decoding algorithm {I}. expected
  complexity,'' \emph{{IEEE} Transactions on Signal Processing}, vol.~53,
  no.~8, pp. 2806--2818, Aug. 2005.

\bibitem{Vikalo_2005:SD_complexity_part_2}
H.~Vikalo and B.~Hassibi, ``On the sphere-decoding algorithm {II}.
  generalizations, second-order statistics, and applications to
  communications,'' \emph{{IEEE} Transactions on Signal Processing}, vol.~53,
  no.~8, pp. 2819--2834, Aug. 2005.

\bibitem{Jalden:SD_complexity_journal}
J.~Jald\'{e}n and B.~Ottersten, ``On the complexity of sphere decoding in
  digital communications,'' \emph{{IEEE} Transactions on Signal Processing},
  vol.~53, no.~4, pp. 1474--1484, Apr. 2005.

\bibitem{Burg_2005:VLSI_depth_first_SD}
A.~Burg, M.~Borgmann, M.~Wenk, M.~Zellweger, W.~Fichtner, and H.~Bolcskei,
  ``{VLSI} implementation of {MIMO} detection using the sphere decoding
  algorithm,'' \emph{{IEEE} Journal of Solid-State Circuits}, vol.~40, no.~7,
  pp. 1566--1577, Jul. 2005.

\bibitem{Wong_2002:K-best_SD_VLSI}
K.-W. Wong, C.-Y. Tsui, R.~S. Cheng, and W.-H. Mow, ``A {VLSI} architecture of
  a {K}-best lattice decoding algorithm for {MIMO} channels,'' in \emph{Proc.
  IEEE International Symposium on Circuits and Systems (ISCAS'02)}, Scottsdale,
  AZ, USA, May 2002, pp. III--273--III--276.

\bibitem{Guo_2006:implementation_K_best_SD_MIMO}
Z.~Guo and P.~Nilsson, ``Algorithm and implementation of the {K}-best sphere
  decoding for {MIMO} detection,'' \emph{{IEEE} Journal on Selected Areas in
  Communications}, vol.~24, no.~3, pp. 491--503, Mar. 2006.

\bibitem{Chen_2007:K-best_VLSI}
S.~Chen, T.~Zhang, and Y.~Xin, ``Relaxed {K}-best {MIMO} signal detector design
  and {VLSI} implementation,'' \emph{IEEE Transactions on Very Large Scale
  Integration (VLSI) Systems}, vol.~15, no.~3, pp. 328--337, 2007.

\bibitem{Studer_2008:soft_SD_implementation}
C.~Studer, A.~Burg, and H.~Bolcskei, ``Soft-output sphere decoding: algorithms
  and {VLSI} implementation,'' \emph{{IEEE} Journal on Selected Areas in
  Communications}, vol.~26, no.~2, pp. 290--300, Feb. 2008.

\bibitem{Murugan_2006:unified_tree_search_sequential_decoder}
A.~R. Murugan, H.~El~Gamal, M.~O. Damen, and G.~Caire, ``A unified framework
  for tree search decoding: rediscovering the sequential decoder,''
  \emph{{IEEE} Transactions on Information Theory}, vol.~52, no.~3, pp.
  933--953, Mar. 2006.

\bibitem{Gowaikar_2007:statistical_pruning_SD_depth_first}
R.~Gowaikar and B.~Hassibi, ``Statistical pruning for near-maximum likelihood
  decoding,'' \emph{{IEEE} Transactions on Signal Processing}, vol.~55, no.~6,
  pp. 2661--2675, Jun. 2007.

\bibitem{Lee_2007:short_path_SD}
K.~Lee and J.~Chun, ``{ML} symbol detection based on the shortest path
  algorithm for {MIMO} systems,'' \emph{{IEEE} Transactions on Signal
  Processing}, vol.~55, no.~11, pp. 5477--5484, Nov. 2007.

\bibitem{Stojnic_2008:speed_up_SD_via_infinity_H_norm}
M.~Stojnic, H.~Vikalo, and B.~Hassibi, ``Speeding up the sphere decoder with
  {$H^{\infty}$} and {SDP} inspired lower bounds,'' \emph{{IEEE} Transactions
  on Signal Processing}, vol.~56, no.~2, pp. 712--726, Feb. 2008.

\bibitem{Kim_2010:best_first_search}
T.-H. Kim and I.-C. Park, ``High-throughput and area-efficient {MIMO} symbol
  detection based on modified {D}ijkstra's search,'' \emph{IEEE Transactions on
  Circuits and Systems I: Regular Papers}, vol.~57, no.~7, pp. 1756--1766, Jul.
  2010.

\bibitem{Chang_2012:best_first_search_A_algorithm}
R.~Y. Chang, W.-H. Chung, and S.-J. Lin, ``A* algorithm inspired
  memory-efficient detection for {MIMO} systems,'' \emph{IEEE Wireless
  Communications Letters}, vol.~1, no.~5, pp. 508--511, Oct. 2012.

\bibitem{Chang_2012:best_first_search}
R.~Y. Chang and W.-H. Chung, ``Best-first tree search with probabilistic node
  ordering for {MIMO} detection: generalization and performance-complexity
  tradeoff,'' \emph{{IEEE} Transactions on Wireless Communications}, vol.~11,
  no.~2, pp. 780--789, Feb. 2012.

\bibitem{Choi:sphere_decoder_look_ahead}
J.~W. Choi, B.~Shim, and A.~C. Singer, ``Efficient soft-input soft-output tree
  detection via an improved path metric,'' \emph{{IEEE} Transactions on
  Information Theory}, vol.~58, no.~3, pp. 1518--1533, Mar. 2012.

\bibitem{Hochwald:SD_near_capacity}
B.~M. Hochwald and S.~ten Brink, ``Achieving near-capacity on a
  multiple-antenna channel,'' \emph{{IEEE} Transactions on Communications},
  vol.~51, no.~3, pp. 389--399, Mar. 2003.

\bibitem{Boutros_2003:SISO_SD_MIMO}
J.~Boutros, N.~Gresset, L.~Brunel, and M.~Fossorier, ``Soft-input soft-output
  lattice sphere decoder for linear channels,'' in \emph{Proc. IEEE Global
  Telecommunications Conference (GLOBECOM'03)}, San Francisco, USA, Dec. 2003,
  pp. 1583--1587.

\bibitem{Vikalo_2004:IDD_SD_MIMO}
H.~Vikalo, B.~Hassibi, and T.~Kailath, ``Iterative decoding for {MIMO} channels
  via modified sphere decoding,'' \emph{{IEEE} Transactions on Wireless
  Communications}, vol.~3, no.~6, pp. 2299--2311, Nov. 2004.

\bibitem{Wang_2006:approaching_MIMO_capacity_hard_SD}
R.~Wang and G.~B. Giannakis, ``Approaching {MIMO} channel capacity with soft
  detection based on hard sphere decoding,'' \emph{{IEEE} Transactions on
  Communications}, vol.~54, no.~4, pp. 587--590, Apr. 2006.

\bibitem{Studer_2010:SISO_single_SD}
C.~Studer and H.~Bolcskei, ``Soft-input sof-output single tree-search sphere
  decoding,'' \emph{{IEEE} Transactions on Information Theory}, vol.~56,
  no.~10, pp. 4827--4842, Oct. 2010.

\bibitem{Rachid_2010:best_first_search}
M.~Rachid and B.~Daneshrad, ``Iterative {MIMO} sphere decoding throughput
  guarantees under realistic channel conditions,'' \emph{{IEEE} Communications
  Letters}, vol.~14, no.~4, pp. 342--344, Apr. 2010.

\bibitem{Barbero:fixed_complexity_SD_journal}
L.~G. Barbero and J.~S. Thompson, ``Fixing the complexity of the sphere decoder
  for {MIMO} detection,'' \emph{{IEEE} Transactions on Wireless
  Communications}, vol.~7, no.~6, pp. 2131--2142, Jun. 2008.

\bibitem{Barbero:soft_fixed_complexity_SD}
------, ``Extending a fixed-complexity sphere decoder to obtain likelihood
  information for {Turbo-MIMO} systems,'' \emph{{IEEE} Transactions on
  Vehicular Technology}, vol.~57, no.~5, pp. 2804--2814, Sep. 2008.

\bibitem{Jalden:FCSD_error_prob}
J.~Jald\'{e}n, L.~G. Barbero, B.~Ottersten, and J.~S. Thompson, ``The error
  probability of the fixed-complexity sphere decoder,'' \emph{{IEEE}
  Transactions on Signal Processing}, vol.~57, no.~7, pp. 2711--2720, Jul.
  2009.

\bibitem{Yao_2002:LR_MIMO_detector}
H.~Yao and G.~W. Wornell, ``Lattice-reduction-aided detectors for {MIMO}
  communication systems,'' in \emph{Proc. IEEE Global Telecommunications
  Conference (GLOBECOM'02)}, Nov. 2002, pp. 424--428.

\bibitem{Fischer_2003:MIMO_precoding_detection}
C.~Windpassinger and R.~F.~H. Fischer, ``Low-complexity near-maximum-likelihood
  detection and precoding for {MIMO} systems using lattice reduction,'' in
  \emph{Proc. IEEE Information Theory Workshop (ITW'03)}, Paris, France, Mar.
  2003, pp. 345--348.

\bibitem{Wubben_2004:LR_reduction_MMSE}
D.~W\"{u}bben, R.~B\"{o}hnke, V.~K\"{u}hn, and K.-D. Kammeyer,
  ``Near-maximum-likelihood detection of {MIMO} systems using {MMSE}-based
  lattice reduction,'' in \emph{Proc. IEEE International Conference on
  Communications (ICC'04)}, Paris, France, Jun. 2004, pp. 798--802.

\bibitem{Windpassinger_2004:LR_precoding_MIMO}
C.~Windpassinger, R.~F.~H. Fischer, and J.~B. Huber, ``Lattice-reduction-aided
  broadcast precoding,'' \emph{{IEEE} Transactions on Communications}, vol.~52,
  no.~12, pp. 2057--2060, Dec. 2004.

\bibitem{Taherzadeh_2007:LLL_LR_receiver_diversity}
M.~Taherzadeh, A.~Mobasher, and A.~K. Khandani, ``{LLL} reduction achieves the
  receive diversity in {MIMO} decoding,'' \emph{{IEEE} Transactions on
  Information Theory}, vol.~53, no.~12, pp. 4801--4805, Dec. 2007.

\bibitem{Ling_2007:LLL_LR}
C.~Ling and N.~Howgrave-Graham, ``Effective {LLL} reduction for lattice
  decoding,'' in \emph{Proc. IEEE International Symposium on Information Theory
  (ISIT'07)}, Nice, France, Jun. 2007, pp. 196--200.

\bibitem{Jalden_2008:complexity_LLL}
J.~Jald\'{e}n, D.~Seethaler, and G.~Matz, ``Worst- and average-case complexity
  of {LLL} lattice reduction in {MIMO} wireless systems,'' in \emph{Proc. IEEE
  International Conference on Acoustics, Speech and Signal Processing
  (ICASSP'08)}, Las Vegas, NV, Mar. 2008, pp. 2685--2688.

\bibitem{Seethaler_2007:SA_LR_MIMO}
D.~Seethaler, G.~Matz, and F.~Hlawatsch, ``Low-complexity {MIMO} data detection
  using {Seysen's} lattice reduction algorithm,'' in \emph{Proc. IEEE
  International Conference on Acoustics, Speech and Signal Processing
  (ICASSP'07)}, vol.~3, Apr. 2007, pp. III--53--III--56.

\bibitem{zhang_2008:analysis}
W.~Zhang, F.~Arnold, and X.~Ma, ``An analysis of {Seysen's} lattice reduction
  algorithm,'' \emph{Signal Processing}, vol.~88, no.~10, pp. 2573--2577, Oct.
  2008.

\bibitem{Burg_2007:LR_Brun_algorithm}
A.~Burg, D.~Seethaler, and G.~Matz, ``{VLSI} implementation of a
  lattice-reduction algorithm for multi-antenna broadcast precoding,'' in
  \emph{Proc. IEEE International Symposium on Circuits and Systems (ISCAS'07)},
  New Orleans, LA, May 2007, pp. 673--676.

\bibitem{Shabany_2008:LR_VLSI_implementation}
M.~Shabany and P.~G. Gulak, ``The application of lattice-reduction to the
  {K}-best algorithm for near-optimal {MIMO} detection,'' in \emph{Proc. IEEE
  International Symposium on Circuits and Systems (ISCAS'08)}, Seattle, WA, May
  2008, pp. 316--319.

\bibitem{Gestner_2008:LR_VLSI_CLLL}
B.~Gestner, W.~Zhang, X.~Ma, and D.~V. Anderson, ``{VLSI} implementation of a
  lattice reduction algorithm for low-complexity equalization,'' in \emph{Proc.
  IEEE International Conference on Circuits and Systems for Communications
  (ICCSC'08)}, Shanghai, China, May 2008, pp. 643--647.

\bibitem{Gan_2005:CLLL_conf}
Y.~H. Gan and W.~H. Mow, ``Complex lattice reduction algorithms for
  low-complexity {MIMO} detection,'' in \emph{Proc. IEEE Global
  Telecommunications Conference (GLOBECOM'05)}, St. Louis, MO, Dec. 2005, pp.
  2953--2957.

\bibitem{Gan_2009:CLLL_MIMO_detection}
Y.~H. Gan, C.~Ling, and W.~H. Mow, ``Complex lattice reduction algorithm for
  low-complexity full-diversity {MIMO} detection,'' \emph{{IEEE} Transactions
  on Signal Processing}, vol.~57, no.~7, pp. 2701--2710, Jul. 2009.

\bibitem{Ma_2008:performance_analysis_CLLL}
X.~Ma and W.~Zhang, ``Performance analysis for {MIMO} systems with
  lattice-reduction aided linear equalization,'' \emph{{IEEE} Transactions on
  Communications}, vol.~56, no.~2, pp. 309--318, Feb. 2008.

\bibitem{Ma_2008:CLLL}
------, ``Fundamental limits of linear equalizers: Diversity, capacity, and
  complexity,'' \emph{{IEEE} Transactions on Information Theory}, vol.~54,
  no.~8, pp. 3442--3456, Aug. 2008.

\bibitem{Silvola_2006:soft_LR_MAP}
P.~Silvola, K.~Hooli, and M.~Juntti, ``Suboptimal soft-output {MAP} detector
  with lattice reduction,'' \emph{{IEEE} Signal Processing Letters}, vol.~13,
  no.~6, pp. 321--324, Jun. 2006.

\bibitem{Qi_2007:soft_LR}
X.-F. Qi and K.~Holt, ``A lattice-reduction-aided soft demapper for high-rate
  coded {MIMO-OFDM} systems,'' \emph{{IEEE} Signal Processing Letters},
  vol.~14, no.~5, pp. 305--308, May 2007.

\bibitem{Ponnampalam_2007:LR_soft_outputs}
V.~Ponnampalam, D.~McNamara, A.~Lillie, and M.~Sandell, ``On generating soft
  outputs for lattice-reduction-aided {MIMO} detection,'' in \emph{Proc. IEEE
  International Conference on Communications (ICC'07)}, Glasgow, UK, Jun. 2007,
  pp. 4144--4149.

\bibitem{Wei_2010:soft_LR_MIMO_detector}
W.~Zhang and X.~Ma, ``Low-complexity soft-output decoding with
  lattice-reduction-aided detectors,'' \emph{{IEEE} Transactions on
  Communications}, vol.~58, no.~9, pp. 2621--2629, Sep. 2010.

\bibitem{Zhou_2013:Lattice_reduction_LS_MIMO_detector}
Q.~Zhou and X.~Ma, ``Element-based lattice reduction algorithms for large
  {MIMO} detection,'' \emph{{IEEE} Journal on Selected Areas in
  Communications}, vol.~31, no.~2, pp. 274--286, Feb. 2013.

\bibitem{Luo:PDA_Sync_CDMA}
J.~Luo, K.~R. Pattipati, P.~K. Willett, and F.~Hasegawa, ``Near optimal
  multiuser detection in synchronous {CDMA} using probabilistic data
  association,'' \emph{{IEEE} Communications Letters}, vol.~5, no.~9, pp.
  361--363, Sep. 2001.

\bibitem{Luo:PDA_thesis}
\BIBentryALTinterwordspacing
J.~Luo, ``Improved multiuser detection in code-division multiple access
  systems,'' Ph.D. dissertation, Univ. of Connecticut, Storrs, May 2002.
  [Online]. Available:
  \url{http://istec.colostate.edu/~rockey/Papers/PhDThesis.pdf}
\BIBentrySTDinterwordspacing

\bibitem{Pham:PDA_Async_CDMA}
D.~Pham, J.~Luo, K.~R. Pattipati, and P.~K. Willett, ``A {PDA}-{Kalman}
  approach to multiuser detection in asynchronous {CDMA},'' \emph{{IEEE}
  Communications Letters}, vol.~6, no.~11, pp. 475--477, Nov. 2002.

\bibitem{Luo_2003:sliding_window_PDA}
J.~Luo, K.~R. Pattipati, and P.~K. Willett, ``A sliding window {PDA} for
  asynchronous {CDMA}, and a proposal for deliberate asynchronicity,''
  \emph{{IEEE} Transactions on Communications}, vol.~51, no.~12, pp.
  1970--1974, 2003.

\bibitem{Penghui_2003:iterative_PDA_MUD}
P.~H. Tan, L.~K. Rasmussen, and J.~Luo, ``Iterative multiuser decoding based on
  probabilistic data association,'' in \emph{Proc. IEEE International Symposium
  on Information Theory (ISIT'03)}, Yokohama, Japan, Jun. 2003, pp. 301--301.

\bibitem{Yin_2004:turbo_equalization_PDA}
Y.~Yin, Y.~Huang, and J.~Zhang, ``Turbo equalization using probabilistic data
  association,'' in \emph{Proc. IEEE Global Telecommunications Conference
  (GLOBECOM'04)}, Dallas, TX, USA, Dec. 2004, pp. 2535--2539.

\bibitem{Huang_2004:generalized_PDA}
Y.~Huang and J.~Zhang, ``A generalized probabilistic data association multiuser
  detector,'' in \emph{Proc. IEEE International Symposium on Information Theory
  (ISIT'04)}, Chicago, IL, USA, Jun. 2004, p. 529.

\bibitem{Pham:GPDA}
D.~Pham, K.~R. Pattipati, P.~K. Willett, and J.~Luo, ``A generalized
  probabilistic data association detector for multiple antenna systems,''
  \emph{{IEEE} Communications Letters}, vol.~8, no.~4, pp. 205--207, Apr. 2004.

\bibitem{Liu:CPDA-apx}
S.~Liu and Z.~Tian, ``Near-optimum soft decision equalization for frequency
  selective {MIMO} channels,'' \emph{{IEEE} Transactions on Signal Processing},
  vol.~52, no.~3, pp. 721--733, Mar. 2004.

\bibitem{Liu:Kalman_PDA_freq_selective}
------, ``A {Kalman}-{PDA} approach to soft-decision equalization for
  frequency-selective {MIMO} channels,'' \emph{{IEEE} Transactions on Signal
  Processing}, vol.~53, no.~10, pp. 3819--3830, Oct. 2005.

\bibitem{Latsoudas_2005:hybrid_PDA_SD}
G.~Latsoudas and N.~D. Sidiropoulos, ``A hybrid probabilistic data
  association-sphere decoding detector for multiple-input-multiple-output
  systems,'' \emph{{IEEE} Signal Processing Letters}, vol.~12, no.~4, pp.
  309--312, Apr. 2005.

\bibitem{Jia_2005:Gaussian_approximation_mixture_reduction_MIMO}
Y.~Jia, C.~Andrieu, R.~J. Piechocki, and M.~Sandell, ``Gaussian approximation
  based mixture reduction for near optimum detection in {MIMO} systems,''
  \emph{{IEEE} Communications Letters}, vol.~9, no.~11, pp. 997--999, Nov.
  2005.

\bibitem{Fricke:Impact_of_Gaussian_approximation}
J.~Fricke, M.~Sandell, J.~Mietzner, and P.~A. Hoeher, ``Impact of the
  {Gaussian} approximation on the performance of the probabilistic data
  association {MIMO} decoder,'' \emph{{EURASIP} Journal on Wireless
  Communications and Networking}, vol. 2005, no.~5, pp. 796--800, 2005.

\bibitem{Shaoqian2005:turbo_PDA}
J.~Wang and S.~Li, ``{MIMO} turbo receiver with new probability data
  association soft interference cancellation,'' in \emph{Proc. International
  Conference on Communications, Circuits and Systems (ICCCAS'05))}, Hong Kong,
  China, May 2005, pp. 232--236.

\bibitem{Penghui_2006:asymptotic_optimum_PDA}
P.~H. Tan and L.~K. Rasmussen, ``Asymptotically optimal nonlinear {MMSE}
  multiuser detection based on multivariate gaussian approximation,''
  \emph{{IEEE} Transactions on Communications}, vol.~54, no.~8, pp. 1427--1428,
  Aug. 2006.

\bibitem{Cai2006:iterative_PDA}
Y.~Cai, X.~Xu, Y.~Cheng, Y.~Xu, and Z.~Li, ``A {SISO} iterative probabilistic
  data association detector for {MIMO} systems,'' in \emph{Proc. 10th
  International Conference on Communication Technology (ICCT'06)}, Guilin,
  China, Nov. 2006, pp. 1--4.

\bibitem{Jia:CPDA}
Y.~Jia, C.~M. Vithanage, C.~Andrieu, and R.~J. Piechocki, ``Probabilistic data
  association for symbol detection in {MIMO} systems,'' \emph{Electronics
  Letters}, vol.~42, no.~1, pp. 38--40, Jan. 2006.

\bibitem{Cao:Relation_of_PDA_and_MMSE-SIC}
F.~Cao, J.~Li, and J.~Yang, ``On the relation between {PDA} and
  {MMSE}-{ISDIC},'' \emph{{IEEE} Signal Processing Letters}, vol.~14, no.~9,
  pp. 597--600, Sep. 2007.

\bibitem{Bavarian_2007:distributed_BS_cooperation_uplink}
S.~Bavarian and J.~K. Cavers, ``Reduced complexity distributed base station
  processing in the uplink of cellular networks,'' in \emph{Proc. IEEE Global
  Telecommunications Conference (GLOBECOM'07)}, Washington, DC, USA, Nov. 2007,
  pp. 4500--4504.

\bibitem{Bavarian_2008:distributed_BS_cooperation_uplink_journal}
------, ``Reduced-complexity belief propagation for system-wide {MUD} in the
  uplink of cellular networks,'' \emph{{IEEE} Journal on Selected Areas in
  Communications}, vol.~26, no.~3, pp. 541--549, Apr. 2008.

\bibitem{Jia_2008:multilevel_SGA_PDA}
Y.~Jia, C.~Andrieu, R.~J. Piechocki, and M.~Sandell, ``Depth-first and
  breadth-first search based multilevel sga algorithms for near optimal symbol
  detection in mimo systems,'' \emph{{IEEE} Transactions on Wireless
  Communications}, vol.~7, no.~3, pp. 1052--1061, Mar. 2008.

\bibitem{Grossmann_2008:turbo_equalization_PDA}
M.~Grossmann and T.~Matsumoto, ``Nonlinear frequency domain {MMSE} turbo
  equalization using probabilistic data association,'' \emph{{IEEE}
  Communications Letters}, vol.~12, no.~4, pp. 295--297, Apr. 2008.

\bibitem{Kim_2008:noncoherent_PDA}
Y.~J.~D. Kim and J.~Bajcsy, ``An iterative receiver for non-coherent {MIMO}
  systems with differential encoding,'' in \emph{Proc. 5th IEEE Consumer
  Communications and Networking Conference (CCNC'08)}, Las Vegas, NV, Jan.
  2008, pp. 46--47.

\bibitem{Shaoshi_2008:PDA_JD}
S.~Yang, T.~Lv, X.~Yun, X.~Su, and J.~Xia, ``A probabilistic data association
  based {MIMO} detector using joint detection of consecutive symbol vectors,''
  in \emph{Proc. 11th IEEE Singapore International Conference on Communication
  Systems (ICCS'08)}, Guangzhou, China, Nov. 2008, pp. 436--440.

\bibitem{Mohammed_2009:PDA_STBC}
S.~K. Mohammed, A.~Chockalingam, and B.~S. Rajan, ``Low-complexity near-{MAP}
  decoding of large non-orthogonal {STBC}s using {PDA},'' in \emph{IEEE
  International Symposium on Information Theory (ISIT'09)}, Seoul, Korea, Jul.
  2009, pp. 1998--2002.

\bibitem{Bavarian:SDE_PDA_freq_selective}
S.~Bavarian and J.~K. Cavers, ``A new framework for soft decision equalization
  in frequency selective {MIMO} channels,'' \emph{{IEEE} Transactions on
  Communications}, vol.~57, no.~2, pp. 415 --422, Feb. 2009.

\bibitem{Shaoshi2011:B_PDA}
S.~Yang, T.~Lv, R.~G. Maunder, and L.~Hanzo, ``Unified bit-based probabilistic
  data association aided {MIMO} detection for high-order {QAM}
  constellations,'' \emph{{IEEE} Transactions on Vehicular Technology},
  vol.~60, no.~3, pp. 981--991, Mar. 2011.

\bibitem{Shaoshi2011:DPDA}
------, ``Distributed probabilistic-data-association-based soft reception
  employing base station cooperation in {MIMO}-aided multiuser multicell
  systems,'' \emph{{IEEE} Transactions on Vehicular Technology}, vol.~60,
  no.~7, pp. 3532--3538, Sep. 2011.

\bibitem{Shaoshi2013:Turbo_AB_Log_PDA}
S.~Yang, L.~Wang, T.~Lv, and L.~Hanzo, ``Approximate {B}ayesian
  probabilistic-data-association-aided iterative detection for {MIMO} systems
  using arbitrary {$M$}-ary modulation,'' \emph{{IEEE} Transactions on
  Vehicular Technology}, vol.~62, no.~3, pp. 1228--1240, Mar. 2013.

\bibitem{Shaoshi_2013:EB_Log_PDA_journal}
S.~Yang, T.~Lv, R.~G. Maunder, and L.~Hanzo, ``From nominal to true \textit{a
  posteriori} probabilities: an exact {B}ayesian theorem based probabilistic
  data association approach for iterative {MIMO} detection and decoding,''
  \emph{IEEE Transactions on Communications}, vol.~61, no.~7, pp. 2782--2793,
  Jul. 2013.

\bibitem{Penghui:SDP_CDMA}
P.~H. Tan and L.~K. Rasmussen, ``The application of semidefinite programming
  for detection in {CDMA},'' \emph{{IEEE} Journal on Selected Areas in
  Communications}, vol.~19, no.~8, pp. 1442--1449, Aug. 2001.

\bibitem{Ma:SDR_CDMA_BPSK}
W.-K. Ma, T.~N. Davidson, K.~M. Wong, Z.-Q. Luo, and P.-C. Ching,
  ``Quasi-maximum-likelihood multiuser detection using semi-definite relaxation
  with application to synchronous {CDMA},'' \emph{{IEEE} Transactions on Signal
  Processing}, vol.~50, no.~4, pp. 912--922, Apr. 2002.

\bibitem{Ma:SDR_CDMA_QPSK}
W.-K. Ma, T.~N. Davidson, K.~M. Wong, and P.-C. Ching, ``A block alternating
  likelihood maximization approach to multiuser detection,'' \emph{{IEEE}
  Transactions on Signal Processing}, vol.~52, no.~9, pp. 2600--2611, Sep.
  2004.

\bibitem{Luo:SDR_performance_analysis}
M.~Kisialiou and Z.-Q. Luo, ``Performance analysis of quasi-maximum-likelihood
  detector based on semi-definite programming,'' in \emph{Proc. IEEE
  International Conference on Acoustics, Speech, and Signal Processing
  (ICASSP'05)}, vol.~3, Philadelphia, PA, Mar. 2005, pp. III/433--III/436.

\bibitem{Jalden:SDR_diversity}
J.~Jald\'{e}n and B.~Ottersten, ``The diversity order of the semidefinite
  relaxation detector,'' \emph{{IEEE} Transactions on Information Theory},
  vol.~54, no.~4, pp. 1406--1422, Apr. 2008.

\bibitem{Luo:SDP_MPSK}
Z.-Q. Luo, X.~Luo, and M.~Kisialiou, ``An efficient quasi-maximum likelihood
  decoder for {PSK} signals,'' in \emph{Proc. IEEE International Conference on
  Acoustics, Speech, and Signal Processing (ICASSP'03)}, vol.~6, Hong Kong,
  China, Apr. 2003, pp. VI/561--VI/564.

\bibitem{Ma:SDR_MPSK}
W.-K. Ma, P.-C. Ching, and Z.~Ding, ``Semidefinite relaxation based multiuser
  detection for {M}-ary {PSK} multiuser systems,'' \emph{{IEEE} Transactions on
  Signal Processing}, vol.~52, no.~10, pp. 2862--2872, Oct. 2004.

\bibitem{Wiesel:PI_SDR_16QAM}
A.~Wiesel, Y.~C. Eldar, and S.~Shamai~(Shitz), ``Semidefinite relaxation for
  detection of 16-{QAM} signaling in {MIMO} channels,'' \emph{{IEEE} Signal
  Processing Letters}, vol.~12, no.~9, pp. 653--656, Sep. 2005.

\bibitem{Yijin:SDR_16QAM_tight}
Y.~Yang, C.~Zhao, P.~Zhou, and W.~Xu, ``{MIMO} detection of 16-{QAM} signaling
  based on semidefinite relaxation,'' \emph{{IEEE} Signal Processing Letters},
  vol.~14, no.~11, pp. 797--800, Nov. 2007.

\bibitem{Sidiropoulos:SDR_HOM}
N.~D. Sidiropoulos and Z.-Q. Luo, ``A semidefinite relaxation approach to
  {MIMO} detection for high-order {QAM} constellations,'' \emph{{IEEE} Signal
  Processing Letters}, vol.~13, no.~9, pp. 525--528, Sep. 2006.

\bibitem{Mobasher:SDR_QAM_journal}
A.~Mobasher, M.~Taherzadeh, R.~Sotirov, and A.~K. Khandani, ``A
  near-maximum-likelihood decoding algorithm for {MIMO} systems based on
  semi-definite programming,'' \emph{{IEEE} Transactions on Information
  Theory}, vol.~53, no.~11, pp. 3869--3886, Nov. 2007.

\bibitem{Mao:SDR_LMR}
Z.~Mao, X.~Wang, and X.~Wang, ``Semidefinite programming relaxation approach
  for multiuser detection of {QAM} signals,'' \emph{{IEEE} Transactions on
  Wireless Communications}, vol.~6, no.~12, pp. 4275--4279, Dec. 2007.

\bibitem{Ma:equivalence_SDR}
W.-K. Ma, C.-C. Su, J.~Jald\'{e}n, T.-H. Chang, and C.-Y. Chi, ``The
  equivalence of semidefinite relaxation {MIMO} detectors for higher-order
  {QAM},'' \emph{IEEE Journal of Selected Topics in Signal Processing}, vol.~3,
  no.~6, pp. 1038--1052, Dec. 2009.

\bibitem{Shaoshi_2013:DVA_SPDR_journal}
S.~Yang, T.~Lv, and L.~Hanzo, ``Semidefinite programming relaxation based
  virtually antipodal detection for {MIMO} systems using {G}ray-coded
  high-order {QAM},'' \emph{{IEEE} Transactions on Vehicular Technology},
  vol.~62, no.~4, pp. 1667--1677, 2013.

\bibitem{Larsson:MIMO_detection_overview}
E.~G. Larsson, ``{MIMO} detection methods: how they work,'' \emph{{IEEE} Signal
  Processing Magazine}, vol.~26, no.~3, pp. 91--95, May 2009.

\bibitem{Lee_1991:overview_CDMA}
W.~C.~Y. Lee, ``Overview of cellular {CDMA},'' \emph{{IEEE} Transactions on
  Vehicular Technology}, vol.~40, no.~2, pp. 291--302, May 1991.

\bibitem{Jung_1993:advantages_of_CDMA_over_TDMA_FDMA}
P.~Jung, P.~W. Baier, and A.~Steil, ``Advantages of {CDMA} and spread spectrum
  techniques over {FDMA} and {TDMA} in cellular mobile radio applications,''
  \emph{{IEEE} Transactions on Vehicular Technology}, vol.~42, no.~3, pp.
  357--364, Aug. 1993.

\bibitem{Viterbi_1995:CDMA_book}
A.~J. Viterbi, \emph{{CDMA}: Principles of Spread Spectrum Communication},
  1st~ed.\hskip 1em plus 0.5em minus 0.4em\relax Addison-Wesley, 1995.

\bibitem{Roy1996:SDMA}
R.~H. Roy and B.~Ottersten, ``Spatial division multiple access wireless
  communication systems,'' U.S. Patent 5 515 378, May 7, 1996.

\bibitem{Gerlach1995:SDMA_thesis}
D.~Gerlach, ``Adaptive transmitting antenna arrays at the base station in
  mobile radio networks,'' Ph.D. dissertation, Department of Electrical
  Engineering, Stanford University, California, USA, 1995.

\bibitem{Ottersten1996:SDMA}
B.~Ottersten, ``Array processing for wireless communications,'' in \emph{Proc.
  8th IEEE Signal Processing Workshop on Statistical Signal and Array
  Processing}, Corfu, Greece, Jun. 1996, pp. 466--473.

\bibitem{Roy1997:SDMA}
R.~H. Roy, ``Spatial division multiple access technology and its application to
  wireless communication systems,'' in \emph{Proc. IEEE 47th Vehicular
  Technology Conference (VTC'97)}, Phoenix, AZ, USA, May 1997, pp. 730--734.

\bibitem{Lotter1998:SDMA}
M.~P. Lotter and P.~van Rooyen, ``Space division multiple access for cellular
  {CDMA},'' in \emph{Proc. IEEE 5th International Symposium on Spread Spectrum
  Techniques and Applications (ISSSTA'98)}, Sun City, South Africa, Sep. 1998,
  pp. 959--964.

\bibitem{vandenameele2001:SDMA}
P.~Vandenameele, L.~Van~der Perre, and M.~Engels, \emph{Space Division Multiple
  Access for Wireless Local Area Networks}.\hskip 1em plus 0.5em minus
  0.4em\relax Norwell, MA, USA: Kluwer, 2001.

\bibitem{Goldsmith_2005:Wireless_Comm}
A.~Goldsmith, \emph{Wireless Communications}, 1st~ed.\hskip 1em plus 0.5em
  minus 0.4em\relax Cambridge University Press, 2005.

\bibitem{Oppenheim:1996:signals_and_systems}
A.~V. Oppenheim, A.~S. Willsky, and S.~H. Nawab, \emph{Signals and Systems},
  2nd~ed.\hskip 1em plus 0.5em minus 0.4em\relax Upper Saddle River, NJ, USA:
  Prentice-Hall, 1996.

\bibitem{Sklar:digital_communications}
B.~Sklar, \emph{Digital Communications: Fundamentals and Applications},
  2nd~ed.\hskip 1em plus 0.5em minus 0.4em\relax Upper Saddle River, NJ, USA:
  Prentice-Hall, 2001.

\bibitem{Rappaport_2002:Wireless_textbook}
T.~S. Rappaport, \emph{Wireless Communications: Principles and Practice},
  2nd~ed.\hskip 1em plus 0.5em minus 0.4em\relax Upper Saddle River, NJ:
  Prentice Hall, 2002.

\bibitem{Proakis_2007:DSP}
J.~G. Proakis, \emph{Digital Signal Processing: Principles, Algorithms and
  Applications}, 4th~ed.\hskip 1em plus 0.5em minus 0.4em\relax Pearson
  Education, 2007.

\bibitem{Haykin_2003:adaptive_filtering}
S.~Haykin, \emph{Adaptive Filter Theory}, 4th~ed.\hskip 1em plus 0.5em minus
  0.4em\relax Upper Saddle River, NJ: Prentice Hall, 2002.

\bibitem{Jamalipour:TDMA}
A.~Jamalipour, T.~Wada, and T.~Yamazato, ``A tutorial on multiple access
  technologies for beyond {3G} mobile networks,'' \emph{{IEEE} Communications
  Magazine}, vol.~43, no.~2, pp. 110--117, Feb. 2005.

\bibitem{Kandukuri_2002:power_control_CDMA}
S.~Kandukuri and S.~Boyd, ``Optimal power control in interference-limited
  fading wireless channels with outage-probability specifications,''
  \emph{{IEEE} Transactions on Wireless Communications}, vol.~1, no.~1, pp.
  46--55, Jan. 2002.

\bibitem{Koffman:OFDMA}
I.~Koffman and V.~Roman, ``Broadband wireless access solutions based on {OFDM}
  access in {IEEE} 802. 16,'' \emph{{IEEE} Communications Magazine}, vol.~40,
  no.~4, pp. 96--103, Apr. 2002.

\bibitem{Morelli_2007:synchronization_OFDMA_tutorial}
M.~Morelli, C.-C.~J. Kuo, and M.-O. Pun, ``Synchronization techniques for
  orthogonal frequency division multiple access ({OFDMA}): A tutorial review,''
  \emph{Proceedings of the {IEEE}}, vol.~95, no.~7, pp. 1394 --1427, Jul. 2007.

\bibitem{Necker:Interference_OFDMA}
M.~Necker, ``Interference coordination in cellular {OFDMA} networks,''
  \emph{{IEEE} Network}, vol.~22, no.~6, pp. 12--19, Nov. 2008.

\bibitem{Benvenuto:SC-FDMA}
N.~Benvenuto, R.~Dinis, D.~Falconer, and S.~Tomasin, ``Single carrier
  modulation with nonlinear frequency domain equalization: An idea whose time
  has come --- again,'' \emph{Proceedings of the {IEEE}}, vol.~98, no.~1, pp.
  69--96, Jan. 2010.

\bibitem{Hara_1997:MC_CDMA}
S.~Hara and R.~Prasad, ``Overview of multicarrier {CDMA},'' \emph{{IEEE}
  Communications Magazine}, vol.~35, no.~12, pp. 126--133, Dec. 1997.

\bibitem{Lieliang_2002:MC_DS_CDMA_Nakagami}
L.-L. Yang and L.~Hanzo, ``Performance of generalized multicarrier {DS-CDMA}
  over {Nakagami-m} fading channels,'' \emph{{IEEE} Transactions on
  Communications}, vol.~50, no.~6, pp. 956--966, Jun. 2002.

\bibitem{Lieliang_2003:MC_DS_CDMA}
------, ``Multicarrier {DS-CDMA}: A multiple access scheme for ubiquitous
  broadband wireless communications,'' \emph{{IEEE} Communications Magazine},
  vol.~41, no.~10, pp. 116--124, Oct. 2003.

\bibitem{Adachi_2005:MC_CDMA}
F.~Adachi, G.~Garg, S.~Takaoka, and K.~Takeda, ``Broadband {CDMA} techniques,''
  \emph{{IEEE} Wireless Communications Magazine}, vol.~12, no.~2, pp. 8--18,
  Apr. 2005.

\bibitem{Hanzo:CDMA_book}
L.~Hanzo, L.-L. Yang, E.-L. Kuan, and K.~Yen, \emph{Single- and Multi-Carrier
  DS-CDMA: Multi-User Detection, Space-Time Spreading, Synchronisation,
  Networking and Standards}.\hskip 1em plus 0.5em minus 0.4em\relax Chichester,
  UK: Wiley-IEEE Press, 2003.

\bibitem{Lieliang_2009:Multicarrier}
L.-L. Yang, \emph{Multicarrier Communications}.\hskip 1em plus 0.5em minus
  0.4em\relax John Wiley \& Sons, 2009.

\bibitem{Chockalingam_2008:HNN_LAS_based_LS_MIMO_detector}
K.~Vardhan, S.~K. Mohammed, A.~Chockalingam, and B.~S. Rajan, ``A
  low-complexity detector for large {MIMO} systems and multicarrier {CDMA}
  systems,'' \emph{{IEEE} Journal on Selected Areas in Communications},
  vol.~26, no.~3, pp. 473--485, Apr. 2008.

\bibitem{Kadir_2015:MIMO_multicarrier_STSK}
M.~Kadir, S.~Sugiura, S.~Chen, and L.~Hanzo, ``Unified {MIMO}-multicarrier
  designs: A space-time shift keying approach,'' \emph{{IEEE} Communications
  Surveys and Tutorials}, vol.~17, no.~2, pp. 550--579, Second Quarter 2015.

\bibitem{Tuchler_2011:turbo_equalisation_overview}
M.~T{\"u}chler and A.~C. Singer, ``Turbo equalization: An overview,''
  \emph{{IEEE} Transactions on Information Theory}, vol.~57, no.~2, pp.
  920--952, Feb. 2011.

\bibitem{Foschini:MIMO}
G.~J. Foschini, ``Layered space-time architecture for wireless communication in
  a fading environment when using multi-element antennas,'' \emph{Bell Labs
  Technical Journal}, vol.~1, no.~2, pp. 41--59, 1996.

\bibitem{Wolniansky:VBLAST}
P.~W. Wolniansky, G.~J. Foschini, G.~D. Golden, and R.~A. Valenzuela,
  ``{V-BLAST}: an architecture for realizing very high data rates over the
  rich-scattering wireless channel,'' in \emph{Proc. URSI International
  Symposium on Signals, Systems, and Electronics (ISSSE'98)}, Pisa, Italy, Sep.
  1998, pp. 295--300.

\bibitem{Golden_1999:VBLAST_first_journal}
G.~D. Golden, G.~J. Foschini, R.~A. Valenzuela, and P.~W. Wolniansky,
  ``Detection algorithm and initial laboratory results using {V-BLAST}
  space-time communication architecture,'' \emph{Electronics Letters}, vol.~35,
  no.~1, pp. 14--16, Jan. 1999.

\bibitem{Sfar_2003:uplink_MU_MIMO_MUD}
S.~Sfar, R.~D. Murch, and K.~B. Letaief, ``Layered space-time multiuser
  detection over wireless uplink systems,'' \emph{{IEEE} Transactions on
  Wireless Communications}, vol.~2, no.~4, pp. 653--668, Jul. 2003.

\bibitem{Serbetli_2004:transceiver_design_MU_MIMO}
S.~Serbetli and A.~Yener, ``Transceiver optimization for multiuser {MIMO}
  systems,'' \emph{{IEEE} Transactions on Signal Processing}, vol.~52, no.~1,
  pp. 214--226, Jan. 2004.

\bibitem{Zhao_2003:cooperative_signal_processing}
F.~Zhao, J.~Liu, J.~Liu, L.~Guibas, and J.~Reich, ``Collaborative signal and
  information processing: an information-directed approach,'' \emph{Proceedings
  of the {IEEE}}, vol.~91, no.~8, pp. 1199--1209, Aug. 2003.

\bibitem{Hunter_2002:cooperative_diversity_coding}
T.~E. Hunter and A.~Nosratinia, ``Cooperation diversity through coding,'' in
  \emph{Proc. IEEE International Symposium on Information Theory (ISIT'02)},
  Lausanne, Switzerland, Jul. 2002, p. 220.

\bibitem{Laneman_2003:distributed_protocol_cooperative_diversity}
J.~N. Laneman and G.~W. Wornell, ``Distributed space-time-coded protocols for
  exploiting cooperative diversity in wireless networks,'' \emph{{IEEE}
  Transactions on Information Theory}, vol.~49, no.~10, pp. 2415--2425, Oct.
  2003.

\bibitem{Laneman_2004:cooperative_diversity}
J.~N. Laneman, D.~N.~C. Tse, and G.~W. Wornell, ``Cooperative diversity in
  wireless networks: efficient protocols and outage behavior,'' \emph{{IEEE}
  Transactions on Information Theory}, vol.~50, no.~12, pp. 3062--3080, Dec.
  2004.

\bibitem{Nosratinia_2004:cooperative_comm}
A.~Nosratinia, T.~E. Hunter, and A.~Hedayat, ``Cooperative communication in
  wireless networks,'' \emph{{IEEE} Communications Magazine}, vol.~42, no.~10,
  pp. 74--80, Oct. 2004.

\bibitem{Janani_2004:coded_cooperation}
M.~Janani, A.~Hedayat, T.~E. Hunter, and A.~Nosratinia, ``Coded cooperation in
  wireless communications: space-time transmission and iterative decoding,''
  \emph{{IEEE} Transactions on Signal Processing}, vol.~52, no.~2, pp.
  362--371, Feb. 2004.

\bibitem{Hunter_2006:diversity_coded_cooperation}
T.~E. Hunter and A.~Nosratinia, ``Diversity through coded cooperation,''
  \emph{{IEEE} Transactions on Wireless Communications}, vol.~5, no.~2, pp.
  283--289, Feb. 2006.

\bibitem{Spencer_2004:ZF_DL_MU_MIMO}
Q.~H. Spencer, A.~L. Swindlehurst, and M.~Haardt, ``Zero-forcing methods for
  downlink spatial multiplexing in multiuser {MIMO} channels,'' \emph{{IEEE}
  Transactions on Signal Processing}, vol.~52, no.~2, pp. 461--471, Feb. 2004.

\bibitem{Spencer:intro_multi-user_MIMO_downlink}
Q.~H. Spencer, C.~B. Peel, A.~L. Swindlehurst, and M.~Haardt, ``An introduction
  to the multi-user {MIMO} downlink,'' \emph{{IEEE} Communications Magazine},
  vol.~42, no.~10, pp. 60--67, Oct. 2004.

\bibitem{Choi_2004:preprocessing_DL_MU_MIMO}
L.-U. Choi and R.~D. Murch, ``A transmit preprocessing technique for multiuser
  {MIMO} systems using a decomposition approach,'' \emph{{IEEE} Transactions on
  Wireless Communications}, vol.~3, no.~1, pp. 20--24, Jan. 2004.

\bibitem{Zhang:CCI_mitigation_downlink}
\BIBentryALTinterwordspacing
H.~Zhang and H.~Dai, ``Cochannel interference mitigation and cooperative
  processing in downlink multicell multiuser {MIMO} networks,'' \emph{EURASIP
  J. Wireless Commun. Netw.}, vol. 2004, no.~2, pp. 222--235,
  doi:10.1155/S1687147204406148, 2004. [Online]. Available:
  \url{http://downloads.hindawi.com/journals/wcn/2004/202654.pdf}
\BIBentrySTDinterwordspacing

\bibitem{Caire_2008:MU_MIMO_partial_cooperation_IA}
G.~Caire, S.~A. Ramprashad, H.~C. Papadopoulos, C.~Pepin, and C.-E.~W.
  Sundberg, ``Multiuser {MIMO} downlink with limited inter-cell cooperation:
  approximate interference alignment in time, frequency and space,'' in
  \emph{Proc. 46th Annual Allerton Conference on Communication, Control, and
  Computing (Allerton'08)}, Urbana-Champaign, IL, USA, Sep. 2008, pp. 730--737.

\bibitem{Caire_2010:MU_MIMO_DL}
G.~Caire, N.~Jindal, M.~Kobayashi, and N.~Ravindran, ``Multiuser {MIMO}
  achievable rates with downlink training and channel state feedback,''
  \emph{{IEEE} Transactions on Information Theory}, vol.~56, no.~6, pp.
  2845--2866, Jun. 2010.

\bibitem{Carleial_1975:earliest_interference_cancellation_idea}
A.~Carleial, ``A case where interference does not reduce capacity,''
  \emph{{IEEE} Transactions on Information Theory}, vol.~21, no.~5, pp.
  569--570, Sep. 1975.

\bibitem{Han_1981:interference_channel_rate_region}
T.~Han and K.~Kobayashi, ``A new achievable rate region for the interference
  channel,'' \emph{{IEEE} Transactions on Information Theory}, vol.~27, no.~1,
  pp. 49--60, Jan. 1981.

\bibitem{Sato_1981:interference_channel_strong_interference}
H.~Sato, ``The capacity of the gaussian interference channel under strong
  interference,'' \emph{{IEEE} Transactions on Information Theory}, vol.~27,
  no.~6, pp. 786--788, Nov. 1981.

\bibitem{Costa_1985:Gaussian_interference_channel}
M.~H.~M. Costa, ``On the gaussian interference channel,'' \emph{{IEEE}
  Transactions on Information Theory}, vol.~31, no.~5, pp. 607--615, Sep. 1985.

\bibitem{Costa_1987:capacity_interference_channel}
M.~H.~M. Costa and A.~El~Gamal, ``The capacity region of the discrete
  memoryless interference channel with strong interference,'' \emph{{IEEE}
  Transactions on Information Theory}, vol.~33, no.~5, pp. 710--711, Sep. 1987.

\bibitem{Sason_2004:rate_regions_Gaussian_interference_channel}
I.~Sason, ``On achievable rate regions for the gaussian interference channel,''
  \emph{{IEEE} Transactions on Information Theory}, vol.~50, no.~6, pp.
  1345--1356, Jun. 2004.

\bibitem{Kramer_2006:review_interference_channel_rate_regions}
G.~Kramer, ``Review of rate regions for interference channels,'' in \emph{Proc.
  International Zurich Seminar on Communications}, Zurich, Switzerland, Feb.
  2006, pp. 162--165.

\bibitem{Jafar_2007:DoF_MIMO_interference_channel}
S.~A. Jafar and M.~J. Fakhereddin, ``Degrees of freedom for the {MIMO}
  interference channel,'' \emph{{IEEE} Transactions on Information Theory},
  vol.~53, no.~7, pp. 2637--2642, Jul. 2007.

\bibitem{Etkin_2008:interference_channel_with_one_bit}
R.~H. Etkin, D.~N.~C. Tse, and H.~Wang, ``Gaussian interference channel
  capacity to within one bit,'' \emph{{IEEE} Transactions on Information
  Theory}, vol.~54, no.~12, pp. 5534--5562, Dec. 2008.

\bibitem{Vishwanath_2003:Z_channel}
S.~Vishwanath, N.~Jindal, and A.~Goldsmith, ``The "{Z}" channel,'' in
  \emph{Proc. IEEE Global Telecommunications Conference (GLOBECOM'03)}, San
  Francisco, CA, USA, Dec. 2003, pp. 1726--1730.

\bibitem{Ali_2008:interference_alignment_journal}
M.~A. Maddah-Ali, A.~S. Motahari, and A.~K. Khandani, ``Communication over
  {MIMO X} channels: interference alignment, decomposition, and performance
  analysis,'' \emph{{IEEE} Transactions on Information Theory}, vol.~54, no.~8,
  pp. 3457--3470, Aug. 2008.

\bibitem{Jafar_2008:DoF_MIMO_X_channel}
S.~A. Jafar and S.~Shamai~(Shitz), ``Degrees of freedom region of the {MIMO X}
  channel,'' \emph{{IEEE} Transactions on Information Theory}, vol.~54, no.~1,
  pp. 151--170, Jan. 2008.

\bibitem{Cadambe2009:IA_for_Wirelss_X_networks}
V.~R. Cadambe and S.~A. Jafar, ``Interference alignment and the degrees of
  freedom of wireless {X} networks,'' \emph{{IEEE} Transactions on Information
  Theory}, vol.~55, no.~9, pp. 3893--3908, Sep. 2009.

\bibitem{Grant:one_dimensional_Wyner_model}
A.~Grant, S.~V. Hanly, J.~S. Evans, and R.~Muller, ``Distributed decoding for
  {Wyner} cellular systems,'' in \emph{Proc. 5th Australian Communications
  Theory Workshop (AusCTW'04)}, Newcastle, Australia, Feb. 2004, pp. 77--81.

\bibitem{Zhang:JT_and_BS_selection}
H.~Zhang, H.~Dai, and Q.~Zhou, ``Base station cooperation for multiuser {MIMO}:
  joint transmission and {BS} selection,'' in \emph{Proc. 38th Annual
  Conference on information sciences and systems (CISS'04)}, Princeton, NJ,
  USA, Mar. 2004, pp. 17--19.

\bibitem{Zhang:CCI_Asynchronous}
H.~Zhang, N.~B. Mehta, A.~F. Molisch, J.~Zhang, and H.~Dai, ``Asynchronous
  interference mitigation in cooperative base station systems,'' \emph{{IEEE}
  Transactions on Wireless Communications}, vol.~7, no.~1, pp. 155--165, Jan.
  2008.

\bibitem{Hadisusanto:BD_and_dual_decomposition}
Y.~Hadisusanto, L.~Thiele, and V.~Jungnickel, ``Distributed base station
  cooperation via block-diagonalization and dual-decomposition,'' in
  \emph{Proc. IEEE Global Telecommunications Conference (GLOBECOM'08)}, New
  Orleans, LO, Nov. 2008, pp. 1--5.

\bibitem{Mayer:Turbo_BS_cooperation_interference_cancellation}
T.~Mayer, H.~Jenkac, and J.~Hagenauer, ``Turbo base-station cooperation for
  intercell interference cancellation,'' in \emph{Proc. IEEE International
  Conference on Communications (ICC'06)}, Istanbul, Turkey, Jun. 2006, pp.
  4977--4982.

\bibitem{Khattak:distributed_max_log_MAP}
\BIBentryALTinterwordspacing
S.~Khattak, W.~Rave, and G.~Fettweis, ``Distributed iterative multiuser
  detection through base station cooperation,'' \emph{EURASIP J. Wireless
  Commun. Netw.}, vol. 2008, Article ID 390489, 15 pages,
  doi:10.1155/2008/390489, 2008. [Online]. Available:
  \url{http://downloads.hindawi.com/journals/wcn/2008/390489.pdf}
\BIBentrySTDinterwordspacing

\bibitem{Aktas:Belief_Propagation_2D_Wyner_model}
E.~Aktas, J.~S. Evans, and S.~V. Hanly, ``Distributed decoding in a cellular
  multiple-access channel,'' \emph{{IEEE} Transactions on Wireless
  Communications}, vol.~7, no.~1, pp. 241--250, Jan. 2008.

\bibitem{Ng:BS_cooperation_downlink_beamforming}
B.~L. Ng, J.~S. Evans, S.~V. Hanly, and D.~Aktas, ``Distributed downlink
  beamforming with cooperative base stations,'' \emph{{IEEE} Transactions on
  Information Theory}, vol.~54, no.~12, pp. 5491--5499, Dec. 2008.

\bibitem{Shaoshi_2011:BS_cooperation_DPDA_conf}
S.~Yang, T.~Lv, and L.~Hanzo, ``Base station cooperation in {MIMO}-aided
  multi-user multi-cell systems employing distributed probabilistic data
  association based soft reception,'' in \emph{Proc. IEEE International
  Conference on Communications (ICC'11)}, Kyoto, Japan, Jun. 2011, pp. 1--5.

\bibitem{Zakhour_2012:BS_cooperation_downlink_large_system}
R.~Zakhour and S.~V. Hanly, ``Base station cooperation on the downlink: large
  system analysis,'' \emph{{IEEE} Transactions on Information Theory}, vol.~58,
  no.~4, pp. 2079--2106, Apr. 2012.

\bibitem{Ali_2006:first_interference_alignment}
M.~A. Maddah-Ali, A.~S. Motahari, and A.~K. Khandani, ``Signaling over {MIMO}
  multi-base systems: combination of multi-access and broadcast schemes,'' in
  \emph{Proc. IEEE International Symposium on Information Theory (ISIT'06)},
  Seattle, WA, Jul. 2006, pp. 2104--2108.

\bibitem{Ali_2006:interference_alignment_report_part_1}
------, ``Communication over {X} channel: signalling and multiplexing gain,''
  Department of Electrical and Computer Engineering, University of Waterloo,
  Waterloo, ON, Canada, Tech. Rep. UW-ECE-2006-12, Jul. 2006.

\bibitem{Ali_2006:interference_alignment_report_part_2}
------, ``Communication over {MIMO X} channel: signalling and performance
  analysis,'' Department of Electrical and Computer Engineering, University of
  Waterloo, Waterloo, ON, Canada, Tech. Rep. UW-ECE-2006-27, Dec. 2006.

\bibitem{Jafar_2007:DoF_MIMO_X_channel_conf}
S.~A. Jafar and S.~Shamai~(Shitz), ``Degrees of freedom of the {MIMO X}
  channel,'' in \emph{Proc. IEEE Global Telecommunications Conference
  (GLOBECOM'07)}, Washington, DC, USA, Nov. 2007, pp. 1632--1636.

\bibitem{Cadambe_Jafar_2008:interference_alignment_journal}
V.~R. Cadambe and S.~A. Jafar, ``Interference alignment and degrees of freedom
  of the {$K$}-user interference channel,'' \emph{{IEEE} Transactions on
  Information Theory}, vol.~54, no.~8, pp. 3425--3441, Aug. 2008.

\bibitem{Jafar_2010:interference_alignment_tutorial}
S.~A. Jafar, ``Interference alignment -- a new look at signal dimensions in a
  communication network,'' \emph{Foundations and Trends in Communications and
  Information Theory}, vol.~7, no.~1, 2010.

\bibitem{Shnidman1967:earliest_ISI_cross_talk_equivalent}
D.~A. Shnidman, ``A generalized {Nyquist} criterion and an optimum linear
  receiver for a pulse modulation system,'' \emph{The Bell System Technical
  Journal}, vol.~46, no.~9, pp. 2163--2177, Nov. 1967.

\bibitem{Kaye_1970:transmit_multiplexed_PAM_over_Multi_channel}
A.~Kaye and D.~George, ``Transmission of multiplexed {PAM} signals over
  multiple channel and diversity systems,'' \emph{IEEE Transactions on
  Communication Technology}, vol. {COM}-18, no.~5, pp. 520--526, Oct. 1970.

\bibitem{Etten_1975:optimum_linear_receiver_for_multi_channel}
W.~van Etten, ``An optimum linear receiver for multiple channel digital
  transmission systems,'' \emph{{IEEE} Transactions on Communications},
  vol.~23, no.~8, pp. 828--834, Aug. 1975.

\bibitem{Etten_1976:ML_receiver_for_multi_channel}
------, ``Maximum likelihood receiver for multiple channel transmission
  systems,'' \emph{{IEEE} Transactions on Communications}, vol.~24, no.~2, pp.
  276--283, Feb. 1976.

\bibitem{Horwood_1975:signal_design_multiple_access}
D.~Horwood and R.~Gagliardi, ``Signal design for digital multiple access
  communications,'' \emph{{IEEE} Transactions on Communications}, vol.~23,
  no.~3, pp. 378--383, Mar. 1975.

\bibitem{Schneider_1979:linear_ZF_CDMA}
K.~S. Schneider, ``Optimum detection of code division multiplexed signals,''
  \emph{{IEEE} Transactions on Aerospace and Electronic Systems}, vol. AES-15,
  no.~1, pp. 181--185, Jan. 1979.

\bibitem{Schneider_1980:crosss_talk_resistant_receiver}
------, ``Crosstalk resistant receiver for {$M$}-ary multiplexed
  communications,'' \emph{{IEEE} Transactions on Aerospace and Electronic
  Systems}, vol. AES-16, no.~4, pp. 426--433, Jul. 1980.

\bibitem{Timor_1980:improved_decoding_CDMA}
U.~Timor, ``Improved decoding scheme for frequency-hopped multilevel {FSK}
  system,'' \emph{The Bell System Technical Journal}, vol.~59, no.~10, pp.
  1839--1855, Dec. 1980.

\bibitem{Timor_1981:multistage_decoding_CDMA}
------, ``Multistage decoding of frequency-hopped {FSK} system,'' \emph{The
  Bell System Technical Journal}, vol.~60, no.~4, pp. 471--483, Apr. 1981.

\bibitem{Verdu_1983:earliest_optimal_MUD_conf_ISIT}
S.~Verd{\'u}, ``Optimum sequence detection of asynchronous multiple-access
  communications,'' in \emph{Abstr. IEEE International Symposium on Information
  Theory (ISIT'83)}, St. Jovite, Canada, Sep. 1983, p.~80.

\bibitem{Verdu_1983:earliest_optimal_MUD_conf_milcom}
------, ``Minimum probability of error for asynchronous multiple access
  communication systems,'' in \emph{Proc. IEEE Military Communications
  Conference (MILCOM'83)}, Washington, DC, Nov. 1983, pp. 213--219.

\bibitem{Verdu_1986:optimal_MUD_asynchronous_CDMA}
------, ``Minimum probability of error for asynchronous {Gaussian}
  multiple-access channels,'' \emph{{IEEE} Transactions on Information Theory},
  vol. IT-32, no.~1, pp. 85--96, Jan. 1986.

\bibitem{Verdu_1986:optimum_MUD_asymptotic_efficiency}
------, ``Optimum multiuser asymptotic efficiency,'' \emph{{IEEE} Transactions
  on Communications}, vol.~34, no.~9, pp. 890--897, Sep. 1986.

\bibitem{Lupas_1989:linear_MUD_synchrohous_CDMA}
R.~Lupas and S.~Verd{\'u}, ``Linear multiuser detectors for synchronous
  code-division multiple-access channels,'' \emph{{IEEE} Transactions on
  Information Theory}, vol.~35, no.~1, pp. 123--136, Jan. 1989.

\bibitem{Lupas_1990:near_far_MUD_asynchronous}
------, ``Near-far resistance of multiuser detectors in asynchronous
  channels,'' \emph{{IEEE} Transactions on Communications}, vol.~38, no.~4, pp.
  496--508, Apr. 1990.

\bibitem{Kohno_1983:PIC_CDMA}
\BIBentryALTinterwordspacing
R.~Kohno and M.~Hatori, ``Cancellation techniques of co-channel interference in
  asynchronous spread spectrum multiple access systems,'' \emph{Electronics and
  Communications in Japan (Part I: Communications)}, vol.~66, no.~5, pp.
  20--29, May 1983. [Online]. Available:
  \url{http://dx.doi.org/10.1002/ecja.4400660504}
\BIBentrySTDinterwordspacing

\bibitem{Kohno_1990:IC_CDMA}
R.~Kohno, H.~Imai, M.~Hatori, and S.~Pasupathy, ``Combinations of an adaptive
  array antenna and a canceller of interference for direct-sequence
  spread-spectrum multiple-access system,'' \emph{{IEEE} Journal on Selected
  Areas in Communications}, vol.~8, no.~4, pp. 675--682, May 1990.

\bibitem{Kohno_1990:PIC_CDMA_journal}
------, ``An adaptive canceller of cochannel interference for spread-spectrum
  multiple-access communication networks in a power line,'' \emph{{IEEE}
  Journal on Selected Areas in Communications}, vol.~8, no.~4, pp. 691--699,
  May 1990.

\bibitem{Varanasi_1990:multistage_detection_asynchronous_CDMA}
M.~K. Varanasi and B.~Aazhang, ``Multistage detection in asynchronous
  code-division multiple-access communications,'' \emph{{IEEE} Transactions on
  Communications}, vol.~38, no.~4, pp. 509--519, Apr. 1990.

\bibitem{Varanasi_1991:multistage_MUD_synchronous_CDMA}
------, ``Near-optimum detection in synchronous code-division multiple-access
  systems,'' \emph{{IEEE} Transactions on Communications}, vol.~39, no.~5, pp.
  725--736, May 1991.

\bibitem{Varanasi_1991:noncoherent_MUD}
------, ``Optimally near-far resistant multiuser detection in differentially
  coherent synchronous channels,'' \emph{{IEEE} Transactions on Information
  Theory}, vol.~37, no.~4, pp. 1006--1018, Jul. 1991.

\bibitem{Yoon_1993:PIC_CDMA_journal}
Y.~C. Yoon, R.~Kohno, and H.~Imai, ``A spread-spectrum multiaccess system with
  cochannel interference cancellation for multipath fading channels,''
  \emph{{IEEE} Journal on Selected Areas in Communications}, vol.~11, no.~7,
  pp. 1067--1075, Sep. 1993.

\bibitem{Divsalar_1998:PIC_MUD_CDMA}
D.~Divsalar, M.~K. Simon, and D.~Raphaeli, ``Improved parallel interference
  cancellation for {CDMA},'' \emph{{IEEE} Transactions on Communications},
  vol.~46, no.~2, pp. 258--268, Feb. 1998.

\bibitem{Buehrer_1996:adaptive_multistage_IC_CDMA}
R.~M. Buehrer and B.~D. Woerner, ``Analysis of adaptive multistage interference
  cancellation for {CDMA} using an improved gaussian approximation,''
  \emph{{IEEE} Transactions on Communications}, vol.~44, no.~10, pp.
  1308--1321, Oct. 1996.

\bibitem{Masamura_1988:earliest_interference_cancellation_SSMA}
T.~Masamura, ``Spread spectrum multiple access system with intrasystem
  interference cancellation,'' \emph{IEICE Transactions}, vol. E71, no.~3, pp.
  224--231, Mar. 1988.

\bibitem{Viterbi_1990:earliest_interference_cancellation_approach_capacity}
A.~J. Viterbi, ``Very low rate convolution codes for maximum theoretical
  performance of spread-spectrum multiple-access channels,'' \emph{{IEEE}
  Journal on Selected Areas in Communications}, vol.~8, no.~4, pp. 641--649,
  May 1990.

\bibitem{Xie_1990:sequential_MUD_for_async_CDMA}
Z.~Xie, C.~K. Rushforth, and R.~T. Short, ``Multiuser signal detection using
  sequential decoding,'' \emph{{IEEE} Transactions on Communications}, vol.~38,
  no.~5, pp. 578--583, May 1990.

\bibitem{Xie_1990:Linear_MMSE_WLS_MUD}
Z.~Xie, R.~T. Short, and C.~K. Rushforth, ``A family of suboptimum detectors
  for coherent multiuser communications,'' \emph{{IEEE} Journal on Selected
  Areas in Communications}, vol.~8, no.~4, pp. 683--690, May 1990.

\bibitem{Xie_1993:joint_signal_detection_estimation}
Z.~Xie, C.~K. Rushforth, R.~T. Short, and T.~K. Moon, ``Joint signal detection
  and parameter estimation in multiuser communications,'' \emph{{IEEE}
  Transactions on Communications}, vol.~41, no.~8, pp. 1208--1216, Aug. 1993.

\bibitem{Hallen_1993:decorrelating_DFD_synch_CDMA}
A.~Duel-Hallen, ``Decorrelating decision-feedback multiuser detector for
  synchronous code-division multiple-access channel,'' \emph{{IEEE}
  Transactions on Communications}, vol.~41, no.~2, pp. 285--290, Feb. 1993.

\bibitem{Hallen:1995:DFD_asynchronous_CDMA}
------, ``A family of multiuser decision-feedback detectors for asynchronous
  code-division multiple-access channels,'' \emph{{IEEE} Transactions on
  Communications}, vol.~43, no. 2/3/4, pp. 421--434, Feb./Mar./Apr. 1995.

\bibitem{Patel_1994:SIC_CDMA}
P.~R. Patel and J.~M. Holtzman, ``Analysis of a simple successive interference
  cancellation scheme in a {DS/CDMA} system,'' \emph{{IEEE} Journal on Selected
  Areas in Communications}, vol.~12, no.~5, pp. 796--807, Jun. 1994.

\bibitem{Varanasi:DFD}
M.~K. Varanasi, ``Decision feedback multiuser detection: a systematic
  approach,'' \emph{{IEEE} Transactions on Information Theory}, vol.~45, no.~1,
  pp. 219--240, Jan. 1999.

\bibitem{Hui_1998:SIC_asynch_CDMA}
A.~L.~C. Hui and K.~B. Letaief, ``Successive interference cancellation for
  multiuser asynchronous {DS/CDMA} detectors in multipath fading links,''
  \emph{{IEEE} Transactions on Communications}, vol.~46, no.~3, pp. 384--391,
  Mar. 1998.

\bibitem{Verdu_1999:spectral_efficiency_CDMA}
S.~Verd{\'u} and S.~Shamai~(Shitz), ``Spectral efficiency of {CDMA} with random
  spreading,'' \emph{{IEEE} Transactions on Information Theory}, vol.~45,
  no.~2, pp. 622--640, Mar. 1999.

\bibitem{Wang_2009:Wireless_advanced_reception}
X.~Wang and H.~V. Poor, \emph{Wireless Communication Systems: Advanced
  Techniques for Signal Reception}, 1st~ed.\hskip 1em plus 0.5em minus
  0.4em\relax Upper Saddle River, NJ, USA: Prentice Hall, 2009.

\bibitem{Paulraj1994:MIMO}
A.~Paulraj and T.~Kailath, ``Increasing capacity in wireless broadcast systems
  using distributed transmission/directional reception,'' U.S. Patent 5 345
  599, Sep. 6, 1994.

\bibitem{Telatar1995:MIMO_capacity_report}
E.~Telatar, ``Capacity of multi-antenna gaussian channels,'' Technical Report
  \#BL0112170-950615-07TM, AT \& T Bell Laboratories, 1995.

\bibitem{Telatar_1999:MIMO_capacity}
\BIBentryALTinterwordspacing
------, ``Capacity of multi-antenna gaussian channels,'' \emph{European
  Transactions on Telecommunications}, vol.~10, no.~6, pp. 585--595,
  November-December 1999. [Online]. Available:
  \url{http://dx.doi.org/10.1002/ett.4460100604}
\BIBentrySTDinterwordspacing

\bibitem{Foschini:MIMO_capacity}
G.~J. Foschini and M.~J. Gans, ``On limits of wireless communications in a
  fading environment when using multiple antennas,'' \emph{Wireless Personal
  Communications}, vol.~6, no.~3, pp. 311--335, Mar. 1998.

\bibitem{Tarokh:Space-Time_code}
V.~Tarokh, N.~Seshadri, and A.~R. Calderbank, ``Space-time codes for high data
  rate wireless communication: performance criterion and code construction,''
  \emph{{IEEE} Transactions on Information Theory}, vol.~44, no.~2, pp.
  744--765, Mar. 1998.

\bibitem{Tarokh:STBC}
V.~Tarokh, H.~Jafarkhani, and A.~R. Calderbank, ``Space-time block codes from
  orthogonal designs,'' \emph{{IEEE} Transactions on Information Theory},
  vol.~45, no.~5, pp. 1456--1467, Jul. 1999.

\bibitem{Naguib_1998:STC_modem}
A.~F. Naguib, V.~Tarokh, N.~Seshadri, and A.~R. Calderbank, ``A space-time
  coding modem for high-data-rate wireless communications,'' \emph{{IEEE}
  Journal on Selected Areas in Communications}, vol.~16, no.~8, pp. 1459--1478,
  Oct. 1998.

\bibitem{Tarokh_1999:STBC_performance_results}
V.~Tarokh, H.~Jafarkhani, and A.~R. Calderbank, ``Space-time block coding for
  wireless communications: performance results,'' \emph{{IEEE} Journal on
  Selected Areas in Communications}, vol.~17, no.~3, pp. 451--460, Mar. 1999.

\bibitem{Tarokh_1999:STC_criteria_CE}
V.~Tarokh, A.~F. Naguib, N.~Seshadri, and A.~R. Calderbank, ``Space-time codes
  for high data rate wireless communication: performance criteria in the
  presence of channel estimation errors, mobility, and multiple paths,''
  \emph{{IEEE} Transactions on Communications}, vol.~47, no.~2, pp. 199--207,
  Feb. 1999.

\bibitem{Jafarkhani_2001:quasi_orthogonal_STBC}
H.~Jafarkhani, ``A quasi-orthogonal space-time block code,'' \emph{{IEEE}
  Transactions on Communications}, vol.~49, no.~1, pp. 1--4, Jan. 2001.

\bibitem{Jafarkhani_2003:super_orthogonal_STTC}
H.~Jafarkhani and N.~Seshadri, ``Super-orthogonal space-time trellis codes,''
  \emph{{IEEE} Transactions on Information Theory}, vol.~49, no.~4, pp.
  937--950, Apr. 2003.

\bibitem{Alamouti_1998:Alamouti_code}
S.~Alamouti, ``A simple transmit diversity technique for wireless
  communications,'' \emph{{IEEE} Journal on Selected Areas in Communications},
  vol.~16, no.~8, pp. 1451--1458, Oct. 1998.

\bibitem{Jafarkhani2005:STC_book}
H.~Jafarkhani, \emph{Space-Time Coding: Theory and Practice}.\hskip 1em plus
  0.5em minus 0.4em\relax Cambridge University Press, 2005.

\bibitem{Divsalar_1990:Multiple_symbol_differential_detection}
D.~Divsalar and M.~K. Simon, ``Multiple-symbol differential detection of
  {MPSK},'' \emph{{IEEE} Transactions on Communications}, vol.~38, no.~3, pp.
  300--308, Mar. 1990.

\bibitem{Tarokh_2000:differential_decoding}
V.~Tarokh and H.~Jafarkhani, ``A differential detection scheme for transmit
  diversity,'' \emph{{IEEE} Journal on Selected Areas in Communications},
  vol.~18, no.~7, pp. 1169--1174, Jul. 2000.

\bibitem{Jafarkhani_2001:MSDD_STC}
H.~Jafarkhani and V.~Tarokh, ``Multiple transmit antenna differential detection
  from generalized orthogonal designs,'' \emph{{IEEE} Transactions on
  Information Theory}, vol.~47, no.~6, pp. 2626--2631, Sep. 2001.

\bibitem{Hughes_2000:differential_ST}
B.~L. Hughes, ``Differential space-time modulation,'' \emph{{IEEE} Transactions
  on Information Theory}, vol.~46, no.~7, pp. 2567--2578, Nov. 2000.

\bibitem{Warrier_2002:spectrally_efficient_noncoherent}
D.~Warrier and U.~Madhow, ``Spectrally efficient noncoherent communication,''
  \emph{{IEEE} Transactions on Information Theory}, vol.~48, no.~3, pp.
  651--668, Mar. 2002.

\bibitem{Schober_2002:noncoherent_receiver_differential_STC}
R.~Schober and L.~Lampe, ``Noncoherent receivers for differential space-time
  modulation,'' \emph{{IEEE} Transactions on Communications}, vol.~50, no.~5,
  pp. 768--777, May 2002.

\bibitem{Hassibi_2003:how_much_training_needed}
B.~Hassibi and B.~M. Hochwald, ``How much training is needed in
  multiple-antenna wireless links?'' \emph{{IEEE} Transactions on Information
  Theory}, vol.~49, no.~4, pp. 951--963, Apr. 2003.

\bibitem{Lampe_2005:MSDSD}
L.~Lampe, R.~Schober, V.~Pauli, and C.~Windpassinger, ``Multiple-symbol
  differential sphere decoding,'' \emph{{IEEE} Transactions on Communications},
  vol.~53, no.~12, pp. 1981--1985, Dec. 2005.

\bibitem{hanzo2010:mimo_coherent_vs_noncoherent}
L.~Hanzo, Y.~Akhtman, L.~Wang, and M.~Jiang, \emph{{MIMO-OFDM for LTE, WiFi and
  WiMAX: Coherent versus Non-Coherent and Cooperative
  Turbo-Transceivers}}.\hskip 1em plus 0.5em minus 0.4em\relax Chichester, UK:
  Wiley, 2010.

\bibitem{Stuber_2004:MIMO_OFDM}
G.~L. Stuber, J.~R. Barry, S.~W. McLaughlin, Y.~Li, M.~A. Ingram, and T.~G.
  Pratt, ``Broadband {MIMO-OFDM} wireless communications,'' \emph{Proceedings
  of the {IEEE}}, vol.~92, no.~2, pp. 271--294, Feb. 2004.

\bibitem{Gamal_2003:ST_SF_MIMO_fre_selec}
H.~El~Gamal, A.~R. Hammons, Y.~Liu, M.~P. Fitz, and O.~Y. Takeshita, ``On the
  design of space-time and space-frequency codes for {MIMO} frequency-selective
  fading channels,'' \emph{{IEEE} Transactions on Information Theory}, vol.~49,
  no.~9, pp. 2277--2292, Sep. 2003.

\bibitem{Abe_2003:ST_Turbo_equalization_fre_selec_MIMO}
T.~Abe and T.~Matsumoto, ``Space-time turbo equalization in frequency-selective
  {MIMO} channels,'' \emph{{IEEE} Transactions on Vehicular Technology},
  vol.~52, no.~3, pp. 469--475, May 2003.

\bibitem{Zhu_2004:SF_equalization_MIMO_fre_selet}
X.~Zhu and R.~D. Murch, ``Layered space-frequency equalization in a
  single-carrier {MIMO} system for frequency-selective channels,'' \emph{{IEEE}
  Transactions on Wireless Communications}, vol.~3, no.~3, pp. 701--708, May
  2004.

\bibitem{Ma_2005:optimum_training_MIMO_fre_selec}
X.~Ma, L.~Yang, and G.~B. Giannakis, ``Optimal training for {MIMO}
  frequency-selective fading channels,'' \emph{{IEEE} Transactions on Wireless
  Communications}, vol.~4, no.~2, pp. 453--466, Mar. 2005.

\bibitem{Wang_2000:multicarrier_fourier_meets_shannon}
Z.~Wang and G.~B. Giannakis, ``Wireless multicarrier communications--{W}here
  {F}ourier meets {S}hannon,'' \emph{{IEEE} Signal Processing Magazine},
  vol.~17, no.~3, pp. 29--48, May 2000.

\bibitem{Webb_1991:star_QAM}
W.~T. Webb, L.~Hanzo, and R.~Steele, ``Bandwidth efficient {QAM} schemes for
  {Rayleigh} fading channels,'' \emph{IEE Proceedings I (Communications, Speech
  and Vision)}, vol. 138, no.~3, pp. 169--175, Jun. 1991.

\bibitem{Webb_1995:star_QAM}
W.~T. Webb and R.~Steele, ``Variable rate {QAM} for mobile radio,''
  \emph{{IEEE} Transactions on Communications}, vol.~43, no.~7, pp. 2223--2230,
  Jul. 1995.

\bibitem{Dong_1998:star_QAM_fre_sel}
X.~Dong, T.~T. Tjhung, and F.~Adachi, ``Error probability analysis for 16
  {STAR-QAM} in frequency-selective {Rician} fading with diversity reception,''
  \emph{{IEEE} Transactions on Vehicular Technology}, vol.~47, no.~3, pp.
  924--935, Aug. 1998.

\bibitem{Dong_1999:star_QAM}
X.~Dong, N.~C. Beaulieu, and P.~H. Wittke, ``Error probabilities of
  two-dimensional {M}-ary signaling in fading,'' \emph{{IEEE} Transactions on
  Communications}, vol.~47, no.~3, pp. 352--355, Mar. 1999.

\bibitem{Forney_1998:Gaussian_constellation}
J.~Forney, G.~D. and G.~Ungerboeck, ``Modulation and coding for linear
  {Gaussian} channels,'' \emph{{IEEE} Transactions on Information Theory},
  vol.~44, no.~6, pp. 2384--2415, Oct. 1998.

\bibitem{Nevat_2010:MIMO_detect_Gaussian_constellation}
I.~Nevat, G.~W. Peters, and J.~Yuan, ``Detection of {Gaussian} constellations
  in {MIMO} systems under imperfect {CSI},'' \emph{{IEEE} Transactions on
  Communications}, vol.~58, no.~4, pp. 1151--1160, Apr. 2010.

\bibitem{Knuth_1997:art_of_programming}
D.~E. Knuth, \emph{Seminumerical Algorithms}, 3rd~ed., ser. The Art of Computer
  Programming.\hskip 1em plus 0.5em minus 0.4em\relax Reading, Massachusetts:
  Addison-Wesley, 1997.

\bibitem{Fischer_2003:real_versus_complex_VBLAST}
R.~F.~H. Fischer and C.~Windpassinger, ``Real versus complex-valued
  equalisation in {V-BLAST} systems,'' \emph{Electronics Letters}, vol.~39,
  no.~5, pp. 470--471, Mar. 2003.

\bibitem{Siti_2006:pairwise_real_model}
M.~Siti and M.~P. Fitz, ``A novel soft-output layered orthogonal lattice
  detector for multiple antenna communications,'' in \emph{Proc. IEEE
  International Conference on Communications (ICC'06)}, vol.~4, Istanbul,
  Turkey, Jun. 2006, pp. 1686--1691.

\bibitem{Azzam_2009:pairwise_real_model_SD}
L.~Azzam and E.~Ayanoglu, ``Reduced complexity sphere decoding via a reordered
  lattice representation,'' \emph{{IEEE} Transactions on Communications},
  vol.~57, no.~9, pp. 2564--2569, Sep. 2009.

\bibitem{Liu_2012:real_versus_complex_MIMO_detection}
\BIBentryALTinterwordspacing
T.-H. Liu and C.-N. Chiu, ``On fast preprocessing schemes for the real-valued
  spatially multiplexed {MIMO} detectors,'' \emph{International Journal of
  Communication Systems}, 2012. [Online]. Available:
  \url{http://dx.doi.org/10.1002/dac.2365}
\BIBentrySTDinterwordspacing

\bibitem{Neeser:proper_Gaussian_process}
F.~D. Neeser and J.~L. Massey, ``Proper complex random processes with
  applications to information theory,'' \emph{{IEEE} Transactions on
  Information Theory}, vol.~39, no.~4, pp. 1293--1302, Jul. 1993.

\bibitem{Adali2011:complex_valued_SP}
T.~Adali, P.~Schreier, and L.~Scharf, ``Complex-valued signal processing: The
  proper way to deal with impropriety,'' \emph{{IEEE} Transactions on Signal
  Processing}, vol.~59, no.~11, pp. 5101--5125, Nov. 2011.

\bibitem{Mandic_2009:complex_valued_signal_processing}
D.~Mandic and S.~L. Goh, \emph{Complex Valued Nonlinear Adaptive Filters:
  Noncircularity, Widely Linear and Neural Models}, ser. Adaptive and Learning
  Systems for Signal Processing, Communications and Control Series.\hskip 1em
  plus 0.5em minus 0.4em\relax John Wiley \& Sons, 2009.

\bibitem{Tufts_1965:Nyquist_problem_ISI}
D.~W. Tufts, ``Nyquist's problem -- the joint optimization of transmitter and
  receiver in pulse amplitude modulation,'' \emph{Proceedings of the IEEE},
  vol.~53, no.~3, pp. 248--259, Mar. 1965.

\bibitem{Li_2002:MIMO_OFDM_SIC_detection_CE}
Y.~G. Li, J.~H. Winters, and N.~R. Sollenberger, ``{MIMO-OFDM} for wireless
  communications: signal detection with enhanced channel estimation,''
  \emph{{IEEE} Transactions on Communications}, vol.~50, no.~9, pp. 1471--1477,
  Sep. 2002.

\bibitem{Lu_2002:MIMO_OFDM_LDPC_SIC}
B.~Lu, G.~Yue, and X.~Wang, ``Performance analysis and design optimization of
  {LDPC}-coded {MIMO OFDM} systems,'' \emph{{IEEE} Transactions on Signal
  Processing}, vol.~52, no.~2, pp. 348--361, Feb. 2004.

\bibitem{Zanella_2005:MMSE_SIC_MIMO}
A.~Zanella, M.~Chiani, and M.~Z. Win, ``{MMSE} reception and successive
  interference cancellation for {MIMO} systems with high spectral efficiency,''
  \emph{{IEEE} Transactions on Wireless Communications}, vol.~4, no.~3, pp.
  1244--1253, May 2005.

\bibitem{Chen_2006:MBER_MIMO_Detection}
S.~Chen, A.~Livingstone, and L.~Hanzo, ``Minimum bit-error rate design for
  space-time equalization-based multiuser detection,'' \emph{{IEEE}
  Transactions on Communications}, vol.~54, no.~5, pp. 824--832, May 2006.

\bibitem{Palomar_2005:MBER_MIMO_tranceiver}
D.~P. Palomar, M.~Bengtsson, and B.~Ottersten, ``Minimum {BER} linear
  transceivers for {MIMO} channels via primal decomposition,'' \emph{{IEEE}
  Transactions on Signal Processing}, vol.~53, no.~8, pp. 2866--2882, Aug.
  2005.

\bibitem{Wubben_2003:MMSE_DSNR_ordering}
D.~W{\"u}bben, R.~B{\"o}hnke, V.~K{\"u}hn, and K.-D. Kammeyer, ``{MMSE}
  extension of {V-BLAST} based on sorted {QR} decomposition,'' in \emph{Proc.
  IEEE 58th Vehicular Technology Conference (VTC'03-Fall)}, Orlando, USA, Oct.
  2003, pp. 508--512.

\bibitem{Wai_2000:IMSE_DSNR_ordering_MMSE_SIC}
K.-W. Wong, C.-Y. Tsui, and R.~S. Cheng, ``A low complexity architecture of the
  {V-BLAST} system,'' in \emph{Proc. IEEE Wireless Communications and
  Networking Confernce (WCNC'00)}, Chicago, IL, USA, Sep. 2000, pp. 310--314.

\bibitem{Bohnke_2003:MMSE_SIC_GSNR}
R.~B{\"o}hnke, D.~W{\"u}bben, V.~K{\"u}hn, and K.-D. Kammeyer, ``Reduced
  complexity {MMSE} detection for {BLAST} architectures,'' in \emph{Proc. IEEE
  Global Telecommunications Conference (GLOBECOM'03)}, San Francisco, USA, Dec.
  2003, pp. 2258--2262.

\bibitem{Hassibi_2000:sqrt_algorithm_BLAST_LMSE_ordering}
B.~Hassibi, ``An efficient square-root algorithm for {BLAST},'' in \emph{Proc.
  IEEE International Conference on Acoustics, Speech, and Signal Processing
  (ICASSP'00)}, Istanbul, Turkey, Jun. 2000, pp. 737--740.

\bibitem{Benesty_2003:MMSE_SIC_LMSE_ordering}
J.~Benesty, Y.~Huang, and J.~Chen, ``A fast recursive algorithm for optimum
  sequential signal detection in a {BLAST} system,'' \emph{{IEEE} Transactions
  on Signal Processing}, vol.~51, no.~7, pp. 1722--1730, Jul. 2003.

\bibitem{Liu_2009:MMSE_SIC_LMSE}
T.-H. Liu, ``Some results for the fast {MMSE-SIC} detection in spatially
  multiplexed {MIMO} systems,'' \emph{{IEEE} Transactions on Wireless
  Communications}, vol.~8, no.~11, pp. 5443--5448, Nov. 2009.

\bibitem{Chin_2000:PIC_BLAST}
W.~H. Chin, A.~G. Constantinides, and D.~B. Ward, ``Parallel multistage
  detection for multiple antenna wireless systems,'' \emph{Electronics
  Letters}, vol.~38, no.~12, pp. 597--599, 2002.

\bibitem{Luo_2008:PIC_MIMO}
Z.~Luo, M.~Zhao, S.~Liu, and Y.~Liu, ``Generalized parallel interference
  cancellation with near-optimal detection performance,'' \emph{{IEEE}
  Transactions on Signal Processing}, vol.~56, no.~1, pp. 304--312, Jan. 2008.

\bibitem{Studer_2011:PIC_MIMO_ASIC_implementation}
C.~Studer, S.~Fateh, and D.~Seethaler, ``{ASIC} implementation of soft-input
  soft-output {MIMO} detection using {MMSE} parallel interference
  cancellation,'' \emph{{IEEE} Journal of Solid-State Circuits}, vol.~46,
  no.~7, pp. 1754--1765, Jul. 2011.

\bibitem{Viterbo_1993:SD_conf}
E.~Viterbo and E.~Biglieri, ``A universal decoding algorithm for lattice
  codes,'' in \emph{Proc. GRETSI 14-{\`e}me Colloque}, Juan-les-Pins, France,
  Sep. 1993.

\bibitem{Wu_2008:early_pruning_SD}
Y.~H. Wu, Y.~T. Liu, H.-C. Chang, Y.-C. Liao, and H.-C. Chang, ``Early-pruned
  {K}-best sphere decoding algorithm based on radius constraints,'' in
  \emph{Proc. IEEE International Conference on Communications (ICC'08)},
  Beijing, China, May 2008, pp. 4496--4500.

\bibitem{Fukatani_2004:best_first_search}
T.~Fukatani, R.~Matsumoto, and T.~Uyematsu, ``Two methods for decreasing the
  computational complexity of the {MIMO ML} decoder,'' \emph{IEICE Transactions
  on Fundamentals of Electronics, Communications and Computer Sciences}, vol.
  E87-A, no.~10, pp. 2571--2576, Oct. 2004.

\bibitem{Okawado_2008:best_first_search}
A.~Okawado, R.~Matsumoto, and T.~Uyematsu, ``Near {ML} detection using
  {D}ijkstra's algorithm with bounded list size over {MIMO} channels,'' in
  \emph{Proc. IEEE International Symposium on Information Theory (ISIT'2008)},
  Toronto, ON, Jul. 2008, pp. 2022--2025.

\bibitem{Luo:SDR_simplest}
Z.-Q. Luo, W.-K. Ma, A.~M.-C. So, Y.~Ye, and S.~Zhang, ``Semidefinite
  relaxation of quadratic optimization problems,'' \emph{{IEEE} Signal
  Processing Magazine}, vol.~27, no.~3, pp. 20--34, May 2010.

\bibitem{Luo:Convex_Optimization_for_SP_Comm}
Z.-Q. Luo and W.~Yu, ``An introduction to convex optimization for
  communications and signal processing,'' \emph{{IEEE} Journal on Selected
  Areas in Communications}, vol.~24, no.~8, pp. 1426--1438, Aug. 2006.

\bibitem{Liang_2006:block_iterative_DFEs_massive_MIMO}
Y.-C. Liang, S.~Sun, and C.~K. Ho, ``Block-iterative generalized decision
  feedback equalizers for large {MIMO} systems: algorithm design and asymptotic
  performance analysis,'' \emph{{IEEE} Transactions on Signal Processing},
  vol.~54, no.~6, pp. 2035--2048, Jun. 2006.

\bibitem{Liang_2007:MMSE_large_system_performance}
Y.-C. Liang, G.~Pan, and Z.~D. Bai, ``Asymptotic performance of {MMSE}
  receivers for large systems using random matrix theory,'' \emph{{IEEE}
  Transactions on Information Theory}, vol.~53, no.~11, pp. 4173--4190, Nov.
  2007.

\bibitem{Liang_2008:relation_between_MMSE_SIC_and_BI_GDFE}
Y.-C. Liang, E.~Y. Cheu, L.~Bai, and G.~Pan, ``On the relationship between
  {MMSE-SIC} and {BI-GDFE} receivers for large multiple-input multiple-output
  channels,'' \emph{{IEEE} Transactions on Signal Processing}, vol.~56, no.~8,
  pp. 3627--3637, Aug. 2008.

\bibitem{Chockalingam_2009:RTS_LS_MIMO_detector}
B.~S. Rajan, S.~K. Mohammed, A.~Chockalingam, and N.~Srinidhi, ``Low-complexity
  near-{ML} decoding of large non-orthogonal {STBCs} using reactive tabu
  search,'' in \emph{Proc. IEEE International Symposium on Information Theory
  (ISIT'09}, Seoul, Korea, Jun. 2009, pp. 1993--1997.

\bibitem{Chockalingam_2009:LAS_non_orthogonal_STBC}
S.~K. Mohammed, A.~Zaki, A.~Chockalingam, and B.~S. Rajan, ``High-rate
  space-time coded large-{MIMO} systems: Low-complexity detection and channel
  estimation,'' \emph{{IEEE} Journal of Selected Topics in Signal Processing},
  vol.~3, no.~6, pp. 958--974, Dec 2009.

\bibitem{Chockalingam_2010:low_complexity_LS_MIMO_detection}
A.~Chockalingam, ``Low-complexity algorithms for large-{MIMO} detection,'' in
  \emph{Proc. 4th International Symposium on Communications, Control and Signal
  Processing (ISCCSP'10)}, Limassol, Cyprus, Mar. 2010, pp. 1--6.

\bibitem{Chockalingam_2011:hybrid_RTS_BP_LS_MIMO_detector}
T.~Datta, N.~Srinidhi, A.~Chockalingam, and B.~S. Rajan, ``A hybrid {RTS-BP}
  algorithm for improved detection of large-{MIMO} {M-QAM} signals,'' in
  \emph{Proc. National Conference on Communications (NCC'11)}, Bangalore,
  India, Jan. 2011, pp. 1--5.

\bibitem{Chockalingam_2011:graphical_model_LS_MIMO_detection}
P.~Som, T.~Datta, N.~Srinidhi, A.~Chockalingam, and B.~S. Rajan,
  ``Low-complexity detection in large-dimension {MIMO-ISI} channels using
  graphical models,'' \emph{{IEEE} Journal of Selected Topics in Signal
  Processing}, vol.~5, no.~8, pp. 1497--1511, Dec. 2011.

\bibitem{Chockalingam_2011:randomized_MCMC_and_search_LS_MIMO_detection}
A.~Kumar, S.~Chandrasekaran, A.~Chockalingam, and B.~S. Rajan, ``Near-optimal
  large-{MIMO} detection using randomized {MCMC} and randomized search
  algorithms,'' in \emph{Proc. IEEE International Conference on Communications
  (ICC'11)}, Jun. 2011, pp. 1--5.

\bibitem{Chockalingam_2013:MC_Sampling_receiver_LS_MIMO}
T.~Datta, N.~A. Kumar, A.~Chockalingam, and B.~S. Rajan, ``A novel
  monte-carlo-sampling-based receiver for large-scale uplink multiuser {MIMO}
  systems,'' \emph{{IEEE} Transactions on Vehicular Technology}, vol.~62,
  no.~7, pp. 3019--3038, Sep. 2013.

\bibitem{Chockalingam_2013:LR_detection_LS_MIMO}
K.~A. Singhal, T.~Datta, and A.~Chockalingam, ``Lattice reduction aided
  detection in large-{MIMO} systems,'' in \emph{Proc. IEEE 14th Workshop on
  Signal Processing Advances in Wireless Communications (SPAWC'13)}, Darmstadt,
  Germany, Jun. 2013, pp. 594--598.

\bibitem{Suthisopapan_2013:capacity_approaching_LDPC_MMSE_detection_LS_MIMO_journal}
P.~Suthisopapan, K.~Kasai, A.~Meesomboon, and V.~Imtawil, ``Achieving near
  capacity of non-binary {LDPC} coded large {MIMO} systems with a novel ultra
  low-complexity soft-output detector,'' \emph{{IEEE} Transactions on Wireless
  Communications}, vol.~12, no.~10, pp. 5185--5199, Oct. 2013.

\bibitem{Chockalingam_2014:Channel_harderning_message_passsing}
T.~Lakshmi~Narasimhan and A.~Chockalingam, ``Channel hardening-exploiting
  message passing ({CHEMP}) receiver in large-scale {MIMO} systems,''
  \emph{{IEEE} Journal of Selected Topics in Signal Processing}, vol.~8, no.~5,
  pp. 847 -- 860, Oct. 2014.

\bibitem{Larsson_2014:SUMIS_LS_MIMO_detector}
M.~Cirkic and E.~Larsson, ``{SUMIS}: Near-optimal soft-in soft-out {MIMO}
  detection with low and fixed complexity,'' \emph{{IEEE} Transactions on
  Signal Processing}, vol.~62, no.~12, pp. 3084 -- 3097, Jun. 2014.

\bibitem{Wu_2014:message_passing_soft_iterative_LS_MIMO_detection}
S.~Wu, L.~Kuang, Z.~Ni, J.~Lu, D.~Huang, and Q.~Guo, ``Low-complexity iterative
  detection for large-scale multiuser {MIMO-OFDM} systems using approximate
  message passing,'' \emph{{IEEE} Journal of Selected Topics in Signal
  Processing}, vol.~8, no.~5, pp. 902 -- 915, Oct. 2014.

\bibitem{Studer_2014:implementation_LS_MIMO_detector}
B.~Yin, M.~Wu, G.~Wang, C.~Dick, J.~R. Cavallaro, and C.~Studer, ``A 3.8 {Gb}/s
  large-scale {MIMO} detector for {3GPP LTE-A}dvanced,'' in \emph{Proc. IEEE
  International Conference on Acoustics, Speech, and Signal Processing
  (ICASSP'14)}, Florence, Italy, May 2014, pp. 3879 -- 3883.

\bibitem{Studer_2014:implementation_LS_MIMO_detector_journal}
M.~Wu, B.~Yin, G.~Wang, C.~Dick, J.~Cavallaro, and C.~Studer, ``Large-scale
  {MIMO} detection for {3GPP LTE}: Algorithm and {FPGA} implementation,''
  \emph{{IEEE} Journal of Selected Topics in Signal Processing}, vol.~8, no.~5,
  pp. 916 -- 929, Oct. 2014.

\bibitem{Shinya_2012:CST_near_capacity_transceiver}
S.~Sugiura, S.~Chen, and L.~Hanzo, ``{MIMO}-aided near-capacity turbo
  transceivers: Taxonomy and performance versus complexity,'' \emph{IEEE
  Communications Surveys and Tutorials}, vol.~14, no.~2, pp. 421--442, Second
  Quarter 2012.

\bibitem{Andrews:Interference_Cancellation_overview}
J.~G. Andrews, ``Interference cancellation for cellular systems: a contemporary
  overview,'' \emph{{IEEE} Wireless Communications Magazine}, vol.~12, no.~2,
  p. 2005, Apr. 19-29.

\bibitem{Viterbi_1967:viterbi_algorithm}
A.~J. Viterbi, ``Error bounds for convolutional codes and an asymptotically
  optimum decoding algorithm,'' \emph{{IEEE} Transactions on Information
  Theory}, vol.~13, no.~2, pp. 260--269, Apr. 1967.

\bibitem{Omura_1969:Viterbi_algorithm_new_interpretation}
J.~Omura, ``On the {Viterbi} decoding algorithm,'' \emph{{IEEE} Transactions on
  Information Theory}, vol.~15, no.~1, pp. 177--179, Jan. 1969.

\bibitem{Forney_1972:MLSE_ISI}
G.~D. Forney, ``Maximum-likelihood sequence estimation of digital sequences in
  the presence of intersymbol interference,'' \emph{{IEEE} Transactions on
  Information Theory}, vol.~18, no.~3, pp. 363--378, May 1972.

\bibitem{Forney_1973:Viterbi_algorithm_tutorial}
------, ``The {Viterbi} algorithm,'' \emph{Proceedings of the IEEE}, vol.~61,
  no.~3, pp. 268--278, Mar. 1973.

\bibitem{Viterbi_2006:viterbi_algorithm_history}
A.~J. Viterbi, ``A personal history of the {Viterbi} algorithm,'' \emph{IEEE
  Signal Processing Magazine}, vol.~23, no.~4, pp. 120--142, Jul. 2006.

\bibitem{Qureshi_1985:adaptive_equalization}
S.~U.~H. Qureshi, ``Adaptive equalization,'' \emph{Proceedings of the {IEEE}},
  vol.~73, no.~9, pp. 1349--1387, Sep. 1985.

\bibitem{honig1995:blind_adaptive_MUD}
M.~Honig, U.~Madhow, and S.~Verd{\'u}, ``Blind adaptive multiuser detection,''
  \emph{{IEEE} Transactions on Information Theory}, vol.~41, no.~4, pp.
  944--960, Jul. 1995.

\bibitem{Tong_1994:blind_identification_equalization}
L.~Tong, G.~Xu, and T.~Kailath, ``Blind identification and equalization based
  on second-order statistics: a time domain approach,'' \emph{{IEEE}
  Transactions on Information Theory}, vol.~40, no.~2, pp. 340--349, Mar. 1994.

\bibitem{Wang_1998:blind_equalization_and_MUD_CDMA}
X.~Wang and H.~V. Poor, ``Blind equalization and multiuser detection in
  dispersive cdma channels,'' \emph{{IEEE} Transactions on Communications},
  vol.~46, no.~1, pp. 91--103, Jan. 1998.

\bibitem{Tranter_2007:best_of_best}
W.~H. Tranter, D.~P. Taylor, and R.~E. Ziemer, Eds., \emph{The best of the
  best: fifty years of communications and networking research}.\hskip 1em plus
  0.5em minus 0.4em\relax Hoboken, NJ: John Wiley \& Sons, 2007.

\bibitem{Verdu_1984:Phd_thesis}
S.~Verd{\'u}, ``Optimum multi-user signal detection,'' Ph.D. dissertation,
  Department of Electrical and Computer Engineering, University of Illinois at
  Urbana-Champaign, Urbana, IL, Aug. 1984.

\bibitem{Poor_1988:SUD_for_MU}
H.~V. Poor and S.~Verd{\'u}, ``Single-user detectors for multiuser channels,''
  \emph{{IEEE} Transactions on Communications}, vol.~36, no.~1, pp. 50--60,
  Jan. 1988.

\bibitem{Grant_1998:ML_diversity_order}
S.~J. Grant and J.~K. Cavers, ``Performance enhancement through joint detection
  of cochannel signals using diversity arrays,'' \emph{{IEEE} Transactions on
  Communications}, vol.~46, no.~8, pp. 1038--1049, Aug. 1998.

\bibitem{Grant_2000:further_results_ML_diversity_order}
------, ``Further analytical results on the joint detection of cochannel
  signals using diversity arrays,'' \emph{{IEEE} Transactions on
  Communications}, vol.~48, no.~11, pp. 1788--1792, Nov. 2000.

\bibitem{Awater_2000:ML_MMSE_diversity_order}
R.~van Nee, V.~van Zelst, and G.~Awater, ``Maximum likelihood decoding in a
  space division multiplexing system,'' in \emph{Proc. IEEE 51st Vehicular
  Technology Conference (VTC'00-Spring)}, Tokyo, Japan, May 2000, pp. 6--10.

\bibitem{Murch_2002:ML_performance_analysis}
X.~Zhu and R.~D. Murch, ``Performance analysis of maximum likelihood detection
  in a {MIMO} antenna system,'' \emph{{IEEE} Transactions on Communications},
  vol.~50, no.~2, pp. 187--191, Feb. 2002.

\bibitem{Lee_2006:ML_VER}
M.~Shin, D.~S. Kwon, and C.~Lee, ``Performance analysis of maximum likelihood
  detection for {MIMO} systems,'' in \emph{Proc. IEEE 63rd Vehicular Technology
  Conference (VTC'06-Spring)}, Melbourne, Austria, May 2006, pp. 2154--2158.

\bibitem{Verdu_1997:MUD_progress_misconception}
\BIBentryALTinterwordspacing
S.~Verd{\'u}, ``Demodulation in the presence of multiuser interference:
  progress and misconceptions,'' in \emph{Intelligent Methods in Signal
  Processing and Communications}, D.~Docampo, A.~R. Figueiras-Vidal, and
  F.~P{\'e}rez-Gonz{\'a}lez, Eds.\hskip 1em plus 0.5em minus 0.4em\relax
  Boston: Birkh{\"a}user, 1997, pp. 15--45. [Online]. Available:
  \url{http://dx.doi.org/10.1007/978-1-4612-2018-3_2}
\BIBentrySTDinterwordspacing

\bibitem{Li_2006:MMSE_SINR_distribution}
P.~Li, D.~Paul, R.~Narasimhan, and J.~Cioffi, ``On the distribution of {SINR}
  for the {MMSE MIMO} receiver and performance analysis,'' \emph{{IEEE}
  Transactions on Information Theory}, vol.~52, no.~1, pp. 271--286, Jan. 2006.

\bibitem{Mehana_2012:MMSE_diversity}
A.~H. Mehana and A.~Nosratinia, ``Diversity of {MMSE MIMO} receivers,''
  \emph{{IEEE} Transactions on Information Theory}, vol.~58, no.~11, pp.
  6788--6805, Nov. 2012.

\bibitem{Abend_1968:earliest_MAP_sequence_detector}
K.~Abend, T.~J. Harley, B.~D. Fritchman, and C.~Gumacos, ``On optimum receivers
  for channels having memory,'' \emph{{IEEE} Transactions on Information
  Theory}, vol.~14, no.~6, pp. 819--820, Nov. 1968.

\bibitem{Abend_1970:earliest_MAP_sequence_detection}
K.~Abend and B.~D. Fritchman, ``Statistical detection for communication
  channels with intersymbol interference,'' \emph{Proceedings of the {IEEE}},
  vol.~58, no.~5, pp. 779--785, May 1970.

\bibitem{Bahl_1974:BCJR_algorithm}
L.~Bahl, J.~Cocke, F.~Jelinek, and J.~Raviv, ``Optimal decoding of linear codes
  for minimizing symbol error rate,'' \emph{{IEEE} Transactions on Information
  Theory}, vol.~20, no.~2, pp. 284--287, Mar. 1974.

\bibitem{Verdu_1984:decision_algorithms}
S.~Verd{\'u} and H.~V. Poor, ``Backward, forward and backward-forward dynamic
  programming models under commutativity conditions,'' in \emph{Proc. the 23rd
  IEEE Conference on Decision and Control (CDC'84)}, Las Vegas, NV, Dec. 1984,
  pp. 1081--1086.

\bibitem{Verdu_1987:decision_algorithms}
\BIBentryALTinterwordspacing
------, ``Abstract dynamic programming models under commutativity conditions,''
  \emph{SIAM Journal on Control and Optimization}, vol.~25, no.~4, pp.
  990--1006, Jul. 1987. [Online]. Available:
  \url{http://epubs.siam.org/doi/abs/10.1137/0325054}
\BIBentrySTDinterwordspacing

\bibitem{Garey_1979:complexity_theory}
M.~R. Garey and D.~S. Johnson, \emph{Computers and Intractability: A Guide to
  the Theory of {NP}-Completeness}.\hskip 1em plus 0.5em minus 0.4em\relax San
  Francisco, USA: W. H. Freeman and Co., 1979.

\bibitem{Cormen_2009:introduction_to_algorithm}
T.~H. Cormen, C.~E. Leiserson, R.~L. Rivest, and C.~Stein, \emph{Introduction
  to Algorithms}, 3rd~ed.\hskip 1em plus 0.5em minus 0.4em\relax Cambridge,
  Masachusetts: MIT Press, 2009.

\bibitem{Garrett_2003:first_ML_VLSI_implementation}
D.~C. Garrett, L.~M. Davis, and G.~K. Woodward, ``19.2 {Mbit/s} $4 \times 4$
  {BLAST/MIMO} detector with soft {ML} outputs,'' \emph{Electronics Letters},
  vol.~39, no.~2, pp. 233--235, Jan 2003.

\bibitem{Burg_2003:VLSI_hard_ML}
A.~Burg, N.~Felber, and W.~Fichtner, ``A 50 {Mbps} $4\times 4$ maximum
  likelihood decoder for multiple-input multiple-output systems with {QPSK}
  modulation,'' in \emph{Proc. IEEE International Conference on Electronics,
  Circuits and Systems (ICECS'03)}, Sharjah, United Arab Emirates, Dec. 2003,
  pp. 332--335.

\bibitem{tenBrink2013:massive_MIMO}
J.~Hoydis, S.~ten Brink, and M.~Debbah, ``Massive mimo in the ul/dl of cellular
  networks: How many antennas do we need?'' \emph{{IEEE} Journal on Selected
  Areas in Communications}, vol.~31, no.~2, pp. 160--171, Feb. 2013.

\bibitem{Lupas_1986:linear_MUD_synchrohous_CDMA_conf}
R.~Lupas and S.~Verd{\'u}, ``Asymptotic efficiency of linear multiuser
  detectors,'' in \emph{Proc. the 25th IEEE Conference on Decision and Control
  (CDC'86)}, Dec. 1986, pp. 2094--2100.

\bibitem{Lupas_1988:linear_MUD_asynchrohous_CDMA_conf}
\BIBentryALTinterwordspacing
------, ``Optimum near-far resistance of linear detectors for code-division
  multiple-access channels,'' in \emph{Abstr. IEEE International Symposium on
  Information Theory (ISIT'88)}, Jun. 1988, p.~14. [Online]. Available:
  \url{http://www.princeton.edu/~verdu/reprints/Optimum Near-Far Resistance Of
  Linear Detectors.pdf}
\BIBentrySTDinterwordspacing

\bibitem{Kay_1993:Fundamentals_statistical_signal_processing_vol_1}
S.~M. Kay, \emph{Fundamentals of Statistical Signal Processing, Volume I:
  Estimation Theory}.\hskip 1em plus 0.5em minus 0.4em\relax Prentice Hall,
  1993.

\bibitem{Xie_1989:Linear_MMSE_WLS_MUD_conf}
Z.~Xie, R.~T. Short, and C.~K. Rushforth, ``Suboptimum coherent detection of
  direct-sequence multiple-access signals,'' in \emph{Proc. IEEE Military
  Communications Conference (MILCOM'89)}, Boston, MA, Oct. 1989, pp. 128--133.

\bibitem{Poor_1997:probability_of_error_MMSE_MUD}
H.~V. Poor and S.~Verd{\'u}, ``Probability of error in {MMSE} multiuser
  detection,'' \emph{{IEEE} Transactions on Information Theory}, vol.~43,
  no.~3, pp. 858--871, May 1997.

\bibitem{Lizhong_2003:DMT}
L.~Zheng and D.~N.~C. Tse, ``Diversity and multiplexing: A fundamental tradeoff
  in multiple-antenna channels,'' \emph{{IEEE} Transactions on Information
  Theory}, vol.~49, no.~5, pp. 1073--1096, May 2003.

\bibitem{Tse_2004:DMT}
D.~N.~C. Tse, P.~Viswanath, and L.~Zheng, ``Diversity-multiplexing tradeoff in
  multiple-access channels,'' \emph{{IEEE} Transactions on Information Theory},
  vol.~50, no.~9, pp. 1859--1874, Sep. 2004.

\bibitem{Hedayat_2007:outage_diversity_ZF_MMSE}
A.~Hedayat and A.~Nosratinia, ``Outage and diversity of linear receivers in
  flat-fading {MIMO} channels,'' \emph{{IEEE} Transactions on Signal
  Processing}, vol.~55, no.~12, pp. 5868--5873, Dec. 2007.

\bibitem{Jorswieck_2007:outage_prob_MIMO}
E.~A. Jorswieck and H.~Boche, ``Outage probability in multiple antenna
  systems,'' \emph{European Transactions on Telecommunications}, vol.~18,
  no.~3, pp. 217--233, 2007.

\bibitem{Moustakas_2009:MMSE_massive_MIMO}
A.~L. Moustakas, K.~R. Kumar, and G.~Caire, ``Performance of {MMSE MIMO}
  receivers: A large {$N$} analysis for correlated channels,'' in \emph{Proc.
  IEEE 69th Vehicular Technology Conference (VTC'09 Spring)}, Barcelona, Spain,
  Apr. 2009, pp. 1--5.

\bibitem{Kumar_2009:asymptotic_antenna_SNR_linear_MIMO_receiver}
K.~R. Kumar, G.~Caire, and A.~L. Moustakas, ``Asymptotic performance of linear
  receivers in {MIMO} fading channels,'' \emph{{IEEE} Transactions on
  Information Theory}, vol.~55, no.~10, pp. 4398--4418, Oct. 2009.

\bibitem{Jiang_2011:performance_analysis_ZF_MMSE}
Y.~Jiang, M.~Varanasi, and J.~Li, ``Performance analysis of {ZF} and {MMSE}
  equalizers for {MIMO} systems: An in-depth study of the high {SNR} regime,''
  \emph{{IEEE} Transactions on Information Theory}, vol.~57, no.~4, pp.
  2008--2026, Apr. 2011.

\bibitem{Mandayam_1993:MBER_optical_CDMA_MUD_conf}
N.~B. Mandayam and B.~Aazhang, ``Generalized sensitivity analysis of optical
  code division multiple access systems,'' in \emph{Proc. 27th Annual
  Conference on Information Sciences and Systems (CISS'93)}, Baltimore, MD,
  Mar. 1993, pp. 302--307.

\bibitem{Mandayam_1997:MBER_optical_CDMA_MUD_part_1}
------, ``Gradient estimation for stochastic optimization of optical
  code-division multiple-access systems: Part {I} -- generalized sensitivity
  analysis,'' \emph{{IEEE} Journal on Selected Areas in Communications},
  vol.~15, no.~4, pp. 731--741, May 1997.

\bibitem{Mandayam_1997:MBER_optical_CDMA_MUD_part_2}
------, ``Gradient estimation for stochastic optimization of optical
  code-division multiple-access systems: Part {II} -- adaptive detection,''
  \emph{{IEEE} Journal on Selected Areas in Communications}, vol.~15, no.~4,
  pp. 742--750, May 1997.

\bibitem{Mandayam_1997:MBER_CDMA_MUD}
------, ``Gradient estimation for sensitivity analysis and adaptive multiuser
  interference rejection in code-division multiple-access systems,''
  \emph{{IEEE} Transactions on Communications}, vol.~45, no.~7, pp. 848--858,
  Jul. 1997.

\bibitem{Burg_2006:linear_MMSE_VLSI}
A.~Burg, S.~Haene, D.~Perels, P.~Luethi, N.~Felber, and W.~Fichtner,
  ``Algorithm and {VLSI} architecture for linear {MMSE} detection in
  {MIMO-OFDM} systems,'' in \emph{Proc. IEEE International Symposium on
  Circuits and Systems (ISCAS'06)}, Island of Kos, Greece, May 2006, pp.
  4102--4105.

\bibitem{Yoshizawa_2009:VLSI_implementation_MMSE_MIMO_OFDM}
S.~Yoshizawa and Y.~Miyanaga, ``{VLSI} implementation of a 4 x 4 {MIMO-OFDM}
  transceiver with an 80-{MHz} channel bandwidth,'' in \emph{Proc. IEEE
  International Symposium on Circuits and Systems (ISCAS'09)}, Taipei, Republic
  of China, May 2009, pp. 1743--1746.

\bibitem{Psaromiligkos_1999:MBER_CDMA}
I.~N. Psaromiligkos, S.~N. Batalama, and D.~A. Pados, ``On adaptive minimum
  probability of error linear filter receivers for {DS-CDMA} channels,''
  \emph{{IEEE} Transactions on Communications}, vol.~47, no.~7, pp. 1092--1102,
  Jul. 1999.

\bibitem{Wang_1999:MBER_MUD_conf}
X.~Wang, W.-S. Lu, and A.~Antoniou, ``Constrained minimum-{BER} multiuser
  detection,'' in \emph{Proc. IEEE International Conference on Acoustics,
  Speech, and Signal Processing (ICASSP'99)}, Phoenix, AZ, USA, Mar. 1999, pp.
  2603--2606.

\bibitem{Wang_2000_MBER_MUD}
------, ``Constrained minimum-{BER} multiuser detection,'' \emph{{IEEE}
  Transactions on Signal Processing}, vol.~48, no.~10, pp. 2903--2909, Oct.
  2000.

\bibitem{Yeh_1998:MBER_conf}
C.-C. Yeh, R.~R. Lopes, and J.~R. Barry, ``Approximate minimum bit-error rate
  multiuser detection,'' in \emph{Proc. IEEE Global Telecommunications
  Conference (GLOBECOM'98)}, Sydney, NSW, Australia, Nov. 1998, pp. 3590--3595.

\bibitem{Yeh_2000:MBER_MUD}
C.-C. Yeh and J.~R. Barry, ``Adaptive minimum bit-error rate equalization for
  binary signaling,'' \emph{{IEEE} Transactions on Communications}, vol.~48,
  no.~7, pp. 1226--1235, Jul. 2000.

\bibitem{Chen_2001:adpative_MBER_linear_MUD}
S.~Chen, A.~K. Samingan, B.~Mulgrew, and L.~Hanzo, ``Adaptive minimum-{BER}
  linear multiuser detection for {DS-CDMA} signals in multipath channels,''
  \emph{{IEEE} Transactions on Signal Processing}, vol.~49, no.~6, pp.
  1240--1247, Jun. 2001.

\bibitem{Bergmans_1974:interference_cancellation_earliest}
P.~Bergmans and T.~M. Cover, ``Cooperative broadcasting,'' \emph{{IEEE}
  Transactions on Information Theory}, vol.~20, no.~3, pp. 317--324, May 1974.

\bibitem{Cover_1975:earliest_interference_cancellation_idea}
\BIBentryALTinterwordspacing
T.~M. Cover, ``Some advances in broadcast channels,'' in \emph{Advances in
  Communication Systems}, A.~J. Viterbi, Ed.\hskip 1em plus 0.5em minus
  0.4em\relax New York: Academic Press, Inc., 1975, vol.~4, pp. 229--260.
  [Online]. Available:
  \url{http://www-isl.stanford.edu/~cover/papers/paper34.pdf}
\BIBentrySTDinterwordspacing

\bibitem{Dent_1992:SIC_CDMA}
P.~Dent, B.~Gudmundson, and M.~Ewerbring, ``{CDMA-IC}: a novel code division
  multiple access scheme based on interference cancellation,'' in \emph{Proc.
  3rd IEEE International Symposium on Personal, Indoor and Mobile Radio
  Communications (PIMRC'92)}, Boston, MA, Oct. 1992, pp. 98--102.

\bibitem{Moshavi_1996_multistage_linear_receivers_CDMA}
\BIBentryALTinterwordspacing
S.~Moshavi, E.~G. Kanterakis, and D.~L. Schilling, ``Multistage linear
  receivers for {DS-CDMA} systems,'' \emph{International Journal of Wireless
  Information Networks}, vol.~3, no.~1, pp. 1--17, Jan. 1996. [Online].
  Available: \url{http://dx.doi.org/10.1007/BF02106658}
\BIBentrySTDinterwordspacing

\bibitem{Kubota_1992:SIC_CDMA}
S.~Kubota, S.~Kato, and K.~Feher, ``Inter-channel interference cancellation
  technique for {CDMA} mobile/personal communication systems,'' in \emph{Proc.
  3rd IEEE International Symposium on Personal, Indoor and Mobile Radio
  Communications (PIMRC'92)}, Boston, MA, Oct. 1992, pp. 112--117.

\bibitem{Kubota_1992:SIC_CDMA_base_station}
------, ``Inter-channel interference cancellation technique for {CDMA}
  mobile/personal communication base stations,'' in \emph{Proc. 2nd IEEE
  International Symposium on Spread Spectrum Techniques and Applications
  (ISSTA'92)}, Yokohama, Japan, Nov. 1992, pp. 91--94.

\bibitem{Patel_1993:SIC_CDMA}
P.~R. Patel and J.~M. Holtzman, ``Analysis of a {DS/CDMA} successive
  interference cancellation scheme using correlations,'' in \emph{Proc. IEEE
  Global Telecommunications Conference (GLOBECOM'93)}, Houston, TX, USA, Dec.
  1993, pp. 76--80.

\bibitem{Holtzman_1994:SIC_CDMA}
J.~M. Holtzman, ``{DS/CDMA} successive interference cancellation,'' in
  \emph{Proc. IEEE 3rd International Symposium on Spread Spectrum Techniques
  and Applications (ISSSTA'94)}, Oulu, Finland, Jul. 1994, pp. 69--78.

\bibitem{Holtzman_1994:SIC_CDMA_milcom}
------, ``Successive interference cancellation for direct sequence code
  division multiple access,'' in \emph{Proc. IEEE Military Communications
  Conference (MILCOM'94)}, Fort Monmouth, NJ, Oct. 1994, pp. 997--1001.

\bibitem{Gupta_2007:SIC_constellation_structure}
A.~S. Gupta and A.~C. Singer, ``Successive interference cancellation using
  constellation structure,'' \emph{{IEEE} Transactions on Signal Processing},
  vol.~55, no.~12, pp. 5716--5730, Dec. 2007.

\bibitem{Andrews:power_control_SIC}
J.~G. Andrews and T.~H. Meng, ``Optimum power control for successive
  interference cancellation with imperfect channel estimation,'' \emph{{IEEE}
  Transactions on Wireless Communications}, vol.~2, no.~2, pp. 375--383, Mar.
  2003.

\bibitem{Yoon_1992:PIC_CDMA}
Y.~C. Yoon, R.~Kohno, and H.~Imai, ``A spread-spectrum multi-access system with
  a cascade of co-channel interference cancelers for multipath fading
  channels,'' in \emph{Proc. 2nd IEEE International Symposium on Spread
  Spectrum Techniques and Applications (ISSTA'92)}, Yokohama, Japan, Nov. 1992,
  pp. 87--90.

\bibitem{Buehrer_1999:linear_versus_nonlinear_IC}
R.~M. Buehrer, S.~P. Nicoloso, and S.~Gollamudi, ``Linear versus nonlinear
  interference cancellation,'' \emph{Journal of Communication and Networks},
  vol.~1, no.~2, pp. 118--133, Jun. 1999.

\bibitem{Guo_2000:PIC_CDMA}
D.~Guo, L.~K. Rasmussen, S.~Sun, and T.~J. Lim, ``A matrix-algebraic approach
  to linear parallel interference cancellation in {CDMA},'' \emph{{IEEE}
  Transactions on Communications}, vol.~48, no.~1, pp. 152--161, Jan. 2000.

\bibitem{Varansi_1988:multistage_detector_asynchronous_CDMA}
M.~K. Varanasi and B.~Aazhang, ``An iterative detector for asynchronous
  spread-spectrum multiple-access systems,'' in \emph{Proc. IEEE Global
  Telecommunications Conference (GLOBECOM'88)}, Nov. 1988, pp. 556--560.

\bibitem{Varanasi_1988:multistage_MUD_synchronous_CDMA_conf}
------, ``Probability of error comparison of linear and iterative multiuser
  detectors,'' in \emph{Proc. International Conference on Advances in
  Communications and Control Systems}, Baton Rouge, LA, Oct. 1988, pp. 54--65.

\bibitem{Hallen_1991:DFD_conf}
A.~Duel-Hallen, ``Linear and decision-feedback multiuser detectors,'' in
  \emph{Proc. IEEE International Symposium on Information Theory (ISIT'91)},
  Budapest, Hungary, Jun. 1991, pp. 24--28.

\bibitem{Hallen_1993:DFE_asynchronous_CDMA}
------, ``Performance of multiuser zero-forcing and {MMSE} decision-feedback
  detectors for {CDMA} channels,'' in \emph{Proc. IEEE Global
  Telecommunications Conference (GLOBECOM'93)}, Houston, TX, USA, Nov. 1993,
  pp. 82--86.

\bibitem{Guo_2003:VLSI_VBLAST_detector}
Z.~Guo and P.~Nilsson, ``A {VLSI} implementation of {MIMO} detection for future
  wireless communications,'' in \emph{Proc. IEEE Proceedings on Personal,
  Indoor and Mobile Radio Communications (PIMRC'03)}, Beijing, China, Sep.
  2003, pp. 2852--2856.

\bibitem{Hallen_1989:delayed_DFSE}
A.~Duel-Hallen and C.~Heegard, ``Delayed decision-feedback sequence
  estimation,'' \emph{{IEEE} Transactions on Communications}, vol.~37, no.~5,
  pp. 428--436, May 1989.

\bibitem{Hallen_1992:Equalizers_for_MIMO_PAM}
A.~Duel-Hallen, ``Equalizers for multiple input/multiple output channels and
  {PAM} systems with cyclostationary input sequences,'' \emph{{IEEE} Journal on
  Selected Areas in Communications}, vol.~10, no.~3, pp. 630--639, Apr. 1992.

\bibitem{Verdu_1989:recent_progress_in_MUD}
\BIBentryALTinterwordspacing
S.~Verd{\'u}, ``Recent progress in multiuser detection,'' in \emph{Advances in
  Communications and Signal Processing}, ser. Lecture Notes in Control and
  Information Sciences, W.~A. Porter and S.~C. Kak, Eds.\hskip 1em plus 0.5em
  minus 0.4em\relax Heidelberg: Springer-Verlag, 1989, vol. 129, pp. 27--38.
  [Online]. Available: \url{http://dx.doi.org/10.1007/BFb0042716}
\BIBentrySTDinterwordspacing

\bibitem{Xie_1988:sequential_decoding_MUD_async_CDMA}
Z.~Xie and C.~K. Rushforth, ``Multi-user signal detection using sequential
  decoding,'' in \emph{Proc. IEEE Military Communications Conference
  (MILCOM'88)}, San Diego, CA, USA, Oct. 1988, pp. 983--988.

\bibitem{Xie_1990:tree_search_MUD_conf}
Z.~Xie, C.~K. Rushforth, R.~T. Short, and T.~K. Moon, ``A tree-search algorithm
  for signal detection and parameter estimation in multi-user communications,''
  in \emph{Proc. IEEE Military Communications Conference (MILCOM'90)},
  Monterey, CA, USA, Oct. 1990, pp. 796--800.

\bibitem{Wei_1997:tree-search_MUD_CDMA}
L.~Wei, L.~K. Rasmussen, and R.~Wyrwas, ``Near optimum tree-search detection
  schemes for bit-synchronous multiuser {CDMA} systems over gaussian and
  two-path rayleigh-fading channels,'' \emph{{IEEE} Transactions on
  Communications}, vol.~45, no.~6, pp. 691--700, Jun. 1997.

\bibitem{Anderson_1969:M-algorithm_MSc_thesis}
F.~Jelinek and J.~B. Anderson, \emph{Instrumentable tree encoding of
  information sources}.\hskip 1em plus 0.5em minus 0.4em\relax School of
  Electrical Engineering, Cornell University, Ithaca, NY, USA, Sep. 1969.

\bibitem{Anderson_1971:M-algorithm}
------, ``Instrumentable tree encoding of information sources,'' \emph{{IEEE}
  Transactions on Information Theory}, vol.~17, no.~1, pp. 118--119, Jan. 1971.

\bibitem{Waltmann_1965:T-algorithm}
\BIBentryALTinterwordspacing
W.~L. Waltmann and R.~J. Lambert, ``{T}-algorithm for tridiagonalization,''
  \emph{Journal of the Society for Industrial and Applied Mathematics},
  vol.~13, no.~4, pp. 1069--1078, Dec. 1965. [Online]. Available:
  \url{http://www.jstor.org/stable/2946426}
\BIBentrySTDinterwordspacing

\bibitem{Simmons_1986:T-algorithm_trellis_decoder}
\BIBentryALTinterwordspacing
S.~J. Simmons, ``A reduced-computation trellis decoder with inherent
  parallelism,'' Ph.D. dissertation, Department of Electrical and Computer
  Engineering, Queen’s University, Kingston, Ontario, Canada, Jun. 1986.
  [Online]. Available:
  \url{http://istec.colostate.edu/~rockey/Papers/PhDThesis.pdf}
\BIBentrySTDinterwordspacing

\bibitem{Simmons_1990:T-algorithm_conf}
S.~J. Simmons and P.~Senyshyn, ``Reduced-search trellis decoding of coded
  modulations over {ISI} channels,'' in \emph{Proc. IEEE Global
  Telecommunications Conference (GLOBECOM'90)}, Dec. 1990, pp. 393--396.

\bibitem{Simmons_1990:T-algorithm_trellis_decoder}
S.~J. Simmons, ``Breadth-first trellis decoding with adaptive effort,''
  \emph{{IEEE} Transactions on Communications}, vol.~38, no.~1, pp. 3--12, Jan.
  1990.

\bibitem{Eyuboglu_1988:reduced_state_SE_DF}
M.~V. Eyubo{\v g}lu and S.~U. Qureshi, ``Reduced-state sequence estimation with
  set partitioning and decision feedback,'' \emph{{IEEE} Transactions on
  Communications}, vol.~36, no.~1, pp. 13--20, Jan. 1988.

\bibitem{Fano_1963:fano_algorithm_sequential_decoding}
R.~Fano, ``A heuristic discussion of probabilistic decoding,'' \emph{{IEEE}
  Transactions on Information Theory}, vol.~9, no.~2, pp. 64--74, Apr. 1963.

\bibitem{Zigangirov_1966:stack_algorithm_sequential_decoding}
K.~S. Zigangirov, ``Some sequential decoding procedures,'' \emph{Problemy
  Peredachi Informatsii}, vol.~2, no.~4, pp. 13--25, Oct. 1966.

\bibitem{Jelinek_1969:stack_algorithm_sequential_decoding}
F.~Jelinek, ``Fast sequential decoding algorithm using a stack,'' \emph{IBM
  Journal of Research and Development}, vol.~13, no.~6, pp. 675--685, Nov.
  1969.

\bibitem{Massey_1972:variable_length_codes_Fano_metric}
J.~L. Massey, ``Variable-length codes and the fano metric,'' \emph{{IEEE}
  Transactions on Information Theory}, vol.~18, no.~1, pp. 196--198, Jan. 1972.

\bibitem{Anderson_1984:sequential_decoding_survey}
J.~B. Anderson and S.~Mohan, ``Sequential coding algorithms: a survey and cost
  analysis,'' \emph{{IEEE} Transactions on Communications}, vol.~32, no.~2, pp.
  169--176, Feb. 1984.

\bibitem{Anderson_1989:sequential_decoding_survey}
J.~B. Anderson, ``Limited search trellis decoding of convolutional codes,''
  \emph{{IEEE} Transactions on Information Theory}, vol.~35, no.~5, pp.
  944--955, Sep. 1989.

\bibitem{Pottie_1989:sequential_decoding_survey}
G.~J. Pottie and D.~P. Taylor, ``A comparison of reduced complexity decoding
  algorithms for trellis codes,'' \emph{{IEEE} Journal on Selected Areas in
  Communications}, vol.~7, no.~9, pp. 1369--1380, Dec. 1989.

\bibitem{Pohst_1981:computation_lattice_vectors_minimum}
M.~Pohst, ``On the computation of lattice vectors of minimal length, successive
  minima and reduced bases with applications,'' \emph{ACM SIGSAM Bulletin},
  vol.~15, no.~1, pp. 37--44, Feb. 1981.

\bibitem{Fincke_1985:improved_lattice_vectors_calculation}
U.~Fincke and M.~Pohst, ``Improved methods for calculating vectors of short
  length in a lattice, including a complexity analysis,'' \emph{Mathematics of
  Computation}, vol.~44, no. 170, pp. 463--471, Apr. 1985.

\bibitem{Schnorr_1994:lattice_basis_reduction}
C.~P. Schnorr and M.~Euchner, ``Lattice basis reduction: improved practical
  algorithms and solving subset sum problems,'' \emph{Mathematical
  Programming}, vol.~66, no. 1-3, pp. 181--199, Aug. 1994.

\bibitem{Shen_2012:best_first_tree_search_VLSI}
C.-A. Shen, A.~M. Eltawil, K.~N. Salama, and S.~Mondal, ``A best-first
  soft/hard decision tree searching {MIMO} decoder for a 4 $\times$ 4 64-{QAM}
  system,'' \emph{{IEEE} Transactions on Very Large Scale Integration ({VLSI})
  Systems}, vol.~20, no.~8, pp. 1537--1541, Aug. 2012.

\bibitem{Hassibi_2001:SD_complexity_first}
B.~Hassibi and H.~Vikalo, ``On the expected complexity of sphere decoding,'' in
  \emph{Proc. 35th Annual Asilomar Conference on Signals, Systems and Computers
  (Asilomar'01)}, Pacific Grove, CA, USA, Nov. 2001, pp. 1051--1055.

\bibitem{Hassibi_2002:SD_complexity}
------, ``On the expected complexity of integer least-squares problems,'' in
  \emph{Proc. IEEE IEEE International Conference on Acoustics, Speech, and
  Signal Processing (ICASSP'02)}, vol.~2, Orlando, FL, USA, May 2002, pp.
  II--1497--II--1500.

\bibitem{Hassibi_2003:ML_ILS_complexity}
------, ``Maximum-likelihood decoding and integer least-squares: The expected
  complexity,'' in \emph{DIMACS Series in Discrete Mathematics and Theoretical
  Computer Science: Multiantenna channels: capacity, coding and signal
  processing}, G.~J. Foschini and S.~Verd{\'u}, Eds.\hskip 1em plus 0.5em minus
  0.4em\relax New York: American Mathematical Society, 2003, vol.~62, pp.
  161--192.

\bibitem{Jalden:SD_complexity_conf}
J.~Jald\'{e}n and B.~Ottersten, ``An exponential lower bound on the expected
  complexity of sphere decoding,'' in \emph{Proc. IEEE International Conference
  on Acoustics, Speech, and Signal Processing (ICASSP'04)}, vol.~4, Montreal,
  Canada, May 2004, pp. 393--396.

\bibitem{Hochwald_2001:IDD_LSD_near_capacity_conf}
B.~M. Hochwald and S.~ten Brink, ``Iterative list sphere decoding to attain
  capacity on a multi-antenna link,'' in \emph{Proc. of 39th Annual Allerton
  Conference On Communication, Control and Computing (Allerton'01)},
  Monticello, IL, USA, Oct. 2001, pp. 815--824.

\bibitem{Garrett_2004:VLSI_ML_SD_silicon_complexity}
D.~C. Garrett, L.~M. Davis, S.~ten Brink, B.~Hochwald, and G.~Knagge, ``Silicon
  complexity for maximum likelihood {MIMO} detection using spherical
  decoding,'' \emph{IEEE Journal of Solid-State Circuits}, vol.~39, no.~9, pp.
  1544--1552, Sep. 2004.

\bibitem{Barbero_2006:fixed_complexity_SD}
L.~G. Barbero and J.~S. Thompson, ``A fixed-complexity {MIMO} detector based on
  the complex sphere decoder,'' in \emph{Proc. IEEE 7th Workshop on Signal
  Processing Advances in Wireless Communications (SPAWC'06)}, Cannes, France,
  Jul. 2006, pp. 1--5.

\bibitem{Guo_2004:K_best_implementation}
Z.~Guo and P.~Nilsson, ``A {VLSI} architecture of the {Schnorr-Euchner} decoder
  for {MIMO} systems,'' in \emph{Proc. IEEE 6th Circuits and Systems Symposium
  on Emerging Technologies: Frontiers of Mobile and Wireless Communication},
  Shanghai, China, May 2004, pp. 65--68.

\bibitem{Wenk_2006:K_best_SD_VLSI}
M.~Wenk, M.~Zellweger, A.~Burg, N.~Felber, and W.~Fichtner, ``{K-best MIMO}
  detection {VLSI} architectures achieving up to 424 {Mbps},'' in \emph{Proc.
  IEEE International Symposium on Circuits and Systems (ISCAS'06)}, May 2006,
  pp. 1151--1154.

\bibitem{Patel_2010:K_best_LTE_WiMAX_soft}
D.~Patel, V.~Smolyakov, M.~Shabany, and P.~G. Gulak, ``{VLSI} implementation of
  a {WiMAX/LTE} compliant low-complexity high-throughput soft-output {K}-best
  {MIMO} detector,'' in \emph{Proc. IEEE International Symposium on Circuits
  and Systems (ISCAS'10)}, Paris, France, May 2010, pp. 593--596.

\bibitem{Dijkstra_1959:Dijkstra_algorithm}
\BIBentryALTinterwordspacing
E.~W. Dijkstra, ``A note on two problems in connexion with graphs,''
  \emph{Numerische Mathematik}, vol.~1, no.~1, pp. 269--271, 1959. [Online].
  Available: \url{http://dx.doi.org/10.1007/BF01386390}
\BIBentrySTDinterwordspacing

\bibitem{Xu_2004:best_first_stack_algorithm}
W.~Xu, Y.~Wang, Z.~Zhou, and J.~Wang, ``A computationally efficient exact {ML}
  sphere decoder,'' in \emph{Proc. IEEE Global Telecommunications Conference
  (GLOBECOM'04)}, Dallas, TX, USA, Dec. 2004, pp. 2594--2598.

\bibitem{Mow_1994:lattice_MLSE}
W.~H. Mow, ``Maximum likelihood sequence estimation from the lattice
  viewpoint,'' \emph{{IEEE} Transactions on Information Theory}, vol.~40,
  no.~5, pp. 1591--1600, Sep. 1994.

\bibitem{Wubben_2011:LR_magazine}
D.~W\"{u}bben, D.~Seethaler, J.~Jald\'{e}n, and G.~Matz, ``Lattice reduction,''
  \emph{{IEEE} Signal Processing Magazine}, vol.~28, no.~3, pp. 70--91, May
  2011.

\bibitem{Daude_1994:Gaussian_reduction}
H.~Daud{\'e}, P.~Flajolet, and B.~Vall{\'e}e, ``An analysis of the gaussian
  algorithm for lattice reduction,'' in \emph{Proc. of the 1st International
  Symposium on Algorithmic Number Theory}, Ithaca, NY, 1994, pp. 144--158.

\bibitem{KZ_1873:LR_reduction}
\BIBentryALTinterwordspacing
A.~Korkine and G.~Zolotareff, ``\BIBforeignlanguage{French}{Sur les formes
  quadratiques},'' \emph{\BIBforeignlanguage{French}{Mathematische Annalen}},
  vol.~6, no.~3, pp. 366--389, 1873. [Online]. Available:
  \url{http://dx.doi.org/10.1007/BF01442795}
\BIBentrySTDinterwordspacing

\bibitem{Yao_2003:LR_MIMO_detector}
H.~Yao, ``Efficient signal, code, and receiver designs for {MIMO} communication
  systems,'' Ph.D. dissertation, 2003.

\bibitem{Zhang_2009:wireless_MIMO_receiver_design}
W.~Z. Zhang, ``Wireless receiver designs: From information theory to {VLSI}
  implementation,'' Ph.D. dissertation, 2009.

\bibitem{SE_1994:LR_reduction}
C.-P. Schnorr and M.~Euchner, ``Lattice basis reduction: Improved practical
  algorithms and solving subset sum problems,'' \emph{Mathematical
  programming}, vol.~66, no. 1-3, pp. 181--199, 1994.

\bibitem{LLL_1982:LLL_LR_algorithm}
\BIBentryALTinterwordspacing
A.~K. Lenstra, J.~Lenstra, H.~W., and L.~Lov\'{a}sz, ``Factoring polynomials
  with rational coefficients,'' \emph{Mathematische Annalen}, vol. 261, no.~4,
  pp. 515--534, 1982. [Online]. Available:
  \url{http://dx.doi.org/10.1007/BF01457454}
\BIBentrySTDinterwordspacing

\bibitem{Seysen_1993:SA_LLR_algorithm}
\BIBentryALTinterwordspacing
M.~Seysen, ``Simultaneous reduction of a lattice basis and its reciprocal
  basis,'' \emph{Combinatorica}, vol.~13, no.~3, pp. 363--376, 1993. [Online].
  Available: \url{http://dx.doi.org/10.1007/BF01202355}
\BIBentrySTDinterwordspacing

\bibitem{Seethaler_2006:Brun_LR_algorithm}
D.~Seethaler and G.~Matz, ``Efficient vector perturbation in multi-antenna
  multi-user systems based on approximate integer relations,'' in \emph{Proc.
  14th European Signal Processing Conference (EUSIPCO'06)}, Florence, Italy,
  Sep. 2006.

\bibitem{Clarkson_1997:Brun_LR_algorithm}
I.~V.~L. Clarkson, ``Approximation of linear forms by lattice points with
  applications to signal processing,'' Ph.D. dissertation, 1997.

\bibitem{Ling_2006:approximate_lattice_decoding}
C.~Ling, ``Approximate lattice decoding: Primal versus dual basis reduction,''
  in \emph{Proc. IEEE International Symposium on Information Theory (ISIT'06)},
  Jul. 2006, pp. 1--5.

\bibitem{Napias_1996:generalization_CLLL}
H.~Napias, ``A generalization of the {LLL}-algorithm over euclidean rings or
  orders,'' \emph{Journal de th{\'e}orie des nombres de Bordeaux}, vol.~8, pp.
  387--396, 1996.

\bibitem{Wubben_2004:LR_reduction_MMSE_WSA}
D.~W\"{u}bben, R.~B\"{o}hnke, V.~K\"{u}hn, and K.-D. Kammeyer, ``{MMSE}-based
  lattice-reduction for near-{ML} detection of {MIMO} systems,'' in \emph{Proc.
  ITG Workshop on Smart Antennas (WSA'04)}, Mar. 2004, pp. 106--113.

\bibitem{Shalom_1975:invention_PDA}
Y.~Bar-Shalom and E.~Tse, ``Tracking in a cluttered environment with
  probabilistic data association,'' \emph{Automatica}, vol.~11, no.~5, pp. 451
  -- 460, 1975.

\bibitem{Fortmann_1980:JPDA_invention_conf}
T.~E. Fortmann, Y.~Bar-Shalom, and M.~Scheffe, ``Multi-target tracking using
  joint probabilistic data association,'' in \emph{19th IEEE Conference on
  Decision and Control including the Symposium on Adaptive Processes},
  Albuquerque, NM, USA, Dec. 1980, pp. 807--812.

\bibitem{Fortmann_1983:JPDA_sonar_tracking_journal}
------, ``Sonar tracking of multiple targets using joint probabilistic data
  association,'' \emph{{IEEE} Journal of Oceanic Engineering}, vol.~8, no.~3,
  pp. 173--184, Jul. 1983.

\bibitem{Shalom:PDA_original}
Y.~Bar-Shalom and X.~R. Li, \emph{Estimation and Tracking: Principles,
  Techniques and Software}.\hskip 1em plus 0.5em minus 0.4em\relax Dedham, MA:
  Artech House, 1993.

\bibitem{Shalom_2004:estimation_applications_to_tracking_navigation}
Y.~Bar-Shalom, X.~R. Li, and T.~Kirubarajan, \emph{Estimation with Applications
  to Tracking and Navigation: Theory, Algorithms and Software}.\hskip 1em plus
  0.5em minus 0.4em\relax John Wiley \& Sons, Inc., 2004.

\bibitem{Shalom_2005:PDA_target_tracking_sonar_radar}
Y.~Bar-Shalom, T.~Kirubarajan, and X.~Lin, ``Probabilistic data association
  techniques for target tracking with applications to sonar, radar and {EO}
  sensors,'' \emph{{IEEE} Aerospace and Electronics Systems Magazine}, vol.~20,
  no.~8, pp. 37--56, 2005.

\bibitem{Shalom2009:PDA_filter}
Y.~Bar-Shalom, F.~Daum, and J.~Huang, ``The probabilistic data association
  filter,'' \emph{{IEEE} Control Systems Magazine}, vol.~29, no.~6, pp.
  82--100, Dec. 2009.

\bibitem{Chang_1984:JPDA_multi_target_tracking}
K.-C. Chang and Y.~Bar-Shalom, ``Joint probabilistic data association for
  multitarget tracking with possibly unresolved measurements and maneuvers,''
  \emph{{IEEE} Transactions on Automatic Control}, vol.~29, no.~7, pp.
  585--594, Jul. 1984.

\bibitem{Chang_1986:JPDA_distributed_sensor_networks}
K.-C. Chang, C.-Y. Chong, and Y.~Bar-Shalom, ``Joint probabilistic data
  association in distributed sensor networks,'' \emph{{IEEE} Transactions on
  Automatic Control}, vol.~31, no.~10, pp. 889--897, Oct. 1986.

\bibitem{Roecker_1993:suboptimal_JPDA}
J.~A. Roecker and G.~L. Phillis, ``Suboptimal joint probabilistic data
  association,'' \emph{{IEEE} Transactions on Aerospace and Electronic
  Systems}, vol.~29, no.~2, pp. 510--517, Apr. 1993.

\bibitem{Musicki_1994:IPDA}
D.~Musicki, R.~J. Evans, and S.~Stankovic, ``Integrated probabilistic data
  association,'' \emph{{IEEE} Transactions on Automatic Control}, vol.~39,
  no.~6, pp. 1237--1241, Jun. 1994.

\bibitem{Kershaw_1997:WSPDA}
D.~J. Kershaw and R.~J. Evans, ``Waveform selective probabilistic data
  association,'' \emph{{IEEE} Transactions on Aerospace and Electronic
  Systems}, vol.~33, no.~4, pp. 1180--1188, Oct. 1997.

\bibitem{Blom_2000:PDA_avoid_track_coalescence}
H.~A.~P. Blom and E.~A. Bloem, ``Probabilistic data association avoiding track
  coalescence,'' \emph{{IEEE} Transactions on Automatic Control}, vol.~45,
  no.~2, pp. 247--259, Feb. 2000.

\bibitem{Kirubarajan_2004:PDA_tracking_clutter}
T.~Kirubarajan and Y.~Bar-Shalom, ``Probabilistic data association techniques
  for target tracking in clutter,'' \emph{Proceedings of the {IEEE}}, vol.~92,
  no.~3, pp. 536--557, Mar. 2004.

\bibitem{Musicki_2004:JIPDA}
D.~Musicki and R.~J. Evans, ``Joint integrated probabilistic data association:
  {JIPDA},'' \emph{{IEEE} Transactions on Aerospace and Electronic Systems},
  vol.~40, no.~3, pp. 1093--1099, Jul. 2004.

\bibitem{Cox_1993:review_statistical_data_association}
\BIBentryALTinterwordspacing
I.~J. Cox, ``A review of statistical data association techniques for motion
  correspondence,'' \emph{International Journal of Computer Vision}, vol.~10,
  no.~1, pp. 53--66, Feb. 1993. [Online]. Available:
  \url{http://dx.doi.org/10.1007/BF01440847}
\BIBentrySTDinterwordspacing

\bibitem{Schulz_2001:tracking_robots}
D.~Schulz, W.~Burgard, D.~Fox, and A.~B. Cremers, ``Tracking multiple moving
  targets with a mobile robot using particle filters and statistical data
  association,'' in \emph{Proc. IEEE International Conference on Robotics and
  Automation (ICRA'01)}, Seoul, Korea, May 2001, pp. 1665--1670.

\bibitem{Schulz_2003:people_tracking_robots}
------, ``People tracking with mobile robots using sample-based joint
  probabilistic data association filters,'' \emph{International Journal of
  Robotics Research}, vol.~22, no.~2, pp. 99--116, Feb. 2003.

\bibitem{Rasmussen_2001:PDA_in_computer_vision}
C.~Rasmussen and G.~D. Hager, ``Probabilistic data association methods for
  tracking complex visual objects,'' \emph{{IEEE} Transactions on Pattern
  Analysis and Machine Intelligence}, vol.~23, no.~6, pp. 560--576, Jun. 2001.

\bibitem{Wang_2003:PDA_CE}
Z.~J. Wang, Z.~Han, and K.~J.~R. Liu, ``{MIMO-OFDM} channel estimation via
  probabilistic data association based {TOA}s,'' in \emph{Proc. IEEE Global
  Telecommunications Conference (GLOBECOM'03)}, San Francisco, CA, USA, Dec.
  2003, pp. 626--630.

\bibitem{Wang_2005:PDA_CE_journal}
------, ``A {MIMO-OFDM} channel estimation approach using time of arrivals,''
  \emph{{IEEE} Transactions on Wireless Communications}, vol.~4, no.~3, pp.
  1207--1213, May 2005.

\bibitem{Shaoshi_2011:BPDA_conf}
S.~Yang, T.~Lv, and L.~Hanzo, ``Unified bit-based probabilistic data
  association aided {MIMO} detection for high-order {QAM},'' in \emph{Proc.
  IEEE Wireless Communications and Networking Conference (WCNC'11)}, Cancun,
  Mexico, Mar. 2011, pp. 1629--1634.

\bibitem{Shaoshi2012gc:Turbo_AB_Log_PDA}
S.~Yang and L.~Hanzo, ``Iterative detection and decoding using approximate
  {B}ayesian theorem based {PDA} method over {MIMO} {N}akagami-$m$ fading
  channels,'' in \emph{Proc. IEEE Global Communications Conference
  (GLOBECOM'12)}, Anaheim, CA, USA, Dec. 2012, pp. 3588--3593.

\bibitem{Shaoshi2013gc:Turbo_EB_Log_PDA}
------, ``Exact {B}ayes' theorem based probabilistic data association for
  iterative {MIMO} detection and decoding,'' in \emph{Proc. IEEE Global
  Communications Conference (GLOBECOM'13)}, Atlanta, GA, USA, Dec. 2013.

\bibitem{Berrou:Turbo_coding_conference}
C.~Berrou, A.~Glavieux, and P.~Thitimajshima, ``Near {Shannon} limit
  error-correcting coding and decoding: Turbo-codes (1),'' in \emph{Proc. IEEE
  International Conference on Communications (ICC'93)}, vol.~2, Geneva,
  Switzerland, May 1993, pp. 1064--1070.

\bibitem{Berrou:Turbo_coding_journal}
C.~Berrou and A.~Glavieux, ``Near optimum error correcting coding and decoding:
  turbo-codes,'' \emph{{IEEE} Transactions on Communications}, vol.~44, no.~10,
  pp. 1261--1271, Oct. 1996.

\bibitem{Gallager:LDPC_code}
R.~Gallager, ``Low-density parity-check codes,'' \emph{IRE Transactions on
  Information Theory}, vol.~8, no.~1, pp. 21--28, Jan. 1962.

\bibitem{MacKay:LDPC_code}
D.~J.~C. MacKay and R.~M. Neal, ``Near {Shannon} limit performance of low
  density parity check codes,'' \emph{Electronics Letters}, vol.~32, no.~18,
  pp. 1645--1646, 1996.

\bibitem{Vandenberg:semidefinite_programming}
L.~Vandenberghe and S.~Boyd, ``Semidefinite programming,'' \emph{SIAM Review},
  vol.~38, no.~1, pp. 49--95, Mar. 1996.

\bibitem{Boyd:Convex_Optimization}
S.~Boyd and L.~Vandenberghe, \emph{Convex Optimization}.\hskip 1em plus 0.5em
  minus 0.4em\relax New York, USA: Cambridge University Press, 2004.

\bibitem{Helmberg:interior_point_algorithm}
C.~Helmberg, F.~Rendl, R.~J. Vanderbei, and H.~Wolkowicz, ``An interior-point
  method for semidefinite programming,'' \emph{SIAM Journal on Optimization},
  vol.~6, pp. 342--361, 1996.

\bibitem{Penghui_2001:SDP_CDMA_conf}
P.~H. Tan, L.~K. Rasmussen, and T.~M. Aulin, ``The application of semidefinite
  programming for detection in {CDMA},'' in \emph{Proc. IEEE International
  Symposium on Information Theory (ISIT'01)}, Washington, DC, Jun. 2001, p.~9.

\bibitem{Ma_2001:SDP_CDMA_conf}
W.-K. Ma, T.~N. Davidson, K.~M. Wong, Z.-Q. Luo, and P.-C. Ching, ``Efficient
  quasi-maximum-likelihood multiuser detection by semi-definite relaxation,''
  in \emph{Proc. IEEE International Conference on Communications (ICC'01)},
  Helsinki, Finland, Jun. 2001, pp. 6--10.

\bibitem{Wang_2001:SDP_CDMA_conf}
X.~Wang, W.-S. Lu, and A.~Antoniou, ``A near-optimal multiuser detector for
  {CDMA} channels using semidefinite programming relaxation,'' in \emph{Proc.
  IEEE International Symposium on Circuits and Systems (ISCAS'01)}, Sydney,
  NSW, Australia, May 2001, pp. 298--301.

\bibitem{Wang_2003:SDP_CDMA_journal}
------, ``A near-optimal multiuser detector for {DS-CDMA} systems using
  semidefinite programming relaxation,'' \emph{{IEEE} Transactions on Signal
  Processing}, vol.~51, no.~9, pp. 2446--2450, Sep. 2003.

\bibitem{luo:soft_SDR}
B.~Steingrimsson, Z.-Q. Luo, and K.~M. Wong, ``Soft quasi-maximum-likelihood
  detection for multiple-antenna wireless channels,'' \emph{{IEEE} Transactions
  on Signal Processing}, vol.~51, no.~11, pp. 2710--2719, Nov. 2003.

\bibitem{Foschini_1999:antenna_correlation_MIMO_capacity}
P.~F. Driessen and G.~J. Foschini, ``On the capacity formula for multiple
  input-multiple output wireless channels: a geometric interpretation,''
  \emph{IEEE Transactions on Communications}, vol.~47, no.~2, pp. 173--176,
  Feb. 1999.

\bibitem{Shiu_2000:fading_correlation_MIMO}
D.-S. Shiu, G.~J. Foschini, M.~J. Gans, and J.~M. Kahn, ``Fading correlation
  and its effect on the capacity of multielement antenna systems,'' \emph{IEEE
  Transactions on Communications}, vol.~48, no.~3, pp. 502--513, Mar. 2000.

\bibitem{Chizhik_2000:antenna_correlation_MIMO}
D.~Chizhik, F.~Rashid-Farrokhi, J.~Ling, and A.~Lozano, ``Effect of antenna
  separation on the capacity of {BLAST} in correlated channels,'' \emph{IEEE
  Communications Letters}, vol.~4, no.~11, pp. 337--339, Nov. 2000.

\bibitem{Tse_2002:MIMO_capacity_correlation}
C.-N. Chuah, D.~N.~C. Tse, J.~M. Kahn, and R.~A. Valenzuela, ``Capacity scaling
  in {MIMO} wireless systems under correlated fading,'' \emph{IEEE Transactions
  on Information Theory}, vol.~48, no.~3, pp. 637--650, Mar. 2002.

\bibitem{Molisch_2002:correlated_MIMO_capacity_measured_channel}
A.~F. Molisch, M.~Steinbauer, M.~Toeltsch, E.~Bonek, and R.~S. Thoma,
  ``Capacity of {MIMO} systems based on measured wireless channels,''
  \emph{IEEE Journal on Selected Areas in Communications}, vol.~20, no.~3, pp.
  561--569, Apr. 2002.

\bibitem{Shin_2003:keyhole_correlation_double_scattering_IT}
H.~Shin and J.~H. Lee, ``Capacity of multiple-antenna fading channels: spatial
  fading correlation, double scattering, and keyhole,'' \emph{IEEE Transactions
  on Information Theory,}, vol.~49, no.~10, pp. 2636--2647, Oct. 2003.

\bibitem{Chizhik_2000:keyhole_letter}
D.~Chizhik, G.~J. Foschini, and R.~A. Valenzuela, ``Capacities of multi-element
  transmit and receive antennas: correlations and keyholes,'' \emph{Electronics
  Letters}, vol.~36, no.~13, pp. 1099--1100, Jun. 2000.

\bibitem{Chizhik_2002:keyhole_effect}
D.~Chizhik, G.~J. Foschini, M.~J. Gans, and R.~A. Valenzuela, ``Keyholes,
  correlations, and capacities of multielement transmit and receive antennas,''
  \emph{IEEE Transactions on Wireless Communications}, vol.~1, no.~2, pp.
  361--368, Apr. 2002.

\bibitem{Gesbert_2000:keyhole}
D.~Gesbert, H.~Bolcskei, D.~Gore, and A.~Paulraj, ``{MIMO} wireless channels:
  capacity and performance prediction,'' in \emph{Proc. IEEE Global
  Telecommunications Conference (GLOBECOM'00)}, San Francisco, CA, US, Nov.
  2000, pp. 1083--1088.

\bibitem{Molisch_2003:keyhole_measurement_letter}
P.~Almers, F.~Tufvesson, and A.~F. Molisch, ``Measurement of keyhole effect in
  a wireless multiple-input multiple-output {MIMO} channel,'' \emph{IEEE
  Communications Letters}, vol.~7, no.~8, pp. 373--375, Aug. 2003.

\bibitem{Molisch_2006:keyhole_MIMO_capacity_measurement}
------, ``Keyhole effect in {MIMO} wireless channels: measurements and
  theory,'' \emph{IEEE Transactions on Wireless Communications}, vol.~5,
  no.~12, pp. 3596--3604, Dec. 2006.

\bibitem{Varanasi_1995:group_detection}
M.~K. Varanasi, ``Group detection for synchronous {G}aussian code-division
  multiple-access channels,'' \emph{IEEE Transactions on Information Theory},
  vol.~41, no.~4, pp. 1083--1096, Jul. 1995.

\bibitem{Schlegel_1996:MUD_projection}
C.~Schlegel, S.~Roy, P.~D. Alexander, and Z.-J. Xiang, ``Multiuser projection
  receivers,'' \emph{IEEE Journal on Selected Areas in Communications,},
  vol.~14, no.~8, pp. 1610--1618, Oct. 1996.

\bibitem{Varanasi_2000:diversity_order_group_detection}
E.~A. Fain and M.~K. Varanasi, ``Diversity order gain for narrow-band multiuser
  communications with pre-combining group detection,'' \emph{IEEE Transactions
  on Communications,}, vol.~48, no.~4, pp. 533--536, Apr. 2000.

\bibitem{Varanasi_2003:overloaded_MUD_CDMA}
A.~Kapur and M.~K. Varanasi, ``Multiuser detection for overloaded {CDMA}
  systems,'' \emph{IEEE Transactions on Information Theory}, vol.~49, no.~7,
  pp. 1728--1742, Jul. 2003.

\bibitem{Zarikoff_2007:group_detection_overloaded_MIMO}
B.~Zarikoff, J.~K. Cavers, and S.~Bavarian, ``An iterative groupwise multiuser
  detector for overloaded {MIMO} applications,'' \emph{IEEE Transactions on
  Wireless Communications}, vol.~6, no.~2, pp. 443--447, Feb. 2007.

\bibitem{Krause_2011:Group_detection_MIMO_overloaded}
M.~Krause, D.~P. Taylor, and P.~A. Martin, ``List-based group-wise symbol
  detection for multiple signal communications,'' \emph{IEEE Transactions on
  Wireless Communications,}, vol.~10, no.~5, pp. 1636--1644, May 2011.

\bibitem{Dayal_2003:afast_SD_rank_deficient}
P.~Dayal and M.~K. Varanasi, ``A fast generalized sphere decoder for optimum
  decoding of under-determined {MIMO} systems,'' in \emph{Proc. 41st Annual
  Allerton Conference On Communication, Control and Computing (Allerton'03)},
  Montecello, IL, US, Oct. 2003, pp. 1216--1225.

\bibitem{Cui_2005:SD_rank_deficient_letter}
T.~Cui and C.~Tellambura, ``An efficient generalized sphere decoder for
  rank-deficient {MIMO} systems,'' \emph{IEEE Communications Letters}, vol.~9,
  no.~5, pp. 423--425, May 2005.

\bibitem{Yang_2005:SD_rank_deficient}
Z.~Yang, C.~Liu, and J.~He, ``A new approach for fast generalized sphere
  decoding in {MIMO} systems,'' \emph{IEEE Signal Processing Letters}, vol.~12,
  no.~1, pp. 41--44, Jan. 2005.

\bibitem{Hanzo_2006:SD_rank_deficient}
A.~Wolfgang, J.~Akhtman, S.~Chen, and L.~Hanzo, ``Iterative {MIMO} detection
  for rank-deficient systems,'' \emph{IEEE Signal Processing Letters}, vol.~13,
  no.~11, pp. 699--702, Nov. 2006.

\bibitem{Paulraj_2007:SD_rank_deficient}
K.-K. Wong, A.~Paulraj, and R.~D. Murch, ``Efficient high-performance decoding
  for overloaded {MIMO} antenna systems,'' \emph{IEEE Transactions on Wireless
  Communications}, vol.~6, no.~5, pp. 1833--1843, May 2007.

\bibitem{Wang2008:center_shifting_SD}
L.~Wang, L.~Xu, S.~Chen, and L.~Hanzo, ``Generic iterative
  search-centre-shifting {K}-best sphere detection for rank-deficient
  {SDM-OFDM} systems,'' \emph{Electronics Letters}, vol.~44, no.~8, pp.
  552--553, 2008.

\bibitem{Tian_2009:overloaded_MIMO_detection}
Z.~Tian, G.~Leus, and V.~Lottici, ``Detection of sparse signals under
  finite-alphabet constraints,'' in \emph{Proc. IEEE International Conference
  on Acoustics, Speech and Signal Processing (ICASSP'2009)}, Taipei, Republic
  of China, Apr. 2009, pp. 2349--2352.

\bibitem{Kanaras_2010:SD_ill_conditioned}
I.~Kanaras, A.~Chorti, M.~R.~D. Rodrigues, and I.~Darwazeh, ``A fast
  constrained sphere decoder for ill conditioned communication systems,''
  \emph{IEEE Communications Letters}, vol.~14, no.~11, pp. 999--1001, Nov.
  2010.

\bibitem{Cui_2006:SDR_overloaded}
T.~Cui, T.~Ho, and C.~Tellambura, ``Polynomial moment relaxation for {MIMO}
  detection,'' in \emph{Proc. IEEE International Conference on Communications
  (ICC'06)}, Istanbul, Turkey, Jun. 2006, pp. 3129--3134.

\bibitem{Juntti_1997:genetic_algorithm_MUD_CDMA}
M.~J. Juntti, T.~Schlosser, and J.~O. Lilleberg, ``Genetic algorithms for
  multiuser detection in synchronous {CDMA},'' in \emph{Proc. IEEE
  International Symposium on Information Theory (ISIT'97)}, Ulm, Germany, Jul.
  1997, p. 492.

\bibitem{Ergun_1998:genetic_algorithm_MUD}
C.~Erg{\"u}n and K.~Hacioglu, ``Application of a genetic algorithm to
  multi-stage detection in {CDMA} systems,'' in \emph{Proc. 9th Mediterranean
  Electrotechnical Conference (MELECON'98)}, Tel-Aviv, Israel, May 1998, pp.
  846--850.

\bibitem{Ergun_2000:genetic_algorithm_MUD_journal}
------, ``Multiuser detection using a genetic algorithm in {CDMA}
  communications systems,'' \emph{{IEEE} Transactions on Communications},
  vol.~48, no.~8, pp. 1374--1383, Aug. 2000.

\bibitem{Wang_1998:genetic_algorithm_MUD}
X.~Wang, W.-S. Lu, and A.~Antoniou, ``A genetic-algorithm-based multiuser
  detector for multiple-access communications,'' in \emph{Proc. IEEE
  International Symposium on Circuits and Systems (ISCAS'98)}, Monterey, CA,
  USA, Jun. 1998, pp. 534--537.

\bibitem{Yen_2001:genetic_algorithm_joint_MUD_CE}
K.~Yen and L.~Hanzo, ``Genetic algorithm assisted joint multiuser symbol
  detection and fading channel estimation for synchronous {CDMA} systems,''
  \emph{{IEEE} Journal on Selected Areas in Communications}, vol.~19, no.~6,
  pp. 985--998, Jun. 2001.

\bibitem{Yen_2003:genetic_algorithm_MUD_synch}
------, ``Antenna-diversity-assisted genetic-algorithm-based multiuser
  detection schemes for synchronous {CDMA} systems,'' \emph{{IEEE} Transactions
  on Communications}, vol.~51, no.~3, pp. 366--370, Mar. 2003.

\bibitem{Yen_2004:genetic_algorithm_MUD_asynch}
------, ``Genetic-algorithm-assisted multiuser detection in asynchronous {CDMA}
  communications,'' \emph{{IEEE} Transactions on Vehicular Technology},
  vol.~53, no.~5, pp. 1413--1422, Sep. 2004.

\bibitem{Colman_2008:GA_overloaded}
G.~W.~K. Colman and T.~J. Willink, ``Overloaded array processing using genetic
  algorithms with soft-biased initialization,'' \emph{IEEE Transactions on
  Vehicular Technology}, vol.~57, no.~4, pp. 2123--2131, Jul. 2008.

\bibitem{Dorigo_2006:ACO}
M.~Dorigo, M.~Birattari, and T.~Stutzle, ``Ant colony optimization,''
  \emph{IEEE Computational Intelligence Magazine}, vol.~1, no.~4, pp. 28--39,
  Nov 2006.

\bibitem{Sun_2004:PSO}
J.~Sun, W.~Xu, and B.~Feng, ``A global search strategy of quantum-behaved
  particle swarm optimization,'' in \emph{Proc. IEEE Conference on Cybernetics
  and Intelligent Systems (CIS'04)}, vol.~1, Dec. 2004, pp. 111--116.

\bibitem{Hijazi_2004:ACO_MUD}
S.~L. Hijazi and B.~Natarajan, ``Novel low-complexity {DS-CDMA} multiuser
  detector based on ant colony optimization,'' in \emph{Proc. IEEE 60th
  Vehicular Technology Conference (VTC'04-Fall)}, vol.~3, Sep. 2004, pp.
  1939--1943.

\bibitem{Xu_2008:ACO_MUD}
C.~Xu, B.~Hu, L.-L. Yang, and L.~Hanzo, ``Ant-colony-based multiuser detection
  for multifunctional-antenna-array-assisted {MC DS-CDMA} systems,''
  \emph{{IEEE} Transactions on Vehicular Technology}, vol.~57, no.~1, pp.
  658--663, Jan. 2008.

\bibitem{Xu_2009:ACO_MUD_letter}
C.~Xu, R.~G. Maunder, L.-L. Yang, and L.~Hanzo, ``Near-optimum multiuser
  detectors using soft-output ant-colony-optimization for the {DS-CDMA}
  uplink,'' \emph{IEEE Signal Processing Letters}, vol.~16, no.~2, pp.
  137--140, Feb. 2009.

\bibitem{Lain_2010:ACO_overloaded}
J.-K. Lain and J.-Y. Chen, ``{Near-MLD MIMO} detection based on a modified ant
  colony optimization,'' \emph{IEEE Communications Letters}, vol.~14, no.~8,
  pp. 722--724, August 2010.

\bibitem{Tasneem_2012:overloaded_ACO_MIMO}
K.~T. Tasneem, P.~A. Martin, and D.~P. Taylor, ``Iterative soft detection of
  cochannel signals using ant colony optimization,'' in \emph{Proc. 23rd IEEE
  International Symposium on Personal, Indoor and Mobile Radio Communications
  (PIMRC'12)}, Sydney, Australia, Sep. 2012, pp. 1617--1621.

\bibitem{Haris_2012:heuristics_MIMO_overloaded}
P.~A. Haris, E.~Gopinathan, and C.~K. Ali, ``Artificial bee colony and tabu
  search enhanced {TTCM} assisted {MMSE} multi-user detectors for rank
  deficient {SDMA-OFDM} system,'' \emph{Wireless Personal Communications},
  vol.~65, no.~2, pp. 425--442, 2012.

\bibitem{Yang_2004:PSO}
S.~Yang, M.~Wang, and L.~Jiao, ``A quantum particle swarm optimization,'' in
  \emph{Proc. Congress on Evolutionary Computation (CEC'04)}, vol.~1, Jun.
  2004, pp. 320--324.

\bibitem{Zhao_2004:PSO_detection}
Y.~Zhao and J.~Zheng, ``Particle swarm optimization algorithm in signal
  detection and blind extraction,'' in \emph{Proc. 7th International Symposium
  on Parallel Architectures, Algorithms and Networks (ISPAN'04)}, Hong Kong,
  China, May 2004, pp. 37--41.

\bibitem{Chen_2010:PSO_MIMO_detection}
S.~Chen, W.~Yao, H.~R. Palally, and L.~Hanzo, ``Particle swarm optimisation
  aided {MIMO} transceiver designs,'' in \emph{Computational Intelligence in
  Expensive Optimization Problems}, ser. Adaptation Learning and Optimization,
  Y.~Tenne and C.-K. Goh, Eds.\hskip 1em plus 0.5em minus 0.4em\relax Springer
  Berlin Heidelberg, 2010, vol.~2, pp. 487--511.

\bibitem{Taufik_2010:SA_MUD}
T.~Abr{\~a}o, L.~D. de~Oliveira, F.~Ciriaco, B.~A. Ang{\'e}lico, P.~E.
  Jeszensky, and F.~J. Casadevall~Palacio, ``{S/MIMO MC-CDMA} heuristic
  multiuser detectors based on single-objective optimization,'' \emph{Wireless
  Personal Communications}, vol.~53, no.~4, pp. 529--553, 2010.

\bibitem{Xia_2008:SA_MUD}
J.~Xia, T.~Lv, X.~Yun, X.~Su, and S.~Yang, ``Simulated annealing based
  multiuser detection for synchronous {SDMA} system,'' in \emph{Proc. 11th IEEE
  Singapore International Conference on Communication Systems (ICCS'08)}, Nov.
  2008, pp. 441--445.

\bibitem{Datta_2012:rank_deficient_LS_MIMO_detection}
T.~Datta, N.~Srinidhi, A.~Chockalingam, and B.~S. Rajan, ``Low-complexity
  near-optimal signal detection in underdetermined large-{MIMO} systems,'' in
  \emph{Proc. National Conference on Communications (NCC'12)}, Kharagpur,
  India, Feb. 2012, pp. 1--5.

\bibitem{Hagenauer:iterative_decoding_concatenated_codes}
J.~Lodge, R.~Young, P.~A. Hoeher, and J.~Hagenauer, ``Separable {MAP} 'filters'
  for the decoding of product and concatenated codes,'' in \emph{Proc. IEEE
  International Conference on Communications (ICC'93)}, vol.~3, Geneva,
  Switzerland, May 1993, pp. 1740--1745.

\bibitem{Hagenauer1997:turbo_principle}
J.~Hagenauer, ``The {Turbo} principle: tutorial introduction and state of the
  art,'' in \emph{Proc. 1st International Symposium on Turbo Codes and Related
  Topics}, Brest, France, Sep. 1997, pp. 1--11.

\bibitem{Hagenauer1996:iterative_decoding_block_conv_codes}
J.~Hagenauer, E.~Offer, and L.~Papke, ``Iterative decoding of binary block and
  convolutional codes,'' \emph{{IEEE} Transactions on Information Theory},
  vol.~42, no.~2, pp. 429--445, Mar. 1996.

\bibitem{Moher_1997:turbo_MUD}
M.~Moher, ``Turbo-based multiuser detection,'' in \emph{Proc. IEEE
  International Symposium on Information Theory (ISIT'97)}, Ulm, Germany, Jun.
  1997, p. 195.

\bibitem{Moher_1998:earliest_iterative_MUD_MAP}
------, ``An iterative multiuser decoder for near-capacity communications,''
  \emph{{IEEE} Transactions on Communications}, vol.~46, no.~7, pp. 870--880,
  Jul. 1998.

\bibitem{Reed1997:IDD_conf_PIMRC}
M.~C. Reed, P.~D. Alexander, J.~A. Asenstorfer, and C.~B. Schlegel, ``Near
  single user performance using iterative multi-user detection for {CDMA} with
  turbo-code decoders,'' in \emph{Proc. 8th IEEE Symposium on Personal, Indoor
  and Mobile Radio Communications (PIMRC'97)}, vol.~2, Helsinki, Finland, Sep.
  1997, pp. 740--744.

\bibitem{Reed_1998:iterative_MUD_coded_CDMA}
M.~C. Reed, C.~B. Schlegel, P.~D. Alexander, and J.~A. Asenstorfer, ``Iterative
  multiuser detection for {CDMA} with {FEC}: Near-single-user performance,''
  \emph{{IEEE} Transactions on Communications}, vol.~46, no.~12, pp.
  1693--1699, 1998.

\bibitem{Tarkoy_1997:IDD_MUD}
F.~Tark{\"o}y, ``Iterative multi-user decoding for asynchronous users,'' in
  \emph{Proc. IEEE International Symposium on Information Theory (ISIT'97)},
  Ulm, Germany, Jun. 1997, p.~30.

\bibitem{Nelson_1996:EM_MUD}
L.~Nelson and H.~V. Poor, ``Iterative multiuser receivers for {CDMA} channels:
  an {EM}-based approach,'' \emph{{IEEE} Transactions on Communications},
  vol.~44, no.~12, pp. 1700--1710, Dec. 1996.

\bibitem{Wang_Xiaodong_1999:iterative_detection}
X.~Wang and H.~V. Poor, ``Iterative (turbo) soft interference cancellation and
  decoding for coded {CDMA},'' \emph{{IEEE} Transactions on Communications},
  vol.~47, no.~7, pp. 1046--1061, Jul. 1999.

\bibitem{Gamal_2000:iterative_MUD_coded_CDMA}
H.~El~Gamal and E.~Geraniotis, ``Iterative multiuser detection for coded {CDMA}
  signals in awgn and fading channels,'' \emph{{IEEE} Journal on Selected Areas
  in Communications}, vol.~18, no.~1, pp. 30--41, Jan. 2000.

\bibitem{Lee_2006:IDD_vblast}
H.~Lee, B.~Lee, and I.~Lee, ``Iterative detection and decoding with an improved
  {V-BLAST} for {MIMO-OFDM} systems,'' \emph{IEEE Journal on Selected Areas in
  Communications}, vol.~24, no.~3, pp. 504--513, Mar. 2006.

\bibitem{DeJong_2005:iterative_tree_search}
Y.~De~Jong and T.~Willink, ``Iterative tree search detection for {MIMO}
  wireless systems,'' \emph{IEEE Transactions on Communications}, vol.~53,
  no.~6, pp. 930--935, Jun. 2005.

\bibitem{Nekuii:without_list_soft_SDR_QPSK_journal}
M.~Nekuii, M.~Kisialiou, T.~Davidson, and Z.-Q. Luo, ``Efficient soft-output
  demodulation of {MIMO} {QPSK} via semidefinite relaxation,'' \emph{{IEEE}
  Journal of Selected Topics in Signal Processing}, vol.~5, no.~8, pp.
  1426--1437, Dec. 2011.

\bibitem{Pauli_2006:turbo_DPSK_MSDSD}
V.~Pauli, L.~Lampe, and R.~Schober, ``{Turbo DPSK} using soft multiple-symbol
  differential sphere decoding,'' \emph{{IEEE} Transactions on Information
  Theory}, vol.~52, no.~4, pp. 1385--1398, Apr. 2006.

\bibitem{ten_brink2001:EXIT_chart}
S.~ten Brink, ``Convergence behavior of iteratively decoded parallel
  concatenated codes,'' \emph{{IEEE} Transactions on Communications}, vol.~49,
  no.~10, pp. 1727--1737, Oct. 2001.

\bibitem{hagenauer2004:EXIT_chart}
J.~Hagenauer, ``The {EXIT} chart--{I}ntroduction to extrinsic information
  transfer in iterative processing,'' in \emph{Proc. 12th European Signal
  Processing Conference (EUSIPCO'04)}, Vienna, Austria, Sep. 2004, pp.
  1541--1548.

\bibitem{Ariyavisitakul_2000:turbo_ST_processing}
S.~L. Ariyavisitakul, ``Turbo space-time processing to improve wireless channel
  capacity,'' \emph{IEEE Transactions on Communications}, vol.~48, no.~8, pp.
  1347--1359, Aug. 2000.

\bibitem{Baro_2003:LISS_iterative_detection}
S.~Baro, J.~Hagenauer, and M.~Witzke, ``Iterative detection of {MIMO}
  transmission using a list-sequential ({LISS}) detector,'' in \emph{Proc. IEEE
  International Conference on Communications (ICC'03)}, Anchorage, Alaska, US,
  May 2003, pp. 2653--2657.

\bibitem{Haykin_2004:turbo_MIMO}
S.~Haykin, M.~Sellathurai, Y.~De~Jong, and T.~Willink, ``{Turbo-MIMO} for
  wireless communications,'' \emph{IEEE Communications Magazine}, vol.~42,
  no.~10, pp. 48--53, Oct. 2004.

\bibitem{Hagenauer_2007:LISS_algorithm}
J.~Hagenauer and C.~Kuhn, ``The list-sequential ({LISS}) algorithm and its
  application,'' \emph{{IEEE} Transactions on Communications}, vol.~55, no.~5,
  pp. 918--928, May 2007.

\bibitem{Peel_2005:vector_perturbation_near_capacity_part_1}
C.~B. Peel, B.~M. Hochwald, and A.~L. Swindlehurst, ``A vector-perturbation
  technique for near-capacity multiantenna multiuser communication -- part {I}:
  channel inversion and regularization,'' \emph{{IEEE} Transactions on
  Communications}, vol.~53, no.~1, pp. 195--202, Jan. 2005.

\bibitem{Hochwald_2005:vector_perturbation_near_capacity_part_2}
B.~M. Hochwald, C.~B. Peel, and A.~L. Swindlehurst, ``A vector-perturbation
  technique for near-capacity multiantenna multiuser communication - part {II}:
  perturbation,'' \emph{{IEEE} Transactions on Communications}, vol.~53, no.~3,
  pp. 537--544, Mar. 2005.

\bibitem{Alamri_2009:near_capacity_three_stage_iterative_detection}
O.~Alamri, J.~Wang, S.~X. Ng, L.-L. Yang, and L.~Hanzo, ``Near-capacity
  three-stage turbo detection of irregular convolutional coded joint
  sphere-packing modulation and space-time coding,'' \emph{IEEE Transactions on
  Communications}, vol.~57, no.~5, pp. 1486--1495, May 2009.

\bibitem{Hanzo_2011:near_capacity_proceeding_IEEE}
L.~Hanzo, M.~El-Hajjar, and O.~Alamri, ``Near-capacity wireless transceivers
  and cooperative communications in the {MIMO} era: Evolution of standards,
  waveform design, and future perspectives,'' \emph{Proceedings of the IEEE},
  vol.~99, no.~8, pp. 1343--1385, Aug. 2011.

\bibitem{Suthisopapan_2012:capacity_approaching_LDPC_MMSE_detection_LS_MIMO}
P.~Suthisopapan, K.~Kasai, V.~Imtawil, and A.~Meesomboon, ``Approaching
  capacity of large {MIMO} systems by non-binary {LDPC} codes and {MMSE}
  detection,'' in \emph{Proc. IEEE International Symposium on Information
  Theory (ISIT'12)}, Cambridge, MA, Jul. 2012, pp. 1712--1716.

\bibitem{Hanzo2009:near_capacity_multi_func_MIMO}
L.~Hanzo, O.~Alamri, M.~El-Hajjar, and N.~Wu, \emph{Near-Capacity
  Multi-Functional MIMO Systems: Sphere-Packing, Iterative Detection and
  Cooperation}.\hskip 1em plus 0.5em minus 0.4em\relax John Wiley \& Sons,
  2009.

\bibitem{Hanzo_2011:near_capacity_book}
L.~Hanzo, T.~H. Liew, B.~L. Yeap, R.~Y.~S. Tee, and S.~X. Ng, \emph{Turbo
  Coding, Turbo Equalisation and Space-Time Coding: EXIT-Chart-Aided
  Near-Capacity Designs for Wireless Channels}.\hskip 1em plus 0.5em minus
  0.4em\relax John Wiley \& Sons, 2011.

\bibitem{Hanzo:2011:near_capacity_variable_length_coding}
L.~Hanzo, R.~G. Maunder, J.~Wang, and L.-L. Yang, \emph{Near-capacity
  variable-length coding: regular and {EXIT}-chart-aided irregular
  designs}.\hskip 1em plus 0.5em minus 0.4em\relax John Wiley \& Sons, 2011.

\bibitem{Wymeersch2007:iterative_receiver_design}
H.~Wymeersch, \emph{Iterative receiver design}.\hskip 1em plus 0.5em minus
  0.4em\relax Cambridge University Press, 2007.

\bibitem{Sampath_2001:MIMO_precoding}
H.~Sampath, P.~Stoica, and A.~Paulraj, ``Generalized linear precoder and
  decoder design for {MIMO} channels using the weighted {MMSE} criterion,''
  \emph{IEEE Transactions on Communications}, vol.~49, no.~12, pp. 2198--2206,
  Dec. 2001.

\bibitem{Palomar_2003:joint_TX_RX_beamforming}
D.~P. Palomar, J.~M. Cioffi, and M.-A. Lagunas, ``Joint {Tx-Rx} beamforming
  design for multicarrier {MIMO} channels: a unified framework for convex
  optimization,'' \emph{IEEE Transactions on Signal Processing}, vol.~51,
  no.~9, pp. 2381--2401, Sep. 2003.

\bibitem{Love_2005:limited_feedback_precoding}
D.~J. Love and R.~W. Heath, ``Limited feedback unitary precoding for spatial
  multiplexing systems,'' \emph{IEEE Transactions on Information Theory},
  vol.~51, no.~8, pp. 2967--2976, Aug. 2005.

\bibitem{Wiesel_2006:linear_precoder_MIMO}
A.~Wiesel, Y.~C. Eldar, and S.~Shamai, ``Linear precoding via conic
  optimization for fixed {MIMO} receivers,'' \emph{IEEE Transactions on Signal
  Processing}, vol.~54, no.~1, pp. 161--176, Jan. 2006.

\bibitem{Sadek_2007:precoding_MU_MIMO}
M.~Sadek, A.~Tarighat, and A.~H. Sayed, ``A leakage-based precoding scheme for
  downlink multi-user {MIMO} channels,'' \emph{IEEE Transactions on Wireless
  Communications}, vol.~6, no.~5, pp. 1711--1721, May 2007.

\bibitem{Paulraj_2007:MIMO_precoding_magazine}
M.~Vu and A.~Paulraj, ``Mimo wireless linear precoding,'' \emph{IEEE Signal
  Processing Magazine}, vol.~24, no.~5, pp. 86--105, Sep. 2007.

\bibitem{Heath_2009:network_MIMO_precoding}
J.~Zhang, R.~Chen, J.~G. Andrews, A.~Ghosh, and R.~W. Heath, ``Networked {MIMO}
  with clustered linear precoding,'' \emph{IEEE Transactions on Wireless
  Communications}, vol.~8, no.~4, pp. 1910--1921, Apr. 2009.

\bibitem{Winters_1994:ZF_MMSE_diversity_order}
J.~H. Winters, J.~Salz, and R.~D. Gitlin, ``The impact of antenna diversity on
  the capacity of wireless communication systems,'' \emph{{IEEE} Transactions
  on Communications}, vol.~42, no. 2/3/4, pp. 1740--1751, Feb./Mar./Apr. 1994.

\bibitem{Varanasi_2004:DFD_performance_analysis}
N.~Prasad and M.~K. Varanasi, ``Analysis of decision feedback detection for
  {MIMO} rayleigh-fading channels and the optimization of power and rate
  allocations,'' \emph{{IEEE} Transactions on Information Theory}, vol.~50,
  no.~6, pp. 1009--1025, Jun. 2004.

\bibitem{Loyka_2006:VBLAST_without_ordering_performance}
S.~Loyka and F.~Gagnon, ``{V-BLAST} without optimal ordering: Analytical
  performance evaluation for {Rayleigh} fading channels,'' \emph{{IEEE}
  Transactions on Communications}, vol.~54, no.~6, pp. 1109--1120, Jun. 2006.

\bibitem{Jiang_2005:Asymptotic_VBLAST}
Y.~Jiang, X.~Zheng, and J.~Li, ``Asymptotic performance analysis of
  {V-BLAST},'' in \emph{Proc. IEEE Global Telecommunications Conference
  (GLOBECOM'05)}, St. Louis, US, Dec. 2005, pp. 3882--3886.

\bibitem{Loyka_2004:VBLAST_performance_ordering}
S.~Loyka and F.~Gagnon, ``Performance analysis of the {V-BLAST} algorithm: An
  analytical approach,'' \emph{{IEEE} Transactions on Wireless Communications},
  vol.~3, no.~4, pp. 1326--1337, Jul. 2004.

\bibitem{Seethaler_2010:SD_infinity_norm}
D.~Seethaler and H.~Bolcskei, ``Performance and complexity analysis of
  infinity-norm sphere-decoding,'' \emph{{IEEE} Transactions on Information
  Theory}, vol.~56, no.~3, pp. 1085--1105, March 2010.

\bibitem{Seethaler_2011:SD_complexity_distribution}
D.~Seethaler, J.~J.~Jald\'{e}n, C.~Studer, and H.~Bolcskei, ``On the complexity
  distribution of sphere decoding,'' \emph{{IEEE} Transactions on Information
  Theory}, vol.~57, no.~9, pp. 5754--5768, Sep. 2011.

\bibitem{Jalden_2012:SD_complexity_exponent}
J.~Jald\'{e}n and P.~Elia, ``Sphere decoding complexity exponent for decoding
  full-rate codes over the quasi-static {MIMO} channel,'' \emph{{IEEE}
  Transactions on Information Theory}, vol.~58, no.~9, pp. 5785--5803, Sep.
  2012.

\bibitem{Jalden:fixed_complexity_SD}
J.~Jald\'{e}n, L.~G. Barbero, B.~Ottersten, and J.~S. Thompson, ``Full
  diversity detection in {MIMO} systems with a fixed-complexity sphere
  decoder,'' in \emph{Proc. IEEE International Conference on Acoustics, Speech,
  and Signal Processing (ICASSP'07)}, vol.~3, Honolulu, HI, Apr. 2007, pp.
  II--49--III--52.

\bibitem{Ling_2006:LR_performance_analysis}
C.~Ling, ``Towards characterizing the performance of approximate lattice
  decoding in {MIMO} communications,'' in \emph{Proc. 4th International
  Symposium on Turbo Codes Related Topics/6th International ITG-Conference on
  Source and Channel Coding}, Munich, Germany, Apr. 2006, pp. 1--6.

\bibitem{Taherzadeh_2010:LLL_LR_limitation}
M.~Taherzadeh and A.~K. Khandani, ``On the limitations of the naive lattice
  decoding,'' \emph{{IEEE} Transactions on Information Theory}, vol.~56,
  no.~10, pp. 4820--4826, Oct. 2010.

\bibitem{Ling_2011:LR_performance_analysis}
C.~Ling, ``On the proximity factors of lattice reduction-aided decoding,''
  \emph{{IEEE} Transactions on Signal Processing}, vol.~59, no.~6, pp.
  2795--2808, Jun. 2011.

\bibitem{Jalden_2010:DMT_optimality_LRA}
J.~Jald\'{e}n and P.~Elia, ``{DMT} optimality of {LR}-aided linear decoders for
  a general class of channels, lattice designs, and system models,''
  \emph{{IEEE} Transactions on Information Theory}, vol.~56, no.~10, pp.
  4765--4780, Oct. 2010.

\bibitem{Singh_2012:vanishing_gap_LR}
A.~K. Singh, P.~Elia, and J.~Jald\'{e}n, ``Achieving a vanishing {SNR} gap to
  exact lattice decoding at a subexponential complexity,'' \emph{{IEEE}
  Transactions on Information Theory}, vol.~58, no.~6, pp. 3692--3707, June
  2012.

\bibitem{Orlik_2013:PDA_performance_analysis}
A.~Yellepeddi, K.~J. Kim, C.~Duan, and P.~Orlik, ``On probabilistic data
  association for achieving near-exponential diversity over fading channels,''
  in \emph{Proc. IEEE International Conference on Communications (ICC'13)},
  Budapest, Hungary, Jun. 2013, pp. 5409--5414.

\bibitem{Jalden:SDR_optimality_conditions_BPSK}
J.~Jald\'{e}n, C.~Martin, and B.~Ottersten, ``Semidefinite programming for
  detection in linear systems - optimality conditions and space-time
  decoding,'' in \emph{Acoustics, Speech, and Signal Processing, 2003.
  Proceedings. (ICASSP '03). 2003 IEEE International Conference on}, vol.~4,
  Apr. 2003, pp. IV--9 -- IV--12.

\bibitem{So:SDR_performance_low_SNR_QAM}
A.~M.-C. So, ``On the performance of semidefinite relaxation mimo detectors for
  qam constellations,'' in \emph{Proc. IEEE International Conference on
  Acoustics, Speech and Signal Processing (ICASSP'09)}, Taipei, Republic of
  China, Apr. 2009, pp. 2449--2452.

\bibitem{Luo_2009:efficient_implementation_SDR}
M.~Kisialiou, X.~Luo, and Z.-Q. Luo, ``Efficient implementation of quasi-
  maximum-likelihood detection based on semidefinite relaxation,'' \emph{{IEEE}
  Transactions on Signal Processing}, vol.~57, no.~12, pp. 4811--4822, Dec.
  2009.

\bibitem{}


\bibitem{Marzetta2010:massive_MIMO}
T.~L. Marzetta, ``Noncooperative cellular wireless with unlimited numbers of
  base station antennas,'' \emph{{IEEE} Transactions on Wireless
  Communications}, vol.~9, no.~11, pp. 3590--3600, Nov. 2010.

\bibitem{Anzhong_2014:massive_MIMO_source_localization}
A.~Hu, T.~Lv, H.~Gao, Z.~Zhang, and S.~Yang, ``An {ESPRIT}-based approach for
  {2-D} localization of incoherently distributed sources in massive {MIMO}
  systems,'' \emph{IEEE Journal of Selected Topics in Signal Processing},
  vol.~8, no.~5, pp. 996--1011, Oct. 2014.

\bibitem{Jiankang_2014:massive_MIMO_pilot_decontamination}
J.~Zhang, B.~Zhang, S.~Chen, X.~Mu, M.~El-Hajjar, and L.~Hanzo, ``Pilot
  contamination elimination for large-scale multiple-antenna aided {OFDM}
  systems,'' \emph{IEEE Journal of Selected Topics in Signal Processing},
  vol.~8, no.~5, pp. 759 -- 772, Oct. 2014.

\bibitem{Haas2011:indoor_optical_wireless}
H.~Elgala, R.~Mesleh, and H.~Haas, ``Indoor optical wireless communication:
  potential and state-of-the-art,'' \emph{{IEEE} Communications Magazine},
  vol.~49, no.~9, pp. 56--62, Sep. 2011.

\bibitem{Hanzo2012:myth_wireless}
L.~Hanzo, H.~Haas, S.~Imre, D.~O'Brien, M.~Rupp, and L.~Gyongyosi, ``Wireless
  myths, realities, and futures: from {3G/4G} to optical and quantum
  wireless,'' \emph{Proceedings of the {IEEE}}, vol. 100, no. Special
  Centennial Issue, pp. 1853--1888, May 2012.

\bibitem{Hoydis2013:small_cell_massive_MIMO}
J.~Hoydis, K.~Hosseini, S.~ten Brink, and M.~Debbah, ``Making smart use of
  excess antennas: massive {MIMO}, small cells, and {TDD},'' \emph{Bell Labs
  Technical Journal}, vol.~18, no.~2, pp. 5--21, Sep. 2013.

\bibitem{Kan_2014:massive_MIMO_CM_survey}
K.~Zheng, S.~Ou, and X.~Yin, ``Massive {MIMO} channel models: A survey,''
  \emph{International Journal of Antennas and Propagation}, vol. 2014, pp. 1 --
  10, Jun. 2014.

\bibitem{Tulino2004:random_matrix_wireless_comm}
\BIBentryALTinterwordspacing
A.~M. Tulino and S.~Verd{\'u}, ``Random matrix theory and wireless
  communications,'' \emph{Foundations and Trends\textregistered in
  Communications and Information Theory}, vol.~1, no.~1, pp. 1--182, 2004.
  [Online]. Available: \url{http://dx.doi.org/10.1561/0100000001}
\BIBentrySTDinterwordspacing

\bibitem{mehta2004:random_matrices}
M.~L. Mehta, \emph{Random Matrices}, 3rd~ed., ser. Pure and Applied
  Mathematics.\hskip 1em plus 0.5em minus 0.4em\relax Elsevier Science, 2004.

\bibitem{Marvcenko1967:distribution_eigenvalues_random_matrix}
V.~A. Mar{\v{c}}enko and L.~A. Pastur, ``Distribution of eigenvalues for some
  sets of random matrices,'' \emph{Math USSR Shornik}, vol.~1, no.~4, pp.
  457--483, 1967.

\bibitem{chockalingam2014:large_MIMO_systems}
A.~Chockalingam and B.~S. Rajan, \emph{Large MIMO Systems}, ser. Large MIMO
  Systems.\hskip 1em plus 0.5em minus 0.4em\relax Cambridge University Press,
  2014.

\bibitem{Marzetta2006:massive_MIMO_single_cell}
T.~L. Marzetta, ``How much training is required for multiuser {MIMO}?'' in
  \emph{Proc. 40th Asilomar Conference on Signals, Systems and Computers
  (ACSSC'06)}, Pacific Grove, CA, Oct. 2006, pp. 359--363.

\bibitem{Sun_1998:HNN_MUD}
Y.~Sun, ``Eliminating-highest-error and fastest-metric-descent criteria and
  iterative algorithms for bit-synchronous {CDMA} multiuser detection,'' in
  \emph{Proc. IEEE International Conference on Communications (ICC'98)},
  vol.~3, Atlanta, GA, Jun. 1998, pp. 1576--1580.

\bibitem{Sun_2000:HNN_MUD_CDMA}
------, ``A family of linear complexity likelihood ascent search detectors for
  {CDMA} multiuser detection,'' in \emph{Proc. IEEE 6th International Symposium
  on Spread Spectrum Techniques and Applications}, vol.~2, Parsippany, NJ, Sep.
  2000, pp. 713--717.

\bibitem{Sun_2009:LAS_MUD_journal}
------, ``A family of likelihood ascent search multiuser detectors: approaching
  optimum performance via random multicodes with linear complexity,''
  \emph{{IEEE} Transactions on Communications}, vol.~57, no.~8, pp. 2215--2220,
  Aug. 2009.

\bibitem{Sun_2000:HNN_image_restoration_part_I}
------, ``Hopfield neural network based algorithms for image restoration and
  reconstruction -- part {I}: Algorithms and simulations,'' \emph{{IEEE}
  Transactions on Signal Processing}, vol.~48, no.~7, pp. 2105--2118, Jul.
  2000.

\bibitem{Sun_2000:HNN_image_restoration_part_II}
------, ``Hopfield neural network based algorithms for image restoration and
  reconstruction -- part {II}: Performance analysis,'' \emph{{IEEE}
  Transactions on Signal Processing}, vol.~48, no.~7, pp. 2119--2131, Jul.
  2000.

\bibitem{Paris_1988:neural_network_MUD_conf}
B.-P. Paris, G.~Orsak, M.~K. Varanasi, and B.~Aazhang, ``Neural net receivers
  in multiple access-communications,'' in \emph{IEEE Conference on Neural
  Information Processing Systems (NIPS'88)}, Denver, CO., Nov. 1988, pp.
  272--280.

\bibitem{Aazhang_1992:neural_networks_MUD_CDMA}
B.~Aazhang, B.-P. Paris, and G.~C. Orsak, ``Neural networks for multiuser
  detection in code-division multiple-access communications,'' \emph{{IEEE}
  Transactions on Communications}, vol.~40, no.~7, pp. 1212--1222, Jul. 1992.

\bibitem{Mitra_1992:neural_networks_MUD_CDMA}
U.~Mitra and H.~V. Poor, ``Adaptive receiver algorithms for near-far resistant
  {CDMA},'' in \emph{Proc. 3rd IEEE International Symposium on Personal, Indoor
  and Mobile Radio Communications (PIMRC'92)}, Boston, MA, Oct. 1992, pp.
  639--644.

\bibitem{Mitra_1994:neural_network_adaptive_MUD}
------, ``Neural network techniques for adaptive multiuser demodulation,''
  \emph{{IEEE} Journal on Selected Areas in Communications}, vol.~12, no.~9,
  pp. 1460--1470, Dec. 1994.

\bibitem{Mitra_1995:neutal_networks_near_far_CDMA}
------, ``Adaptive receiver algorithms for near-far resistant {CDMA},''
  \emph{{IEEE} Transactions on Communications}, vol.~43, no. 2/3/4, pp.
  1713--1724, Feb./Mar./Apr. 1995.

\bibitem{Kechriotis_1996:hopfield_neural_network_MUD}
G.~Kechriotis and E.~S. Manolakos, ``Hopfield neural network implementation of
  the optimal {CDMA} multiuser detector,'' \emph{{IEEE} Transactions on Neural
  Networks}, vol.~7, no.~1, pp. 131--141, Jan. 1996.

\bibitem{Murch_2010:LAS_LS_MIMO_detection}
P.~Li and R.~D. Murch, ``Multiple output selection-{LAS} algorithm in large
  {MIMO} systems,'' \emph{{IEEE} Communications Letters}, vol.~14, no.~5, pp.
  399--401, May 2010.

\bibitem{Mitola_1999:CR}
J.~Mitola~III and G.~Q. Maguire~Jr., ``Cognitive radio: Making software radios
  more personal,'' \emph{IEEE Personal Communications}, vol.~6, no.~4, pp.
  13--18, Aug. 1999.

\bibitem{Mitola_2000:Phd_thesis}
J.~Mitola~III, ``Cognitive radio: An integrated agent architecture for software
  defined radio,'' Ph.D. dissertation, Royal Institute of Technology (KTH),
  Sweden, May 2000.

\bibitem{Haykin_2005:cognitive_radio}
S.~Haykin, ``Cognitive radio: Brain-empowered wireless communications,''
  \emph{{IEEE} Journal on Selected Areas in Communications}, vol.~23, no.~2,
  pp. 201--220, Feb. 2005.

\bibitem{Rui_2008:MIMO_cognitive_radio}
R.~Zhang and Y.-C. Liang, ``Exploiting multi-antennas for opportunistic
  spectrum sharing in cognitive radio networks,'' \emph{IEEE Journal of
  Selected Topics in Signal Processing}, vol.~2, no.~1, pp. 88--102, Feb. 2008.

\bibitem{Vishwanath_2008:MIMO_cognitive_radio_capacity}
S.~Sridharan and S.~Vishwanath, ``On the capacity of a class of {MIMO}
  cognitive radios,'' \emph{IEEE Journal of Selected Topics in Signal
  Processing}, vol.~2, no.~1, pp. 103--117, Feb. 2008.

\bibitem{Palomar_2008:MIMO_cognitive_radio}
G.~Scutari, D.~P. Palomar, and S.~Barbarossa, ``Cognitive {MIMO} radio,''
  \emph{{IEEE} Signal Processing Magazine}, vol.~25, no.~6, pp. 46--59, Nov.
  2008.

\bibitem{Letaief_2009:MIMO_cognitive_radio}
K.~Hamdi, W.~Zhang, and K.~Letaief, ``Opportunistic spectrum sharing in
  cognitive {MIMO} wireless networks,'' \emph{{IEEE} Transactions on Wireless
  Communications}, vol.~8, no.~8, pp. 4098--4109, Aug. 2009.

\bibitem{Gan_2009:MIMO_cognitive_radio_beamforming}
G.~Zheng, K.-K. Wong, and B.~Ottersten, ``Robust cognitive beamforming with
  bounded channel uncertainties,'' \emph{{IEEE} Transactions on Signal
  Processing}, vol.~57, no.~12, pp. 4871--4881, Dec. 2009.

\bibitem{Palomar_2010:MIMO_cognitive_radio_game_theory}
G.~Scutari and D.~P. Palomar, ``{MIMO} cognitive radio: A game theoretical
  approach,'' \emph{{IEEE} Transactions on Signal Processing}, vol.~58, no.~2,
  pp. 761--780, Feb. 2010.

\bibitem{Debbah_2010:MIMO_cognitive_radio}
S.~M. Perlaza, N.~Fawaz, S.~Lasaulce, and M.~Debbah, ``From spectrum pooling to
  space pooling: Opportunistic interference alignment in {MIMO} cognitive
  networks,'' \emph{{IEEE} Transactions on Signal Processing}, vol.~58, no.~7,
  pp. 3728--3741, Jul. 2010.

\bibitem{GBG_2011:MIMO_cognitive_radio}
S.-J. Kim and G.~B. Giannakis, ``Optimal resource allocation for {MIMO} ad hoc
  cognitive radio networks,'' \emph{{IEEE} Transactions on Information Theory},
  vol.~57, no.~5, pp. 3117--3131, May 2011.

\bibitem{Zhang_2011:MIMO_cognitive_radio_SDP}
Y.~J.~A. Zhang and A.~M.-C. So, ``Optimal spectrum sharing in {MIMO} cognitive
  radio networks via semidefinite programming,'' \emph{{IEEE} Journal on
  Selected Areas in Communications}, vol.~29, no.~2, pp. 362--373, Feb. 2011.

\bibitem{Palomar_2011:robust_MIMO_cogntive_radio_game_theory}
J.~Wang, G.~Scutari, and D.~P. Palomar, ``Robust {MIMO} cognitive radio via
  game theory,'' \emph{{IEEE} Transactions on Signal Processing}, vol.~59,
  no.~3, pp. 1183--1201, Mar. 2011.

\bibitem{Varanasi_2012:MIMO_cognitive_radio_capacity}
C.~S. Vaze and M.~K. Varanasi, ``The degree-of-freedom regions of {MIMO}
  broadcast, interference, and cognitive radio channels with no {CSIT},''
  \emph{{IEEE} Transactions on Information Theory}, vol.~58, no.~8, pp.
  5354--5374, Aug. 2012.

\bibitem{Sun_2015:MIMO_SDR_SDN}
S.~Sun, M.~Kadoch, L.~Gong, and B.~Rong, ``Integrating network function
  virtualization with {SDR} and {SDN} for {4G/5G} networks,'' \emph{{IEEE}
  Network}, vol.~29, no.~3, pp. 54--59, May 2015.

\bibitem{fishler2004_MIMO_radar}
E.~Fishler, A.~Haimovich, R.~S. Blum, D.~Chizhik, L.~J. Cimini, and R.~A.
  Valenzuela, ``{MIMO} radar: an idea whose time has come,'' in
  \emph{Proceedings of the IEEE Radar Conference}, Philadelphia, Pennsylvania,
  USA, Apr. 2004, pp. 71--78.

\bibitem{fishler2006_MIMO_radar}
E.~Fishler, A.~Haimovich, R.~S. Blum, L.~J. Cimini, D.~Chizhik, and R.~A.
  Valenzuela, ``Spatial diversity in radars-models and detection performance,''
  \emph{{IEEE} Transactions on Signal Processing}, vol.~54, no.~3, pp.
  823--838, Mar. 2006.

\bibitem{Li_2007:MIMO_radar}
J.~Li and P.~Stoica, ``{MIMO} radar with colocated antennas,'' \emph{{IEEE}
  Signal Processing Magazine}, vol.~24, no.~5, pp. 106--114, Sep. 2007.

\bibitem{haimovich_2008:MIMO_radar}
A.~M. Haimovich, R.~S. Blum, and L.~J. Cimini, ``{MIMO} radar with widely
  separated antennas,'' \emph{{IEEE} Signal Processing Magazine}, vol.~25,
  no.~1, pp. 116--129, Jan. 2008.

\bibitem{Li_2009:MIMO_radar_signal_processing}
J.~Li and P.~Stoica, Eds., \emph{{MIMO} radar signal processing}.\hskip 1em
  plus 0.5em minus 0.4em\relax John Wiley \& Sons, 2009.

\bibitem{Maio_2007:MIMO_radar_detector}
A.~De~Maio and M.~Lops, ``Design principles of {MIMO} radar detectors,''
  \emph{IEEE Transactions on Aerospace and Electronic Systems}, vol.~43, no.~3,
  pp. 886--898, Jul. 2007.

\bibitem{Li2008:signal_detection_MIMO_radar}
J.~Li, P.~Stoica, and X.~Zheng, ``Signal synthesis and receiver design for
  {MIMO} radar imaging,'' \emph{{IEEE} Transactions on Signal Processing},
  vol.~56, no.~8, pp. 3959--3968, Aug. 2008.

\bibitem{Shah_2005:coherent_MIMO}
A.~R. Shah, R.~C.~J. Hsu, A.~Tarighat, A.~H. Sayed, and B.~Jalali, ``Coherent
  optical {MIMO} ({COMIMO}),'' \emph{Journal of Lightwave Technology}, vol.~23,
  no.~8, pp. 2410--2419, Aug. 2005.

\bibitem{Tarighat2007:fundamentals_challenges_MIMO_fibre}
A.~Tarighat, R.~C.~J. Hsu, A.~Shah, A.~H. Sayed, and B.~Jalali, ``Fundamentals
  and challenges of optical multiple-input multiple-output multimode fiber
  links,'' \emph{{IEEE} Communications Magazine}, vol.~45, no.~5, pp. 57--63,
  May 2007.

\bibitem{Randel_2011_optical_fibre_MIMO}
S.~Randel, R.~Ryf, A.~Sierra, P.~J. Winzer, A.~H. Gnauck, C.~A. Bolle, R.-J.
  Essiambre, D.~W. Peckham, A.~McCurdy, R.~Lingle \emph{et~al.}, ``6$\times$
  56-gb/s mode-division multiplexed transmission over 33-km few-mode fiber
  enabled by 6$\times$ 6 {MIMO} equalization,'' \emph{Optics Express}, vol.~19,
  no.~17, pp. 16\,697--16\,707, Aug. 2011.

\bibitem{Ryf_2011:optical_fibre_MIMO_conf}
R.~Ryf, S.~Randel, A.~H. Gnauck, C.~Bolle, R.-J. Essiambre, P.~J. Winzer, D.~W.
  Peckham, A.~McCurdy, and R.~Lingle, ``Space-division multiplexing over 10 km
  of three-mode fiber using coherent 6 $\times$ 6 {MIMO} processing,'' in
  \emph{Proceedings of Optical Fiber Communication Conference / National Fiber
  Optic Engineers Conference}, Los Angeles, California, USA, Mar. 2011, p.
  PDPB10.

\bibitem{Ryf_2011:optical_fibre_MIMO}
R.~Ryf, S.~Randel, A.~H. Gnauck, C.~Bolle, A.~Sierra, S.~Mumtaz,
  M.~Esmaeelpour, E.~C. Burrows, R.-J. Essiambre, P.~J. Winzer \emph{et~al.},
  ``Mode-division multiplexing over 96 km of few-mode fiber using coherent 6
  $\times$ 6 {MIMO} processing,'' \emph{Journal of Lightwave Technology},
  vol.~30, no.~4, pp. 521--531, Feb. 2012.

\bibitem{Essiambre_2012:MIMO_optical_capacity}
R.~Essiambre and R.~W. Tkach, ``Capacity trends and limits of optical
  communication networks,'' \emph{Proceedings of the IEEE}, vol. 100, no.~5,
  pp. 1035--1055, May 2012.

\bibitem{Arik_2014:MIMO_optic_overview}
S.~Arik, J.~M. Kahn, and K.-P. Ho, ``{MIMO} signal processing for mode-division
  multiplexing: An overview of channel models and signal processing
  architectures,'' \emph{{IEEE} Signal Processing Magazine}, vol.~31, no.~2,
  pp. 25 -- 34, Mar. 2014.

\end{thebibliography}
